\documentclass{aa}

\usepackage{graphicx}
\usepackage{txfonts}
\usepackage{natbib}
\usepackage{lscape}
\usepackage{subfig}                              

\newcommand{\kms}{km\,s$^{-1}$}

\begin{document}

\title{Spot evolution on the red giant star XX~Triangulum\thanks{Based on data obtained with the
STELLA robotic telescopes in Tenerife, an AIP facility jointly operated with IAC.}}

\subtitle{A starspot-decay analysis based on time-series Doppler imaging}

\author{A. K\"unstler\and
        T. A. Carroll\and
        K. G. Strassmeier}

\institute{Leibniz Institute for Astrophysics Potsdam (AIP),
           An der Sternwarte 16, 14482 Potsdam, Germany\\
           \email{akuenstler@aip.de; tcarroll@aip.de; kstrassmeier@aip.de}}

\date{Received ??? / Accepted ???}


\abstract
{Solar spots appear to decay linearly proportional to their size. The decay rate of solar spots is
directly related to magnetic diffusivity, which itself is a key quantity for the length of a
magnetic-activity cycle. Is a linear spot decay also seen on other stars, and is this in agreement
with the large range of solar and stellar activity cycle lengths?}
{We investigate the evolution of starspots on the rapidly-rotating ($P_{\rm rot}$\;$\approx$\;24~d)
K0 giant XX~Tri, using consecutive time-series Doppler images. Our aim is to obtain a well-sampled 
movie of the stellar surface over many years, and thereby detect and quantify a starspot decay law
for further comparison with the Sun.}
{We obtained continuous high-resolution and phase-resolved spectroscopy with the 1.2-m robotic
STELLA telescope on Tenerife over six years, and these observations are ongoing. For each observing
season, we obtained between 5 to 7 independent Doppler images, one per stellar rotation, making up a
total of 36 maps. All images were reconstructed with our line-profile inversion code \textit{iMap}.
A wavelet analysis was implemented for denoising the line profiles. To quantify starspot area decay
and growth, we match the observed images with simplified spot models based on a Monte Carlo
approach.}
{It is shown that the surface of XX~Tri is covered with large high-latitude and even polar spots and
with occasional small equatorial spots. Just over the course of six years, we see a systematically
changing spot distribution with various timescales and morphology, such as spot fragmentation and
spot merging as well as spot decay and formation. An average linear decay of
$D$\;=\;$-$0.022\,$\pm$\,0.002~SH/day is inferred. We found evidence of an active longitude in
phase toward the (unseen) companion star. Furthermore, we detect a weak solar-like differential
rotation with a surface shear of $\alpha$\;=\;0.016\,$\pm$\,0.003. From the decay rate, we
determine a turbulent diffusivity of $\eta_T$\;=\;(6.3\,$\pm$\,0.5)\,$\times$\,10$^{14}$~cm$^2$/s
and predict a magnetic activity cycle of $\approx$\;26\,$\pm$\,6~years. Finally, we present a
short movie of the spatially resolved surface of XX~Tri.}
{}

\keywords{Stars: activity, starspots, stars: late-type, stars: individual: XX~Tri, technique:
Doppler imaging}

\authorrunning{A. K\"unstler\and T. A. Carroll\and K. G. Strassmeier}
\titlerunning{Spot evolution on the red giant star XX~Triangulum}

\maketitle


\section{Introduction}
\label{sec:introduction}

Sunspots enable us to trace differential rotation and meridional circulation on the solar surface to
extreme precision as well as to measure in detail a surface magnetic field in full three dimensions
\citep[e.g.,][]{solanki2003,moradi2010}. Sunspots are understood as the emergence of magnetic flux
tubes originating from a dynamo process in the interior \citep{babcock1961,leighton1964,stix1989},
in mean-field terms referred to as the so-called $\alpha\Omega$-dynamo. As sunspots host strong
magnetic fields, their study may provide indirect information about the internal dynamo activity,
reminding us that the latter is still very controversial
\citep[e.g.,][]{kapyla2011,kapyla2012,jabbari2014,yadav2015}.

Even the decay of sunspots, and thus the decay of the surface magnetic flux, is still not fully
understood \citep{ruediger2000,martinez2002}. Analyzing a subset of the Greenwich Photoheliographic
Results (GPR), \cite{bumba1963} proposed a linear decay law of form $dA/dt$\;=\;$D$ (where $A$ is the
area of the sunspot usually given in units of millionths of the solar hemisphere,
1~MSH\;=\;3.05~Mm$^2$) for recurrent sunspots (those with two or more disk passages) with a mean
value for $D$ of $-$4.2~MSH/day and an exponential law for nonrecurrent spot groups.
\cite{moreno1988} analyzed the GPR data from 1874-1939 and obtained a parabolic decay law. Using the
more accurate Debrecen data, \cite{petrovay1997} obtained results supporting a parabolic decay law.
\cite{martinez1993} analyzed sunspot decay rates using the GPR data from 1874-1976 and found a
lognormal distribution as well as some evidence for weak nonlinearities in the decay process of
isolated spots. A linear area decay law has been studied from a theoretical point of view
\citep{gokhale1972,meyer1974,krause1975}. \cite{meyer1974} obtained a constant area decrease rate
through a process of diffusion of magnetic field over the entire area of the spot. This diffusion
model predicts a linear area (and magnetic flux) decay, implying that $dA/dt$ is proportional to the
turbulent diffusivity $\eta_T$. \cite{krause1975} proposed a similar model based on turbulence,
where the turbulent diffusion was related to the flux decay by $d\Phi/dt$\;$\propto$\;$\eta_T$.
\cite{ruediger2000} used a mean-field formulation of diffusivity quenching and produced
quasi-linear decays for both the spot area as well as the magnetic flux.

\begin{figure*}[!ht]
{\bf (a)} \\
\includegraphics[trim=30 10 10 20, angle=0, width=\textwidth, clip]
{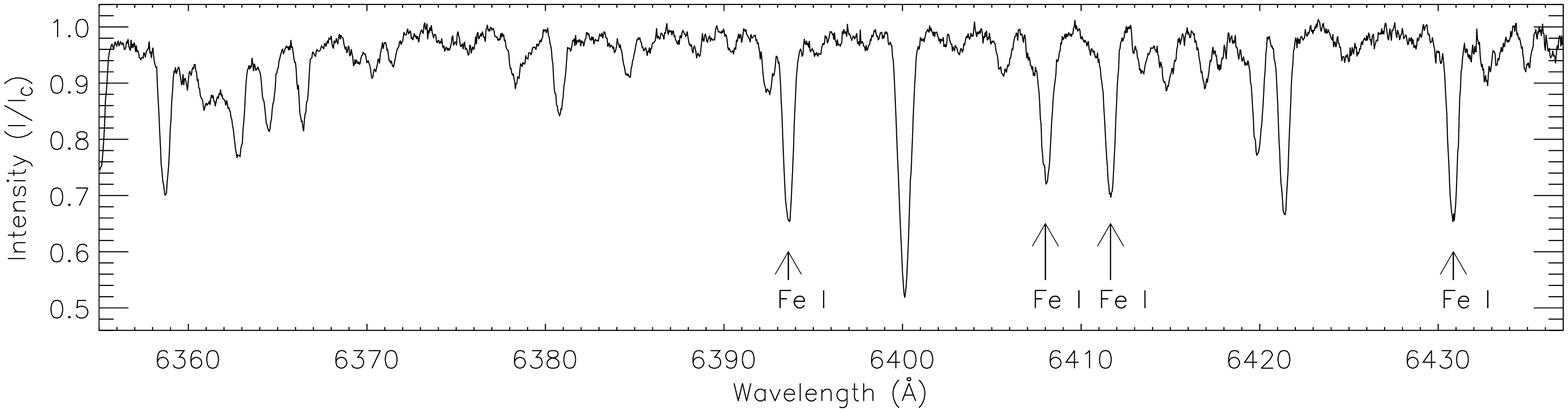}
{\bf (b)} \\
\includegraphics[trim=30 10 10 20, angle=0, width=\textwidth, clip]
{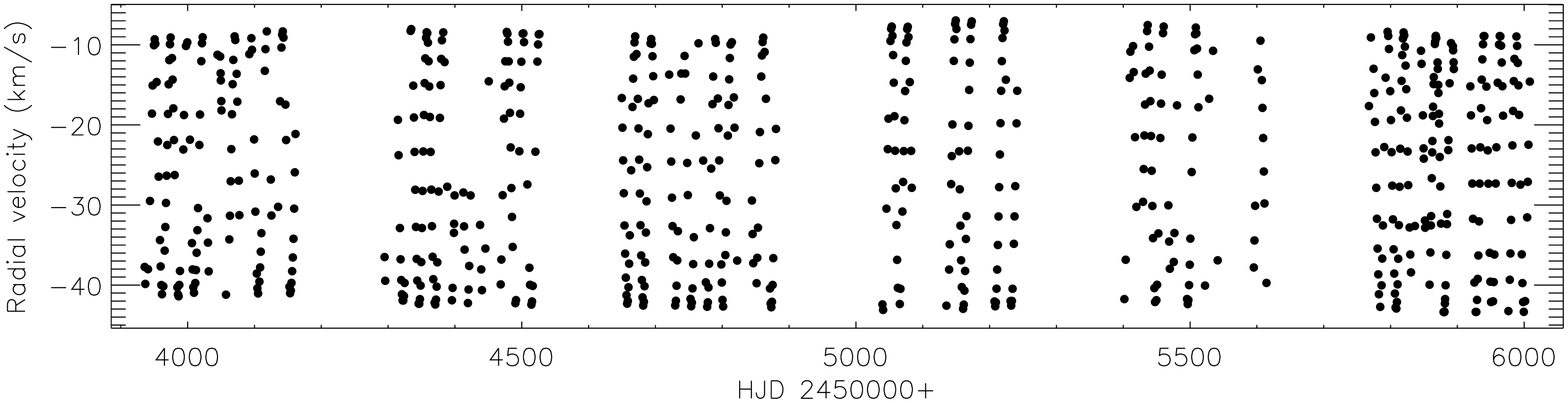}
{\bf (c)} \\
\includegraphics[trim=30 10 10 20, angle=0, width=\textwidth, clip]
{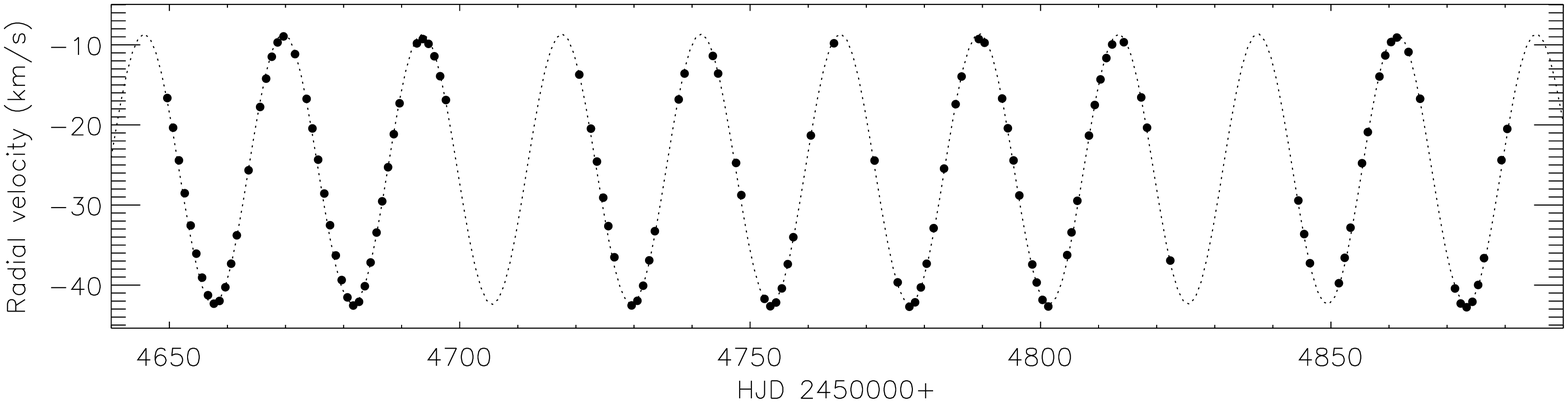}
\caption{STELLA spectroscopy of XX~Tri. a) A section of a representative single spectrum showing a
subset of the spectral lines used for the Doppler imaging inversions. b) The full time-series
radial velocities from a cross-correlation analysis. It shows the rotational phase sampling for the
STELLA spectra. Each dot is from a full echelle spectrum, covering the wavelength range 390-880~nm.
c) A section of b) referring to the observational season 2008/09. The dotted line represents
a radial-velocity fit.}
\label{fig:spectrum_radvel}
\end{figure*}

Since the modern rediscovery of starspots \citep[][and the many references
therein]{hall1972,strassmeier2009}, the study of spots on other stars introduced an extra line of
research to better understand the Sun. Despite this, it is generally not possible to resolve stellar
surfaces directly, with a few recent exceptions \citep[e.g.,][]{kloppenborg2010}, clever
mathematical methods and observing techniques were introduced to resolve stellar surfaces indirectly
\citep[e.g.,][]{vogt1983,rice2002, carroll2012}. These techniques, commonly referred to as Doppler
imaging, became the most advanced tool for the study of starspots. Unfortunately, the needed
time series of well-sampled, high-resolution spectra are difficult to obtain, even for a single
Doppler image, given that intrinsic stellar surface variations confine the data gathering process to
a single or possibly two consecutive stellar rotations. For a star like XX~Tri with a rotation period
of 24~days, this is obviously not a simple task. Sampling the evolution of starspots requires repeated
visits of a target, or preferably even continuous decade-long time series of spectra. In the present
paper, we present for the first time, such long-term, highly-sampled, phase-resolved spectroscopic
data for Doppler imaging, made possible by the use of STELLA robotic telescopes
\citep{strassmeier2010a}.

There have been several other attempts to monitor active stars by means of Doppler imaging. For
example, for the rapidly rotating single late-type giant FK~Comae, a series of snapshots consisting
of 25 Doppler maps between 1993-2003 exists, see \cite{korhonen2007b}. The RS~CVn binary II~Peg was
monitored by \cite{berdyugina1998b,berdyugina1999} between 1992-1999, resulting in 15 Doppler maps.
These maps were partially reanalyzed by \cite{lindborg2011}, including six new maps.
\cite{hackman2012} added 12 more Doppler maps using the same spectrograph (SOFIN at the Nordic
Optical Telescope), resulting in a total of 28 Doppler maps between 1994-2010. Another RS~CVn
binary, IM~Peg, was monitored by \cite{berdyugina2000} who obtained eight Doppler maps between
1996-1999. \cite{marsden2007} obtained a movie from 31 Doppler maps for IM~Peg during a monitoring
campaign between 2004-2007. Another starspot movie was published earlier for HR~1099 (V711~Tau)
from 37 Doppler maps taken in 1996 \citep{strassmeier2000}. Table~2 in \cite{strassmeier2009} lists
all Doppler-imaging attempts and their general results until 2008.

Our target is the spotted, red giant XX~Tri (HD~12545), a member of the RS~CVn class of magnetically
active components of close binaries. This red giant is a synchronized SB1-type system with a period
of $P_{\rm rot}$\;$\approx$\;$P_{\rm orb}$\;$\approx$\;24~days. We have obtained 667 usable spectra
between July 2006 and April 2012, which cover 36 rotational periods. This star is famous for its
detected superspot with a linear extension of $12$\,$\times$\,20 solar radii
\citep{strassmeier1999}. In the mid-1980's XX~Tri became one of the most attractive targets for
photometrists because of its unusually large lightcurve variations. \cite{bidelman1985} observed strong
Ca\,{\sc ii}~H\&K emission lines in the spectrum of XX~Tri and suggested that it might also be a
photometric variable. Photometric variability of RS~CVn stars is generally attributed to the
presence of large, cool starspots moving in and out of view as the star rotates. The presence of an
extremely active chromosphere on XX~Tri was confirmed by \cite{strassmeier1990}, showing that the
Ca\,{\sc ii}~H\&K emission intensity was two to three times that of the local continuum. The first
published photometric observations dated back to 1986-87 and showed a fairly scattered light curve
\citep{hooten1990}. Three years later, a remarkably high $V$ amplitude of $0\fm6$ was observed by
\cite{nolthenius1991}. Multicolor photometry, obtained in early 1991, showing a large amplitude of
$0\fm5$ in $V$ and $0\fm12$ in $V-I$, and allowed \cite{strassmeier1992} to derive a precise
temperature difference between spot and photosphere of 1100\,$\pm$\,35~K, which suggests a spot
coverage of approximately 20\,\% of the entire stellar surface. Furthermore, the $U-B$ and $B-V$
values suggest a K0III classification rather than the G5IV spectral type previously reported. Using
spectroscopic data, \cite{bopp1993} confirmed the K0III spectral type and determined an orbital
period of $P_{\rm orb}$\;=\;23.97~days and $v\sin i$\;=\;17\,$\pm$\,2~\kms. A record light-curve
amplitude of $0\fm63$ in $V$ and $0\fm17$ in $V-I$ in January 1998 was reported by
\cite{strassmeier1999}. At this state of high activity, spectroscopic observations were obtained at
Kitt Peak National Observatory (KPNO) that yielded the first Doppler image of XX~Tri
\citep{strassmeier1999}. It showed a gigantic, cool, high-latitude spot of elliptical shape with a
temperature of around 1300~K below the photospheric temperature. Beside this superspot, a smaller,
cool spot with a temperature of around 4000~K and a warm equatorial spot on the adjacent hemisphere
were reconstructed. A 28-yr data set of photometric observations \citep{olah2014}, including about
20~years of APT photometry (see \citealt{strassmeier1997} for details about APT), revealed a
long-term modulation in $V$ of $1\fm05$ from the deepest minimum to the overall maximum with a
length comparable to the length of the data set.

Most spotted stars investigated so far show solar-like differential rotation, i.e., the equator
rotates faster than the pole \citep[e.g.,][]{barnes2005,reiners2006}. In the last decade, several
stars with antisolar differential rotation were detected \citep[see, e.g.,][and references
therein]{kovari2015}. As the differential rotation holds important information about the stellar
dynamo working beneath the surface, theoretical developments need continuous feedback from
observations. It is particularly important to measure stars of different types because it is still
not fully understood how stellar dynamos work. In close binaries, such as the RS~CVn-systems, tidal
effects are thought to play an important role, as they help maintain the fast rotation and magnetic
activity on high levels \citep[cf.][]{scharlemann1981,scharlemann1982,schrijver1991,holzwarth2002}.

In this paper, we present a series of starspot distributions with different morphology, such as spot
fragmentation and spot merging, and with apparently a large range of variability timescales. We
found evidence for starspot decay and active longitudes, and we estimate a magnetic cycle timescale.
The paper is structured as follows. Sect.~\ref{sec:observation} describes the STELLA observations.
Sect.~\ref{sec:parameter} redetermines the absolute astrophysical parameters of XX~Tri. Among these
parameters is a revised orbit for the binary system, revised atmospheric abundances, and better
constrained values for mass, radius, and age. Sect.~\ref{sec:imaging} presents our Doppler imagery
of a total of 36 maps over the course of six years. In this section, firstly we describe our
Doppler-imaging code \textit{iMap} and the parallel line-profile denoising algorithm. Secondly, we
present and apply a phase-refilling scheme that uses data from the previous or the following stellar
rotation. Next, we present several tests of this scheme with partially artificial, partially real
data. It includes inversion tests with phase gaps like in the real data, e.g., due to bad weather.
Thirdly, we define the term spot area on the basis of photometric image analysis that fits a spot
model to each Doppler image. These images are then analyzed in detail in Sect.~\ref{sec:results} and
discussed in Sect.~\ref{sec:discussion}.


\section{Spectroscopic observations}
\label{sec:observation}

Time-series, high-resolution echelle spectroscopy of XX~Tri was taken with the 1.2-m STELLA
telescopes and the STELLA Echelle Spectrograph (SES) on a nightly basis between July 2006 and April
2012. The STELLA observatory is located at the Izana Observatory on Tenerife and operates fully
robotic with two 1.2-m telescopes \citep{strassmeier2004,strassmeier2010a}. The SES is a white-pupil
spectrograph with an R2 grating with two off-axis collimators, a prism cross disperser and a folded
Schmidt camera with an e2v 2k$\times$2k CCD as the detector, the latter two items were replaced by a
fully refractive camera and an e2v 4k$\times$4k CCD in mid-2012. In 2010, the SES fiber was moved to
the prime focus of the second STELLA telescope (STELLA-II), while STELLA-I now hosts the Wide-Field
STELLA Imaging Photometer (WiFSIP). Further details of the performance of the system were reported
by \cite{granzer2010} and \cite{weber2012}.

\begin{table}
\centering
\caption{Astrophysical properties of XX~Tri (HD~12545).}
\label{tab:xxtri_par}
\begin{tabular}{l l l}
\hline\noalign{\smallskip}
Parameter                   & Value                & Based on \\
\noalign{\smallskip}\hline\noalign{\smallskip}
Classification, MK          & K0III                & \cite{strassmeier1999} \\
Distance, pc                & $160_{-22}^{+32}$    & \cite{vanleeuwen2007} \\
$V_{\rm max}$, mag          & 7.76                 & \cite{olah2014} \\
$V-I_C$, mag                & 1.18                 & \cite{olah2014} \\
Rotation period, d          & $\approx$\;24.0      & \cite{strassmeier1999} \\
Orbital period, d           & 23.9674$^1$          & radial velocities \\
Inclination, deg            & 60\,$\pm$\,10        & \cite{strassmeier1999} \\
$v\sin i$, \kms             & 19.9\,$\pm$\,0.7     & spectrum synthesis \\
Temperature, K              & 4,620\,$\pm$\,30     & spectrum synthesis \\
Log gravity, cgs            & 2.82\,$\pm$\,0.04    & spectrum synthesis \\
Metallicity, [Fe/H]$_\odot$ & $-$0.13\,$\pm$\,0.04 & spectrum synthesis \\
Microturb., \kms            & 1.5                  & spectrum synthesis \\
Macroturb., \kms            & 3.0                  & spectrum synthesis \\
Radius, R$_\odot$           & 10.9\,$\pm$\,1.2     & from $R\sin i$ and $i$ \\
Luminosity, L$_\odot$       & $30_{-8}^{+13}$      & from $M_{\rm bol}$ \\
Mass, M$_\odot$             & 1.26\,$\pm$\,0.15    & evolutionary tracks \\
Age, Gyr                    & 7.7\,$\pm$\,3.1      & evolutionary tracks \\
\noalign{\smallskip}\hline
\end{tabular}
\tablefoot{Values not cited in the third column were obtained in this paper. Note that the errors
for spectrum-synthesis related parameters are internal errors.~$^1$\,$\pm$\,0.0005, see text.}
\end{table}

We obtained a total of 667 usable spectra from six observational seasons. Spectra cover the
wavelength range from 388-882~nm with increasing inter-order gaps near the red end starting at
734~nm toward 882~nm before the camera and CCD exchange. The resolving power is $R$\;=\;55,000
corresponding to a spectral resolution of 0.12~\AA\ at 650~nm.

We set the integration time to 7200~s because of the relative faintness of the target for a 1m-class
telescope. Depending on weather conditions, the averaged signal-to-noise (S/N) ratios are between
50-300:1 for each spectrum, but typically 150:1. The SES spectra are automatically reduced using the
IRAF-based STELLA-SES data-reduction pipeline \citep{weber2008}. We corrected the images for bad
pixels and cosmic-ray impacts. We removed bias levels by subtracting the average overscan from
each image followed by the subtraction of the mean of the (already overscan subtracted) master bias
frame. We flattened the target spectra with a nightly master flat, which itself is constructed from
around 50 individual flats observed during dust, dawn, and around midnight. After removal of
scattered light, the one-dimensional spectra were extracted using an optimal-extraction algorithm.
We then removed the blaze function from the target spectra, followed by a wavelength calibration
using consecutively recorded Th-Ar spectra. Finally, the extracted spectral orders were continuum
normalized by dividing with a flux-normalized synthetic spectrum of the same spectral classification
as XX~Tri.

Fig.~\ref{fig:spectrum_radvel}a shows an example spectrum (Echelle order~\#89) from
HJD~2,454,146.37 while Fig.~\ref{fig:spectrum_radvel}b gives an overview of the time and phase
sampling of all 667 STELLA spectra.


\section{Astrophysical parameters of XX Tri}
\label{sec:parameter}

The revised reduction of the \emph{Hipparcos} data \citep{vanleeuwen2007} yielded a parallax of
6.24\,$\pm$\,1.02~mas and fixed the distance of XX~Tri (HIP~9630) to $\approx$\;160~pc. With an
apparent maximum visual magnitude of $7\fm76$ \citep{olah2014}, the absolute visual magnitude of
XX~Tri is $M_V$\;=\;$1\fm58_{-0.40}^{+0.32}$. We took interstellar absorption into account with
$0\fm1$ per 100~pc \citep{strassmeier1999}. We note that \cite{olah2014} estimated the reddening of
XX~Tri from all-sky infrared imaging and found a color excess $E(B-V)$ of $\approx$\;$0\fm05$,
which leads to the same value of extinction as above. With a bolometric correction of $-0\fm517$
\citep{flower1996}, the bolometric magnitude of XX~Tri is $1\fm06$ and, with an absolute magnitude
for the Sun of $M_{\rm bol,\odot}$\;=\;$4\fm75$, the luminosity must be approximately
$30_{-8}^{+13}$~L$_\odot$.

\begin{figure}[!ht]
\begin{minipage}{1.0\textwidth}
\subfloat{\includegraphics[trim=20 40 20 40, angle=0, width=0.25\textwidth, clip]
{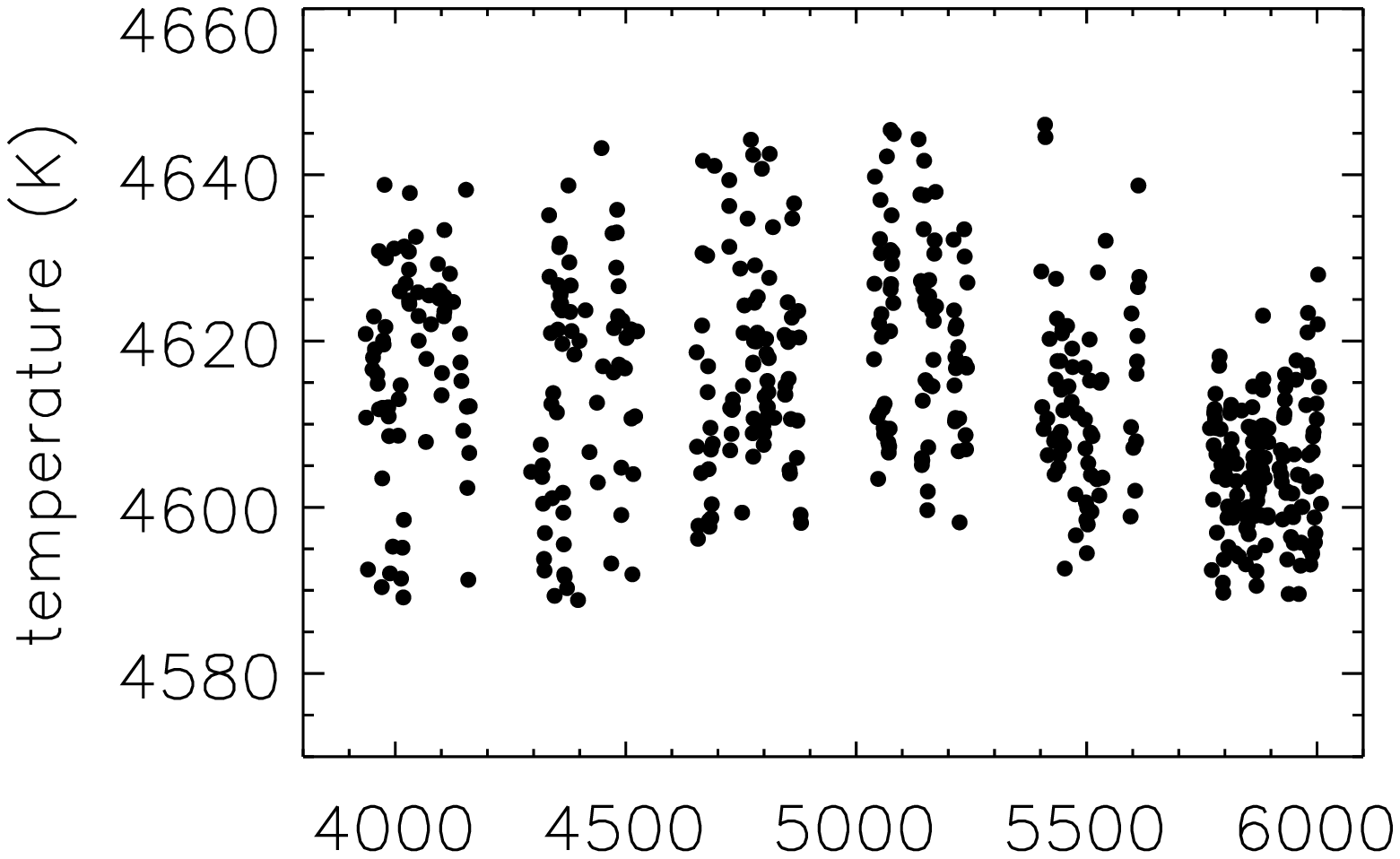}}
\subfloat{\includegraphics[trim=30 40 10 40, angle=0, width=0.25\textwidth, clip]
{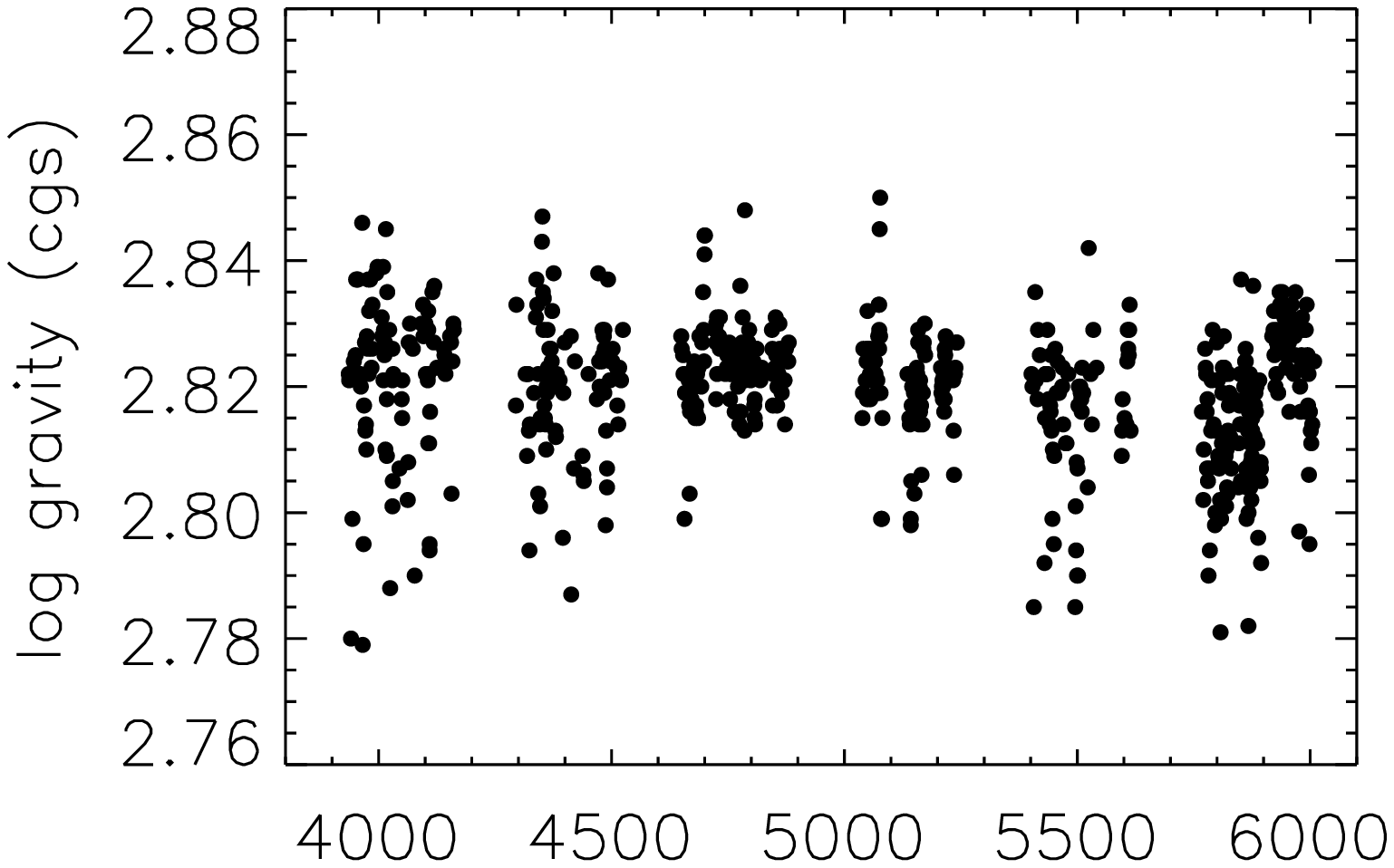}}
\end{minipage}\hspace{1ex}
\begin{minipage}{1.0\textwidth}
\subfloat{\includegraphics[trim=20 10 20 40, angle=0, width=0.25\textwidth, clip]
{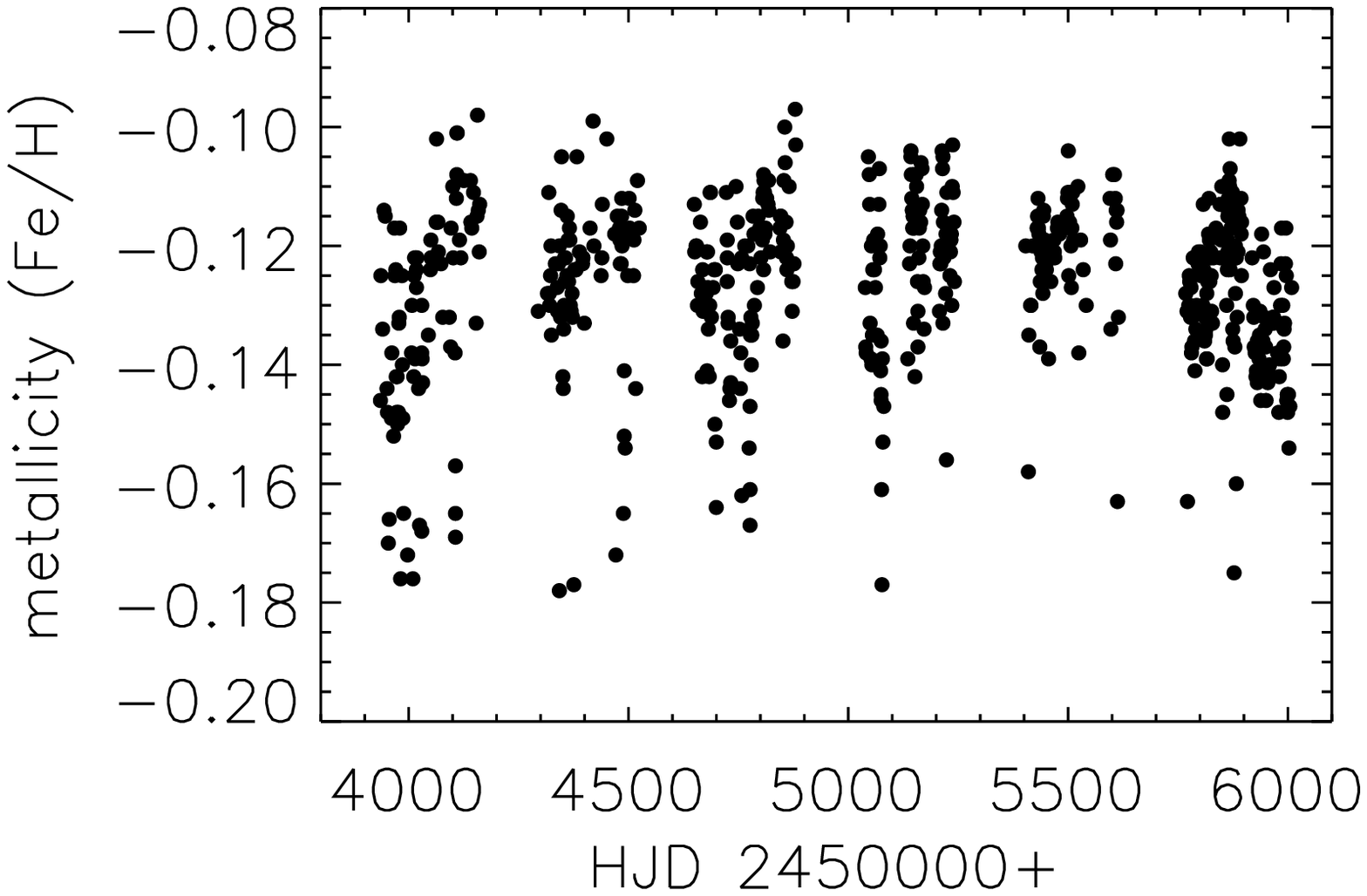}}
\subfloat{\includegraphics[trim=30 10 10 40, angle=0, width=0.25\textwidth, clip]
{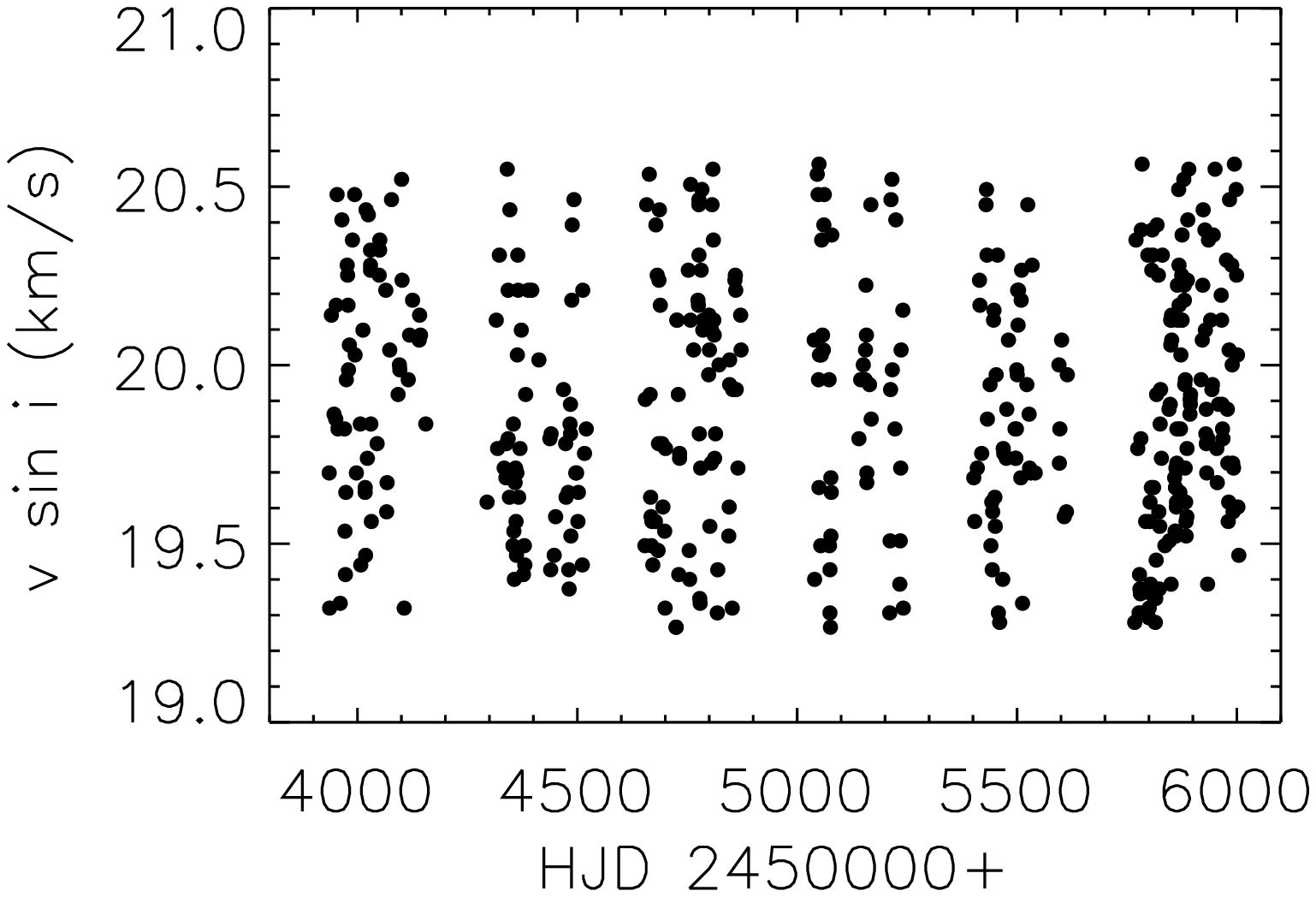}}
\end{minipage}\hspace{1ex}
\caption{Stellar parameter determinations with PARSES. Shown are the obtained values for effective
temperature, log gravity, metallicity, and $v\sin i$ for all spectra excluding 3-$\sigma$ outliers.}
\label{fig:parses_parameters}
\end{figure}

Atmospheric surface stellar parameters (effective temperature, log gravity, metallicity, and $v\sin
i$) were determined with the program PARSES \citep{allende2004,jovanovic2013}, which is included in
the SES data reduction pipeline. It fits synthetic spectra to a defined spectral region, in our case
most echelle orders between 480-750~nm, using MARCS model atmospheres \citep{gustafsson2008}. We
verified this approach by applying it to the ELODIE library \citep{prugniel2001}, and used linear
regressions to the offsets with respect to the literature values to correct the zero point of our
PARSES results. In Fig.~\ref{fig:parses_parameters} these corrected values are shown for all spectra
(excluding 3-$\sigma$ outliers). For more details of this procedure, we refer to previous
applications \citep[e.g.,][]{strassmeier2010b,strassmeier2012}. The mean values are $T_{\rm eff}$ of
4620\,$\pm$\,30~K, a gravity $\log g$ of 2.82\,$\pm$\,0.04, a $v\sin i$ of 19.9\,$\pm$\,0.7~\kms,
and a metallicity of $-$0.13\,$\pm$\,0.04~dex relative to the Sun. Its errors are internal errors
based on the rms of the entire time series. The above effective temperature is close to
what \cite{olah2014} obtained from $VI$ photometry during maximum photometric brightness, but 
the metallicity differs by a factor of two.

Fig.~\ref{fig:parses_parameters} also shows that the plotted values are systematically variable. In
particular the effective temperature and line broadening clearly vary with the rotational period
of the star and from season to season. There may be a similar trend in gravity, i.e., an
increase during the first three observing seasons and a decrease during the last three seasons. A
possible explanation could be related to the assumption that a higher surface temperature means
fewer cool spots, which implies a weaker internal magnetic field (and therefore magnetic pressure),
and hence that the star contracts a bit and thus the surface gravity increases. We will
investigate this behavior in a forthcoming paper in more detail where we compare the observed
broadband light curves with the photometric predictions from our Doppler maps along with the
spectrum-integrated values from PARSES.

\begin{figure}[!ht]
\includegraphics[trim=40 10 0 10, angle=0, width=0.5\textwidth, clip]
{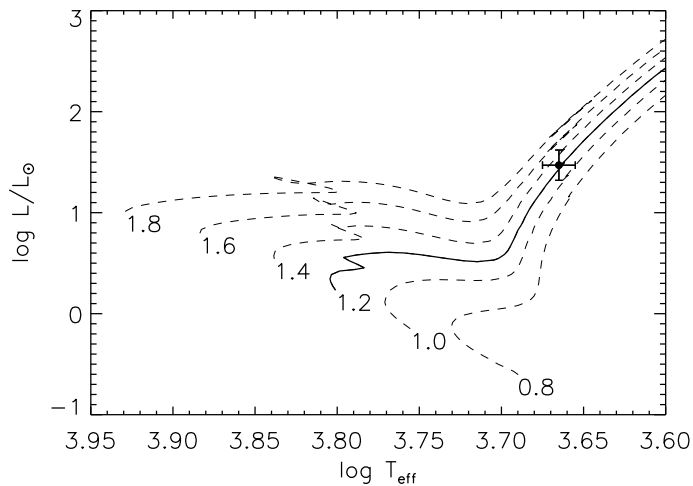}
\caption{Stellar evolutionary tracks of low-mass stars from \cite{bertelli2008}
together with the position of XX~Tri. Each track is interpolated to a metallicity of $-0.13$ and
covers the range from the ZAMS up to the RGB. The stellar mass (in M$_\odot$) is indicated
at the initial point of evolution.}
\label{fig:evol_tracks}
\end{figure}

To determine the mass and age of XX~Tri, a trilinear interpolation between stellar evolutionary
tracks \citep{bertelli2008} based on a Monte Carlo (MC) method \citep{kuenstler2008} was used.
Within the three-dimensional space ($L$, $T_{\rm eff}$, [Fe/H]) 10,000 random positions were
generated, taking Gaussian errors into account. For each generated position, we calculated the mass
and age. The obtained mean values are a mass of 1.26\,$\pm$\,0.15~M$_{\odot}$ and an age of
7.7\,$\pm$\,3.1~Gyrs. Fig.~\ref{fig:evol_tracks} shows the evolutionary tracks interpolated to the
metallicity of XX~Tri including the star's position.

A rotation period of 24.0 days together with a projected rotational velocity of $v\sin
i$\;=\;19.9\,$\pm$\,0.7~\kms\ yields a minimum radius of $R\sin i$\;=\;9.4\,$\pm$\,0.3~R$_{\odot}$.
With an inclination of $i$\;$\approx$\;60\,$\pm$\,10$^{\circ}$ \citep{strassmeier1999}, the stellar
radius is $R$\;=\,10.9\,$\pm$\,1.2~R$_{\odot}$. The unprojected equatorial rotational velocity would
then be $v_{\mathrm{eq}}$\;=\;23.0~\kms.

We use preliminary revised orbital elements from our radial-velocity fit of the STELLA data, see
Fig.~\ref{fig:spectrum_radvel}c: $P_{\mathrm{orb}}$\;=\;23.9674\,$\pm$\,0.0005~days,
$\gamma$\;=\;$-$25.389\,$\pm$\,0.031~\kms, $K$\;=\;16.772\,$\pm$\,0.044~\kms,
$a \sin i$\;=\;5.528\,$\pm$\,0.014\,$\times$\,10$^6$~km, $e$\;=\;0 (adopted), and
$f(M)$\;=\;0.0117\,$\pm$\,0.0001. Phase is always computed from a time of maximum positive radial
velocity with the revised orbital period,
\begin{equation}
HJD = 2,453,926.6663 + E \times 23.9674 \ .
\label{eqn:period_orb}
\end{equation}
Final orbital elements will be presented in our forthcoming paper where we remove the
radial-velocity jitter due to spots and add three more years of data.

The revised mass function, together with the primary mass of 1.26~M$_\odot$ and an orbital
inclination of $i$\;$\approx$\;60$^{\circ}$, suggests a low-mass secondary star with a mass of
$\approx$\;0.36~M$_{\odot}$. Because the secondary is not seen in the spectrum, the most likely
secondary star is then a red dwarf of spectral type M.
The most important astrophysical properties of XX~Tri are summarized in Table~\ref{tab:xxtri_par}.

\begin{table}
\centering
\caption{Spectral lines used in the inversion process.}
\label{tab:linelist}
\begin{tabular}{l l | l l}
\hline \noalign{\smallskip}
Ion & $\lambda$ (\AA) & Ion & $\lambda$ (\AA) \\
\noalign{\smallskip}\hline\noalign{\smallskip}
Fe\,{\sc i} & 5049.820 & Fe\,{\sc i} & 5576.089 \\
Cu\,{\sc i} & 5105.537 & Ca\,{\sc i} & 5581.965 \\
Fe\,{\sc i} & 5198.711 & Ca\,{\sc i} & 5601.277 \\
Fe\,{\sc i} & 5232.940 & Fe\,{\sc i} & 6020.169 \\
Fe\,{\sc i} & 5302.300 & Fe\,{\sc i} & 6024.058 \\
Fe\,{\sc i} & 5307.361 & Fe\,{\sc i} & 6065.482 \\
Fe\,{\sc i} & 5324.179 & Ca\,{\sc i} & 6122.217 \\
Cr\,{\sc i} & 5345.796 & Fe\,{\sc i} & 6173.334 \\
Cr\,{\sc i} & 5348.315 & Fe\,{\sc i} & 6219.281 \\
Fe\,{\sc i} & 5367.466 & Fe\,{\sc i} & 6254.258 \\
Fe\,{\sc i} & 5383.369 & Fe\,{\sc i} & 6265.132 \\
Fe\,{\sc i} & 5393.167 & Fe\,{\sc i} & 6322.685 \\
Mn\,{\sc i} & 5394.677 & Fe\,{\sc i} & 6393.600 \\
Mn\,{\sc i} & 5420.355 & Fe\,{\sc i} & 6408.018 \\
Fe\,{\sc i} & 5434.524 & Fe\,{\sc i} & 6411.648 \\
Fe\,{\sc i} & 5445.042 & Fe\,{\sc i} & 6421.350 \\
Fe\,{\sc i} & 5497.516 & Fe\,{\sc i} & 6430.845 \\
Fe\,{\sc i} & 5501.465 & Ca\,{\sc i} & 6439.075 \\
Fe\,{\sc i} & 5506.779 & Ca\,{\sc i} & 6717.681 \\
Fe\,{\sc i} & 5569.618 & Fe\,{\sc i} & 6750.152 \\
\noalign{\smallskip}\hline
\end{tabular}
\end{table}


\section{Doppler Imaging}
\label{sec:imaging}

The spectral resolution of 55,000 combined with the $v\sin i$ of just $\approx$\;20~\kms\ and the
relative faintness of the star for a 1.2-m telescope, places XX~Tri close to our limit for Doppler
imaging. Significant effort is thus put to remove instrumental noise and other systematics from the
data and to denoise the line profiles to prepare for the inversion.


\subsection{The iMap code}
\label{sec:imap}

All maps were computed with our Doppler Imaging (DI) and (Zeeman Doppler Imaging) ZDI-code
\textit{iMap} \citep{carroll2007,carroll2008,carroll2009,carroll2012}. Here we give just a brief
description of the code, for further details see \cite{carroll2012}. The code performs a multiline
inversion of a large number of photospheric line profiles simultaneously. For the local line profile
calculation, the code utilizes a full (polarized) radiative transfer solver \citep{carroll2008}. The
atomic line parameters are taken from the VALD database \citep{kupka1999}. We used Kurucz model
atmospheres \citep{castelli2004} which are interpolated for each desired temperature, gravity, and
metallicity during the course of the inversion. Additional input parameters are $v\sin i$, micro-
and macroturbulence.

Because of the typical ill-posed nature of the problem, an iterative regularization based on a Landweber
algorithm \citep{carroll2012} is implemented, having the advantage that no additional constraints
are imposed in the image reconstruction. For all temperature maps, the surface segmentation is set
to a $5^{\circ}$\,$\times$\,$5^{\circ}$ equal-degree partition, resulting in 2592 segments. Because of
the inclination of $60^{\circ}$ a total of 432 segments are hidden and therefore only 2160 segments
are included during the inversion process. The code calculates the full radiative transfer of all
involved line profiles for each surface segment depending on the current effective temperature and
atmospheric model. The surface temperature of each segment is adjusted according to the local
(temperature) gradient information. The line profile discrepancy is reduced until a minimum $\chi^2$
is obtained.


\subsection{Line profile denoising}
\label{sec:denoising}

We included 40 well-defined absorption lines simultaneously in our inversion, which are listed in
Table~\ref{tab:linelist}. These lines were chosen individually by investigating the stellar spectra
and VALD database and several other criteria, such as having a minimum line depth of 0.75~$I/I_C$,
being almost blend-free, and having a good continuum stratification above 0.9~$I/I_C$. Additionally,
all blends within $\pm$\,1~\AA\ of each extracted line profile and a minimum line depth of 0.1 are
included in the inversion. As we have to deal with relatively low S/N ratios, a wavelet analysis
based on the \textit{$\grave{a}$ trous}-algorithm \citep[][chapter 1.4.4 and references
therein]{starck1998} is implemented for further denoising. \cite{starck1997} showed that for noisy
data the wavelet transform is a powerful signal processing technique for spectral analysis. Each
line profile, in our case the mean profile out of the 40 individual lines, is split into so-called
wavelet scales $w_j$ and a smoothed array $c_p$, whereas their sum represents the original spectrum
$c_0(\lambda)$\;=\;$c_p(\lambda)$\,+\,$\sum_{j=1}^{p}$\,$w_j(\lambda)$. For each wavelet scale the
standard deviation is determined and only signals above 3~$\sigma$ are overtaken in the
recomposition of the spectral line.

\begin{figure}[!t]
\begin{minipage}{1.0\textwidth}
\captionsetup[subfigure]{labelfont=bf,textfont=bf,singlelinecheck=off,justification=raggedright,
position=top}
\subfloat[]{\includegraphics[width=250pt]{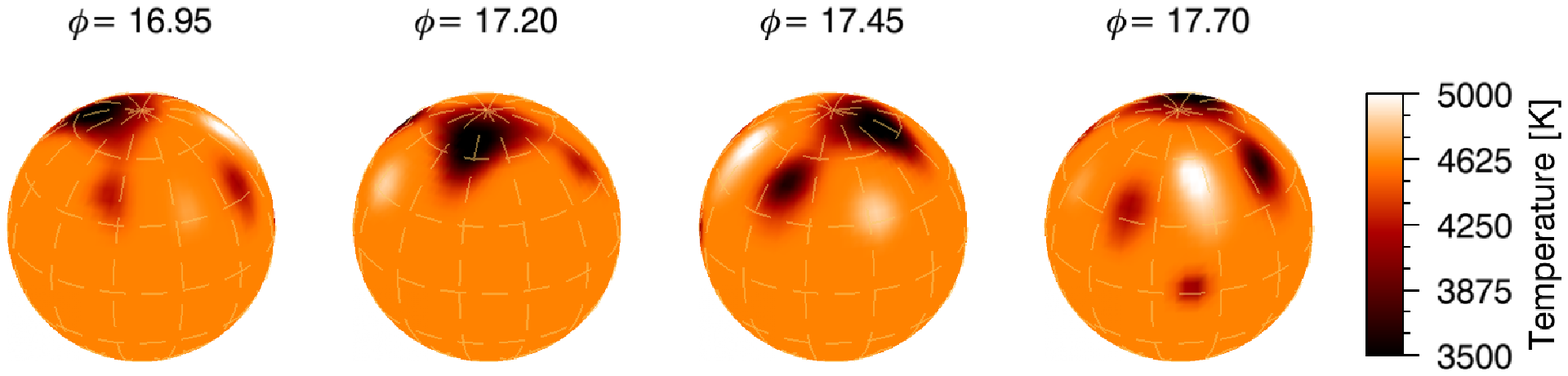}}
\end{minipage}\hspace{1ex}
\begin{minipage}{1.0\textwidth}
\captionsetup[subfigure]{labelfont=bf,textfont=bf,singlelinecheck=off,justification=raggedright,
position=top}
\subfloat[]{\includegraphics[width=250pt]{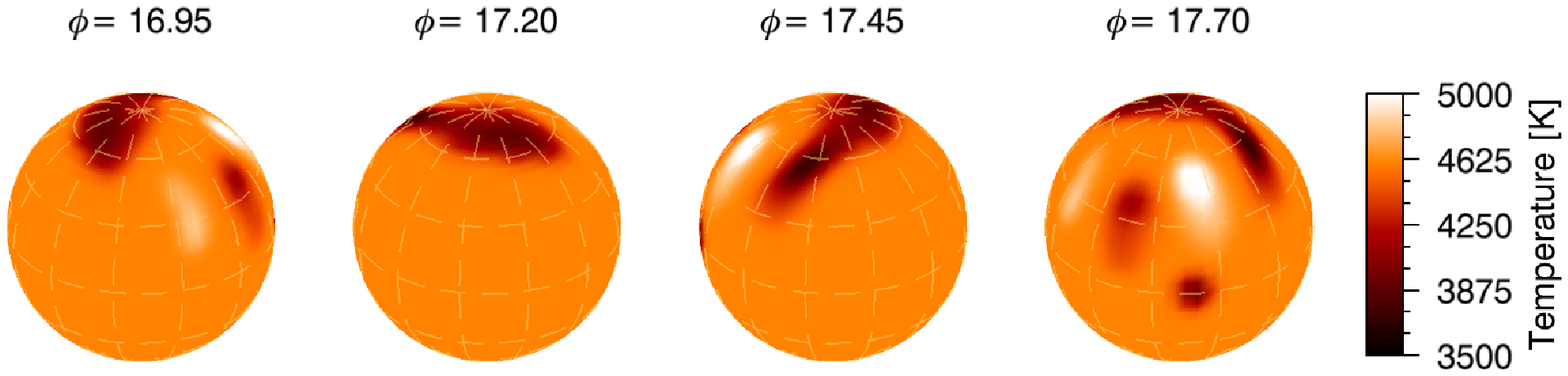}}
\end{minipage}\hspace{1ex}
\begin{minipage}{1.0\textwidth}
\captionsetup[subfigure]{labelfont=bf,textfont=bf,singlelinecheck=off,justification=raggedright,
position=top}
\subfloat[]{\includegraphics[width=250pt]{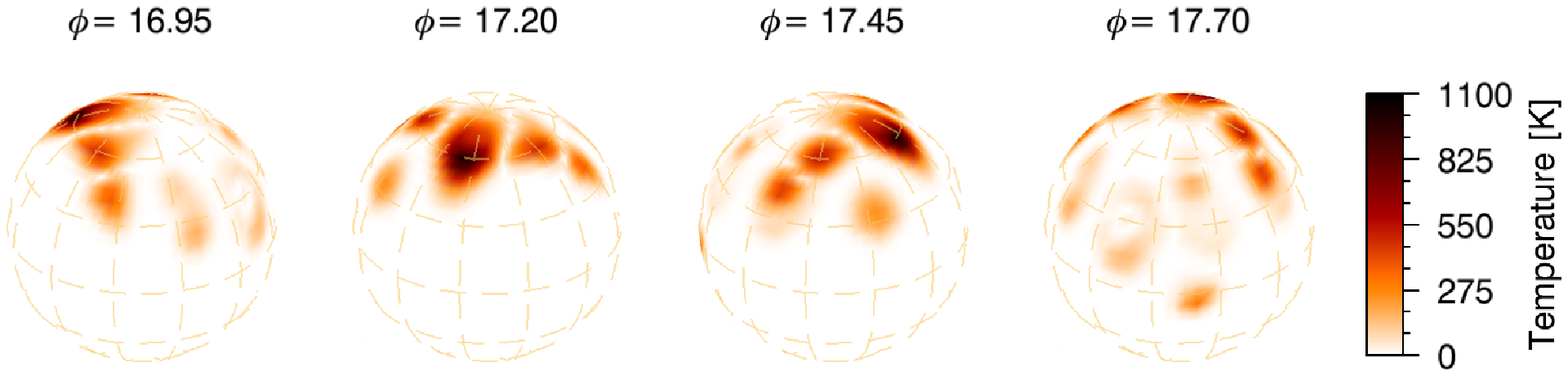}}
\end{minipage}\hspace{1ex}
\begin{minipage}{1.0\textwidth}
\captionsetup[subfigure]{labelfont=bf,textfont=bf,singlelinecheck=off,justification=raggedright,
position=top}
\subfloat[]{\includegraphics[width=250pt]{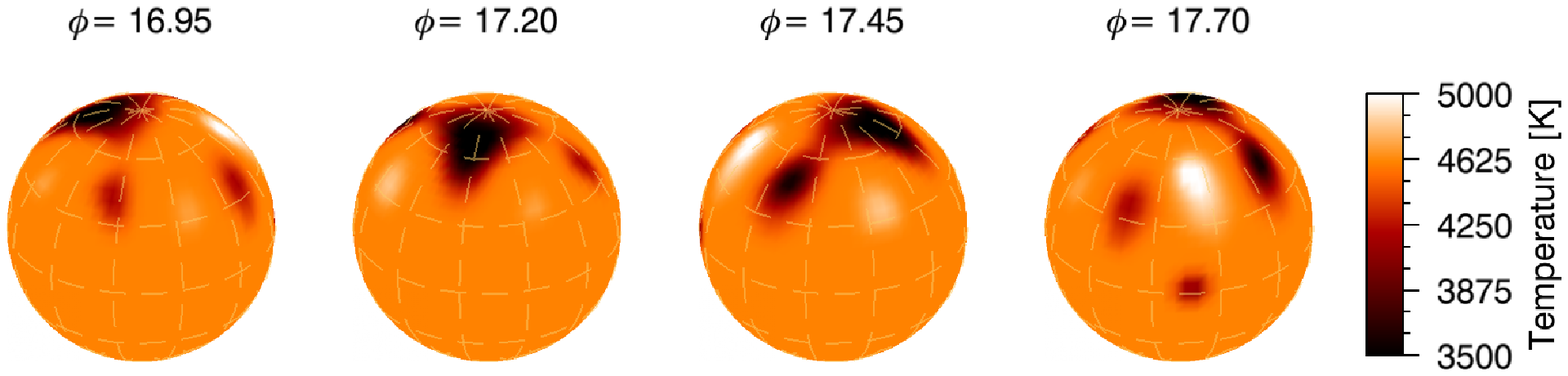}}
\end{minipage}\hspace{1ex}
\begin{minipage}{1.0\textwidth}
\captionsetup[subfigure]{labelfont=bf,textfont=bf,singlelinecheck=off,justification=raggedright,
position=top}
\subfloat[]{\includegraphics[width=250pt]{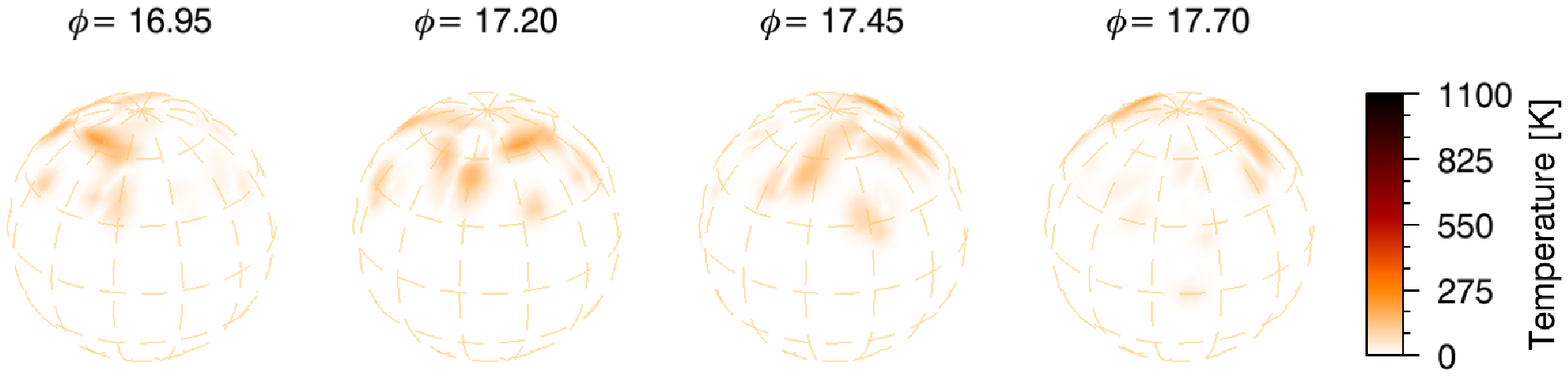}}
\end{minipage}\hspace{1ex}
\caption{Test \#1 of the influence of phase gaps and filled gaps. Each image is shown in four
spherical projections $90^{\circ}$ apart. a) Input map. This map is identical to the first
reconstruction from season 2007/08 (DI~\#8 from 2007.67) with a total of 21 phases. b) The
reconstruction when ignoring phases no. 4-9 in the inversion process. c) The (absolute) difference
$a-b$. d) The reconstruction when the  phase gap is filled with phases from the following
stellar rotation. e) The (absolute) difference $a-d$.}
\label{fig:di_phase_gaps_1}
\end{figure}

\begin{figure}[!t]
\begin{minipage}{1.0\textwidth}
\captionsetup[subfigure]{labelfont=bf,textfont=bf,singlelinecheck=off,justification=raggedright,
position=top}
\subfloat[]{\includegraphics[width=250pt]{figures/season2_map1_original.eps}}
\end{minipage}\hspace{1ex}
\begin{minipage}{1.0\textwidth}
\captionsetup[subfigure]{labelfont=bf,textfont=bf,singlelinecheck=off,justification=raggedright,
position=top}
\subfloat[]{\includegraphics[width=250pt]{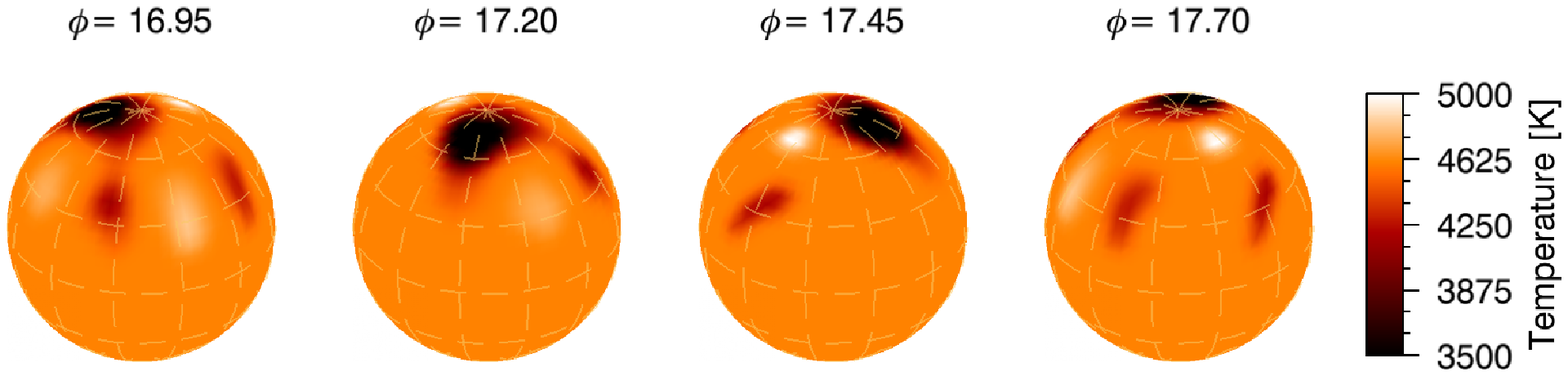}}
\end{minipage}\hspace{1ex}
\begin{minipage}{1.0\textwidth}
\captionsetup[subfigure]{labelfont=bf,textfont=bf,singlelinecheck=off,justification=raggedright,
position=top}
\subfloat[]{\includegraphics[width=250pt]{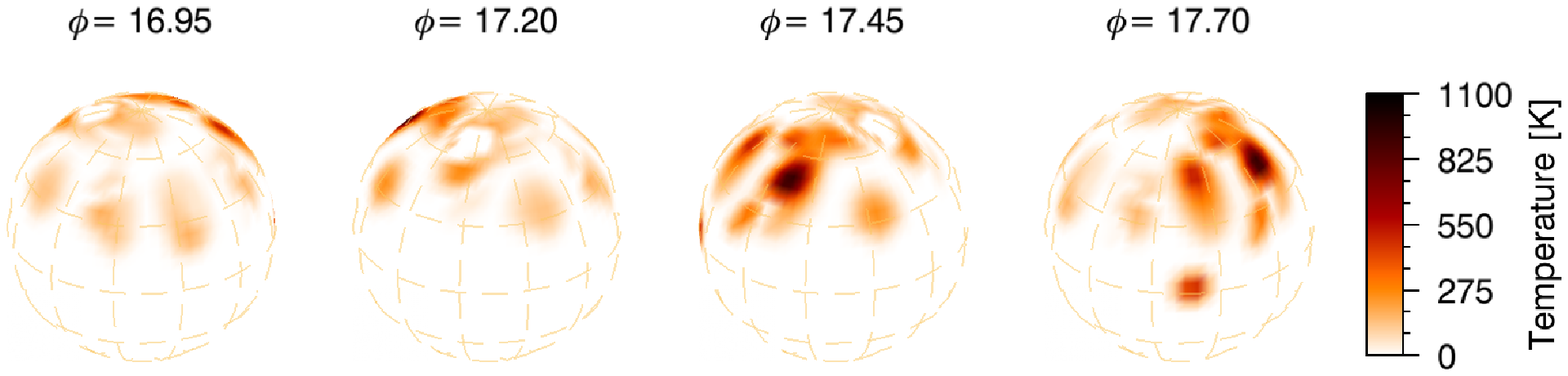}}
\end{minipage}\hspace{1ex}
\begin{minipage}{1.0\textwidth}
\captionsetup[subfigure]{labelfont=bf,textfont=bf,singlelinecheck=off,justification=raggedright,
position=top}
\subfloat[]{\includegraphics[width=250pt]{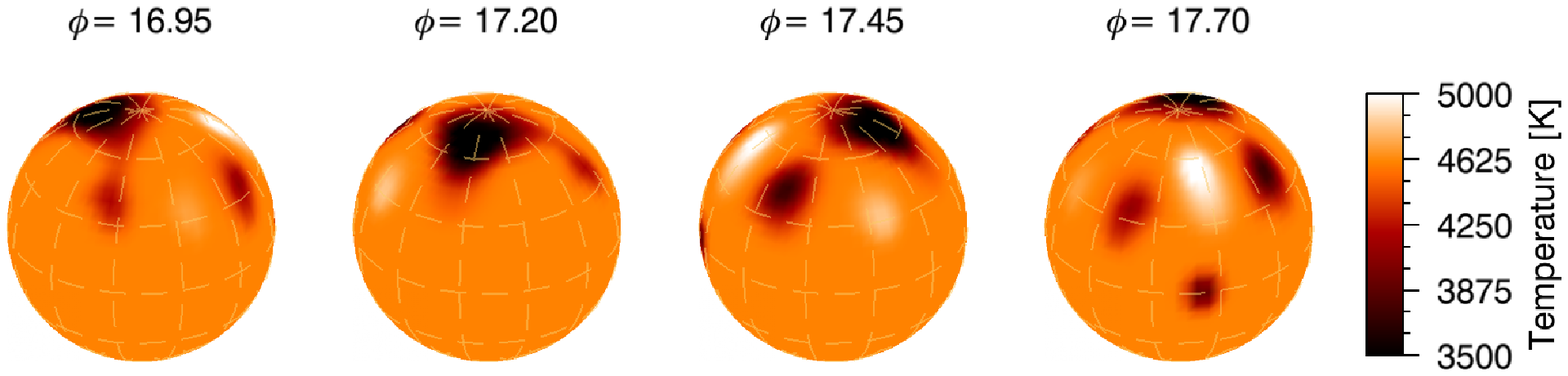}}
\end{minipage}\hspace{1ex}
\begin{minipage}{1.0\textwidth}
\captionsetup[subfigure]{labelfont=bf,textfont=bf,singlelinecheck=off,justification=raggedright,
position=top}
\subfloat[]{\includegraphics[width=250pt]{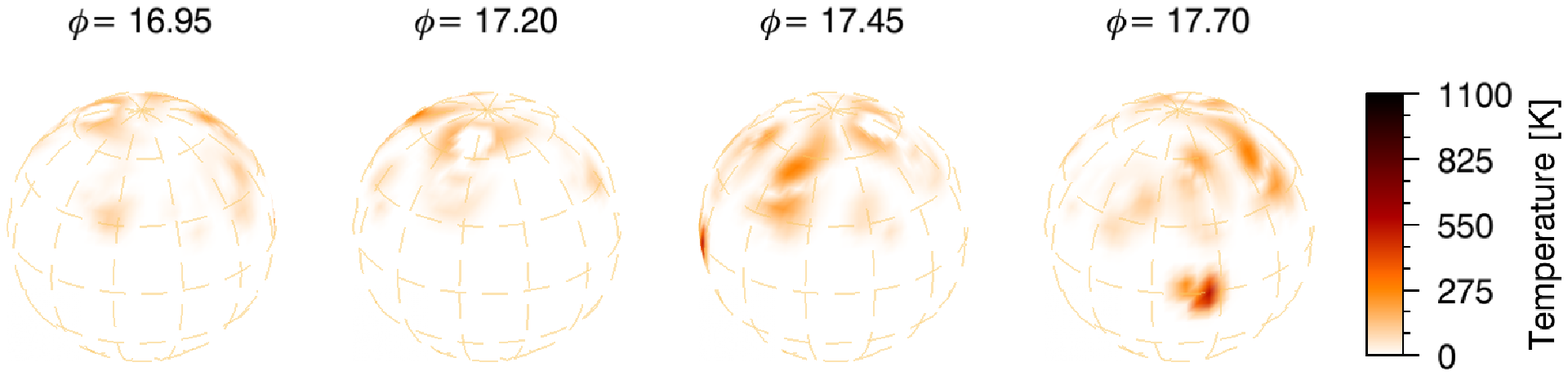}}
\end{minipage}\hspace{1ex}
\caption{Test \#2 of the influence of phase gaps and filled gaps. Each image is shown in four
spherical projections $90^{\circ}$ apart. a) Input map. This map is identical to the first
reconstruction from season 2007/08 (DI~\#8 from 2007.67) with a total of 21 phases. b) The
reconstruction when ignoring phases no. 10-15 in the inversion process. c) The (absolute) difference
$a-b$. d) The reconstruction when the  phase gap is filled with phases from the following
stellar rotation. e) The (absolute) difference $a-d$.}
\label{fig:di_phase_gaps_2}
\end{figure}


\subsection{Phase selection and gap filling}
\label{sec:phasing}

To later quantify the continuous evolution of spots based on consecutive Doppler maps, we
first deal with the inherent limitations of our phase coverage. In certain circumstances it is
difficult to compare consecutive maps that had different phase coverage and that even contained some
larger observational gaps (several tenths of a phase) at different rotational phases at different
times. The effect of phase gaps on the recovery of individual spots had been simulated by many
authors in the past \citep[e.g.,][and references therein]{rice2000}. Generally, Doppler imaging is
very robust against small phase gaps but large phase gaps may introduce spurious spots at surface
locations not covered by the data.

Fig.~\ref{fig:di_phase_gaps_1} and \ref{fig:di_phase_gaps_2} show our simulations with \textit{iMap}
based on real data of XX~Tri. During the season 2007/08, STELLA has covered two consecutive stellar
rotations completely with one observation per night (DI~\#8 from 2007.67 and DI~\#9 from
2007.73, amounting to 21 phases from 23 nights per rotation). From DI~\#8, we removed six consecutive
phases to create an artificial phase gap of $90^{\circ}$ (0\fp25), and compared the resulting map
with the original map (Fig.~\ref{fig:di_phase_gaps_1}a-c). The two darkest and biggest spots at
phases around 17\fp25 and 17\fp55 could not be separated anymore. The larger spot loses a big part
of its area, which is seen in the difference map in Fig.~\ref{fig:di_phase_gaps_1}c with a
temperature similar to the difference between photospheric and spot temperature. In the next step,
we filled these gaps with observations from the following stellar rotation (DI~\#9) and again
compared the resulting map with the original map (Fig.~\ref{fig:di_phase_gaps_1}d-e). All
individual spots are now reconstructed with no changes of their size or temperature exceeding the
expected errors driven by the S/N of the data. Fig.~\ref{fig:di_phase_gaps_2}a-c shows another
simulation of the same data with an artificial phase gap of $90^{\circ}$ but at a different
rotational phase. Here, the smaller spot at phase around 17\fp55 has almost completely vanished,
whereas it is recovered for the case with gap filling (Fig.~\ref{fig:di_phase_gaps_2}d-e). We
further verified this method on a few other Doppler images with the same result (not shown).

We conclude that large ($\approx$\;90$\degr$) phase gaps have a strong impact on the recovery of the
global stellar spot distribution. It affects not only their size and shape but also their location.
For smaller spots located within or near the missing phases it affects even their existence.
However, when we compare the original map with the maps where the missing phases were filled by
phases from one rotation earlier or later, we see a significantly better agreement. This is the
case even if the spot distribution had evolved in the meantime. Based on these simulations, we have
a simple method to minimize the systematics due to phase gaps. This is only a first-order
approximation because the spot evolution within the missing surface area remains unknown.
Fig.~\ref{fig:phase_coverage} shows the phase coverage and gap filling of each Doppler image for all
seasons, if applicable. Real phase gaps ranged between 1-13~d or 0\fp04-0\fp54 at the extremes, but
with typical values more near $\approx$\;4~d or 0\fp17.

For two out of the 36 Doppler images, we had to use phases from two rotations earlier and/or later,
as there were no observations closer in time available. In the first case (DI~\#5) we filled a gap
of 5~d (0\fp21) with two phases. To minimize a possible smearing effect, one phase was taken from
two rotations earlier and one from two rotations later. In the case of DI~\#20, we had to deal with
a small number of existing phases in addition. In general, we would not have taken this
rotation into account, but in this particular case, we would have only four maps for the observing
season 2007/08, two consecutive maps at the beginning and two consecutive maps at the end of the
season, which would have severely limited our coverage for this season. We will rediscuss this
explicitly in Sect.~\ref{sec:spot_evol} in terms of spot-area evolution.

Table~\ref{tab:maps_log} summarizes our Doppler-image log. The year indicates the mid-time of the
Doppler image, followed by its Heliocentric Julian Date (HJD) range, the time $\Delta t$ elapsed in
days, the number of spectra $N$, the largest phase gap in number of consecutive days (the rotation
period is 24~d), and the number of observations from following or preceding stellar rotations that
were used to fill phase gaps. In total, we obtained 36 individual Doppler images, each with
between 11 to 24 spectra.


\subsection{Image analysis: definition of the spot area}
\label{sec:image_analysis}

One parameter to be extracted from each image is the surface area of a spot. Its measurement depends
on the definition where a spot ends and where the undisturbed photosphere begins. Our inversion code
is completely free in the reconstruction and usually recovers an irregular spot morphology.
Furthermore, the images contain small-scale structures, which seem to appear or disappear from one
rotation to another and/or move to different longitudes/latitudes on timescales that are likely
shorter than a stellar rotation. These structures often appear as hot spots or as a pair of a hot
and a cool spot (but not necessarily along the same isoradial line which would indicate an
artifact). The appearance of elongated appendages to the polar spot, or spots that appear connected
to another spot, are further complications for determining a spot's area. Because of inherent
limitations of the spatial surface resolution due to the low $v\sin i$, as well as to a lesser
extent also due to the choice of regularization during the inversion, these spot configurations
cannot easily be separated from each other anymore. Simple area integration of the disk within a
certain temperature difference would thus lead to erroneous spot areas.

We define the spot area by fitting artificial spot models to the maps. Our artificial spots
have circular shapes of arbitrary size, but a constant temperature of 3500~K. From photometric
spot-modeling with spot temperature as free parameter the derived temperature difference $\Delta
T$\;=\;$T_{\rm phot}$\,$-$\,$T_{\rm spot}$ were determined to 1100~K \citep{strassmeier1992} and
1280~K \citep{eker1995} as well as in the range of 650-1200~K \citep{hampton1996} for various
epochs. Furthermore, the superspot on the first Doppler image had a temperature difference of
1300~K \citep{strassmeier1999}. These values are in agreement with the derived spot temperatures of
$\approx$\;3500~K from our Doppler images for long-lived spot structures. As our focus lies in
the evolution of starspots, i.e., their decay or growth, we investigate mainly large-scale spot
structures that are repeatedly reconstructed from one stellar rotation to the next.

Our method is based on the spot-modeling procedure used in light-curve analysis
\citep[e.g.,][]{ribarik2003}, where an appropriate number of spots with circular shape and a defined
temperature is taken as input. An initial guess of the spot's location and radius was taken directly
from the observed Doppler images. With these starting values, we calculated the best fit with an
area- and temperature-weighted Monte Carlo (MC) method and thus extracted a definition-dependent,
best-effort spot location and area. Within the three-parameter space (longitude, latitude, radius)
10,000 random positions were generated, using a range of $15^{\circ}$ for each parameter, and then
cross correlated with the original Doppler map. From the best 100 correlation maps (which
corresponds to $\sqrt{N}$), the mean values and their standard deviations are determined. In the
following, spot area always refers to the spot area deduced from this analysis.

The area of the individual spots from the spot-model fits is summarized in
Table~\ref{tab:maps_log} in units of solar hemispheres (SH). To estimate the quality of the
spot-model fit, we compared the total spotted area between the spot model and the Doppler image.
To determine the total spotted area of the Doppler image, a temperature weighting by analogy to the
MC method was used. We obtain the total spotted area such that each spotted segment $i$ attributes
a certain fraction of its area $a_i$ to the total spotted area dependent on its temperature given by
\begin{equation}
A_{\rm total} = \sum_{i} a_i \frac{(T_{\rm eff} - T_i)}{(\Delta T)_{\rm max}} \ ,
\label{eqn:spot_area_orig_di}
\end{equation}
where $(\Delta T)_{\rm max}$\;=\;$(T_{\rm phot}$\,$-$\,$T_{\rm spot})_{\rm max}$\;=\;1120~K. The
small-scale surface structures, cool and hot spots, were not counted into the total spotted
area. The quality of the spot-model fits is given in Table~\ref{tab:maps_log} in terms of the
uncertainty of the spot models, where all our spot models lie within 1~$\sigma$. The absolute scale
of the area units in m$^2$ is set by the stellar radius of 10.9~R$_{\odot}$. Formally, the stellar
surface of XX~Tri is 724~Gm$^2$ and thus 118.8 times the surface of the Sun or
1~SH\;$\approx$\;0.8\,\%\ of an hemisphere of XX~Tri.

\begin{figure}[!t]
\begin{minipage}{1.0\textwidth}
\captionsetup[subfigure]{labelfont=bf,textfont=bf,singlelinecheck=off,justification=raggedright,
position=top}
\subfloat[]{\includegraphics[width=250pt]{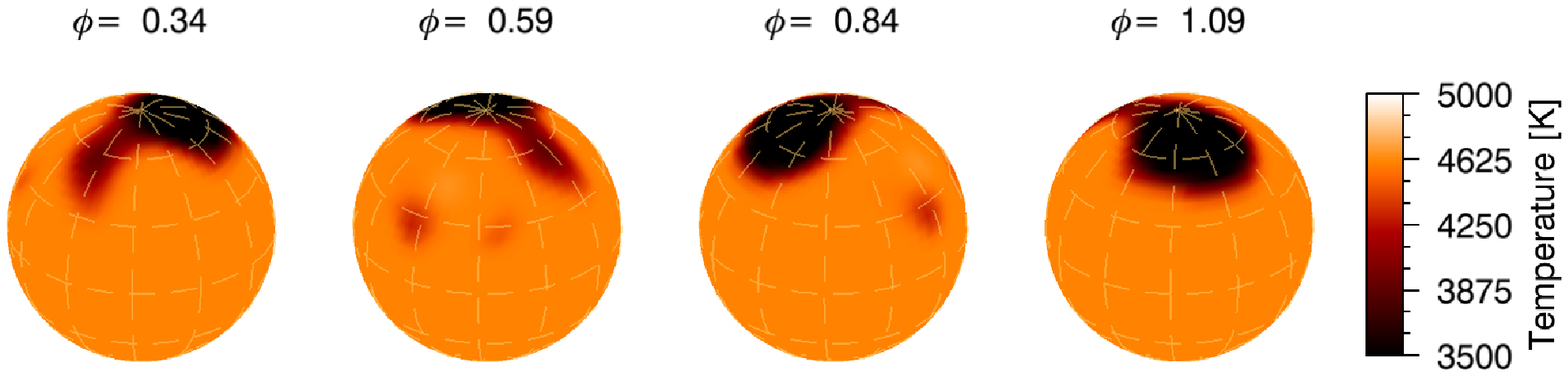}}
\end{minipage}\hspace{1ex}
\begin{minipage}{1.0\textwidth}
\captionsetup[subfigure]{labelfont=bf,textfont=bf,singlelinecheck=off,justification=raggedright,
position=top}
\subfloat[]{\includegraphics[width=250pt]{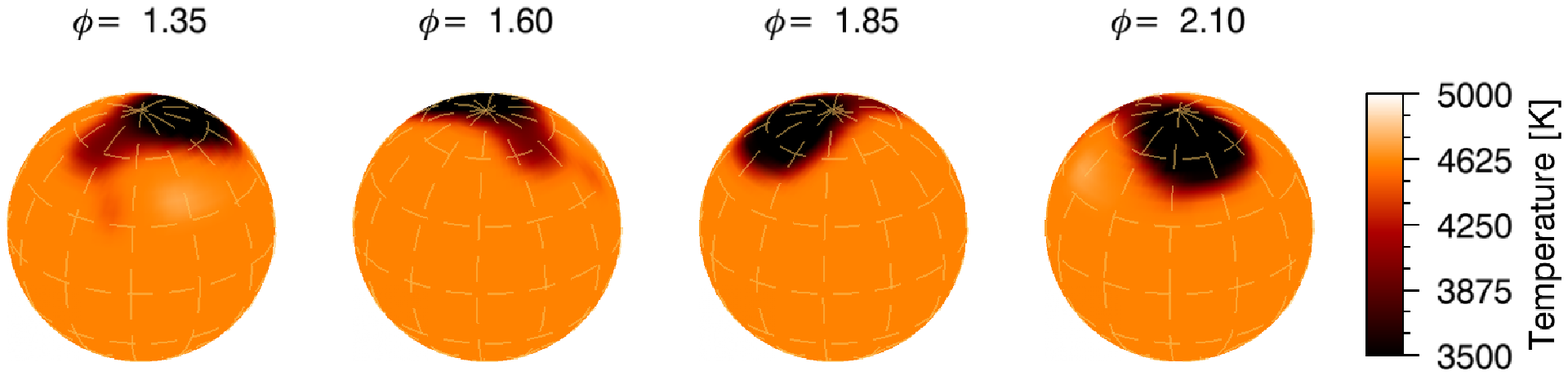}}
\end{minipage}\hspace{1ex}
\begin{minipage}{1.0\textwidth}
\captionsetup[subfigure]{labelfont=bf,textfont=bf,singlelinecheck=off,justification=raggedright,
position=top}
\subfloat[]{\includegraphics[width=250pt]{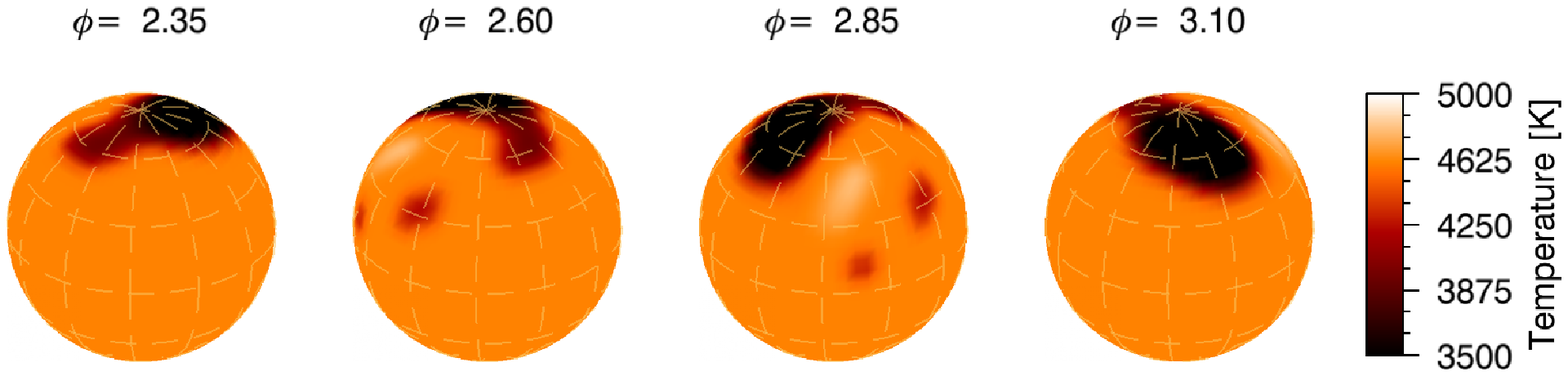}}
\end{minipage}\hspace{1ex}
\begin{minipage}{1.0\textwidth}
\captionsetup[subfigure]{labelfont=bf,textfont=bf,singlelinecheck=off,justification=raggedright,
position=top}
\subfloat[]{\includegraphics[width=250pt]{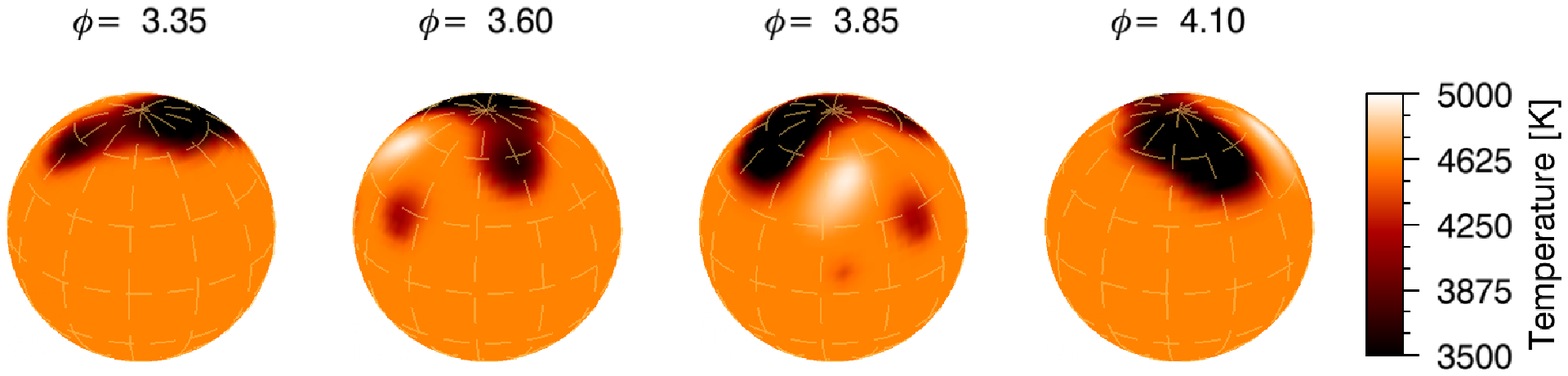}}
\end{minipage}\hspace{1ex}
\begin{minipage}{1.0\textwidth}
\captionsetup[subfigure]{labelfont=bf,textfont=bf,singlelinecheck=off,justification=raggedright,
position=top}
\subfloat[]{\includegraphics[width=250pt]{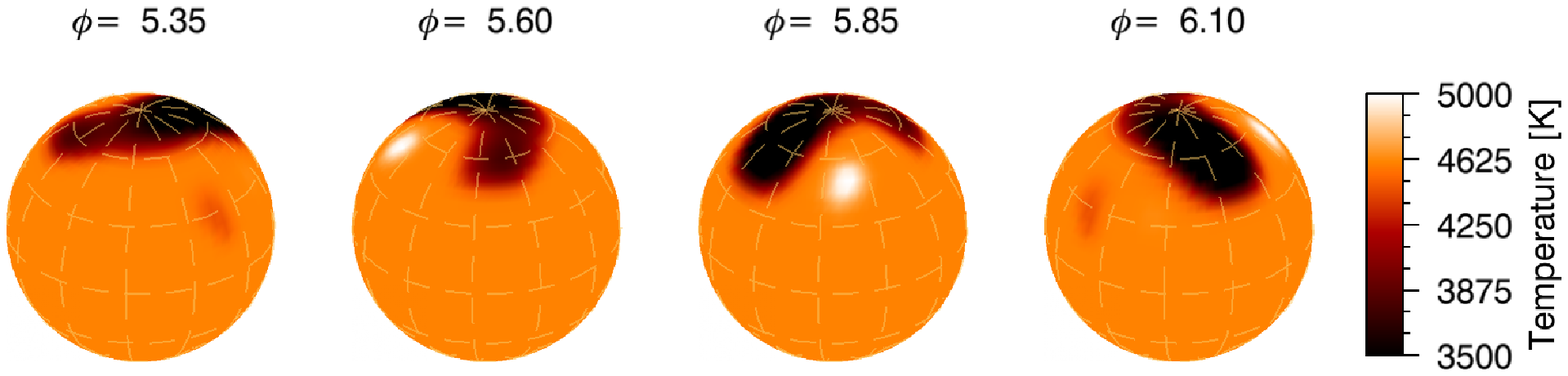}}
\end{minipage}\hspace{1ex}
\begin{minipage}{1.0\textwidth}
\captionsetup[subfigure]{labelfont=bf,textfont=bf,singlelinecheck=off,justification=raggedright,
position=top}
\subfloat[]{\includegraphics[width=250pt]{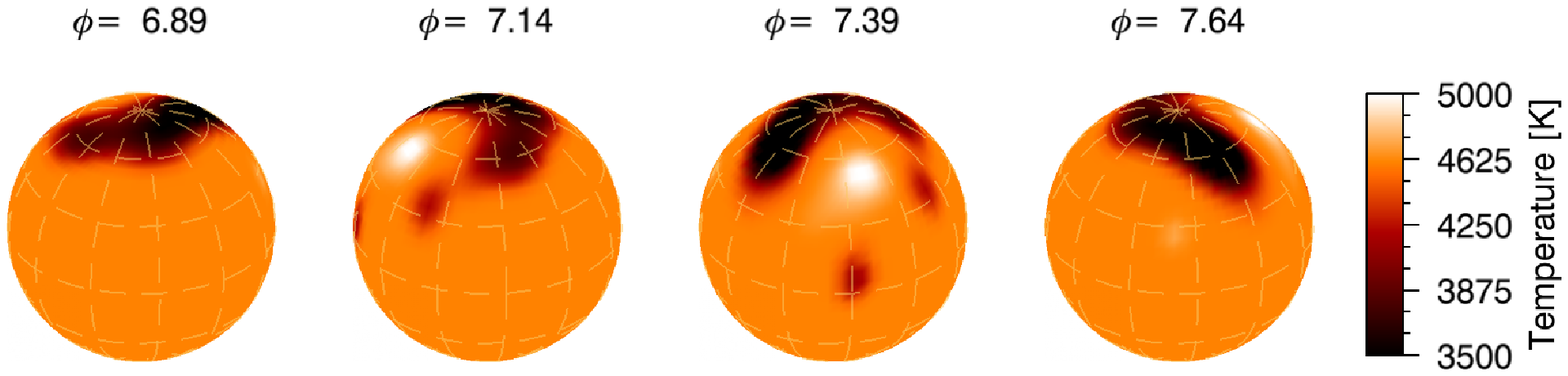}}
\end{minipage}\hspace{1ex}
\begin{minipage}{1.0\textwidth}
\captionsetup[subfigure]{labelfont=bf,textfont=bf,singlelinecheck=off,justification=raggedright,
position=top}
\subfloat[]{\includegraphics[width=250pt]{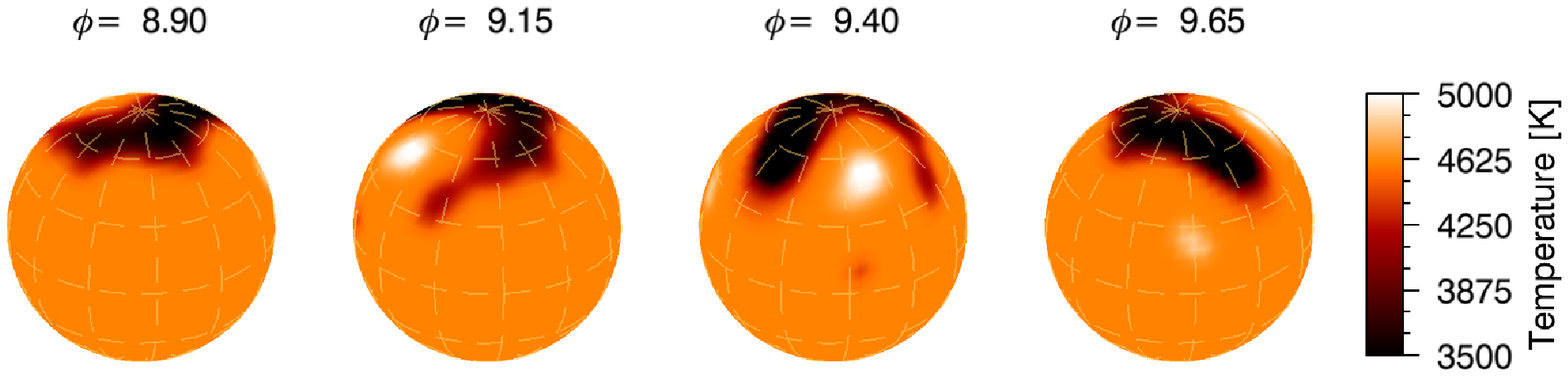}}
\end{minipage}\hspace{1ex}
\caption{Doppler images of XX~Tri for the observing season 2006/07. Each image is shown in four
spherical projections separated by $90^{\circ}$. The rotational shift between consecutive images is
corrected, i.e., the stellar orientation remains the same from map to map and from season to season.
The time difference between each Doppler image is indicated in units of rotational phase $\phi$.}
\label{fig:di_season_1}
\end{figure}

\begin{figure}[!t]
\begin{minipage}{1.0\textwidth}
\captionsetup[subfigure]{labelfont=bf,textfont=bf,singlelinecheck=off,justification=raggedright,
position=top}
\subfloat[]{\includegraphics[width=250pt]{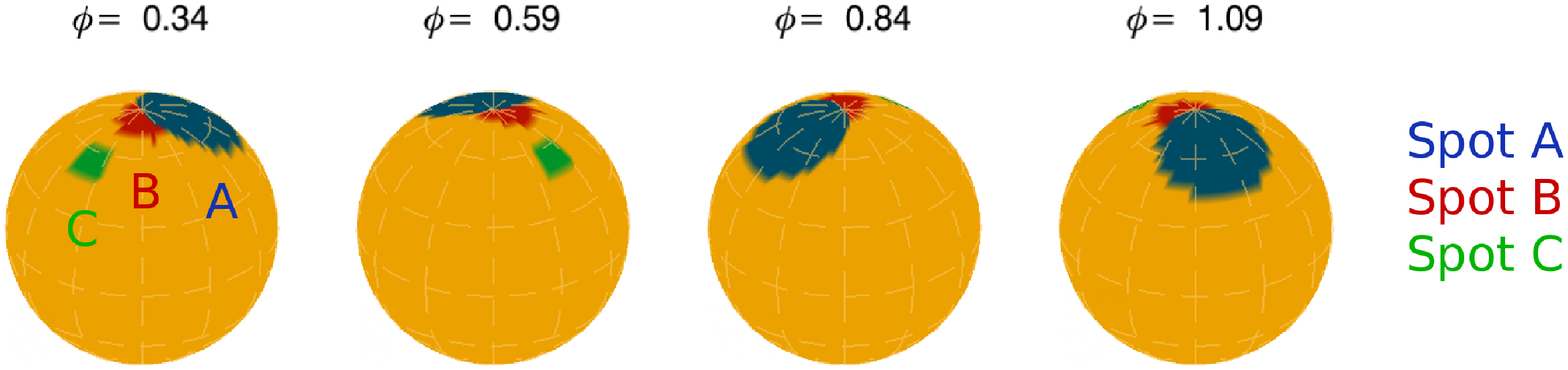}}
\end{minipage}\hspace{1ex}
\begin{minipage}{1.0\textwidth}
\captionsetup[subfigure]{labelfont=bf,textfont=bf,singlelinecheck=off,justification=raggedright,
position=top}
\subfloat[]{\includegraphics[width=250pt]{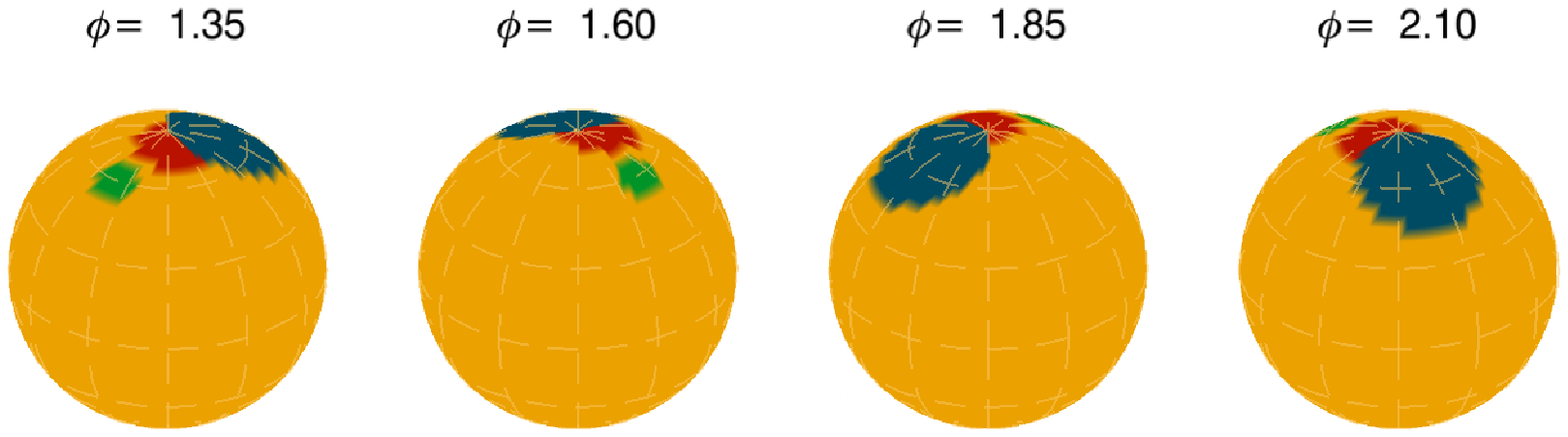}}
\end{minipage}\hspace{1ex}
\begin{minipage}{1.0\textwidth}
\captionsetup[subfigure]{labelfont=bf,textfont=bf,singlelinecheck=off,justification=raggedright,
position=top}
\subfloat[]{\includegraphics[width=250pt]{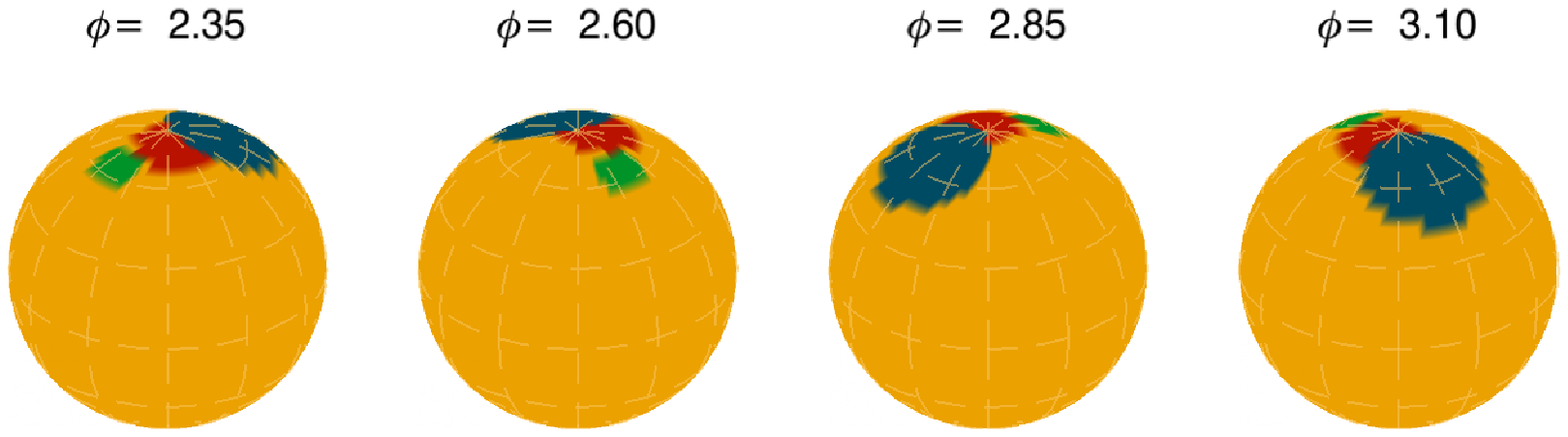}}
\end{minipage}\hspace{1ex}
\begin{minipage}{1.0\textwidth}
\captionsetup[subfigure]{labelfont=bf,textfont=bf,singlelinecheck=off,justification=raggedright,
position=top}
\subfloat[]{\includegraphics[width=250pt]{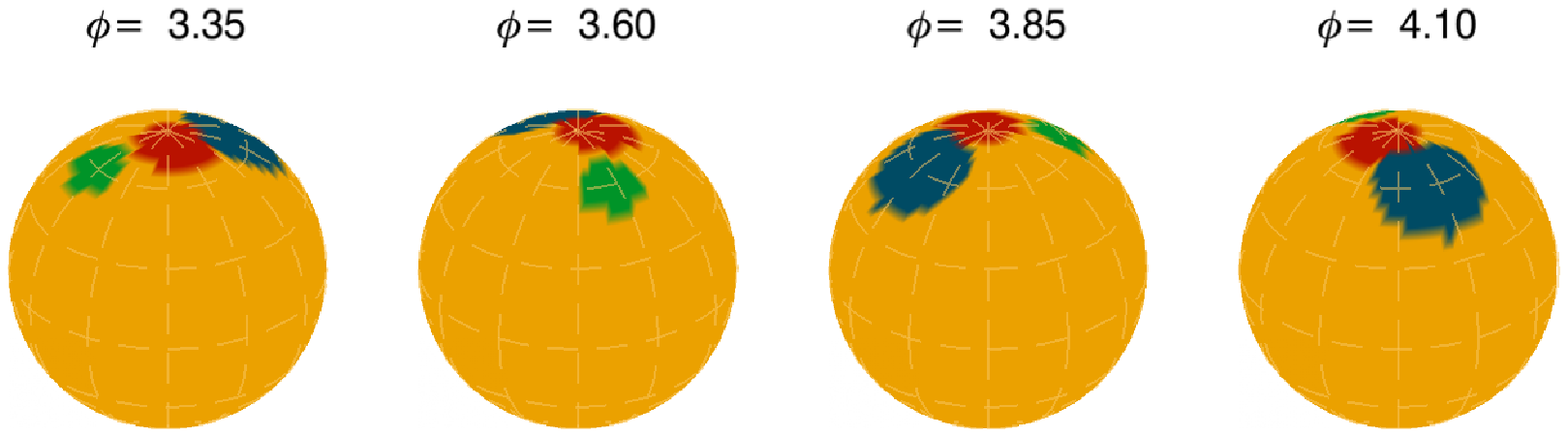}}
\end{minipage}\hspace{1ex}
\begin{minipage}{1.0\textwidth}
\captionsetup[subfigure]{labelfont=bf,textfont=bf,singlelinecheck=off,justification=raggedright,
position=top}
\subfloat[]{\includegraphics[width=250pt]{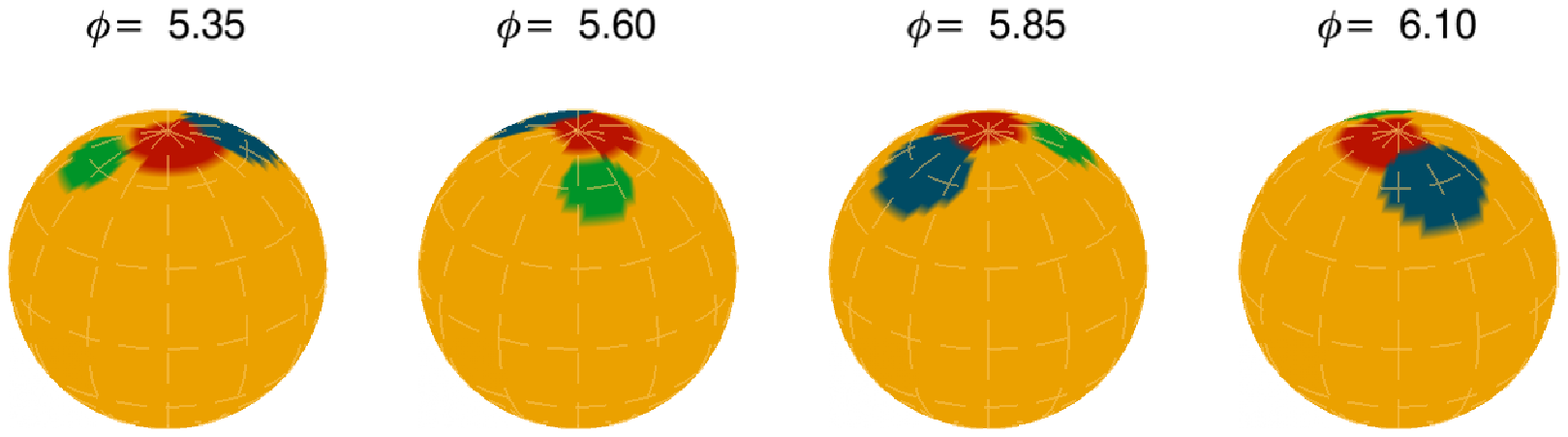}}
\end{minipage}\hspace{1ex}
\begin{minipage}{1.0\textwidth}
\captionsetup[subfigure]{labelfont=bf,textfont=bf,singlelinecheck=off,justification=raggedright,
position=top}
\subfloat[]{\includegraphics[width=250pt]{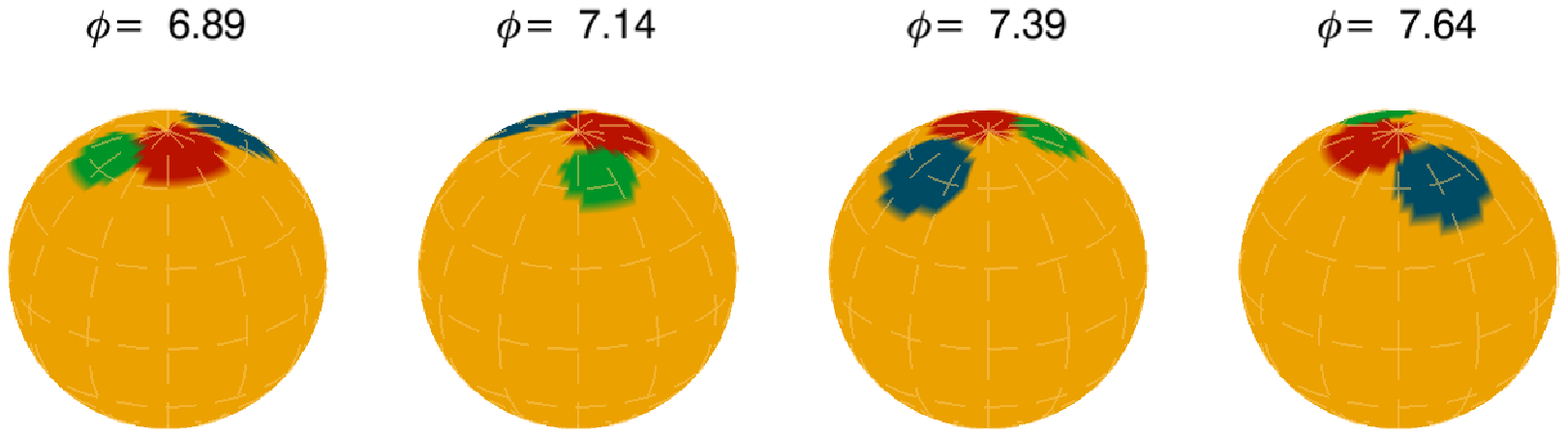}}
\end{minipage}\hspace{1ex}
\begin{minipage}{1.0\textwidth}
\captionsetup[subfigure]{labelfont=bf,textfont=bf,singlelinecheck=off,justification=raggedright,
position=top}
\subfloat[]{\includegraphics[width=250pt]{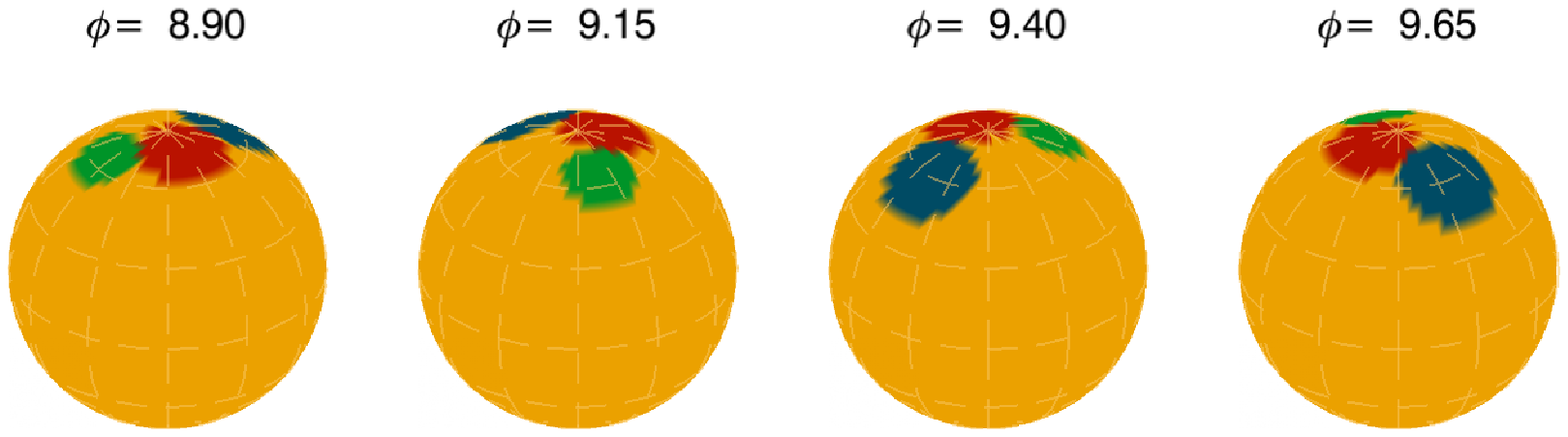}}
\end{minipage}\hspace{1ex}
\caption{Spot-model fits of the Doppler images in Fig.~\ref{fig:di_season_1}. Each spot is shown
with different color/contrast for better visualization.}
\label{fig:di_season_1_fit}
\end{figure}


\subsection{Doppler images and spot models}
\label{sec:doppler_maps}

Fig.~\ref{fig:di_season_1} is a representative figure for our results and shows all maps for the
observing season 2006/07. Seven consecutive Doppler images are reconstructed from a total of
$\approx$\;9.5 stellar rotations. All maps show a large polar spot with a temperature of
$\approx$\;3500~K. During this season, the polar spot drifted apart and changed its morphology from
almost circular to an elongated spot form. This drift could be a sign of differential rotation and
is investigated further in Section~\ref{sec:diff_rot}. Furthermore, a smaller high-latitude spot
with a temperature of around 3800~K is reconstructed. It is seen that the larger spot approached
the smaller spot. Because of the inherent technical limitations of Doppler imaging, the surface
resolution near the rotational pole is poor and thus spot separations not well constrained.
Therefore, one cannot say whether the large spot is a monolithic structure or being a conglomerate
of several smaller spots. Besides the polar spot, scattered small cool and/or hot spots are visible
at latitudes between 0-60$^{\circ}$ with absolute temperatures between 4200 to 5000~K.

\begin{landscape}
\centering
\begin{table}
\caption{Doppler image log, fit quality, surface temperature, and spot areas. Detailed information is
given in Section~\ref{sec:phasing}-\ref{sec:doppler_maps}.}
\label{tab:maps_log}
\begin{tabular}{l l l l l l l l l l l l l l l l l}
\hline\noalign{\smallskip}
DI & Year & HJD range & $\Delta t$ & $N$ & $\phi$-gap & $N$-fill\tablefootmark{a} &
RMS\tablefootmark{b} & RMS\tablefootmark{c} & \multicolumn{2}{c}{Temperature (K)} &
\multicolumn{6}{c}{Spot area (SH)\tablefootmark{d}} \\
\noalign{\smallskip}
\# &   & (2450000+) & (d) &   & (d) &   &   & ($\sigma_{\rm spot}$) & $T_{\rm max}$ & $T_{\rm mean}$
& Spot A & Spot B & Spot C & Spot D & Spot E & Spot Total \\
\noalign{\smallskip}\hline\noalign{\smallskip}
1  & 2006.58 & 3935-3963 & 28  & 15 & 4  & 2 & 0.0048 & 0.6 & 4667 & 4470 & $9.2\pm0.3$ &
$1.1\pm0.6$ & $1.3\pm0.8$ & --- & --- & $11.5\pm1.0$ \\
2  & 2006.64 & 3959-3980 & 21  & 19 & 3  & 0 & 0.0046 & 0.6 & 4731 & 4477 & $8.5\pm0.5$ &
$1.5\pm0.5$ & $1.2\pm0.7$ & --- & --- & $11.2\pm0.9$ \\
3  & 2006.71 & 3966-4015 & 49  & 14 & 5  & 3 & 0.0047 & 0.7 & 4796 & 4489 & $8.1\pm0.6$ &
$2.1\pm0.6$ & $1.6\pm0.9$ & --- & --- & $11.8\pm1.0$ \\
4  & 2006.78 & 4007-4050 & 43  & 18 & 7  & 3 & 0.0047 & 0.7 & 4961 & 4476 & $7.2\pm0.7$ &
$2.6\pm0.5$ & $2.4\pm0.7$ & --- & --- & $12.2\pm0.8$ \\
5  & 2006.91 & 4011-4108 & 97\tablefootmark{e}  & 17 & 5  & 3 & 0.0047 & 0.8 & 5059 & 4477 &
$6.0\pm0.5$ & $3.3\pm0.5$ & $3.0\pm1.1$ & --- & --- & $12.3\pm1.1$ \\
6  & 2007.01 & 4092-4138 & 46  & 16 & 5  & 2 & 0.0044 & 0.5 & 5052 & 4492 & $5.4\pm0.6$ &
$3.7\pm0.6$ & $3.1\pm1.2$ & --- & --- & $12.2\pm1.3$ \\
7  & 2007.14 & 4124-4161 & 37  & 19 & 5  & 2 & 0.0043 & 0.7 & 5125 & 4486 & $5.4\pm0.7$ &
$3.9\pm0.6$ & $3.2\pm1.1$ & --- & --- & $12.5\pm1.3$ \\
\noalign{\smallskip}\hline\noalign{\smallskip}
8  & 2007.67 & 4333-4356 & 23  & 21 & 3  & 0 & 0.0040 & 0.4 & 5129 & 4541 & $5.5\pm0.5$ &
$1.6\pm0.4$ & ---         & --- & --- &  $7.1\pm0.8$ \\
9  & 2007.73 & 4357-4380 & 23  & 21 & 2  & 0 & 0.0035 & 0.4 & 5027 & 4537 & $5.2\pm0.4$ &
$1.8\pm0.7$ & ---         & --- & --- &  $7.0\pm0.9$ \\
10 & 2007.87 & 4362-4475 & 113\tablefootmark{e} & 15 & 13 & 8 & 0.0033 & 0.7 & 4912 & 4538 &
$4.1\pm0.7$ & $2.6\pm0.3$ & ---         & --- & --- &  $6.7\pm0.7$ \\
11 & 2008.05 & 4468-4512 & 44  & 19 & 4  & 1 & 0.0037 & 0.6 & 4685 & 4534 & $4.3\pm0.7$ &
$2.5\pm0.4$ & ---         & --- & --- &  $6.8\pm0.7$ \\
12 & 2008.12 & 4481-4516 & 35  & 15 & 5  & 2 & 0.0038 & 0.6 & 5188 & 4524 & $4.1\pm0.5$ &
$2.6\pm0.5$ & ---         & --- & --- &  $6.6\pm0.4$ \\
\noalign{\smallskip}\hline\noalign{\smallskip}
13 & 2008.53 & 4649-4671 & 22  & 20 & 2  & 0 & 0.0040 & 0.5 & 4903 & 4503 & $4.8\pm0.2$ &
$3.2\pm0.3$ & ---         & --- & --- &  $8.0\pm0.4$ \\
14 & 2008.60 & 4673-4696 & 23  & 22 & 3  & 0 & 0.0037 & 0.5 & 4839 & 4501 & $5.5\pm0.3$ &
$3.0\pm0.4$ & ---         & --- & --- &  $8.5\pm0.5$ \\
15 & 2008.73 & 4722-4744 & 22  & 14 & 5  & 0 & 0.0038 & 0.7 & 4845 & 4501 & $6.4\pm0.4$ &
$2.0\pm0.2$ & ---         & --- & --- &  $8.4\pm0.4$ \\
16 & 2008.80 & 4726-4790 & 64  & 15 & 7  & 5 & 0.0034 & 0.5 & 4745 & 4497 & $7.5\pm0.5$ &
$1.6\pm0.4$ & ---         & --- & --- &  $9.1\pm0.6$ \\
17 & 2008.88 & 4775-4798 & 23  & 16 & 3  & 0 & 0.0035 & 0.4 & 5057 & 4503 & $8.0\pm0.7$ &
$0.7\pm0.2$ & ---         & --- & --- &  $8.7\pm0.8$ \\
18 & 2008.94 & 4799-4822 & 23  & 15 & 4  & 0 & 0.0032 & 0.3 & 5004 & 4508 & $8.1\pm0.6$ &
$0.6\pm0.3$ & ---         & --- & --- &  $8.8\pm0.7$ \\
19 & 2009.07 & 4844-4873 & 29  & 16 & 5  & 2 & 0.0035 & 0.9 & 5193 & 4493 & $8.2\pm0.4$ &
$1.2\pm0.3$ & ---         & --- & --- &  $9.4\pm0.5$ \\
\noalign{\smallskip}\hline\noalign{\smallskip}
20 & 2009.60 & 5039-5066 & 27  & 18 & 5  & 1 & 0.0033 & 0.3 & 4918 & 4522 & $7.8\pm1.2$ &
$2.9\pm0.4$ & ---         & --- & --- & $10.7\pm1.3$ \\
21 & 2009.67 & 5039-5083 & 44  & 17 & 6  & 2 & 0.0035 & 0.7 & 4825 & 4500 & $8.1\pm1.0$ &
$3.3\pm0.4$ & ---         & --- & --- & $11.3\pm1.1$ \\
22 & 2009.86 & 5135-5161 & 26  & 17 & 4  & 1 & 0.0039 & 0.9 & 5061 & 4495 & $8.3\pm1.1$ &
$4.7\pm0.5$ & ---         & --- & --- & $13.0\pm1.1$ \\
23 & 2009.93 & 5154-5173 & 19  & 16 & 11 & 3 & 0.0036 & 0.4 & 5001 & 4511 & $7.4\pm1.1$ &
$4.8\pm0.3$ & ---         & --- & --- & $12.2\pm1.1$ \\
24 & 2010.06 & 5207-5224 & 17  & 17 & 7  & 0 & 0.0035 & 0.6 & 5146 & 4506 & $6.8\pm0.7$ &
$4.9\pm0.6$ & ---         & --- & --- & $11.6\pm1.0$ \\
\noalign{\smallskip}\hline\noalign{\smallskip}
25 & 2010.59 & 5401-5447 & 46  & 12 & 6  & 4 & 0.0029 & 0.6 & 4935 & 4535 & $4.5\pm0.7$ &
$3.3\pm0.3$ & $0.9\pm0.6$ & --- & --- &  $8.7\pm0.9$ \\
26 & 2010.67 & 5429-5452 & 23  & 19 & 2  & 0 & 0.0031 & 0.4 & 5088 & 4538 & $4.0\pm0.7$ &
$3.1\pm0.4$ & $1.1\pm0.6$ & --- & --- &  $8.1\pm0.9$ \\
27 & 2010.74 & 5439-5496 & 57  & 13 & 7  & 3 & 0.0029 & 0.9 & 4967 & 4532 & $3.2\pm0.4$ &
$2.4\pm0.6$ & $1.5\pm0.7$ & --- & --- &  $7.2\pm0.9$ \\
28 & 2010.84 & 5467-5512 & 45  & 17 & 7  & 2 & 0.0030 & 0.9 & 4943 & 4522 & $2.6\pm0.4$ &
$1.5\pm0.6$ & $2.4\pm0.6$ & --- & --- &  $6.5\pm0.7$ \\
29 & 2011.12 & 5595-5614 & 19  & 11 & 5  & 0 & 0.0021 & 0.5 & 4620 & 4513 & ---         & ---
  & ---         & $5.1\pm0.5$ & $2.3\pm0.4$ & $7.4\pm0.5$ \\
\noalign{\smallskip}\hline\noalign{\smallskip}
30 & 2011.61 & 5774-5797 & 23  & 18 & 3  & 0 & 0.0033 & 0.3 & 4758 & 4493 & $4.7\pm0.6$ &
$2.9\pm0.4$ & $1.2\pm0.6$ & --- & --- &  $8.8\pm0.7$ \\
31 & 2011.68 & 5799-5822 & 23  & 24 & 1  & 0 & 0.0032 & 0.4 & 4756 & 4501 & $4.5\pm0.7$ &
$2.5\pm0.8$ & $1.7\pm0.6$ & --- & --- &  $8.8\pm0.8$ \\
32 & 2011.81 & 5830-5879 & 49  & 17 & 7  & 2 & 0.0029 & 0.8 & 4649 & 4509 & $3.7\pm0.7$ &
$1.4\pm0.7$ & $3.1\pm0.4$ & --- & --- &  $8.2\pm1.0$ \\
33 & 2011.88 & 5871-5894 & 23  & 16 & 4  & 0 & 0.0025 & 0.9 & 4722 & 4501 & $3.0\pm0.5$ &
$2.7\pm0.8$ & $3.4\pm0.3$ & --- & --- &  $9.1\pm0.8$ \\
34 & 2012.01 & 5919-5940 & 21  & 20 & 3  & 0 & 0.0027 & 0.4 & 4664 & 4486 & $3.6\pm0.7$ &
$3.4\pm0.9$ & $2.6\pm0.4$ & --- & --- &  $9.6\pm1.0$ \\
35 & 2012.08 & 5944-5967 & 23  & 18 & 3  & 0 & 0.0026 & 0.7 & 4620 & 4506 & $3.3\pm0.7$ &
$4.2\pm0.9$ & $1.7\pm0.4$ & --- & --- &  $9.2\pm1.1$ \\
36 & 2012.19 & 5983-6006 & 23  & 17 & 3  & 0 & 0.0031 & 0.3 & 4767 & 4477 & $3.6\pm0.7$ &
$5.8\pm0.6$ & $0.6\pm0.4$ & --- & --- & $10.0\pm0.8$ \\
\noalign{\smallskip}\hline
\end{tabular}
\tablefoot{\tablefoottext{a}{Number of phases/observations borrowed from the following and/or
preceding stellar rotation.}\tablefoottext{b}{Deviation between observed and calculated line
profiles.}\tablefoottext{c}{Deviation of spot area between spot models and Doppler image normalized
by the standard deviation of the total spotted area.}\tablefoottext{d}{In units of solar hemispheres
(SH).}\tablefoottext{e}{Including observations from \emph{two} stellar rotations earlier and/or
later.}}
\end{table}
\end{landscape}

Comparing our new Doppler images with the first image published by \cite{strassmeier1999} shows a
very good agreement of spot locations and spot temperatures, despite $\approx$\;10-year time
difference. In both cases, the largest spot appears near the pole with a temperature of around
3500~K. Smaller cool/hot spots appear on lower latitudes, again with comparable temperature
differences. The super spot from 1998 with an area of approximately 11\,\%\ of the entire stellar
surface is much larger than the largest recovered spots that we present. This is expected
because the star appeared to be in its brightest stage ever in 2009. In parallel, the
rotationally-modulated photometric light curves show a small amplitude compared to that in 1998.
During the season 1998/99 a record amplitude of $\approx$\;$0\fm6$ in $V$ was observed in comparison
to an amplitude of $<$\;$0\fm4$ during the seasons from 2006/07 to 2011/12 \citep[see][]{olah2014}.

In Fig.~\ref{fig:di_season_1_fit} the best spot-model fits for observing season 2006/07 are shown.
Three artificial spots were sufficient to reach a high correlation for all Doppler images in this
season. Two spots were utilized to represent the large polar spot and to model its drift during
almost ten stellar rotations. The third spot represented the smaller high-latitude spot on the
opposite hemisphere. All other cool and/or hot spots at lower latitudes with no or only very short
continuous appearance were ignored and therefore not implemented in the spot-model analysis.

Doppler images and spot models for the observing seasons 2007-12 are shown in
Fig.~\ref{fig:di_season_2}-\ref{fig:di_season_6_fit}. The observed and inverted line profiles of
each Doppler image are given in Fig.~\ref{fig:stokes_profiles_1} and \ref{fig:stokes_profiles_2}.
Table~\ref{tab:maps_log} lists the fit quality between observed ($x_{i,O}$) and calculated
($x_{i,C}$) profiles,
\begin{equation}
RMS = \sqrt{\frac{1}{n}\sum_{i=1}^{n}(x_{i,O}-x_{i,C})^2} \ ,
\label{eqn:rms_obs_calc}
\end{equation}
where $n$ is the total number of points of all line profiles used during the inversion.
Table~\ref{tab:maps_log} also tabulates the maximum and mean surface temperatures of each Doppler
image.


\section{Results}
\label{sec:results}

\subsection{Spot area evolution}
\label{sec:spot_evol}

\begin{figure*}[!t]
\begin{minipage}{1.0\textwidth}
\captionsetup[subfigure]{labelfont=bf,textfont=bf,singlelinecheck=off,justification=raggedright,
position=top}
\subfloat[]{\includegraphics[width=175pt]{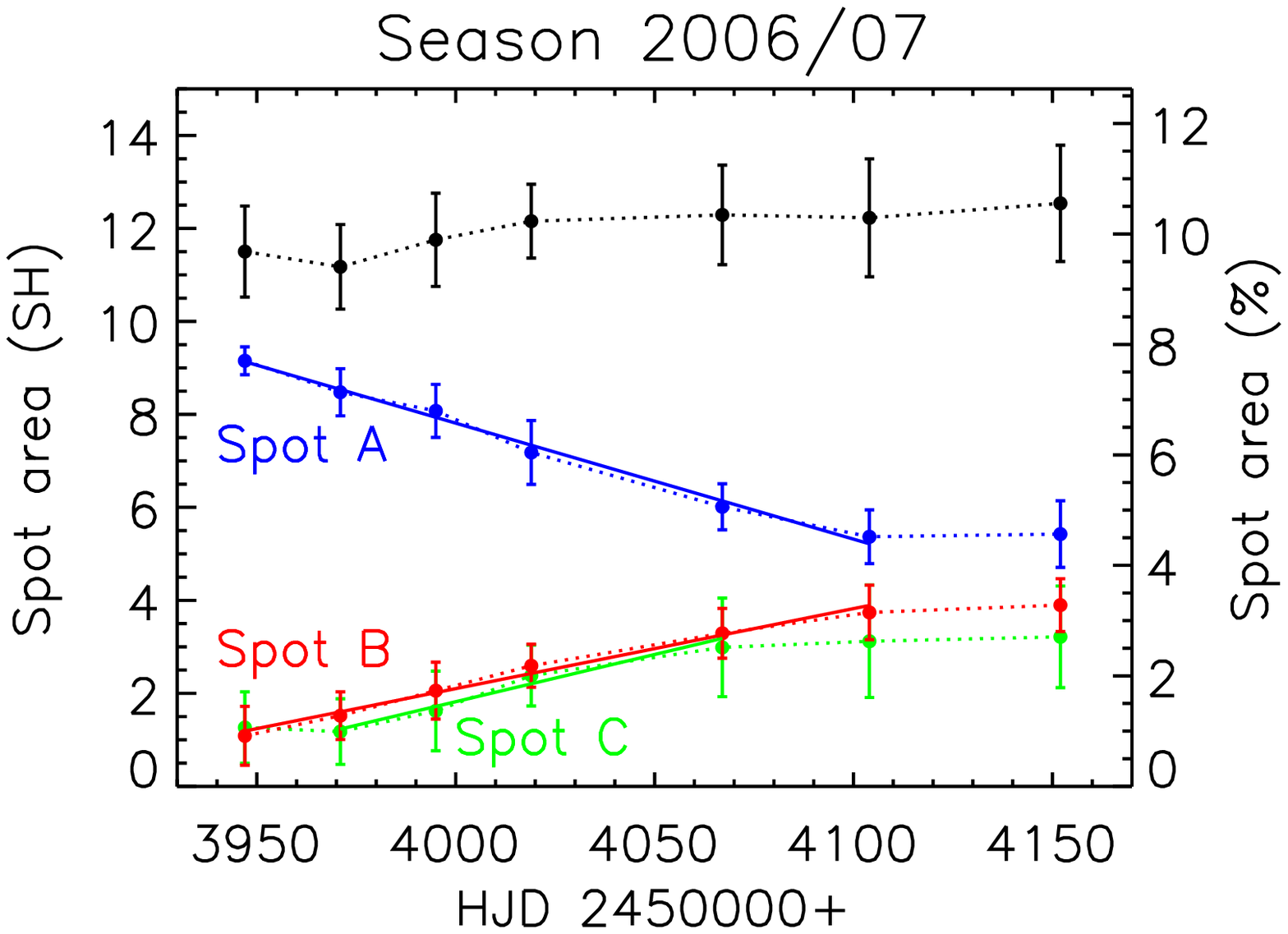}}
\subfloat[]{\includegraphics[width=175pt]{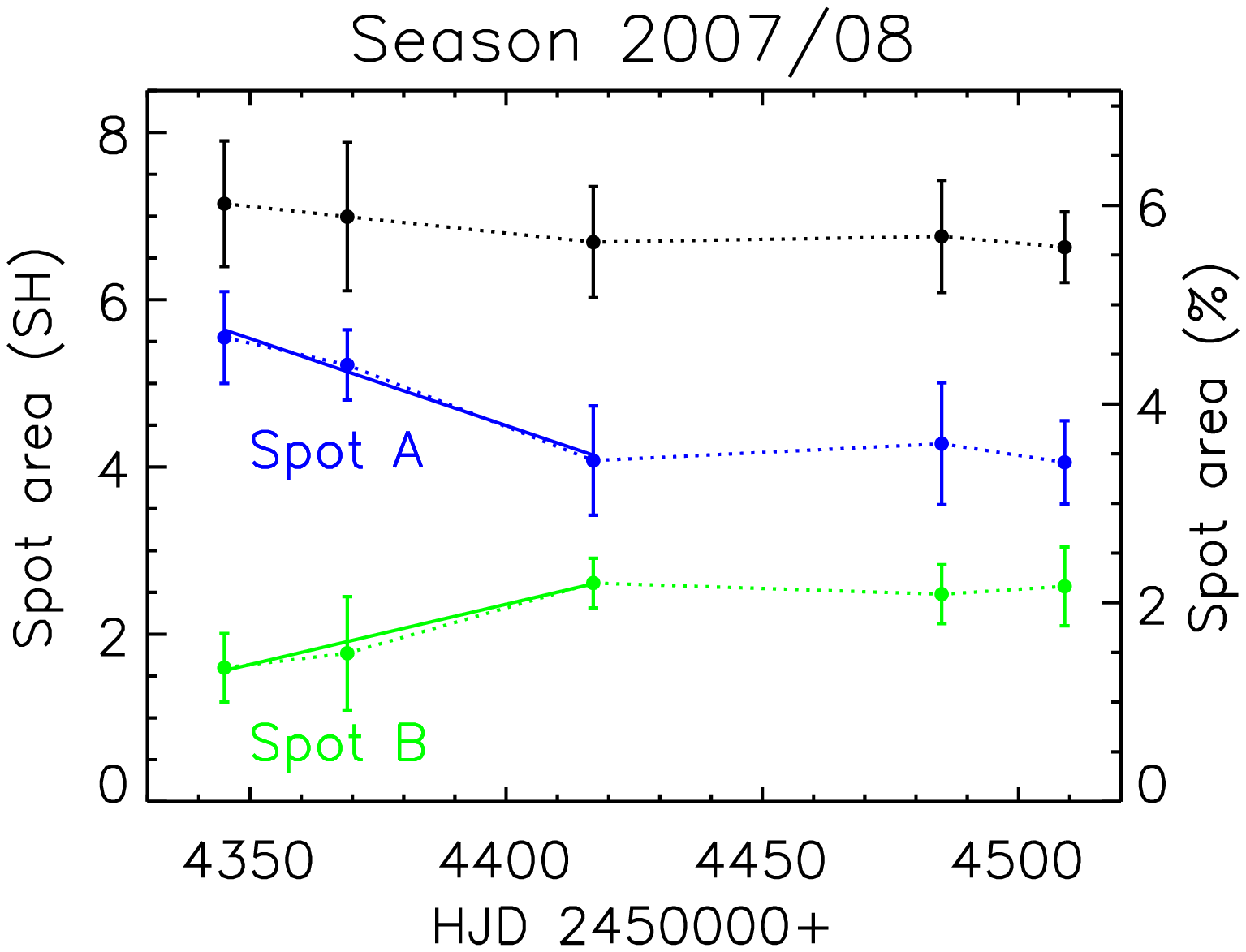}}
\subfloat[]{\includegraphics[width=175pt]{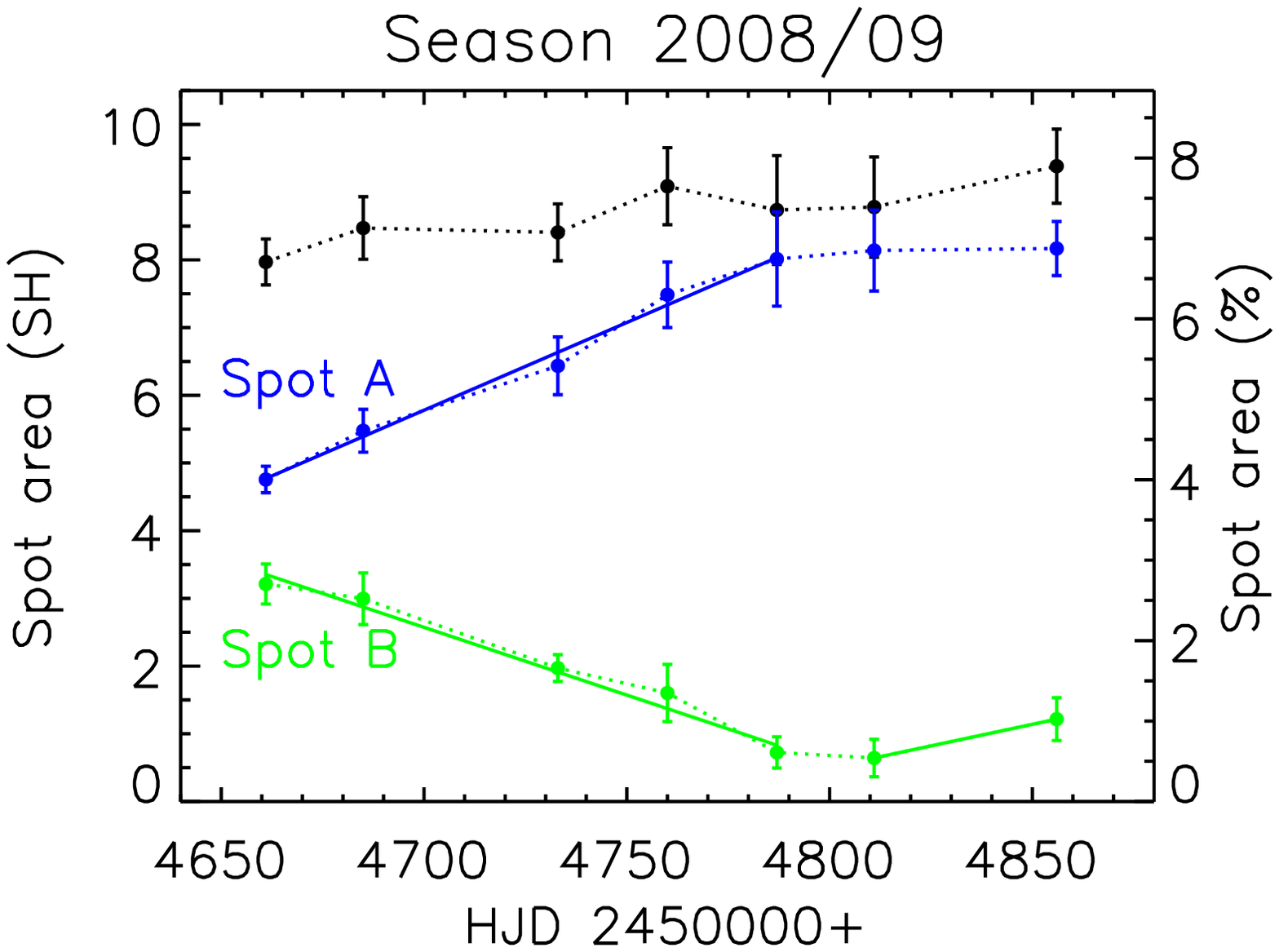}}
\end{minipage}\hspace{1ex}
\begin{minipage}{1.0\textwidth}
\captionsetup[subfigure]{labelfont=bf,textfont=bf,singlelinecheck=off,justification=raggedright,
position=top}
\subfloat[]{\includegraphics[width=175pt]{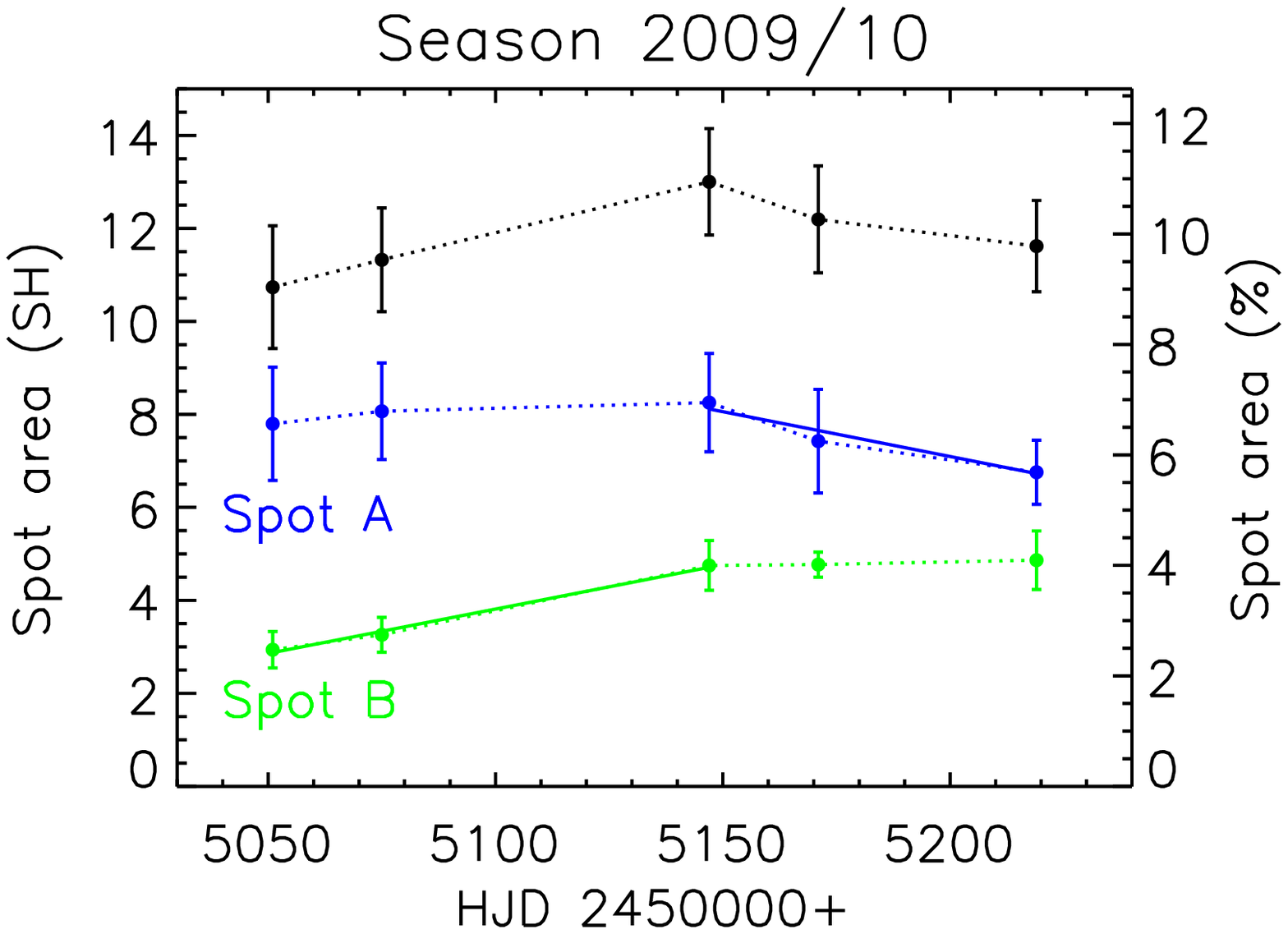}}
\subfloat[]{\includegraphics[width=175pt]{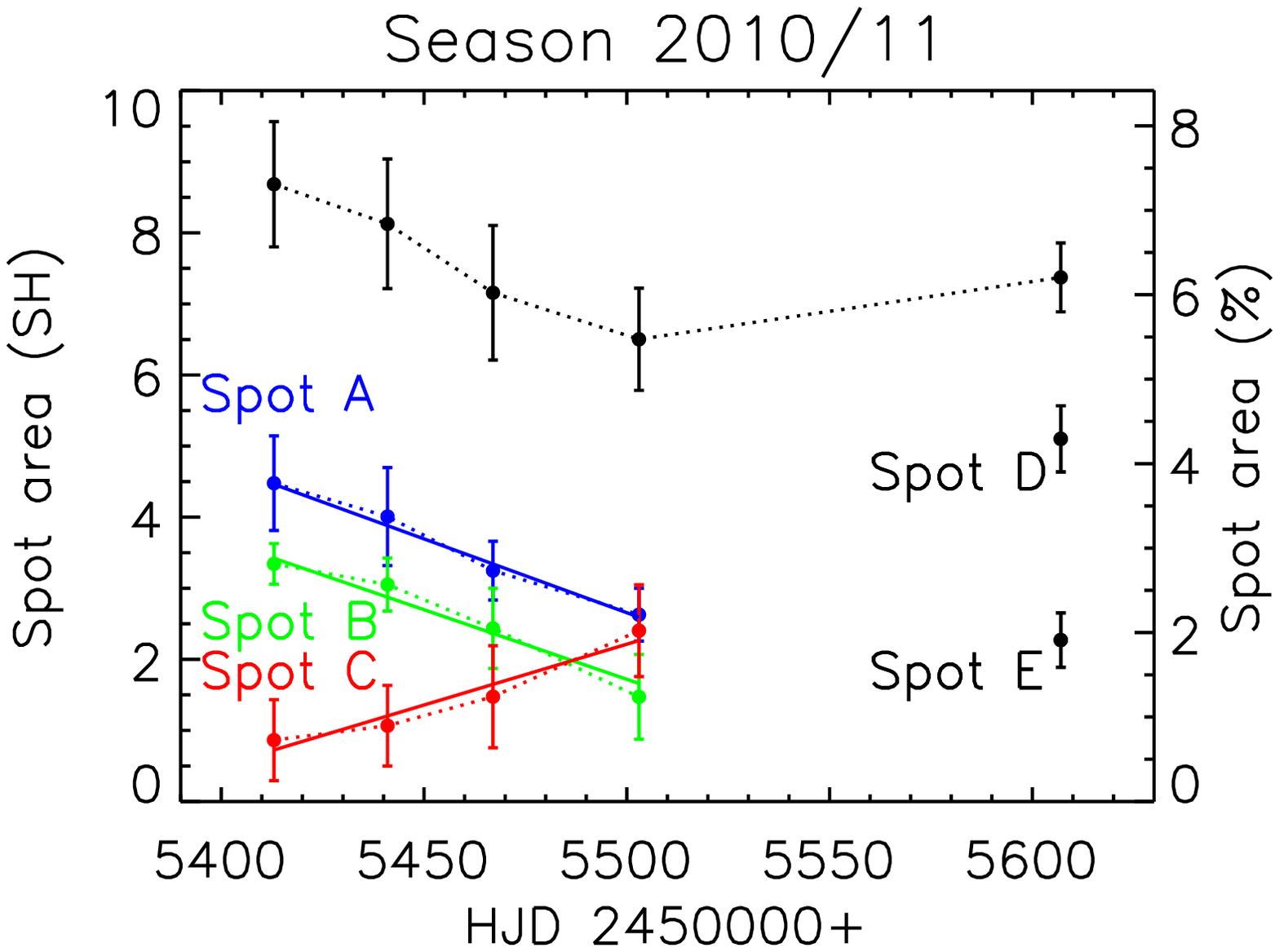}}
\subfloat[]{\includegraphics[width=175pt]{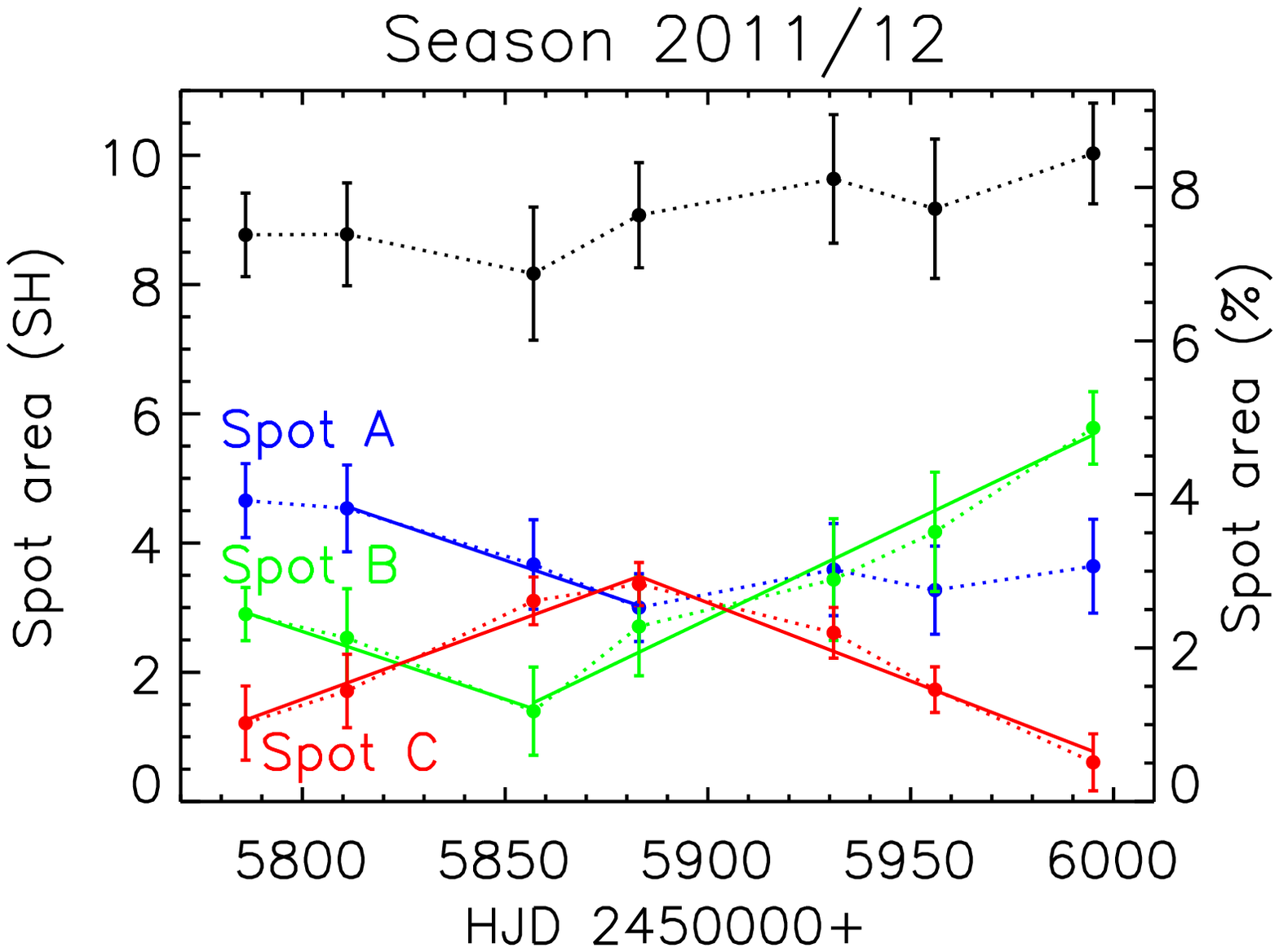}}
\end{minipage}\hspace{1ex}
\caption{Spot area evolution on XX~Tri from 2006 to 2012. Shown are the seasonal evolution of
the individual spots (dotted colored lines) for each observing season. The solid (colored)
lines represent linear fits to the decay or growth of a spot. The black dotted line represents the
total spotted area. The spot area is given in solar hemispheres on the left axis
(1~SH\;=\;3.05~$\mathrm{Gm^2}$) and relative to the total area of a stellar hemisphere of XX~Tri on
the right axis.}
\label{fig:spot_area_season_all}
\end{figure*}

In each observing season, we see both decay and growth of individual spots. From these spots we
infer a linear area decay/formation law of form
\begin{equation}
dA(t)/dt = D \ .
\end{equation}
As known from sunspot-decay studies \citep[e.g.,][]{martinez2002}, the application of a linear law
is the most appropriate way to describe sunspot decay as well as sunspot growth rates.
Fig.~\ref{fig:spot_area_season_all} shows the area evolution of the total spotted area of XX~Tri as
well as of the individual spots from the spot models for each season. The numerical values of $D$
for each spot together with the overall mean are summarized in Table~\ref{tab:spot_decay_par}.

\emph{Season 2006/07.} Three artificial spots are sufficient to match the observed spot distribution
in this season. The three spots appear partly merged and make up for the elongated polar-spot
appendage in the Doppler images. In the case of overlapping spots, the overlapped area refers to the
larger spot. The two overlapping spots, A and B, fragment, while the larger spot A shrinks and the
smaller spot B waxes. Between DI~\#1-6 ($\Delta t$\;$\approx$\;6.5~$P_{\rm rot}$) spot A loses
around 40\,\% of its area from 9.2 to 5.4~SH ($D$\;=\;$-$0.025\,$\pm$\,0.003~SH/day), while the
smaller spot B increases to more than three times its area from 1.1 to 3.7~SH
($D$\;=\;+0.017\,$\pm$\,0.004~SH/day). It seems that spot A ``feeds'' spot B, suggesting flux
transport between these two spots. Between DI~\#6-7 both spots A and B remain almost constant in
area. Spot C is located on the opposite hemisphere with a longitudinal shift of around
$180^{\circ}$. Between DI~\#2-5 ($\Delta t$\;$\approx$\;5.5~$P_{\rm rot}$) it waxes to almost three
times its area from 1.2 to 3.0~SH ($D$\;=\;+0.020\,$\pm$\,0.012~SH/day). Afterwards, spot C
remains almost constant in area. If we were to exclude DI~\#5 (see Sect.~\ref{sec:phasing}), the
determined values of $D$ would be $-$0.024\,$\pm$\,0.004~SH/day for spot A,
+0.017\,$\pm$\,0.005~SH/day for spot B, and +0.016\,$\pm$\,0.010~SH/day for spot C, respectively.

\emph{Season 2007/08.} We concentrate on the two larger spots (at phases around 17\fp25 and 17\fp55)
for a meaningful analysis of spot area evolution. Both spots A and B are separated by around a
quarter of rotation in DI~\#8. During the observing season they merge, as the larger spot A rotates
much slower near the pole than the smaller spot B, which is located at mid-latitudes. This indicates
that differential rotation is detectable as shown in Sect.~\ref{sec:diff_rot}. Between DI~\#8-10
($\Delta t$\;=\;3~$P_{\rm rot}$) spot A loses around 25\,\% of its area from 5.5 to 4.1~SH
($D$\;=\;$-$0.021\,$\pm$\,0.012~SH/day), while spot B increases to almost two times its area from 1.6
to 2.6~SH ($D$\;=\;+0.014\,$\pm$\,0.007~SH/day). This phenomenon of two spots interacting with each
other, while one spot is decaying and the other is growing, has also been detected in the previous
season. If we were to exclude DI~\#10 (see Sect.~\ref{sec:phasing}), a reliable determination of $D$
for this season would not have been possible because of the time sampling of the Doppler images.
However, removing it from the entire time series has no impact on the mean decay rate.

\emph{Season 2008/09.} Two spots are sufficient to characterize the spot evolution during this
season. The two spots, A and B, are very close and appear connected to each other as seen in DI~\#13.
The large spot A increases from 4.8 to 8.0~SH ($D$\;=\;+0.026\,$\pm$\,0.004~SH/day) between
DI~\#13-17 ($\Delta t$\;$\approx$\;5~$P_{\rm rot}$) and afterwards remains almost constant in area.
Within the same time span, the smaller spot B loses in area from 3.2 to 0.7~SH
($D$\;=\;$-$0.020\,$\pm$\,0.003~SH/day). After an almost complete decay, it starts to increase to two
times its area from 0.6 to 1.2~SH ($D$\;=\;+0.013\,$\pm$\,0.009~SH/day) between DI~\#18-19 ($\Delta
t$\;$\approx$\;2~$P_{\rm rot}$). As in the previous season, indications of differential rotation are
seen. Spot A is located nearer to the pole than spot B and therefore both spots get separated from
each other.

\emph{Season 2009/10.} Again, two spots are adequate to match the observed spot distribution
throughout this season. Both spots A and B are located at opposite hemispheres with a longitudinal
shift of around $180^{\circ}$. Between DI~\#20-22 ($\Delta t$\;=\;4~$P_{\rm rot}$) the smaller spot
B waxes to around 160\,\% its area from 2.9 to 4.7~SH ($D$\;=\;+0.019\,$\pm$\,0.007~SH/day),
whereas the larger spot A remains almost constant in area. Between DI~\#22-24 ($\Delta
t$\;=\;3~$P_{\rm rot}$) the larger spot A loses in area from 8.3 to 6.8~SH
($D$\;=\;$-$0.019\,$\pm$\,0.016~SH/day), whereas the smaller spot B remains almost constant in area.

\emph{Season 2010/11.} During this season three spots are required to characterize the spot
evolution. The two large polar spots, A and B, are located at opposite hemispheres with a longitudinal
shift of around $150^{\circ}$ and are moving toward each other. Between DI~\#25-28 ($\Delta
t$\;$\approx$\;4~$P_{\rm rot}$) both spots A and B lose in area from 4.5 to 2.6~SH
($D$\;=\;$-$0.021\,$\pm$\,0.008~SH/day) and 3.3 to 1.5~SH ($D$\;=\;$-$0.020\,$\pm$\,0.007~SH/day),
respectively. During the same time span, the smaller spot C increases from 0.9 to 2.4~SH
($D$\;=\;+0.017\,$\pm$\,0.009~SH/day) and moves toward higher latitudes. There is a large time gap
of around four rotations between DI~\#28-29. During this time span, the spot configuration obviously
changes to a two-spot-model. As it is not clear whether one can identify these spots with spots
from DI~\#25-28, we decide to regard them separately. These two polar spots, D and E, are located at
opposite hemispheres with a longitudinal shift of around $180^{\circ}$. This spot configuration is
almost identical with the first DI (\#30) of the following observational season.

\emph{Season 2011/12.} Again, three spots are sufficient to match the observed large-scale spot
distribution during this season. As in the first season, the three spots appear partly merged and
make up for the elongated polar-spot appendage in the Doppler images. The two spots, A and B, are
located at opposite hemispheres with a longitudinal shift of around $180^{\circ}$. This spot
distribution is very similar to that at the end of the previous season. The larger spot A fragments
into two smaller spots, A and C, between DI~\#30-32. Spot C is located very close to the pole
and rotates much slower than the other two spots A and B. Between the time of DI~\#33-36 spot C
merges with spot B. Therefore, it again suggests flux transport between spot A and B. Spot A
loses in area from 4.5 to 3.0~SH ($D$\;=\;$-$0.021\,$\pm$\,0.012~SH/day) between DI~\#31-33
($\Delta t$\;$\approx$\;3~$P_{\rm rot}$) and remains almost constant in size afterwards. Spot B
loses in area from 2.9 to 1.4~SH ($D$\;=\;$-$0.021\,$\pm$\,0.011~SH/day) between DI~\#30-32 ($\Delta
t$\;=\;3~$P_{\rm rot}$) and waxes between DI~\#32-36 ($\Delta t$\;$\approx$\;6~$P_{\rm rot}$) up to
5.8~SH ($D$\;=\;+0.030\,$\pm$\,0.006~SH/day). Spot C increases in area up to 3.4~SH
($D$\;=\;+0.023\,$\pm$\,0.006~SH/day) between DI~\#30-33 and decays between DI~\#33-36
($\Delta t$\;$\approx$\;4.5~$P_{\rm rot}$) almost completely
($D$\;=\;$-$0.024\,$\pm$\,0.005~SH/day).

Finally, if we exclude Doppler images \#5 and \#10, our spot area evolution analysis would lead to
an identical mean value for spot decay (<$D$>\;=\;$-$0.022\,$\pm$\,0.002~SH/day) and only a
marginally increased mean value for spot formation (<$D$>\;=\;+0.022\,$\pm$\,0.002~SH/day).
Therefore, we include them in our analysis. Furthermore, we repeated the entire analysis also for
the scattered small cool and/or hot spots at low latitudes. Their respective values of $D$ (for
small cool spots $\pm$\,0.02 for growth and decay, respectively; hot spots $\pm$\,0.03) scatter
within the range of the large-scale spots, but the time sampling is such that we can not determine a
true beginning nor ending of the evolution.

\begin{table}
\centering
\caption{Individual spot decay and growth rates. Derived values represent a linear area decay law of
form $dA(t)/dt$\;=\;$D$. $D$ is given in solar hemispheres (SH) per day
(1~SH\;=\;3.05~$\mathrm{Gm^2}$).}
\label{tab:spot_decay_par}
\centering
\begin{tabular}{l | l l l l}
\hline\noalign{\smallskip}
Season & Spot & DI \# & $D$ (SH/day) & type \\
\noalign{\smallskip}\hline\noalign{\smallskip}
2006/07 & A (blue)  &  1-6  & $-0.025\pm0.003$ & decay \\
        & B (red)   &  1-6  & $+0.017\pm0.004$ & growth \\
        & C (green) &  2-5  & $+0.020\pm0.012$ & growth \\
\noalign{\smallskip}\hline\noalign{\smallskip}
2007/08 & A (blue)  &  8-10 & $-0.021\pm0.012$ & decay \\
        & B (green) &  8-10 & $+0.014\pm0.007$ & growth \\
\noalign{\smallskip}\hline\noalign{\smallskip}
2008/09 & A (blue)  & 13-17 & $+0.026\pm0.004$ & growth \\
        & B (green) & 13-17 & $-0.020\pm0.003$ & decay \\
        & B (green) & 18-19 & $+0.013\pm0.009$ & growth \\
\noalign{\smallskip}\hline\noalign{\smallskip}
2009/10 & A (blue)  & 22-24 & $-0.019\pm0.016$ & decay \\
        & B (green) & 20-22 & $+0.019\pm0.007$ & growth \\
\noalign{\smallskip}\hline\noalign{\smallskip}
2010/11 & A (blue)  & 25-28 & $-0.021\pm0.008$ & decay \\
        & B (green) & 25-28 & $-0.020\pm0.007$ & decay \\
        & C (red)   & 25-28 & $+0.017\pm0.009$ & growth \\
\noalign{\smallskip}\hline\noalign{\smallskip}
2011/12 & A (blue)  & 31-33 & $-0.021\pm0.012$ & decay \\
        & B (green) & 30-32 & $-0.021\pm0.011$ & decay \\
        & B (green) & 32-36 & $+0.030\pm0.006$ & growth \\
        & C (red)   & 30-33 & $+0.023\pm0.006$ & growth \\
        & C (red)   & 33-36 & $-0.024\pm0.005$ & decay \\
\noalign{\smallskip}\hline\noalign{\smallskip}
2006-12 &           &     & <$-0.022\pm0.002$> & decay \\
        &           &     & <$+0.021\pm0.002$> & growth \\
\noalign{\smallskip}\hline
\end{tabular}
\end{table}


\subsection{Active longitudes}
\label{sec:act_lon}

Active longitudes are longitudes on which spots occur preferentially. The analysis of long-term
photometric observations as well as time-series Doppler imaging revealed active longitudes on
several stars. \cite{berdyugina1998a} found permanent active longitudes on four RS~CVn stars. If two
active longitudes, which are typically separated by 180$^\circ$, change their spot activity a
so-called ``flip-flop'' occured. This kind of a phenomenon was first noticed on the late-type giant
FK Com \citep{jetsu1991}. The average time between these phenomena is referred to as flip-flop cycle
and has been observed to be in the range of a few years up to a decade. An observational overview is
given in \cite{berdyugina2007} and \cite{korhonen2007a}. In binary stars, a longitudinal dependence
due to tidal effects is suggested \citep{holzwarth2002}. Observations show that in binaries
preferred longitudes exist on giant components mostly at the substellar points
\citep[e.g.,][]{olah2002}.

\begin{figure*}[!t]
\begin{minipage}{1.0\textwidth}
\captionsetup[subfigure]{labelfont=bf,textfont=bf,singlelinecheck=off,justification=raggedright,
position=top}
\subfloat[]{\includegraphics[width=200pt]
{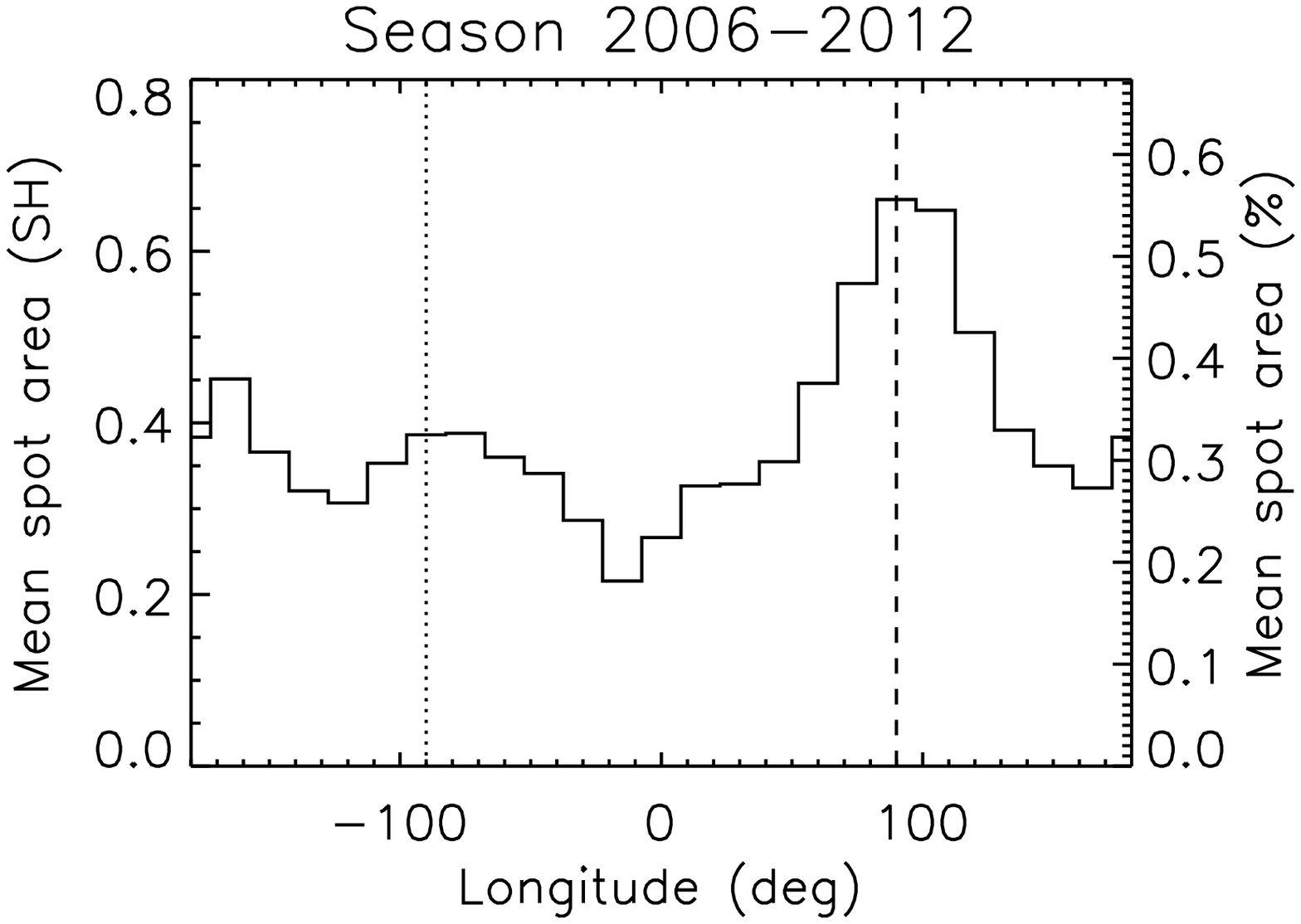}}
\subfloat[]{\includegraphics[width=325pt]
{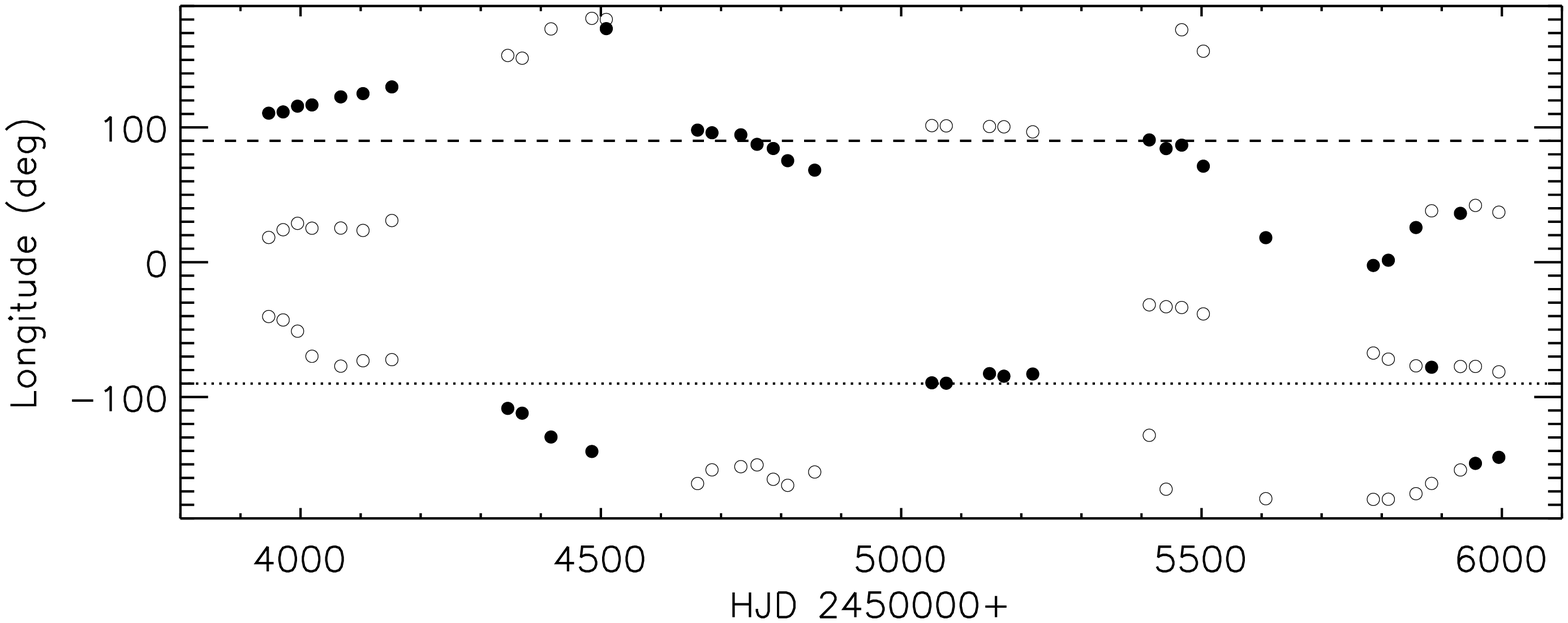}}
\end{minipage}\hspace{1ex}
\caption{Active longitudes on XX~Tri from 2006 to 2012. a)~The overall mean distribution of the spot
area (histogram in bins of 15$^\circ$) is shown. The spot area is given in solar hemispheres on the
left axis (1~SH\;=\;3.05~$\mathrm{Gm^2}$) and relative to the total area of a stellar hemisphere of
XX~Tri on the right axis. The dashed line represents the phase toward the companion star, whereas
the dotted line represents the phase in the opposite direction. b)~Spot longitudes as a function of
time. Spots A-E from all Doppler images are shown. The filled dots represent the larger spot at any
given time. The dashed and dotted lines represent again the phases toward the companion star and in
the opposite direction, respectively.}
\label{fig:active_longitudes}
\end{figure*}

Fig.~\ref{fig:spot_distribution_season_all}a-f shows the mean longitudinal distributions of all
individual spots from our spot-models for each season. There is clear evidence for preferred
longitudes during each season but significantly spread out in location because of individual spot
evolution. A particularly well-defined pair of active longitudes separated by 180$^\circ$ is seen in
season 2009/10. Averaging the longitudinal spot distribution from all 36 Doppler images, we find the
most spotted longitude to appear on average in phase toward the unseen companion star
($\pm$\,90$^\circ$; Fig.~\ref{fig:active_longitudes}a). Fig.~\ref{fig:active_longitudes}b plots the
spot centers from our spot-model fits (spots A-E) as a function of time. The larger spot always
appears in alternating hemispheres at locations either close to the phase toward the companion star
or shifted by around 180$^\circ$. The larger spot appears in between these phases near quadrature
possibly only between DI~\#29-32. Nevertheless, we interpret this behavior as a flip-flop and
estimate a tentative period of around two years.


\subsection{Differential rotation}
\label{sec:diff_rot}

\begin{figure}[!ht]
\begin{minipage}{1.0\textwidth}
\captionsetup[subfigure]{labelfont=bf,textfont=bf,singlelinecheck=off,justification=raggedright,
position=top}
\subfloat[]{\includegraphics[width=270pt]{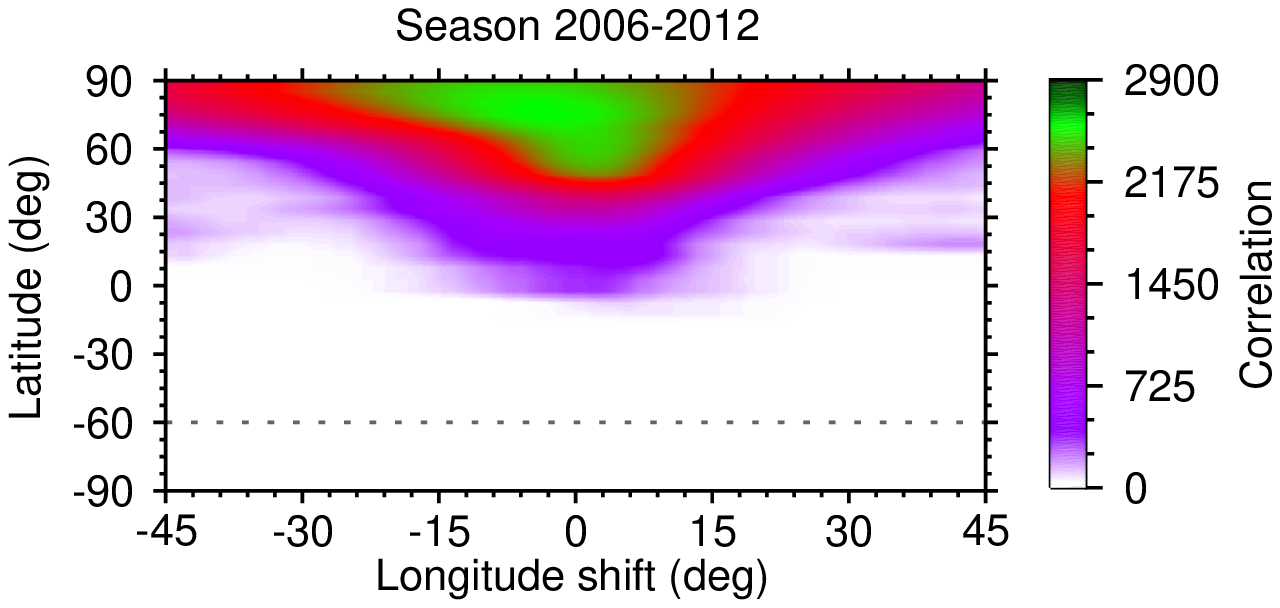}}
\end{minipage}\hspace{1ex}
\begin{minipage}{1.0\textwidth}
\captionsetup[subfigure]{labelfont=bf,textfont=bf,singlelinecheck=off,justification=raggedright,
position=top}
\subfloat[]{\includegraphics[width=250pt]
{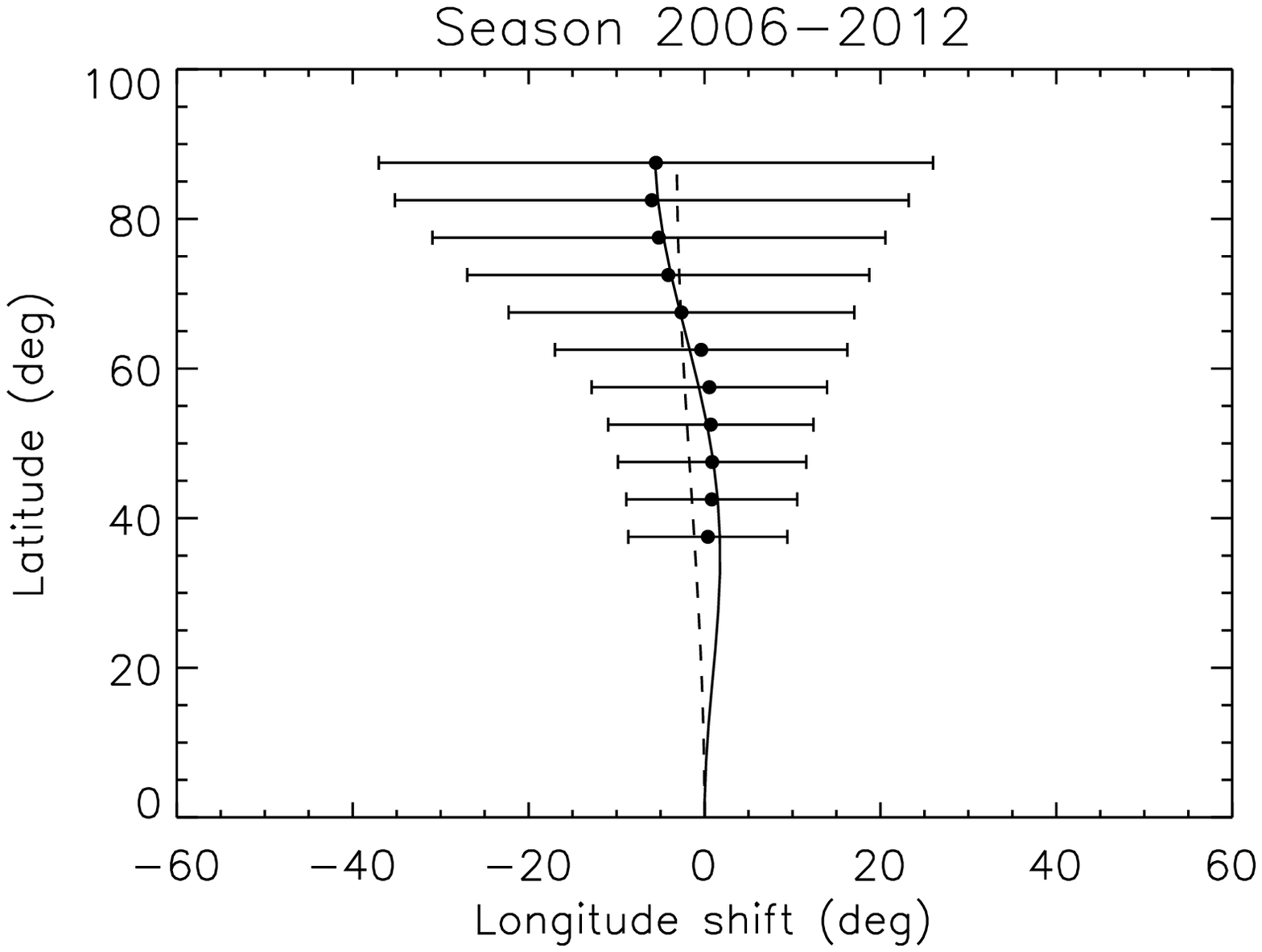}}
\end{minipage}\hspace{1ex}
\caption{a)~Grand average cross-correlation-function map from 2006 to 2012. The map represents the
average ccf map of all seasonal ccf maps from Fig.~\ref{fig:ccf_maps_all}, thus consisting of 29 ccf
maps in total. b)~Global differential rotation signature. The dots are the correlation peaks per
$5^{\circ}$-latitude bin and their error bars are defined as the FWHM of the corresponding Gaussian
fits. The dashed line represents a fit using Eq.~\ref{eqn:diff_rot_eqn_1}, whereas the solid line
represents a fit using Eq.~\ref{eqn:diff_rot_eqn_2}. The parameters for each fit are summarized in
Table~\ref{tab:diff_rot_par}.}
\label{fig:ccf_dr_total}
\end{figure}

Tracking sunspots is a classic technique to measure solar differential rotation and other surface
velocity fields like meridional flows \citep[e.g.,][]{woehl2002,brajvsa2002}. In case of XX~Tri a
large number of temperature surface maps with unprecedented good sampling is available, and
therefore may enable us to reveal a similarly accurate differential rotation law by tracking
individual starspots. Differential rotation has been detected on a number of stars by
cross-correlating consecutive Doppler images in longitudinal direction
\citep[e.g.,][]{donati1997,kovari2007b}. Applying this method to our data, we reconstruct between
three and six cross-correlation-function (ccf) maps per observing season, which we average to
increase their validity. We did not use the ccf map with DI~\#29, the last map in season 2010/11,
because the time span to the previous map (\#28) is comparable large, approximately four rotational
periods. Within such a time span, we expect significant local spot evolution.
Fig.~\ref{fig:ccf_maps_all} shows the average ccf maps for each observational season. The resulting
grand average ccf map, which consists of 29 ccf maps in total, is given in
Fig.~\ref{fig:ccf_dr_total}a.

We determined the correlation peak for each longitudinal stripe of $5^{\circ}$ width with a Gaussian
profile and fitted a standard differential rotation law of the form
\begin{equation}
\Omega(b) = \Omega_{\mathrm{eq}} - \Delta\Omega \sin^2(b) \ ,
\label{eqn:diff_rot_eqn_1}
\end{equation}
which is usually used for differential rotation measurements on stars. The parameter $\Omega(b)$
represents the angular velocity at latitude $b$, while
$\Delta\Omega$\;=\;$\Omega_{\mathrm{eq}}$\,$-$\,$\Omega_{\mathrm{pole}}$ represents the difference
between the angular velocities at the equator and at the pole, respectively. The surface shear
parameter $\alpha$ is defined as $\Delta\Omega/\Omega_{\mathrm{eq}}$, and the lap time as the
reciprocal of the rotational shear, i.e., the time it takes for the equator to do a full lap more
than the pole.

Alternatively, we fit a differential rotation law of the form
\begin{equation}
\Omega(b) = \Omega_{\mathrm{eq}} + \Omega_1 \sin^2(b) + \Omega_2 \sin^4(b) \ ,
\label{eqn:diff_rot_eqn_2}
\end{equation}
which is usually used for differential rotation measurements on the Sun. In this case the angular
velocity at the pole is defined as
$\Omega_{\mathrm{pole}}$\;=\;$\Omega_{\mathrm{eq}}$\,+\,$\Omega_1$\,+\,$\Omega_2$.
Fig.~\ref{fig:diff_rot_fit_all} and Fig.~\ref{fig:ccf_dr_total}b show the observed differential
rotation pattern determined from the ccf maps together with the best fit of the differential
rotation following Eq.~\ref{eqn:diff_rot_eqn_1} and Eq.~\ref{eqn:diff_rot_eqn_2}. In
Table~\ref{tab:diff_rot_par} all seasonal fits for differential rotation are listed.

\begin{table*}
\centering
\caption{Results for differential rotation. Parameters for best fits using
Eq.~\ref{eqn:diff_rot_eqn_1} (middle part) and Eq.~\ref{eqn:diff_rot_eqn_2} (right part). The
parameter $\Omega_{\mathrm{eq}}$ is fixed for $P_{\mathrm{rot}}$\;=\;24.0~d in both cases.}
\label{tab:diff_rot_par}
\centering
\begin{tabular}{l l | l l l | l l l l}
\hline\noalign{\smallskip}
Season & $\#$ ccf maps & $\Delta\Omega$ ($^\circ$/d) & $\alpha$ & lap time (d) & $\Omega_1$
($^\circ$/d) & $\Omega_2$ ($^\circ$/d) & $\alpha$ & lap time (d) \\
\noalign{\smallskip}\hline\noalign{\smallskip}
2006/07 & 6  & 0.05$\pm$0.03 & 0.003$\pm$0.002 & $\approx$6870 & 0.53$\pm$0.04 & --0.68$\pm$0.04 &
0.010$\pm$0.001 & $\approx$2430 \\
2007/08 & 4  & 0.30$\pm$0.08 & 0.020$\pm$0.005 & $\approx$1190 & 0.83$\pm$0.20 & --1.35$\pm$0.23 &
0.035$\pm$0.010 & $\approx$690 \\
2008/09 & 6  & 0.29$\pm$0.06 & 0.019$\pm$0.004 & $\approx$1240 & 0.51$\pm$0.16 & --0.95$\pm$0.18 &
0.030$\pm$0.011 & $\approx$810 \\
2009/10 & 4  & 0.11$\pm$0.04 & 0.007$\pm$0.003 & $\approx$3290 & 0.22$\pm$0.21 & --0.39$\pm$0.24 &
0.011$\pm$0.012 & $\approx$2130 \\
2010/11 & 3  & 0.19$\pm$0.06 & 0.013$\pm$0.004 & $\approx$1870 & --0.20$\pm$0.25 & 0.01$\pm$0.33 &
0.013$\pm$0.058 & $\approx$1890 \\
2011/12 & 6  & 0.00$\pm$0.04 & 0.000$\pm$0.003 & ---           & 0.64$\pm$0.14 & --0.74$\pm$0.17 &
0.007$\pm$0.002 & $\approx$3460 \\
\noalign{\smallskip}\hline\noalign{\smallskip}
2006-12 & 29 & 0.13$\pm$0.04 & 0.009$\pm$0.003 & $\approx$2740 & 0.45$\pm$0.07 & --0.69$\pm$0.09 &
0.016$\pm$0.003 & $\approx$1530 \\
\noalign{\smallskip}\hline
\end{tabular}
\end{table*}

Because most spots on XX~Tri appear at high latitudes, Eq.~\ref{eqn:diff_rot_eqn_2} with its $\sin^4
b$ term leads to a much better fit to the observed shear than Eq.~\ref{eqn:diff_rot_eqn_1}.
Therefore, we favor Eq.~\ref{eqn:diff_rot_eqn_2} over Eq.~\ref{eqn:diff_rot_eqn_1}. All observing
seasons show a solar-like differential rotation law with an overall shear parameter of
$\alpha$\;=\;0.016\,$\pm$\,0.003 and a lap time of $\approx$\;1500~days. The surface shear on XX~Tri
is therefore only around a tenth of the solar value.


\subsection{Stellar cycle prediction}
\label{sec:mag_cycle}

Because turbulent diffusion is believed to be the dominating effect of spot decay, the decay rate of
spot area is directly proportional to the turbulent diffusivity \citep{meyer1974,krause1975},
\begin{equation}
dA/dt = -4\pi\eta_T \ .
\end{equation}
Using the mean decay rate of $D$\;=\;$-$0.022\,$\pm$\,0.002~SH/day from our analysis in
Table~\ref{tab:spot_decay_par} leads to a turbulent diffusivity of
$\eta_T$\;=\;(6.3\,$\pm$\,0.5)\,$\times$\,10$^{14}$~cm$^2$/s. This value is at minimum one order of
magnitude higher than that predicted for solar values, which vary from $10^{10}$~cm$^2$/s
\citep{dikpati1999} to $10^{13}$~cm$^2$/s \citep{ruediger2000}. The diffusion timescale for the
magnetic field inside the convection zone (CZ) is given by
\begin{equation}
\tau = \frac{L_{CZ}^2}{\eta_T} \ ,
\end{equation}
where $L_{CZ}^2$ is the width of the stellar convection zone. Using stellar models calculated with
the Yale Rotational stellar Evolution Code (YREC; see \cite{spada2013} for more details) we estimate
a depth of 0.94~$R_\star$ for the convection zone of XX Tri. Thus, leading to a magnetic cycle of
approximately 26\,$\pm$\,6~years.


\subsection{A starspot movie}
\label{sec:movie}

We merged all 36 Doppler images into an animated {\tt gif} file ({\tt xxtri-di-anim.gif}), which is
available through the A\&A video depository or our group web
page\footnote{http://www.aip.de/en/research/research-area-cmf/
\\cosmic-magnetic-fields/stellar/stellar-activity/news}. It particularly emphasizes surface detail
not covered in the spot-decay analysis and shows the general migration and decay trends. It may
be used for general demonstration purposes.

The movie shows the same four equidistant surface views as, e.g., in Fig.~\ref{fig:di_season_1} but
just as a function of time. Between each Doppler image a time delay of 250~ms is included.


\section{Discussion and summary}
\label{sec:discussion}

Thanks to our robotic STELLA telescopes the present time series of Doppler images resolves 36 single
stellar rotations over a time span of six years. The sample is long enough that it enables, for
the first time, a direct determination of a starspot decay law. Our target is the K0 giant XX~Tri
with a rotation period ($\approx$\;24~d) comparable to that of the Sun but being more massive by
26\,\%\ and significantly older with an age of $\approx$\;8~Gyrs. This combination of parameters
is only possible because the star is a component of a close binary. A comparison with solar
analogies is therefore only for general guidance.

The time series enabled the cartography of a variety of surface activity phenomena, such as active
longitudes, flip-flops, and differential rotation on XX~Tri. However, our main result is a spot decay
law leading to a prediction of a magnetic activity cycle solely based on an observationally
constrained value of the turbulent magnetic diffusivity. The individual spot decay rates,
$dA/dt$\;=\;$D$, scattered between $-0.019$ and $-0.025$~SH/day over the six year observing period
with a mean value of $D$\;=\;$-$0.022\,$\pm$\,0.002~SH/day. The rates for spot growth
were between $+0.013$ and $+0.030$~SH/day with a mean of $D$\;=\;+0.021\,$\pm$\,0.002~SH/day and
thus of nearly the same amount as the decay. In above units, the spot decay on XX~Tri would be of
the order of $10^4$ times faster than what \cite{bumba1963} suggested for sunspots. Bumba proposed a
mean value for $D$ of $-4.2$~MSH/day. However, the areal size of starspots on XX~Tri is also $10^3$
to $10^4$ times larger than the largest observed sunspots ($\approx$\;10$^{-3}$~SH;
\citealt{baumann2005}). From this, we depict a turbulent diffusivity for XX~Tri of
$\approx$\;6\,$\times$\,10$^{14}$~cm$^2$/s, a value between 10 to 10,000 times larger than current
model values for the solar convection zone (from the surface layers to the bottom of the convection
zone). Because the (squared) absolute depth of the convection zone of XX~Tri is about 1,200 times
larger than that of the Sun (200~Mm), the diffusion timescale becomes comparable to that of the Sun.
We obtain an average diffusion time of $\approx$\;26~yr for XX~Tri compared to $\approx$\;12~yr
for the Sun (the latter for an assumed diffusivity of 10$^{12}$~cm$^2$/s).

So far, stellar activity cycles were inferred from long-term chromospheric Ca\,{\sc ii} H\&K or
photospheric $V$-band variations \citep[e.g.,][]{baliunas1995,olah2007} or from repeated detections
of a ``flip-flop'' phenomenon \citep[see][]{hackman2013,korhonen2007a}. Derived timescales for
RS~CVn stars range between 2-50~yr. Just recently, \cite{olah2014} investigated 28 years of
photometry of XX~Tri along with two other over-active K giants and found a long-term sinusoidal
brightness trend with a length comparable to the length of the data set, i.e., $\approx$\;28~yr.
Our diffusivity-based cycle prediction of $\approx$\;26~yr matches this observation surprisingly
well despite the comparable shortness of the photometric coverage. XX~Tri reached its maximum
brightness in 2009 and is in a declining state since then. Removing this overall trend,
\cite{olah2014} found a second shorter-period variation of approximately 6~yr or an integer
multiple of it. No clear explanation for this period, if real, could be given.

As already mentioned in Sect.~\ref{sec:doppler_maps}, the reconstructed giant spots may be
monolithic, but could also be a conglomerate of smaller, unresolved spots. This ``classical''
uncertainty could in principle impact on the interpretation of the observed decay rate, and thus the
cycle length. Assuming that each unresolved spot only decays (and never grows), and does not
interact with an other spot fragment, then we should observe on average the same decay rate as if
the spot were monolithic. If decay and growth of individual fragments coexist, then our determined
decay rate would be just a lower boundary. It then enables a maximum decay rate of
\;$\approx$\;$-0.03$~SH/day, resulting in a cycle length of $\approx$\;19~yr. This cycle period
would be close to our determined 1-$\sigma$ uncertainty. On the contrary, no such cycle period was
detected from broadband photometry. In addition, a possible cycle length of six~years (from
photometry) or two~years (from flip-flop) would suggest a decay rate of $-0.10$ and
$-0.29$~SH/day, respectively; neither of which is supported by our analysis. We conclude that our
predicted cycle of $\approx$\;26~yr, including an uncertainty of 6~yr, appears to be the most
credible.

Active longitudes are a common feature in rapidly-rotating active stars and there is even some
evidence for long-term active longitudes and a 7-yr flip-flop period on the Sun
\citep{berdyugina2003}. Our Doppler imagery provides evidence for a $\approx$\;2-yr flip-flop
period on XX~Tri with a preferred longitude typically facing the (unseen) companion star. This is
significantly shorter than the 6-yr period from photometric data, but could be the true flip-flop
cycle length and thus would identify the 6-yr period just as an alias. From a theoretical
perspective, flip-flops possibly represent the nonaxisymmetric component of a mixed-mode dynamo for
weakly differentially rotating stars \citep{elstner2005,moss2004}. These authors explained
flip-flops as an excited nonaxisymmetric dynamo mode, giving rise to two permanent active
longitudes in opposite stellar hemispheres, but still need an oscillating axisymmetric magnetic
field in parallel. The stability of this kind of a mixed mode is still a matter of discussion.

For XX~Tri, our Doppler images indicate a weak solar-like differential rotation of
$\alpha$\;=\;0.016\,$\pm$\,0.003, which seems to be a typical value for this kind of rapidly
rotating stars. Despite the fact that large-scale mean-field dynamo models predict only a poor tracing
quality for its large spots (\citealt{korhonen2011}; but see also \citealt{czesla2013}), numerous
differential-rotation laws were deduced from cross correlations of cool features from consecutive
Doppler images (at this point we refer to the many references cited in \citealt{korhonen2011}).
Recently, weak solar-like differential rotation was confirmed, e.g., on the K giants IL~Hya
\citep{kovari2014,weber1998} or $\zeta$~And \citep{kovari2012,kovari2007a}, while weak antisolar
differential rotation was confirmed on the K-giant $\sigma$~Gem \citep{kovari2015,kovari2007b}.
Furthermore, weak antisolar differential rotation was claimed for the K giants UZ~Lib
\citep{olah2003}, HD\,31993 \citep{strassmeier2003}, and possibly HU~Vir \citep{strassmeier1994}. All
the latter still in need of further independent verification. For a previous summary on this topic
we refer to \cite{weber2005}. XX~Tri fits into the differentially rotating giants with approximately
ten times weaker surface latitudinal shear when compared to the Sun.


\bibliographystyle{aa}                           
\bibliography{literature}                        

\begin{acknowledgements}
We are grateful to the State of Brandenburg and the German federal ministry for education and
research (BMBF) for their continuous support of the STELLA activities. The STELLA facility is a
collaboration of the AIP in Brandenburg with the IAC in Tenerife. We thank all scientists and
engineers involved, in particular Michael Weber and Thomas Granzer from AIP as well as Ignacio
del Rosario and Miguel Serra from the IAC Tenerife daytime crew. It is our pleasure to thank
G.~R\"udiger, R.~Arlt and C.~Denker for many helpful discussions on solar decay laws. Last but not
least, we like to thank our referee P.~Petit for his many helpful suggestions that improved this
paper.
\end{acknowledgements}

%

\Online

\begin{appendix}

\section{Doppler images for the seasons 2007/08 to 2011/12}

\subsection{Season 2007/08}

\begin{figure}[!t]
\begin{minipage}{1.0\textwidth}
\captionsetup[subfigure]{labelfont=bf,textfont=bf,singlelinecheck=off,justification=raggedright,
position=top}
\subfloat[]{\includegraphics[width=250pt]{figures/season2_map1_original.eps}}
\end{minipage}\hspace{1ex}
\begin{minipage}{1.0\textwidth}
\captionsetup[subfigure]{labelfont=bf,textfont=bf,singlelinecheck=off,justification=raggedright,
position=top}
\subfloat[]{\includegraphics[width=250pt]{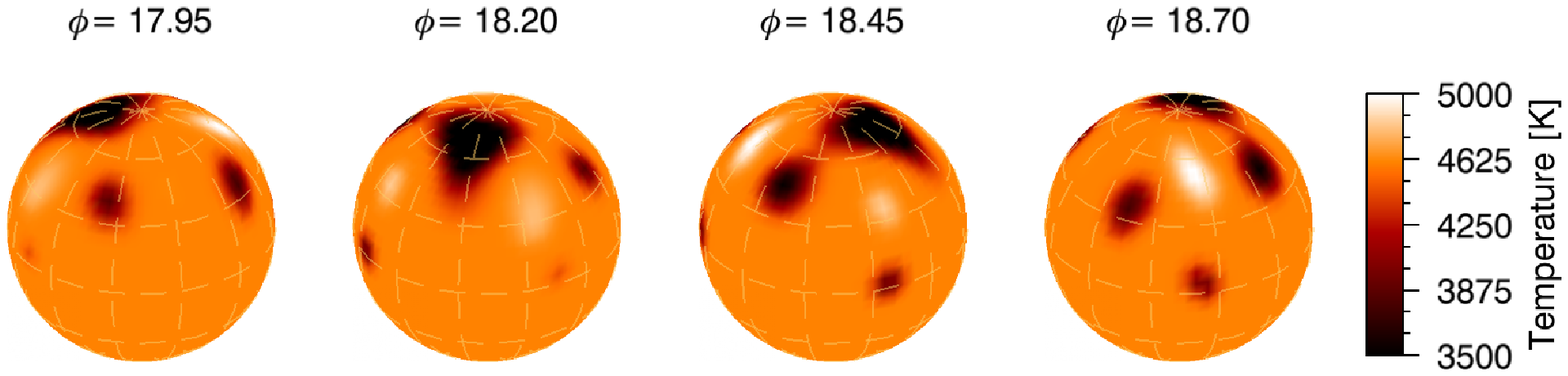}}
\end{minipage}\hspace{1ex}
\begin{minipage}{1.0\textwidth}
\captionsetup[subfigure]{labelfont=bf,textfont=bf,singlelinecheck=off,justification=raggedright,
position=top}
\subfloat[]{\includegraphics[width=250pt]{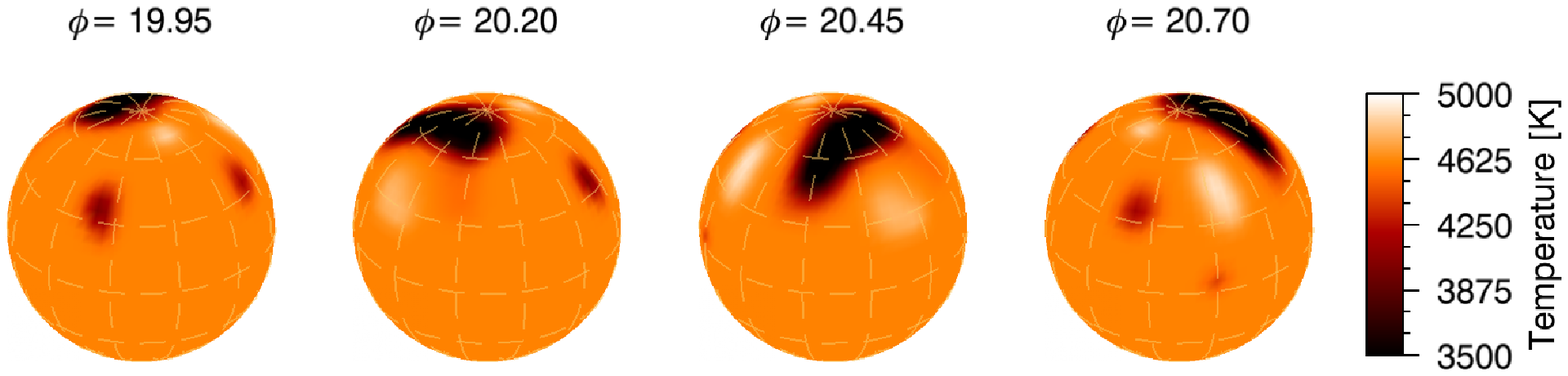}}
\end{minipage}\hspace{1ex}
\begin{minipage}{1.0\textwidth}
\captionsetup[subfigure]{labelfont=bf,textfont=bf,singlelinecheck=off,justification=raggedright,
position=top}
\subfloat[]{\includegraphics[width=250pt]{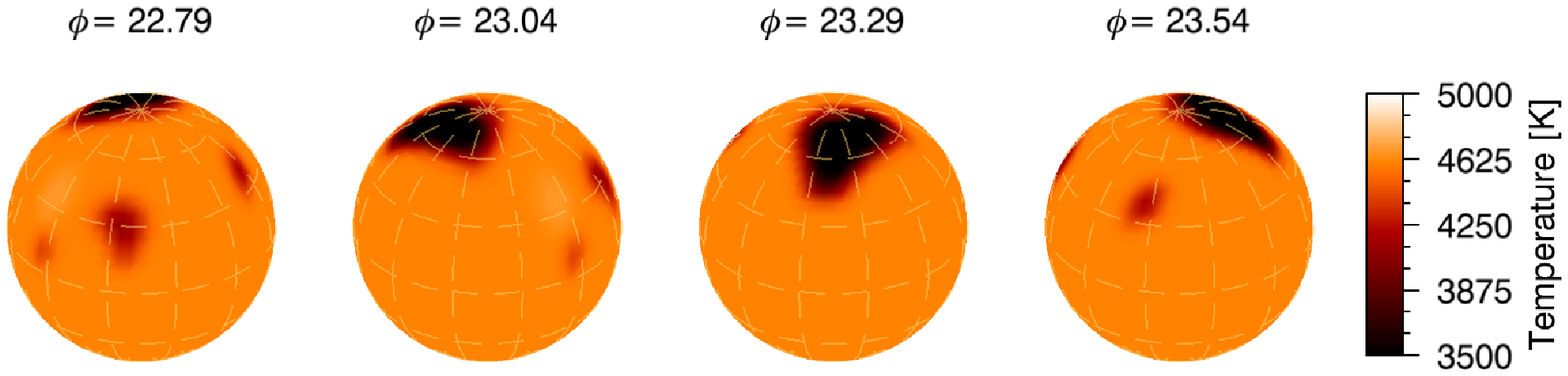}}
\end{minipage}\hspace{1ex}
\begin{minipage}{1.0\textwidth}
\captionsetup[subfigure]{labelfont=bf,textfont=bf,singlelinecheck=off,justification=raggedright,
position=top}
\subfloat[]{\includegraphics[width=250pt]{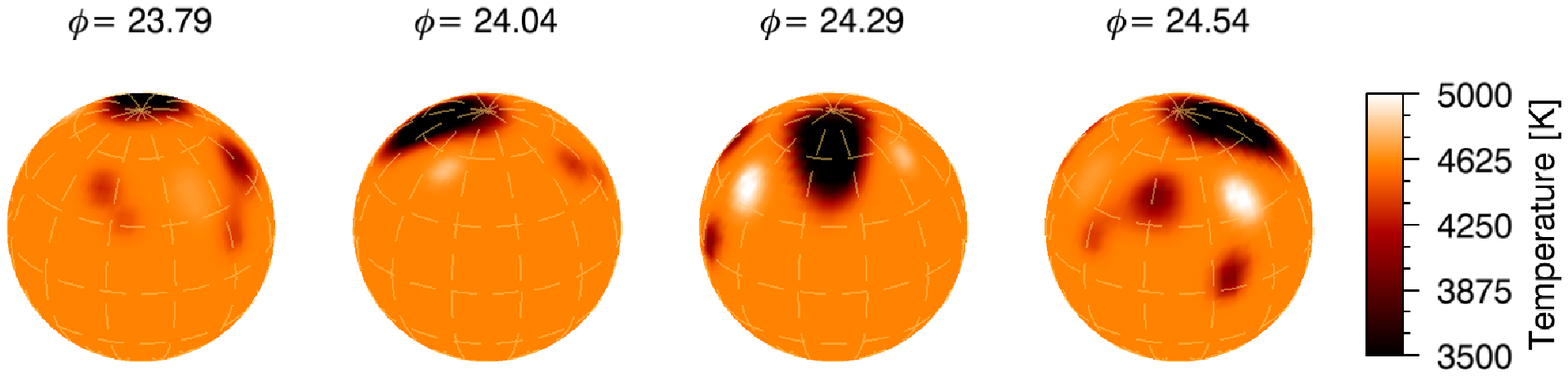}}
\end{minipage}\hspace{1ex}
\caption{Doppler images of XX~Tri for the observing season 2007/08. Each DI is shown in four
spherical projections separated by $90^{\circ}$. The rotational shift between consecutive images is
corrected, i.e., the stellar orientation remains the same from map to map and from season to season.
The time difference between each Doppler image is indicated in units of rotational phase $\phi$.}
\label{fig:di_season_2}
\end{figure}

\begin{figure}[!t]
\begin{minipage}{1.0\textwidth}
\captionsetup[subfigure]{labelfont=bf,textfont=bf,singlelinecheck=off,justification=raggedright,
position=top}
\subfloat[]{\includegraphics[width=250pt]{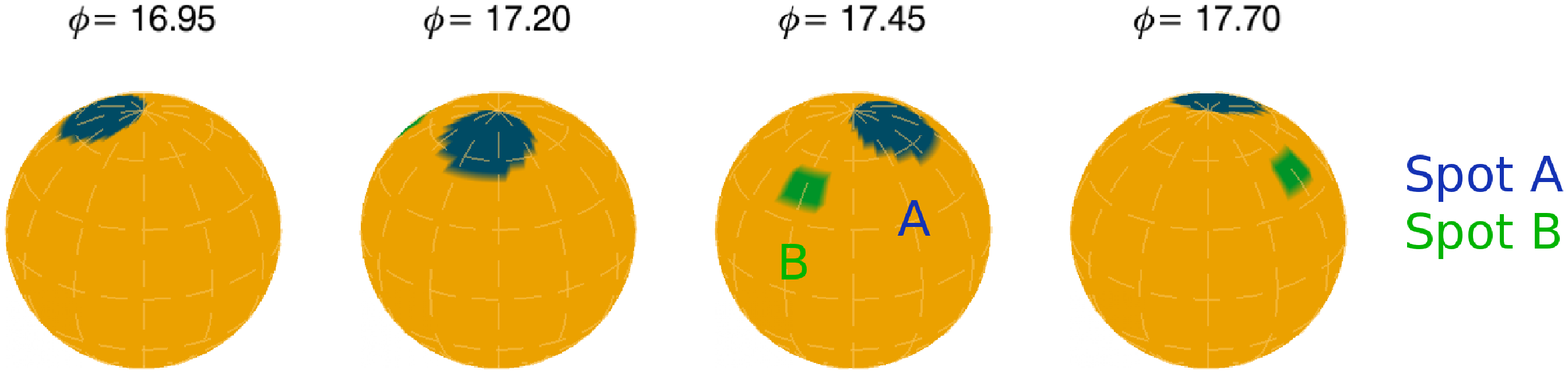}}
\end{minipage}\hspace{1ex}
\begin{minipage}{1.0\textwidth}
\captionsetup[subfigure]{labelfont=bf,textfont=bf,singlelinecheck=off,justification=raggedright,
position=top}
\subfloat[]{\includegraphics[width=250pt]{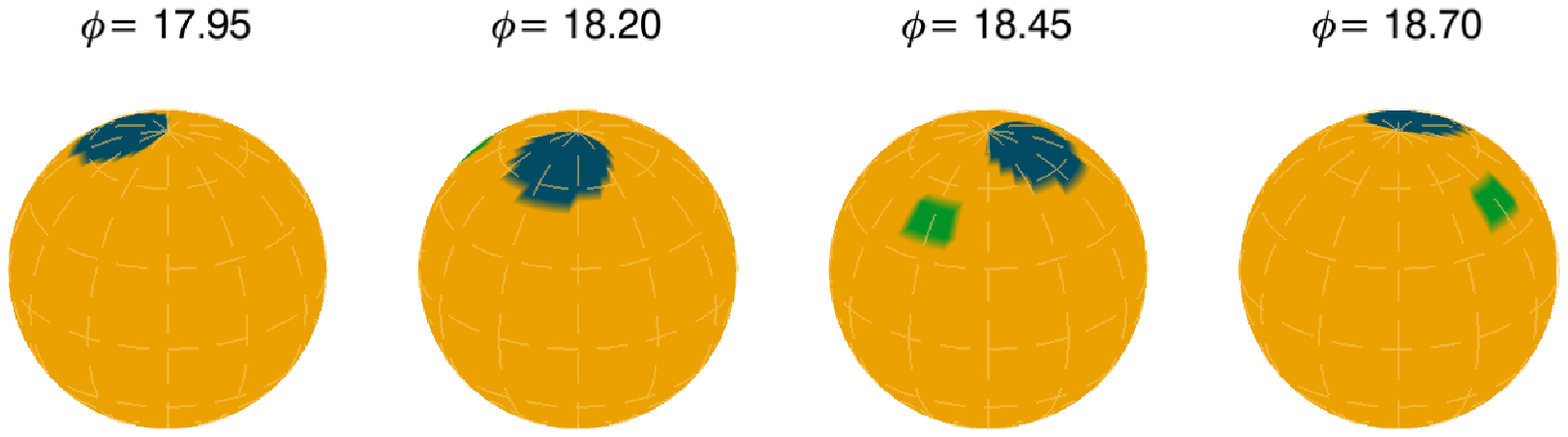}}
\end{minipage}\hspace{1ex}
\begin{minipage}{1.0\textwidth}
\captionsetup[subfigure]{labelfont=bf,textfont=bf,singlelinecheck=off,justification=raggedright,
position=top}
\subfloat[]{\includegraphics[width=250pt]{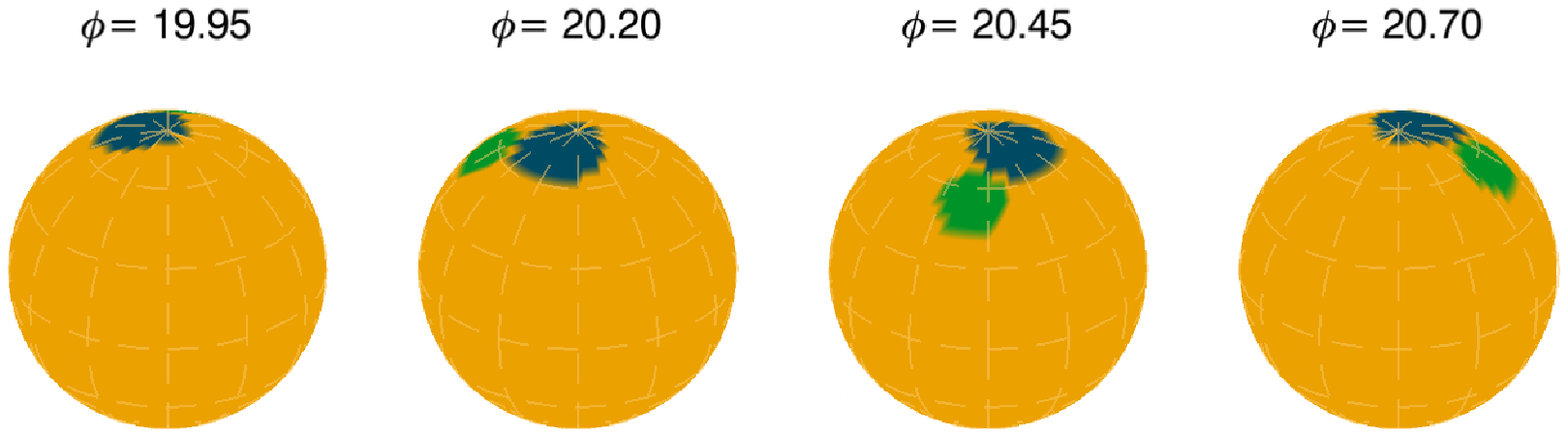}}
\end{minipage}\hspace{1ex}
\begin{minipage}{1.0\textwidth}
\captionsetup[subfigure]{labelfont=bf,textfont=bf,singlelinecheck=off,justification=raggedright,
position=top}
\subfloat[]{\includegraphics[width=250pt]{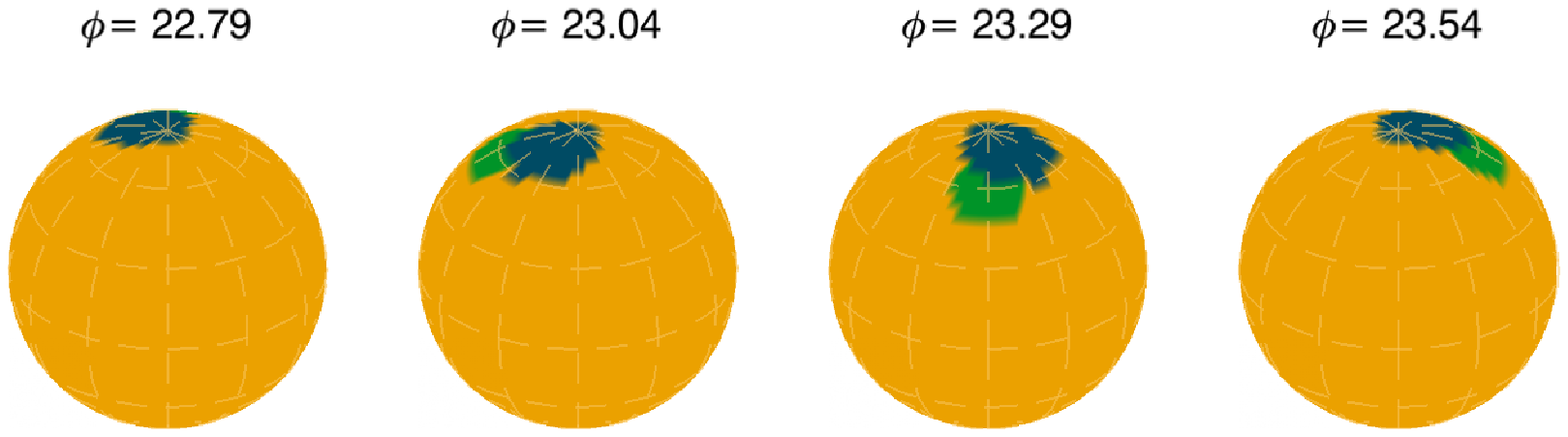}}
\end{minipage}\hspace{1ex}
\begin{minipage}{1.0\textwidth}
\captionsetup[subfigure]{labelfont=bf,textfont=bf,singlelinecheck=off,justification=raggedright,
position=top}
\subfloat[]{\includegraphics[width=250pt]{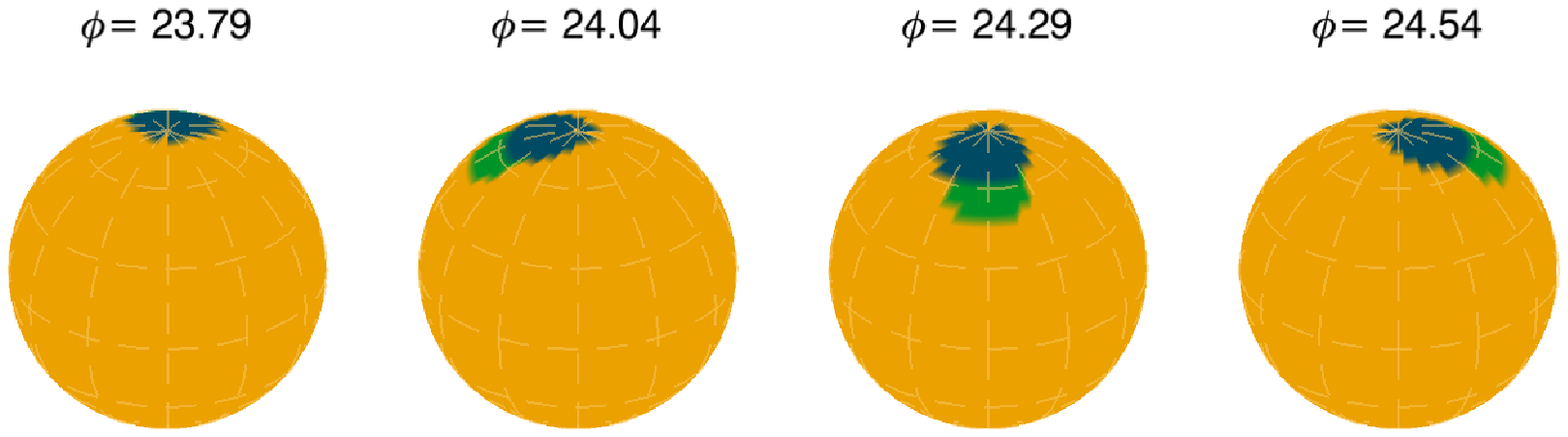}}
\end{minipage}\hspace{1ex}
\caption{Spot-model fits of the Doppler images in Fig.~\ref{fig:di_season_2}. Each spot is shown
with different color/contrast for better visualization.}
\label{fig:di_season_2_fit}
\end{figure}

In Fig.~\ref{fig:di_season_2} five almost consecutive Doppler images are shown, which cover around
eight rotations. In Fig.~\ref{fig:di_season_2_fit} the spot-model fits of the Doppler images are
shown.

\subsection{Season 2008/09}

\begin{figure}[!t]
\begin{minipage}{1.0\textwidth}
\captionsetup[subfigure]{labelfont=bf,textfont=bf,singlelinecheck=off,justification=raggedright,
position=top}
\subfloat[]{\includegraphics[width=250pt]{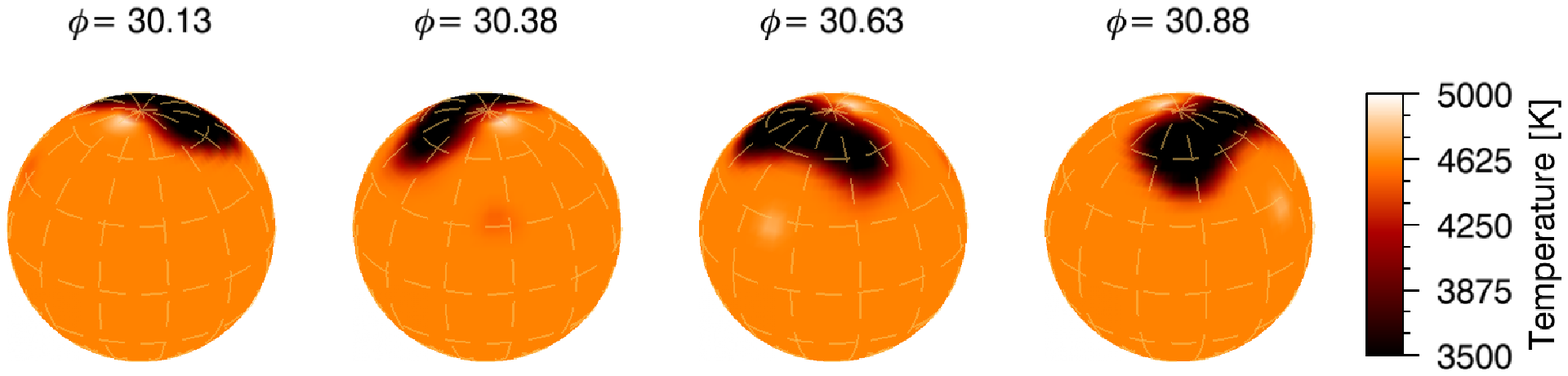}}
\end{minipage}\hspace{1ex}
\begin{minipage}{1.0\textwidth}
\captionsetup[subfigure]{labelfont=bf,textfont=bf,singlelinecheck=off,justification=raggedright,
position=top}
\subfloat[]{\includegraphics[width=250pt]{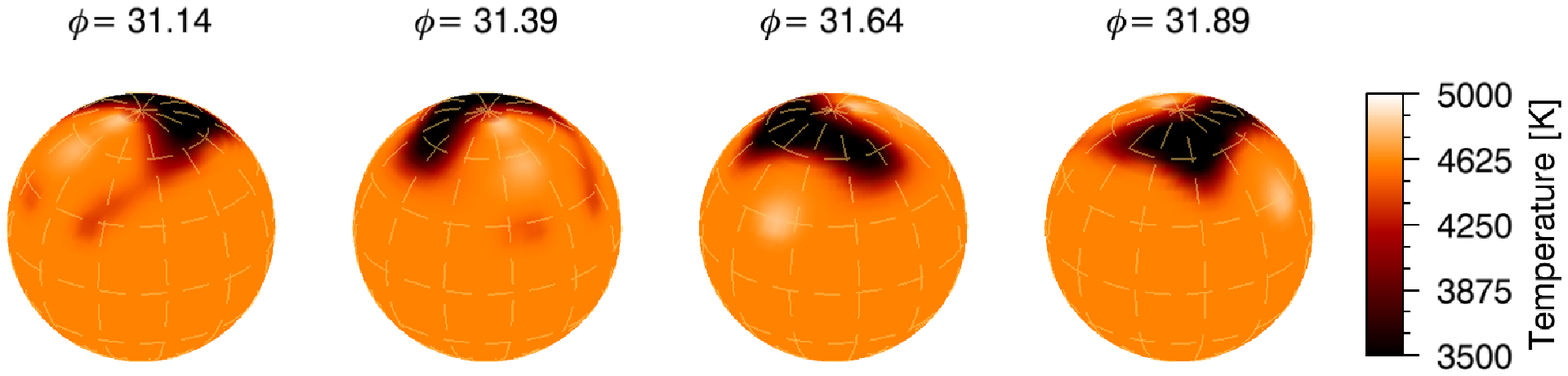}}
\end{minipage}\hspace{1ex}
\begin{minipage}{1.0\textwidth}
\captionsetup[subfigure]{labelfont=bf,textfont=bf,singlelinecheck=off,justification=raggedright,
position=top}
\subfloat[]{\includegraphics[width=250pt]{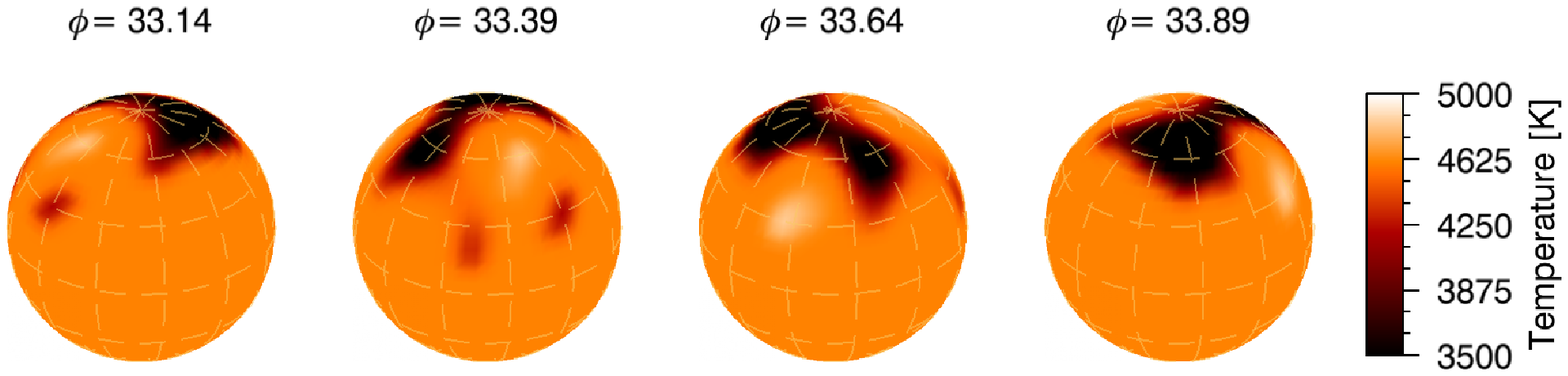}}
\end{minipage}\hspace{1ex}
\begin{minipage}{1.0\textwidth}
\captionsetup[subfigure]{labelfont=bf,textfont=bf,singlelinecheck=off,justification=raggedright,
position=top}
\subfloat[]{\includegraphics[width=250pt]{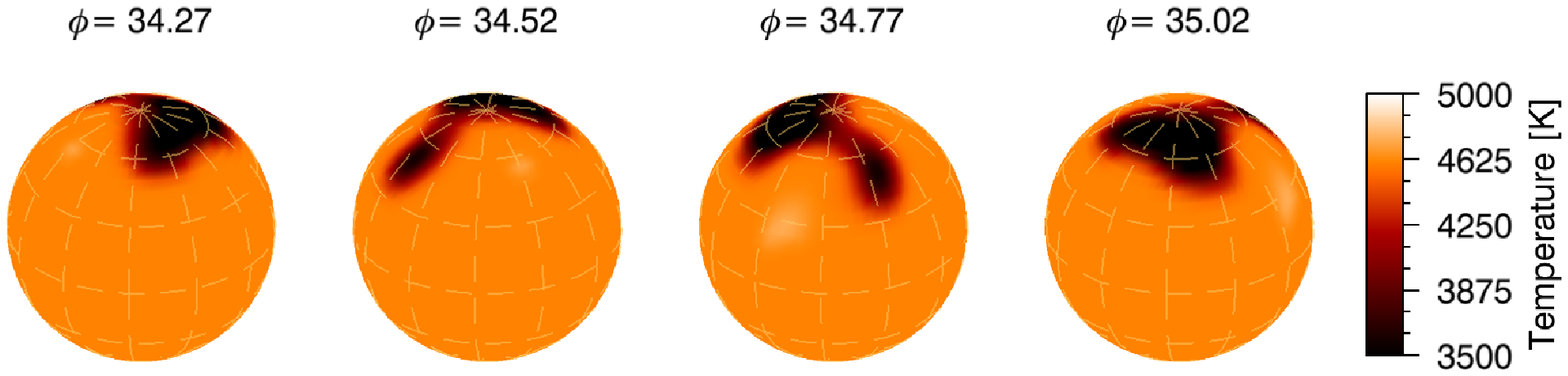}}
\end{minipage}\hspace{1ex}
\begin{minipage}{1.0\textwidth}
\captionsetup[subfigure]{labelfont=bf,textfont=bf,singlelinecheck=off,justification=raggedright,
position=top}
\subfloat[]{\includegraphics[width=250pt]{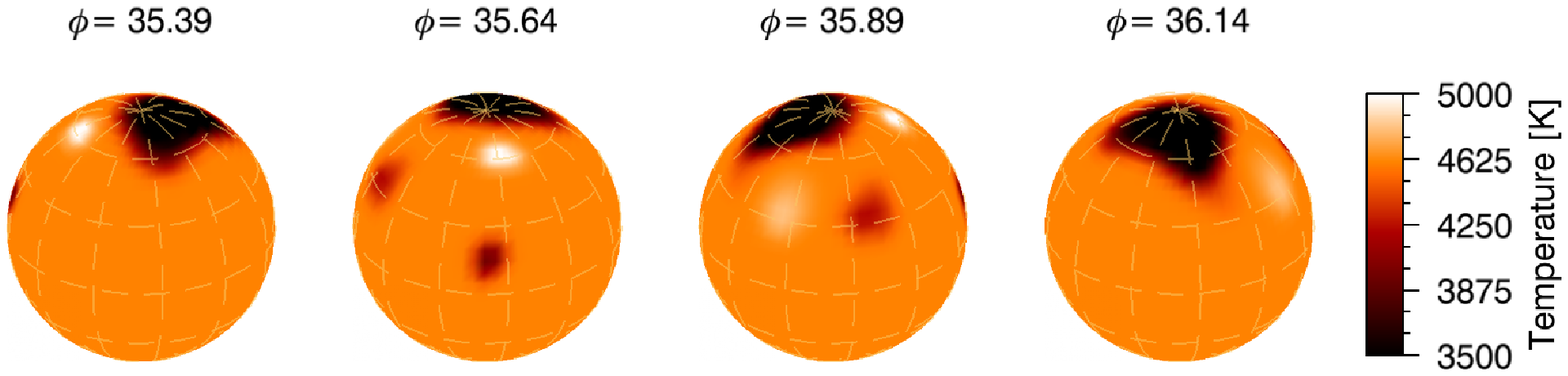}}
\end{minipage}\hspace{1ex}
\begin{minipage}{1.0\textwidth}
\captionsetup[subfigure]{labelfont=bf,textfont=bf,singlelinecheck=off,justification=raggedright,
position=top}
\subfloat[]{\includegraphics[width=250pt]{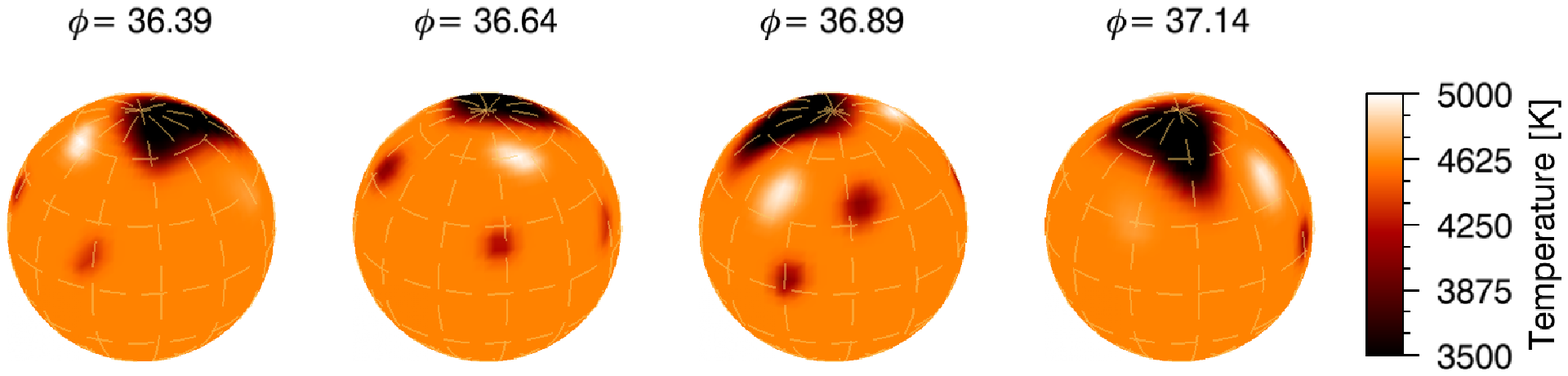}}
\end{minipage}\hspace{1ex}
\begin{minipage}{1.0\textwidth}
\captionsetup[subfigure]{labelfont=bf,textfont=bf,singlelinecheck=off,justification=raggedright,
position=top}
\subfloat[]{\includegraphics[width=250pt]{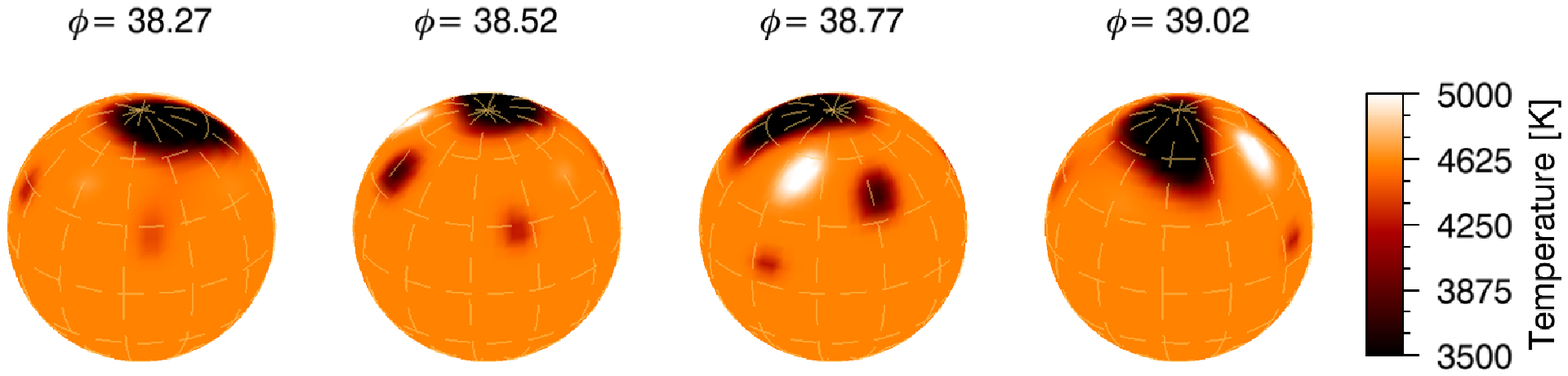}}
\end{minipage}\hspace{1ex}
\caption{Doppler images of XX~Tri for the observing season 2008/09. Otherwise as in
Fig.~\ref{fig:di_season_2}.}
\label{fig:di_season_3}
\end{figure}

\begin{figure}[!t]
\begin{minipage}{1.0\textwidth}
\captionsetup[subfigure]{labelfont=bf,textfont=bf,singlelinecheck=off,justification=raggedright,
position=top}
\subfloat[]{\includegraphics[width=250pt]{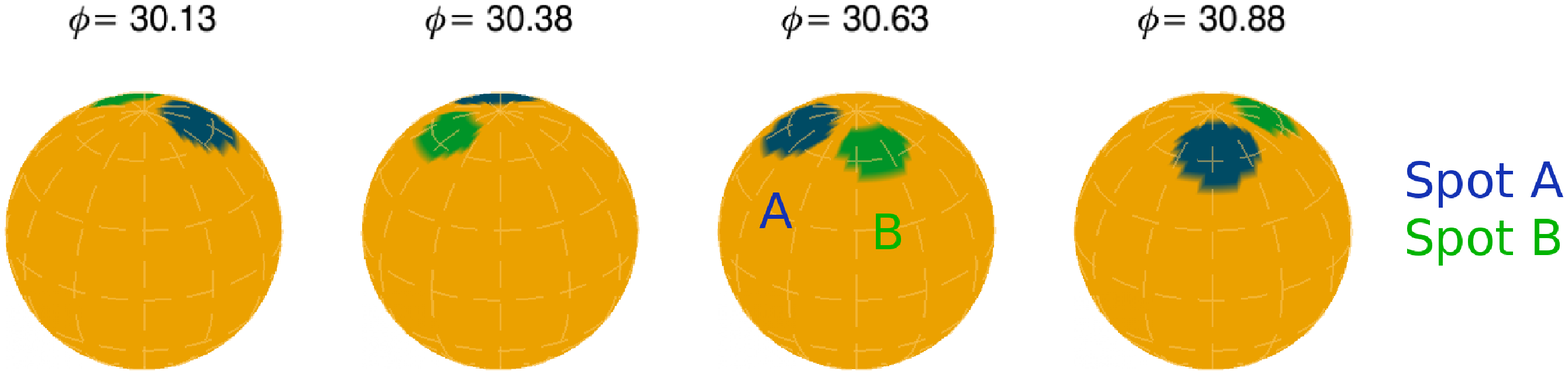}}
\end{minipage}\hspace{1ex}
\begin{minipage}{1.0\textwidth}
\captionsetup[subfigure]{labelfont=bf,textfont=bf,singlelinecheck=off,justification=raggedright,
position=top}
\subfloat[]{\includegraphics[width=250pt]{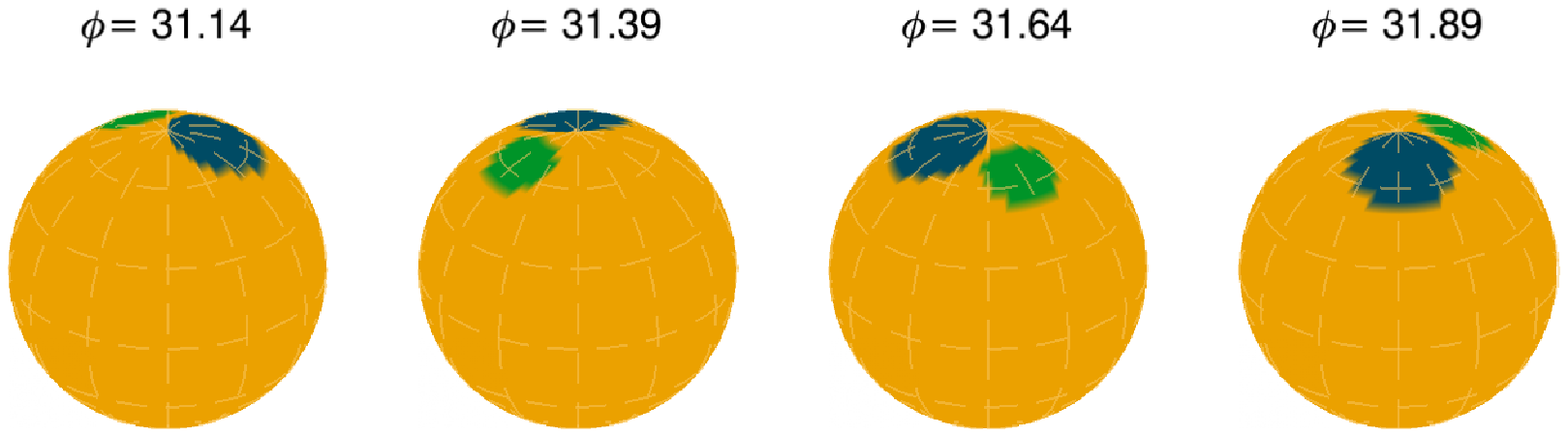}}
\end{minipage}\hspace{1ex}
\begin{minipage}{1.0\textwidth}
\captionsetup[subfigure]{labelfont=bf,textfont=bf,singlelinecheck=off,justification=raggedright,
position=top}
\subfloat[]{\includegraphics[width=250pt]{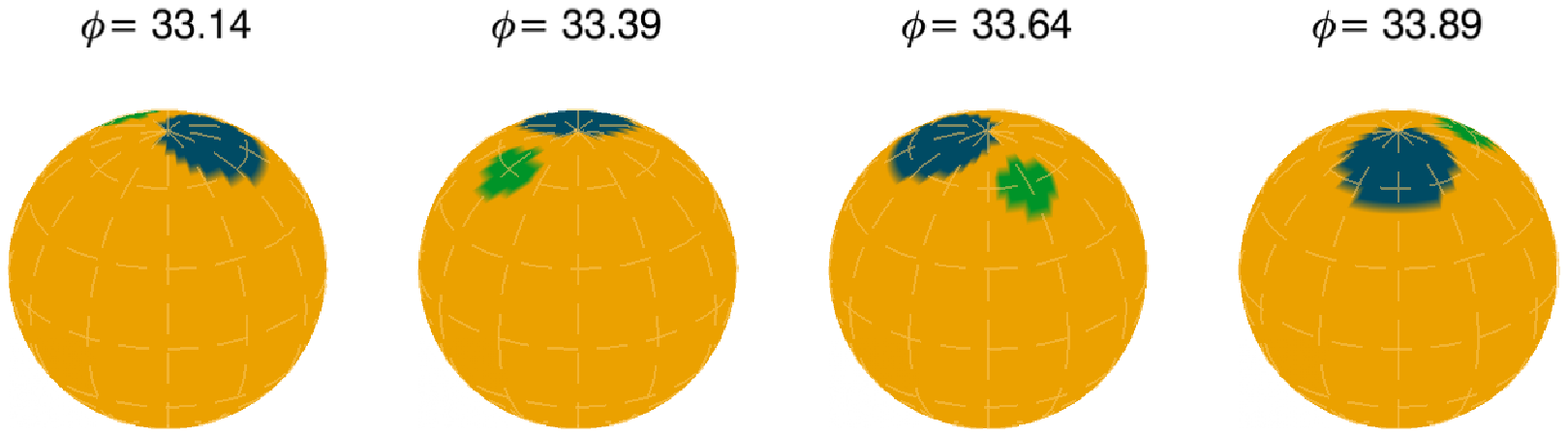}}
\end{minipage}\hspace{1ex}
\begin{minipage}{1.0\textwidth}
\captionsetup[subfigure]{labelfont=bf,textfont=bf,singlelinecheck=off,justification=raggedright,
position=top}
\subfloat[]{\includegraphics[width=250pt]{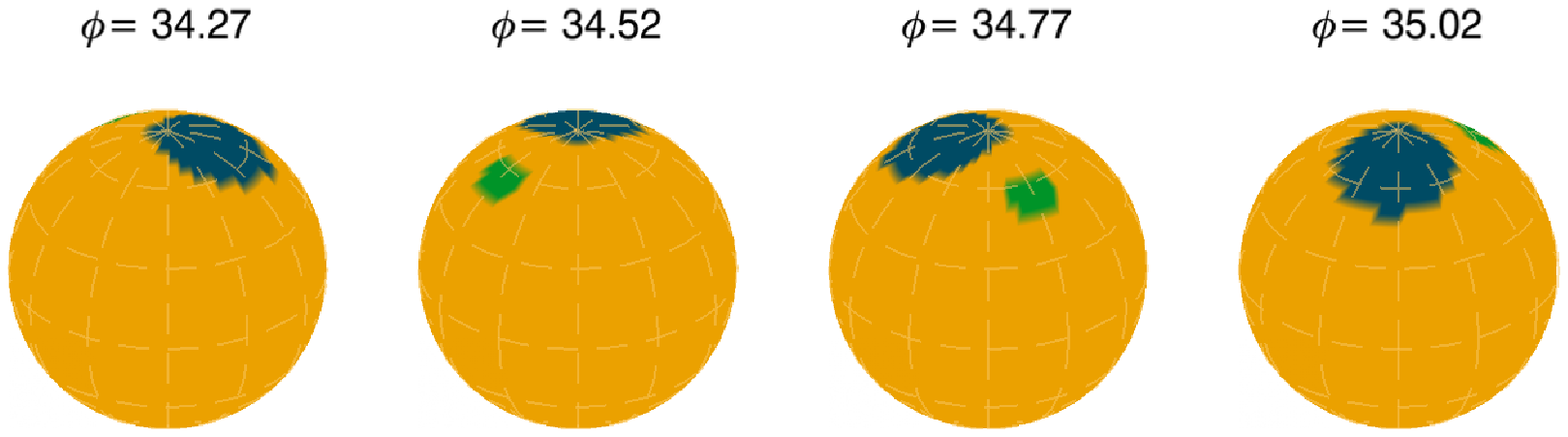}}
\end{minipage}\hspace{1ex}
\begin{minipage}{1.0\textwidth}
\captionsetup[subfigure]{labelfont=bf,textfont=bf,singlelinecheck=off,justification=raggedright,
position=top}
\subfloat[]{\includegraphics[width=250pt]{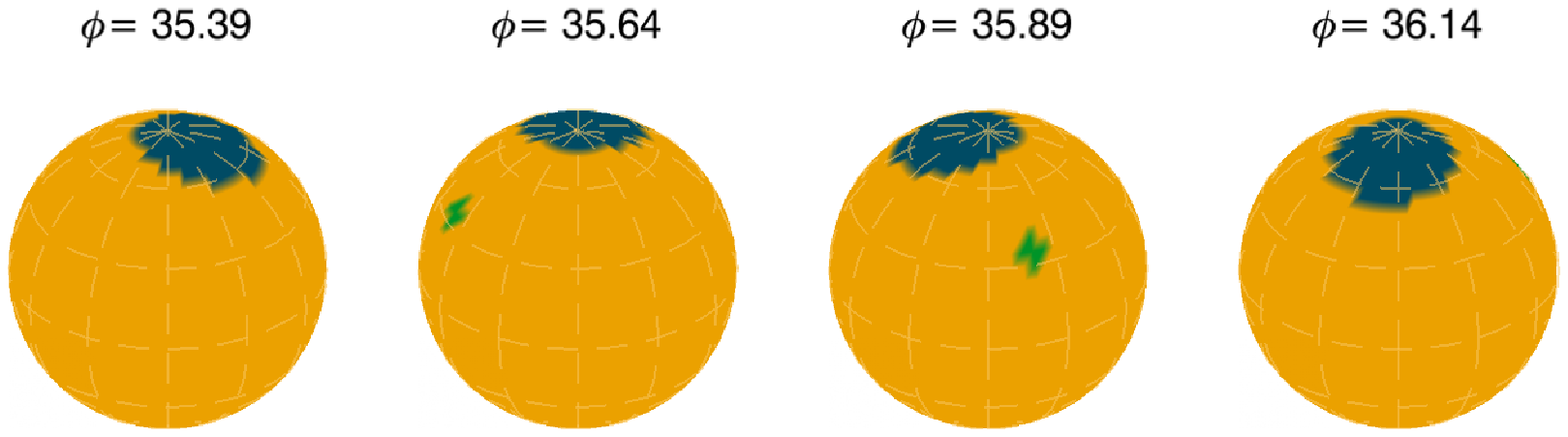}}
\end{minipage}\hspace{1ex}
\begin{minipage}{1.0\textwidth}
\captionsetup[subfigure]{labelfont=bf,textfont=bf,singlelinecheck=off,justification=raggedright,
position=top}
\subfloat[]{\includegraphics[width=250pt]{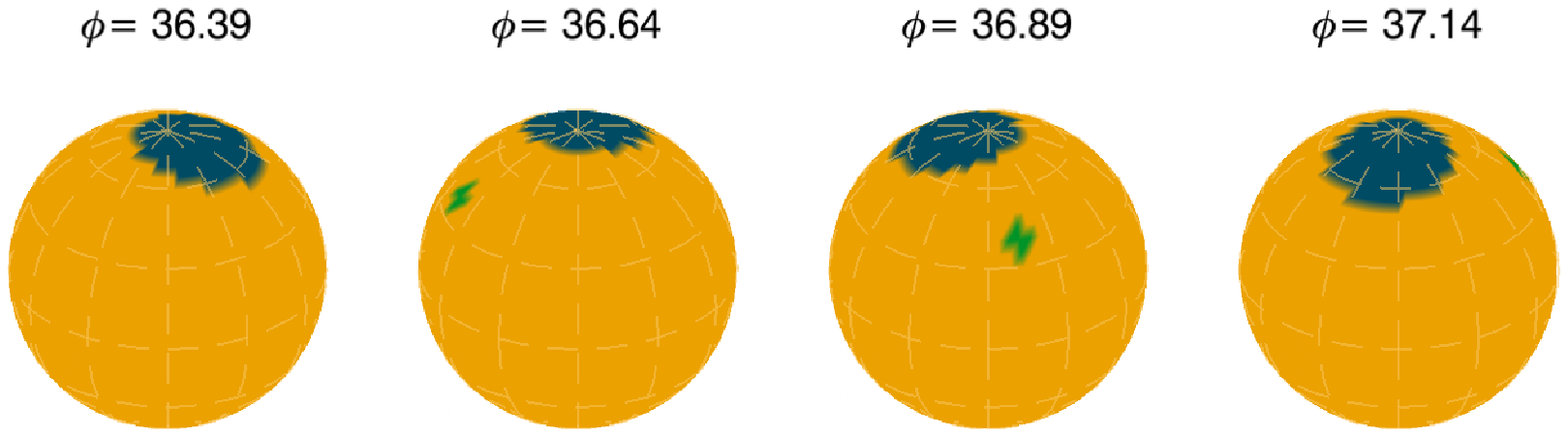}}
\end{minipage}\hspace{1ex}
\begin{minipage}{1.0\textwidth}
\captionsetup[subfigure]{labelfont=bf,textfont=bf,singlelinecheck=off,justification=raggedright,
position=top}
\subfloat[]{\includegraphics[width=250pt]{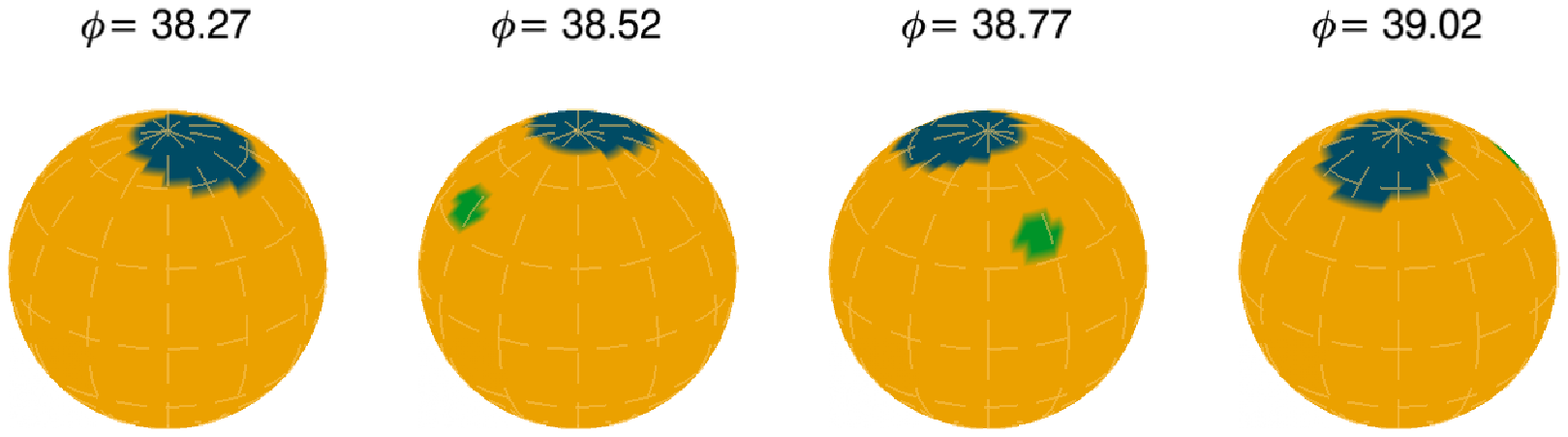}}
\end{minipage}\hspace{1ex}
\caption{Spot-model fits of the Doppler images in Fig.~\ref{fig:di_season_3}. Otherwise as in
Fig.~\ref{fig:di_season_2_fit}.}
\label{fig:di_season_3_fit}
\end{figure}

In Fig.~\ref{fig:di_season_3} seven almost consecutive Doppler images are shown, which cover around
nine rotations. In Fig.~\ref{fig:di_season_3_fit} the spot-model fits of the Doppler images are
shown.

\subsection{Season 2009/10}

\begin{figure}[!t]
\begin{minipage}{1.0\textwidth}
\captionsetup[subfigure]{labelfont=bf,textfont=bf,singlelinecheck=off,justification=raggedright,
position=top}
\subfloat[]{\includegraphics[width=250pt]{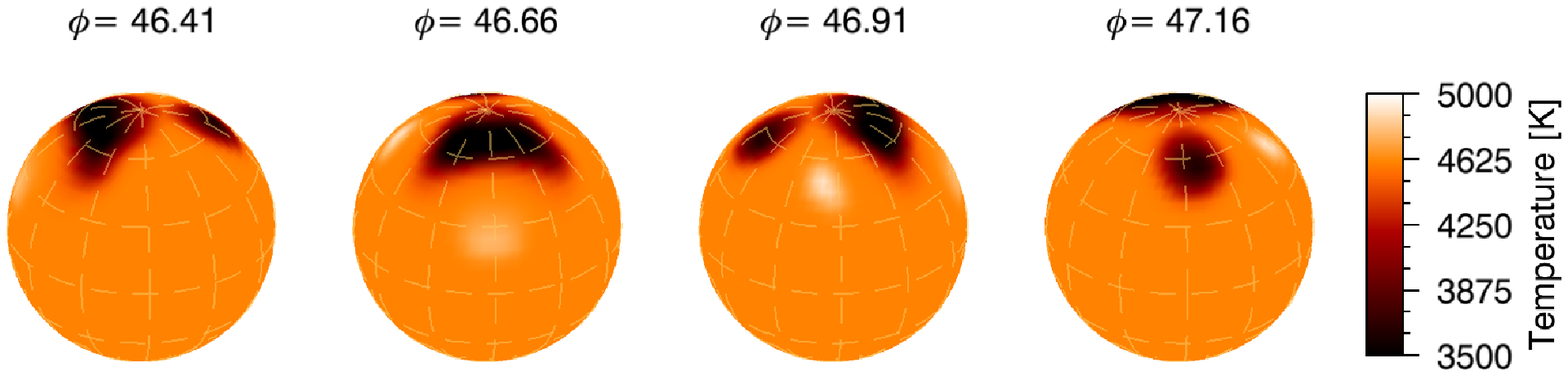}}
\end{minipage}\hspace{1ex}
\begin{minipage}{1.0\textwidth}
\captionsetup[subfigure]{labelfont=bf,textfont=bf,singlelinecheck=off,justification=raggedright,
position=top}
\subfloat[]{\includegraphics[width=250pt]{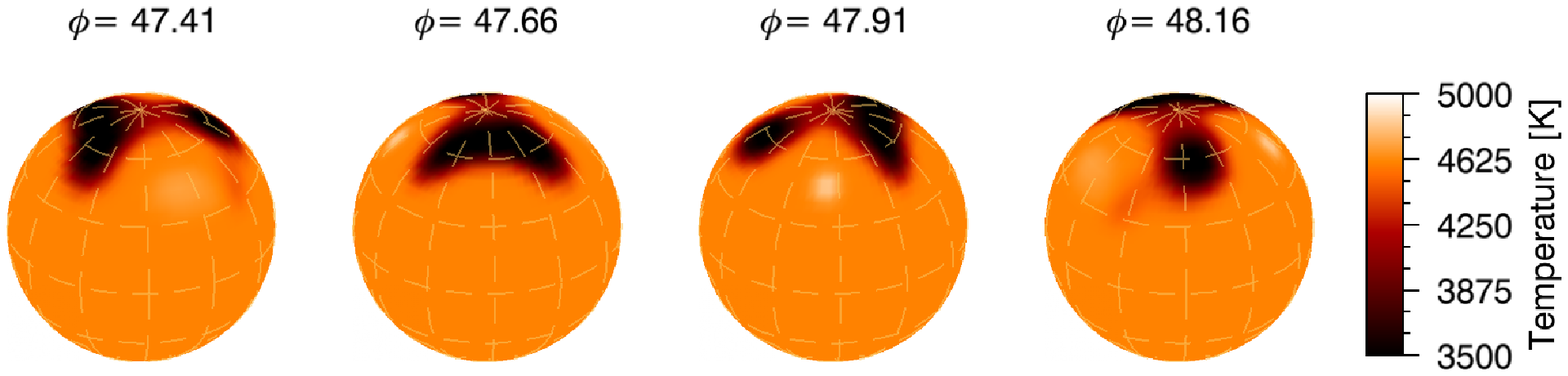}}
\end{minipage}\hspace{1ex}
\begin{minipage}{1.0\textwidth}
\captionsetup[subfigure]{labelfont=bf,textfont=bf,singlelinecheck=off,justification=raggedright,
position=top}
\subfloat[]{\includegraphics[width=250pt]{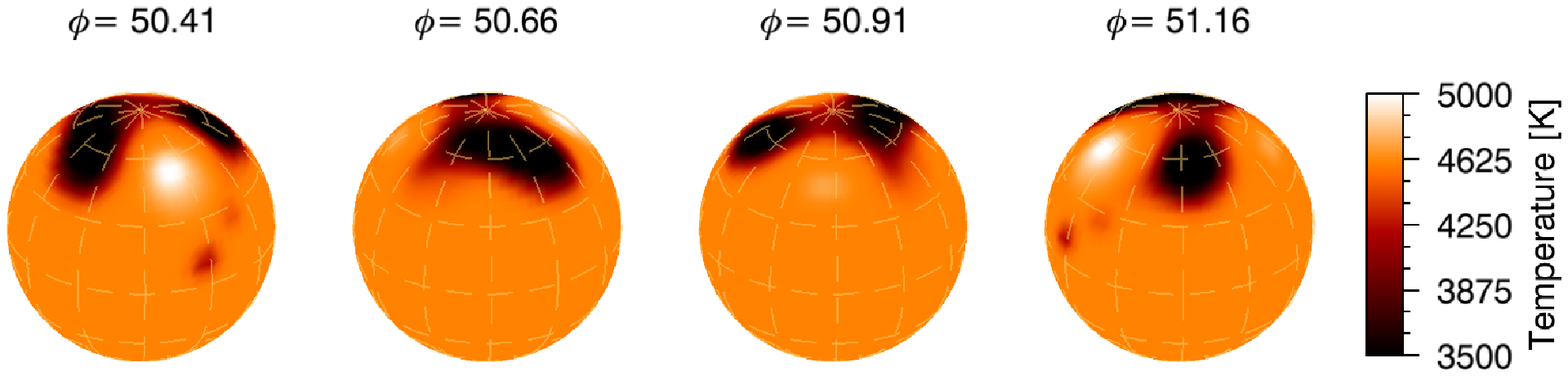}}
\end{minipage}\hspace{1ex}
\begin{minipage}{1.0\textwidth}
\captionsetup[subfigure]{labelfont=bf,textfont=bf,singlelinecheck=off,justification=raggedright,
position=top}
\subfloat[]{\includegraphics[width=250pt]{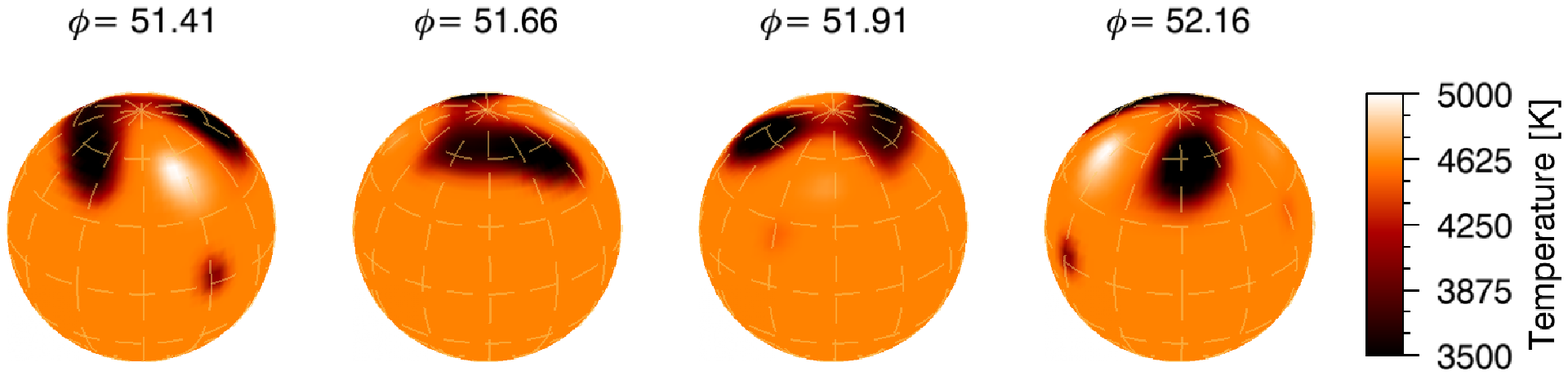}}
\end{minipage}\hspace{1ex}
\begin{minipage}{1.0\textwidth}
\captionsetup[subfigure]{labelfont=bf,textfont=bf,singlelinecheck=off,justification=raggedright,
position=top}
\subfloat[]{\includegraphics[width=250pt]{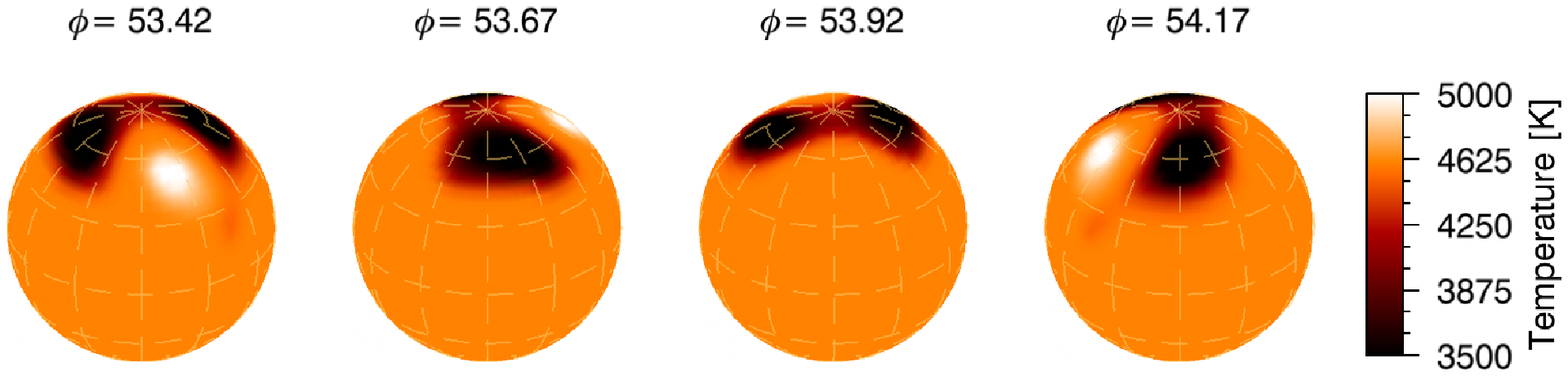}}
\end{minipage}\hspace{1ex}
\caption{Doppler images of XX~Tri for the observing season 2009/10. Otherwise as in
Fig.~\ref{fig:di_season_2}.}
\label{fig:di_season_4}
\end{figure}

\begin{figure}[!t]
\begin{minipage}{1.0\textwidth}
\captionsetup[subfigure]{labelfont=bf,textfont=bf,singlelinecheck=off,justification=raggedright,
position=top}
\subfloat[]{\includegraphics[width=250pt]{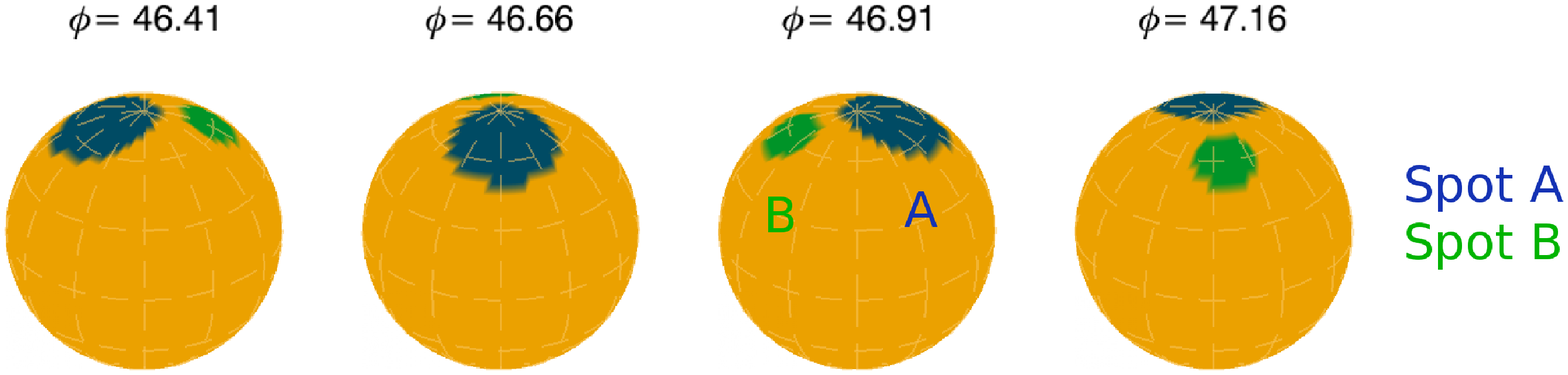}}
\end{minipage}\hspace{1ex}
\begin{minipage}{1.0\textwidth}
\captionsetup[subfigure]{labelfont=bf,textfont=bf,singlelinecheck=off,justification=raggedright,
position=top}
\subfloat[]{\includegraphics[width=250pt]{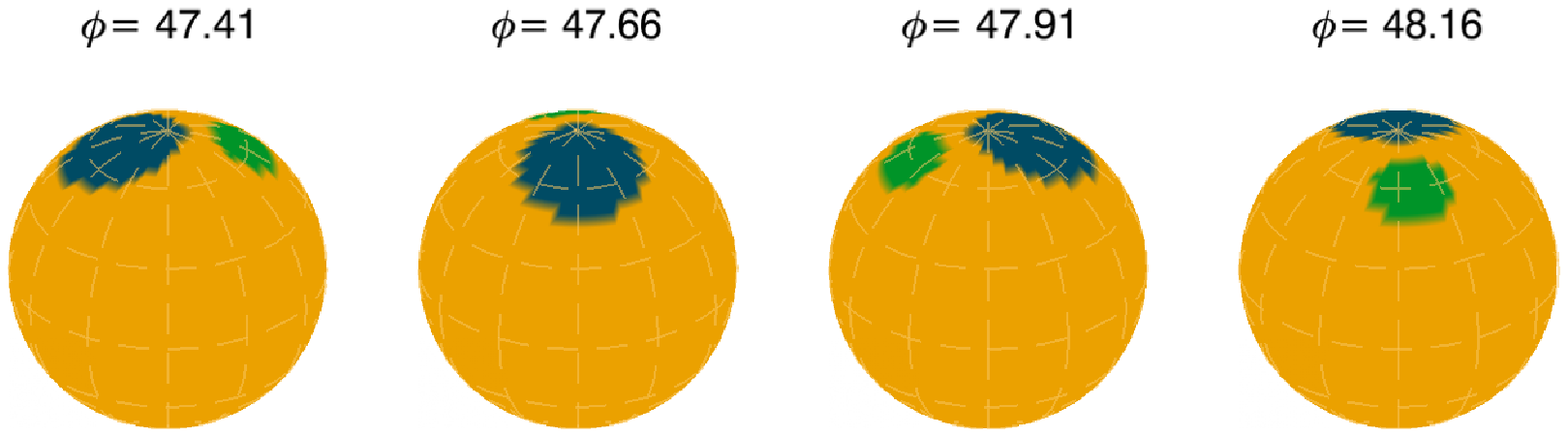}}
\end{minipage}\hspace{1ex}
\begin{minipage}{1.0\textwidth}
\captionsetup[subfigure]{labelfont=bf,textfont=bf,singlelinecheck=off,justification=raggedright,
position=top}
\subfloat[]{\includegraphics[width=250pt]{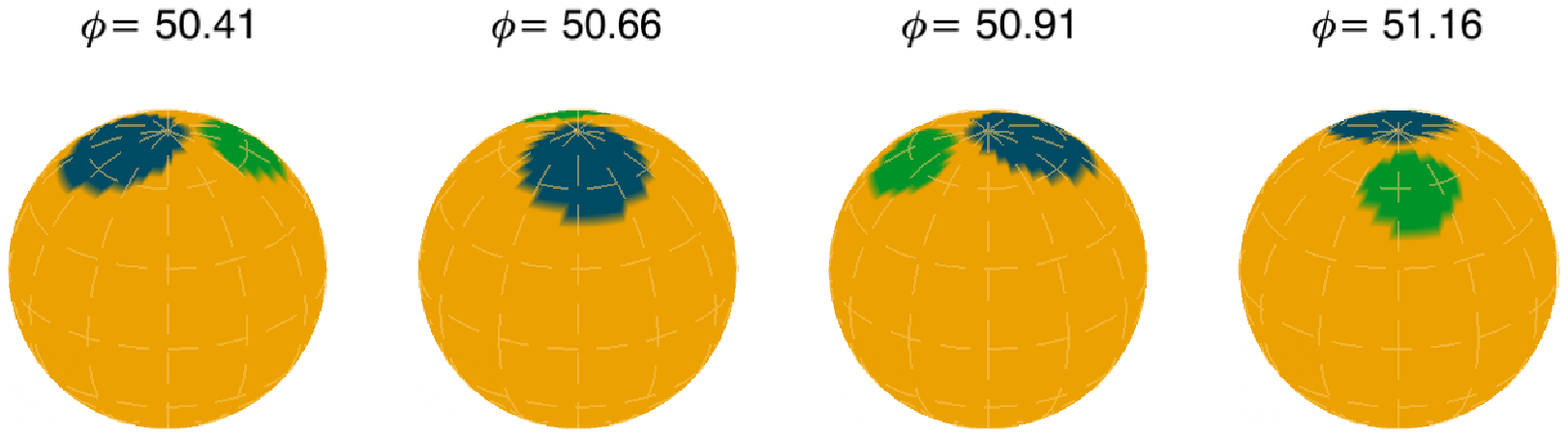}}
\end{minipage}\hspace{1ex}
\begin{minipage}{1.0\textwidth}
\captionsetup[subfigure]{labelfont=bf,textfont=bf,singlelinecheck=off,justification=raggedright,
position=top}
\subfloat[]{\includegraphics[width=250pt]{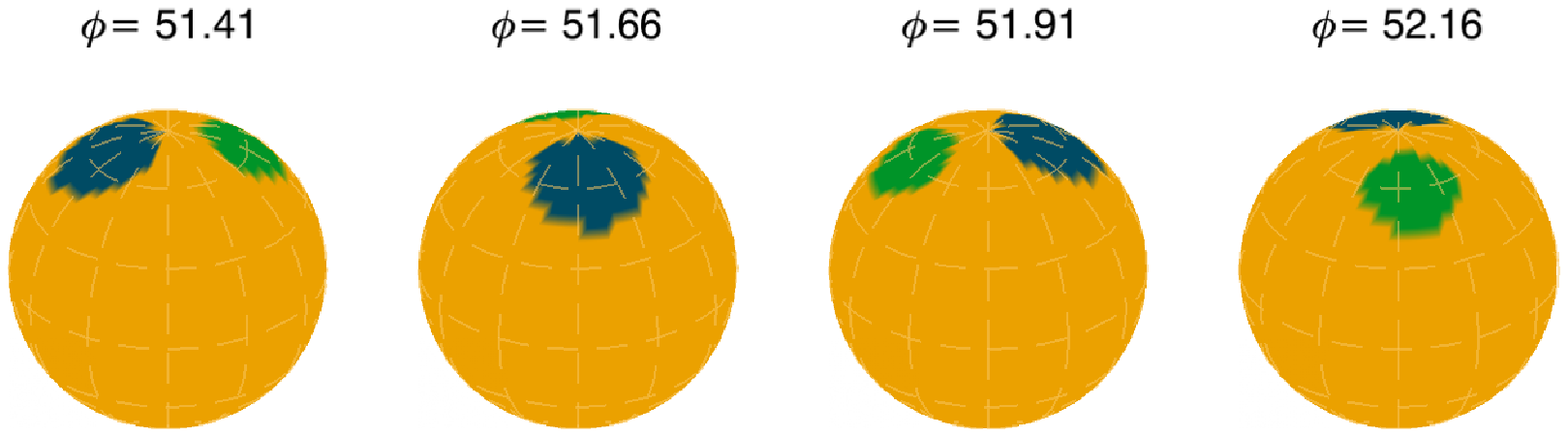}}
\end{minipage}\hspace{1ex}
\begin{minipage}{1.0\textwidth}
\captionsetup[subfigure]{labelfont=bf,textfont=bf,singlelinecheck=off,justification=raggedright,
position=top}
\subfloat[]{\includegraphics[width=250pt]{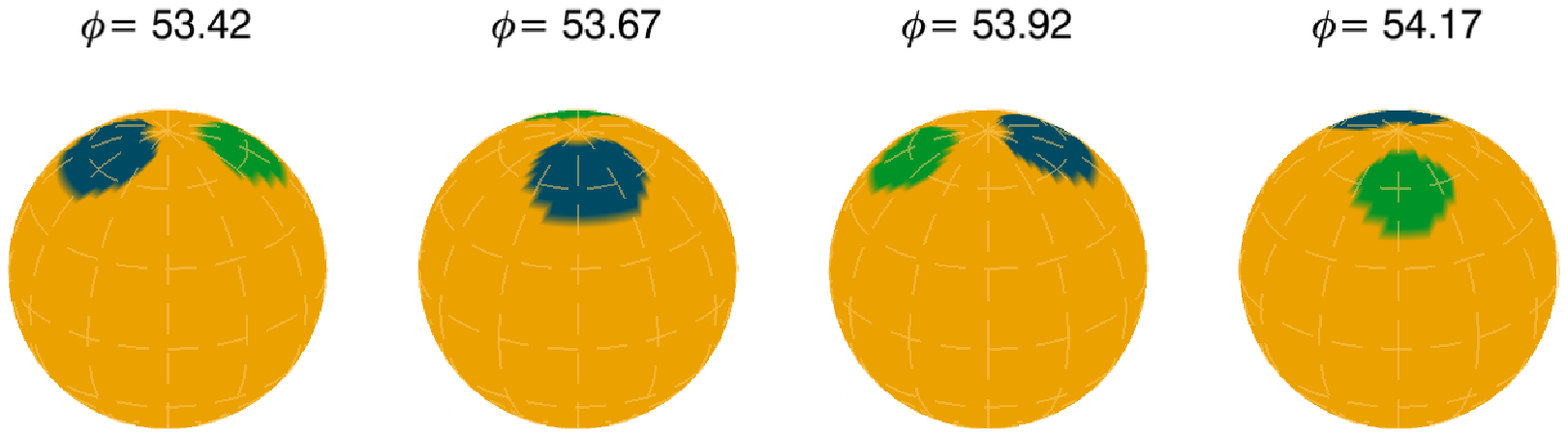}}
\end{minipage}\hspace{1ex}
\caption{Spot-model fits of the Doppler images in Fig.~\ref{fig:di_season_4}. Otherwise as in
Fig.~\ref{fig:di_season_2_fit}.}
\label{fig:di_season_4_fit}
\end{figure}

In Fig.~\ref{fig:di_season_4} five almost consecutive Doppler images are shown, which cover around
eight rotations. In Fig.~\ref{fig:di_season_4_fit} the spot-model fits of the Doppler images are
shown.

\subsection{Season 2010/11}

\begin{figure}[!t]
\begin{minipage}{1.0\textwidth}
\captionsetup[subfigure]{labelfont=bf,textfont=bf,singlelinecheck=off,justification=raggedright,
position=top}
\subfloat[]{\includegraphics[width=250pt]{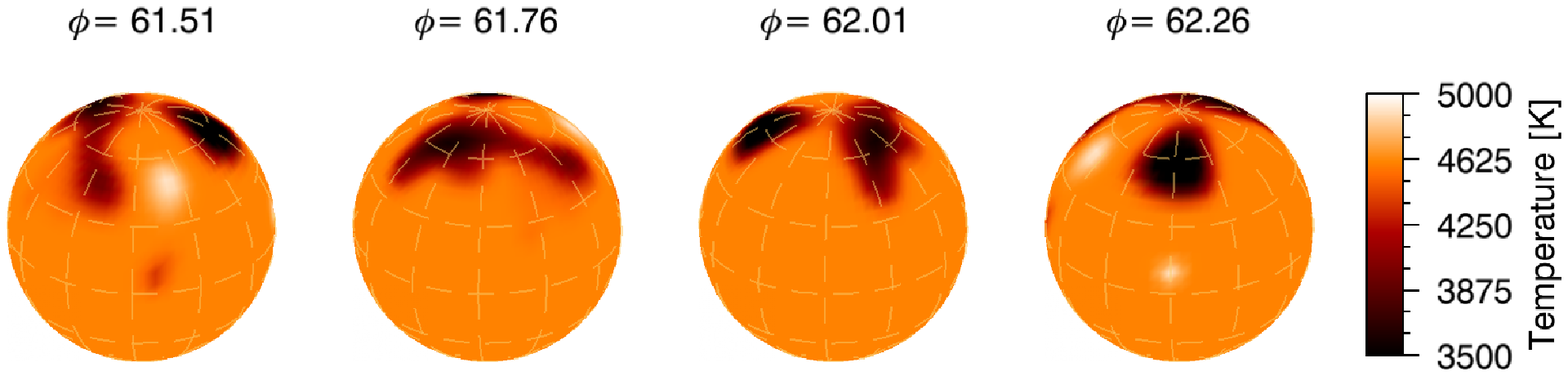}}
\end{minipage}\hspace{1ex}
\begin{minipage}{1.0\textwidth}
\captionsetup[subfigure]{labelfont=bf,textfont=bf,singlelinecheck=off,justification=raggedright,
position=top}
\subfloat[]{\includegraphics[width=250pt]{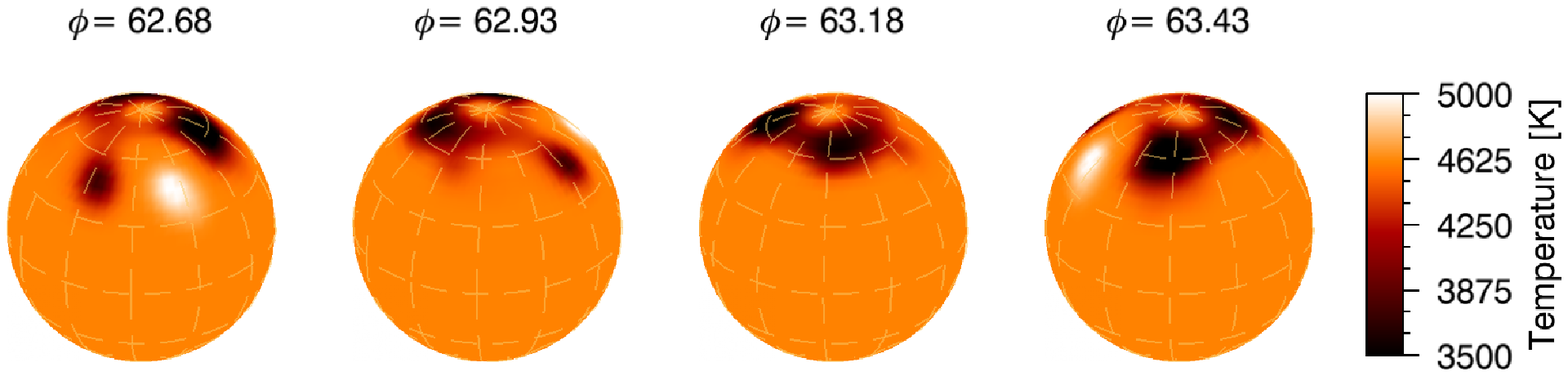}}
\end{minipage}\hspace{1ex}
\begin{minipage}{1.0\textwidth}
\captionsetup[subfigure]{labelfont=bf,textfont=bf,singlelinecheck=off,justification=raggedright,
position=top}
\subfloat[]{\includegraphics[width=250pt]{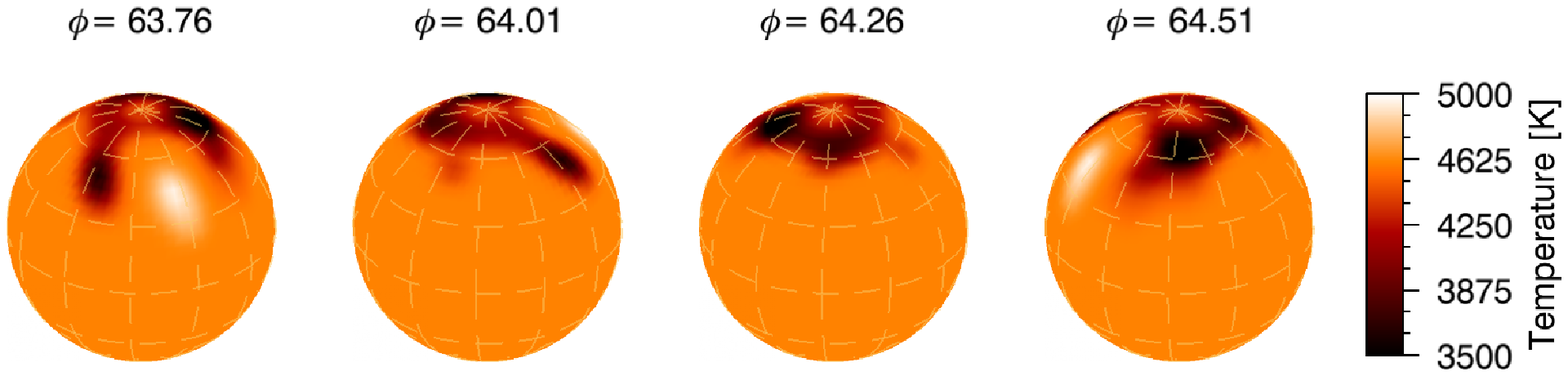}}
\end{minipage}\hspace{1ex}
\begin{minipage}{1.0\textwidth}
\captionsetup[subfigure]{labelfont=bf,textfont=bf,singlelinecheck=off,justification=raggedright,
position=top}
\subfloat[]{\includegraphics[width=250pt]{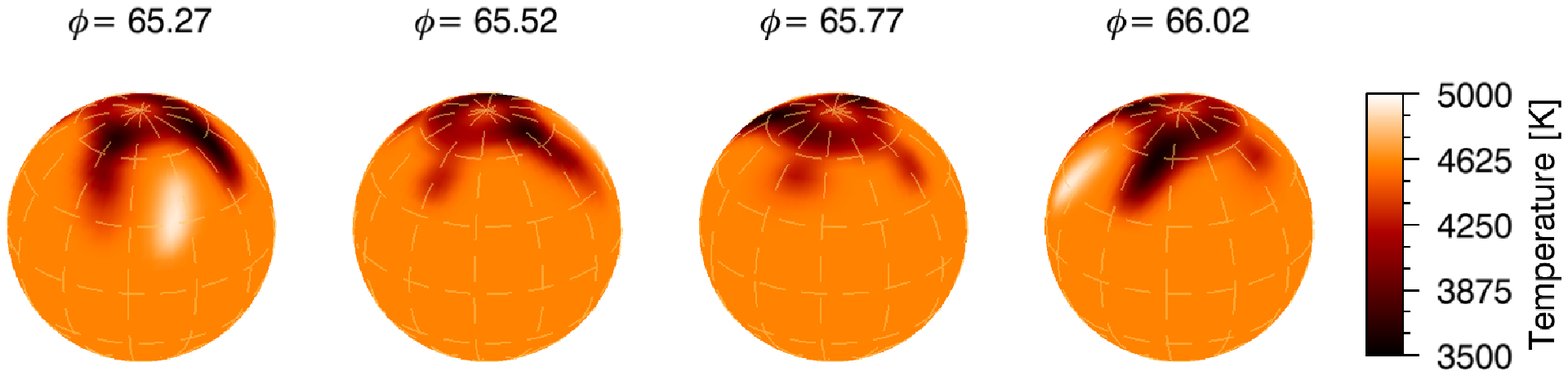}}
\end{minipage}\hspace{1ex}
\begin{minipage}{1.0\textwidth}
\captionsetup[subfigure]{labelfont=bf,textfont=bf,singlelinecheck=off,justification=raggedright,
position=top}
\subfloat[]{\includegraphics[width=250pt]{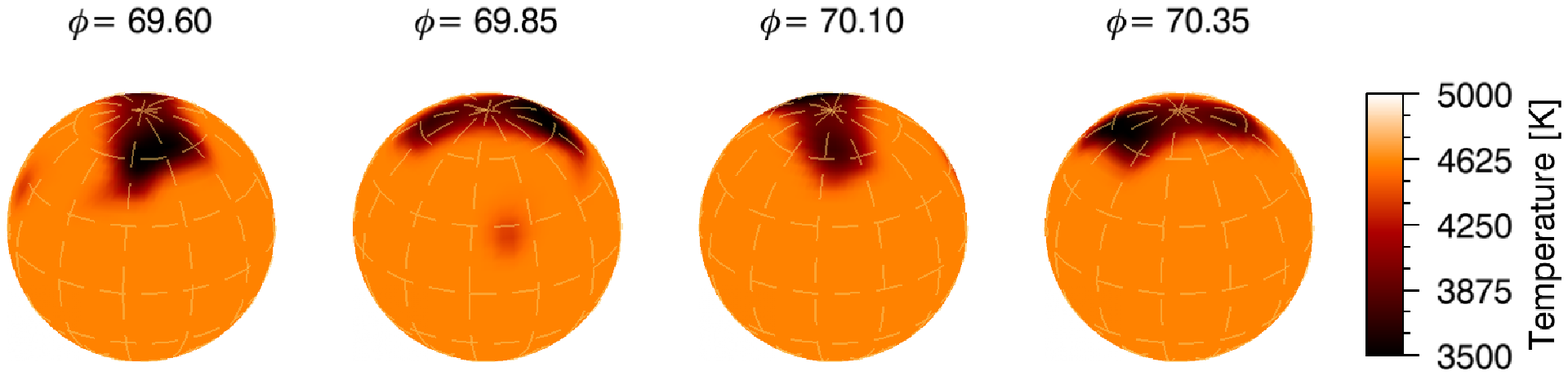}}
\end{minipage}\hspace{1ex}
\caption{Doppler images of XX~Tri for the observing season 2010/11. Otherwise as in
Fig.~\ref{fig:di_season_2}.}
\label{fig:di_season_5}
\end{figure}

\begin{figure}[!t]
\begin{minipage}{1.0\textwidth}
\captionsetup[subfigure]{labelfont=bf,textfont=bf,singlelinecheck=off,justification=raggedright,
position=top}
\subfloat[]{\includegraphics[width=250pt]{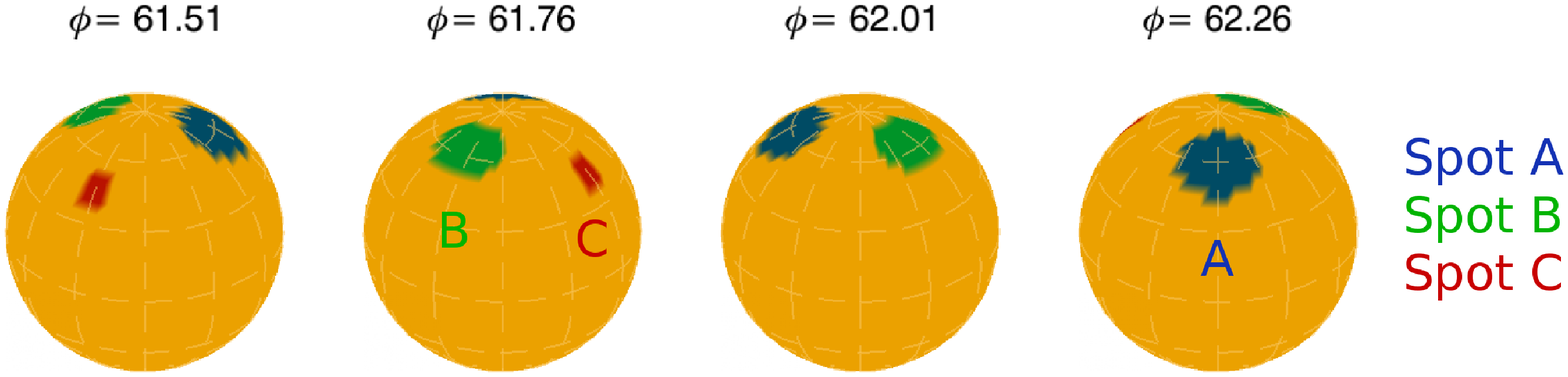}}
\end{minipage}\hspace{1ex}
\begin{minipage}{1.0\textwidth}
\captionsetup[subfigure]{labelfont=bf,textfont=bf,singlelinecheck=off,justification=raggedright,
position=top}
\subfloat[]{\includegraphics[width=250pt]{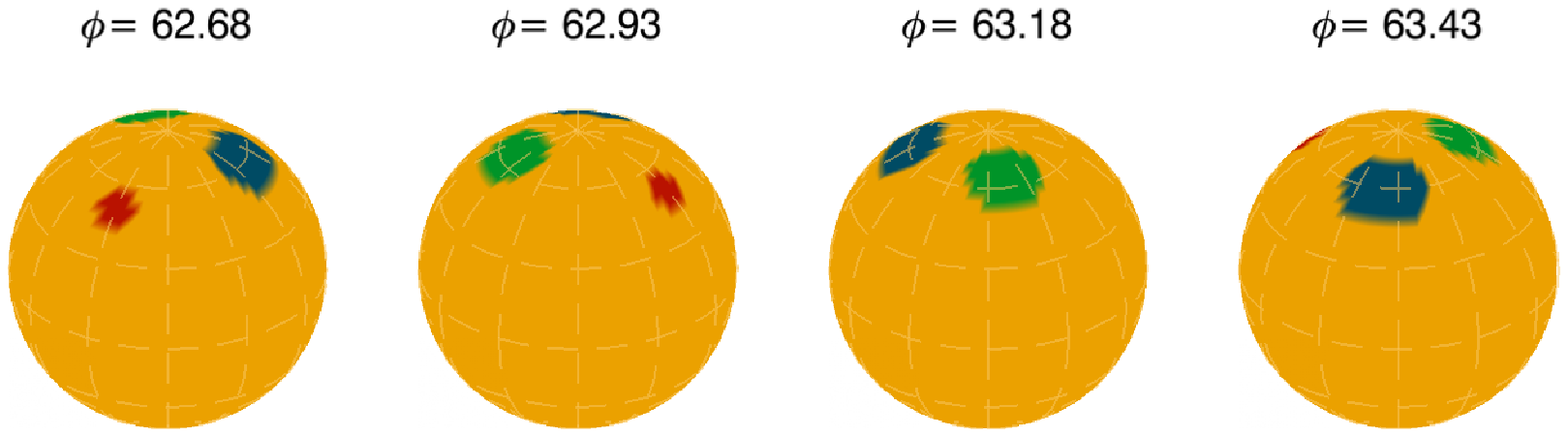}}
\end{minipage}\hspace{1ex}
\begin{minipage}{1.0\textwidth}
\captionsetup[subfigure]{labelfont=bf,textfont=bf,singlelinecheck=off,justification=raggedright,
position=top}
\subfloat[]{\includegraphics[width=250pt]{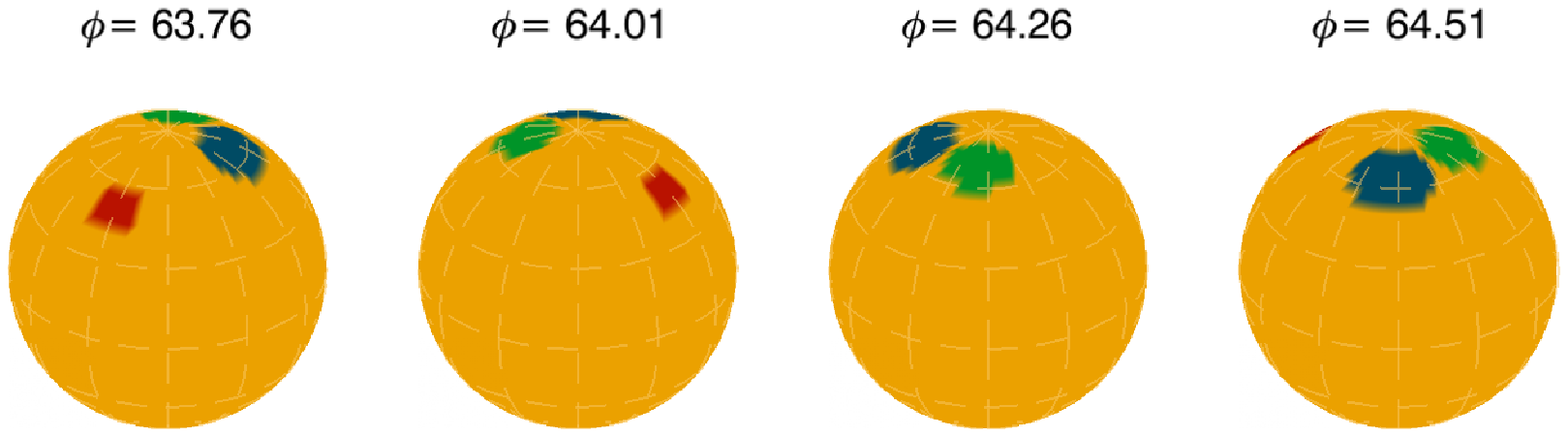}}
\end{minipage}\hspace{1ex}
\begin{minipage}{1.0\textwidth}
\captionsetup[subfigure]{labelfont=bf,textfont=bf,singlelinecheck=off,justification=raggedright,
position=top}
\subfloat[]{\includegraphics[width=250pt]{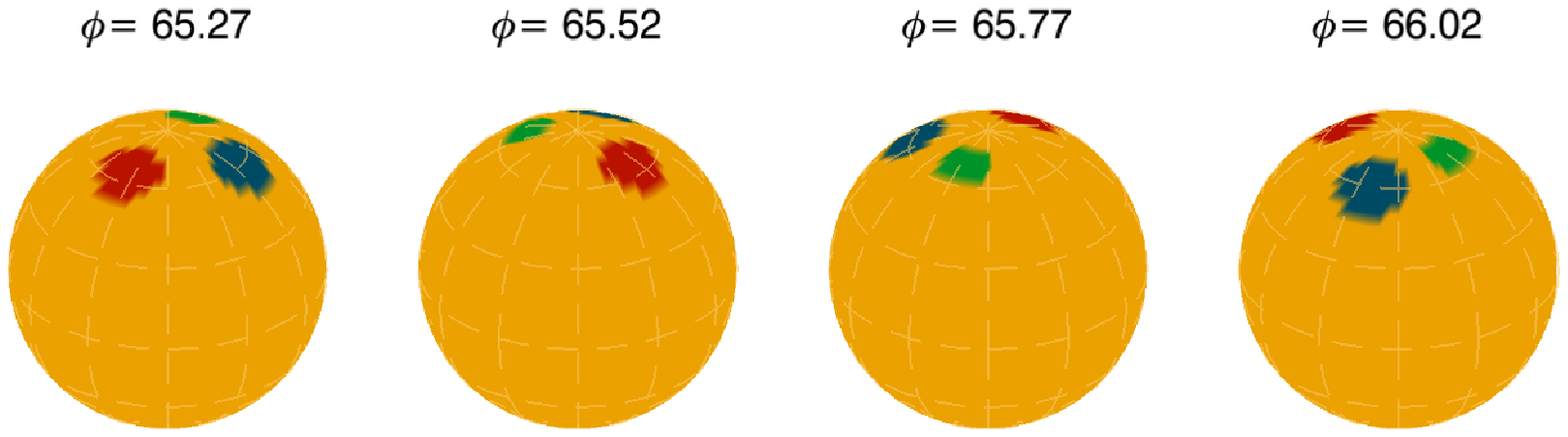}}
\end{minipage}\hspace{1ex}
\begin{minipage}{1.0\textwidth}
\captionsetup[subfigure]{labelfont=bf,textfont=bf,singlelinecheck=off,justification=raggedright,
position=top}
\subfloat[]{\includegraphics[width=250pt]{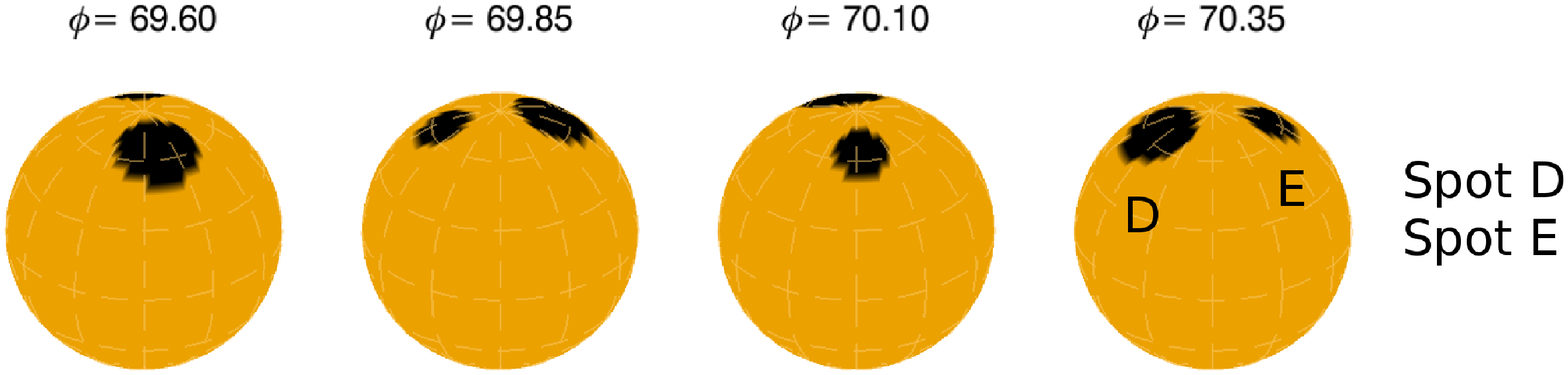}}
\end{minipage}\hspace{1ex}
\caption{Spot-model fits of the Doppler images in Fig.~\ref{fig:di_season_5}. Otherwise as in
Fig.~\ref{fig:di_season_2_fit}.}
\label{fig:di_season_5_fit}
\end{figure}

In Fig.~\ref{fig:di_season_5} five almost consecutive Doppler images are shown, which cover around
nine rotations. In Fig.~\ref{fig:di_season_5_fit} the spot-model fits of the Doppler images are
shown.

\subsection{Season 2011/12}

\begin{figure}[!t]
\begin{minipage}{1.0\textwidth}
\captionsetup[subfigure]{labelfont=bf,textfont=bf,singlelinecheck=off,justification=raggedright,
position=top}
\subfloat[]{\includegraphics[width=250pt]{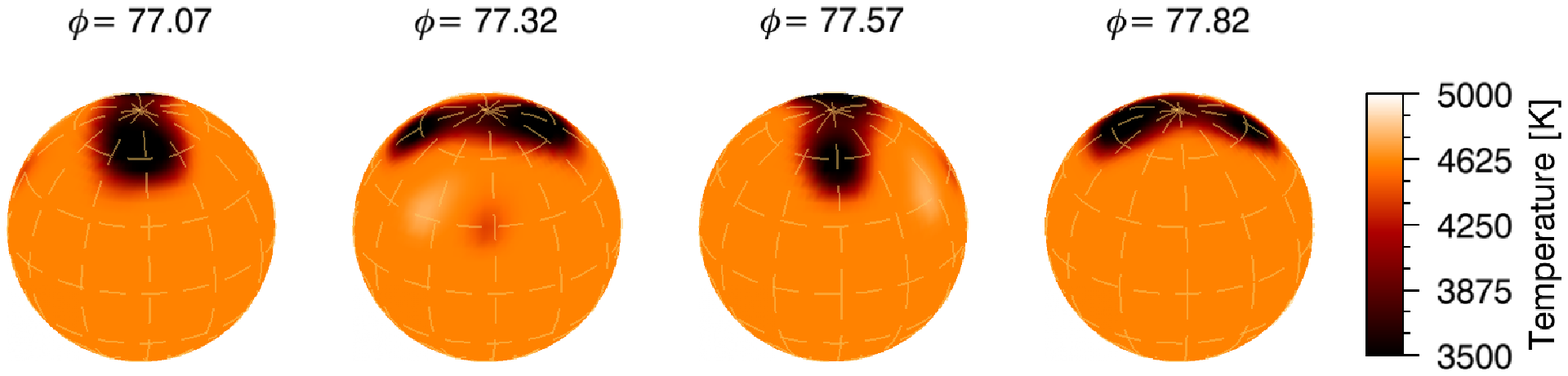}}
\end{minipage}\hspace{1ex}
\begin{minipage}{1.0\textwidth}
\captionsetup[subfigure]{labelfont=bf,textfont=bf,singlelinecheck=off,justification=raggedright,
position=top}
\subfloat[]{\includegraphics[width=250pt]{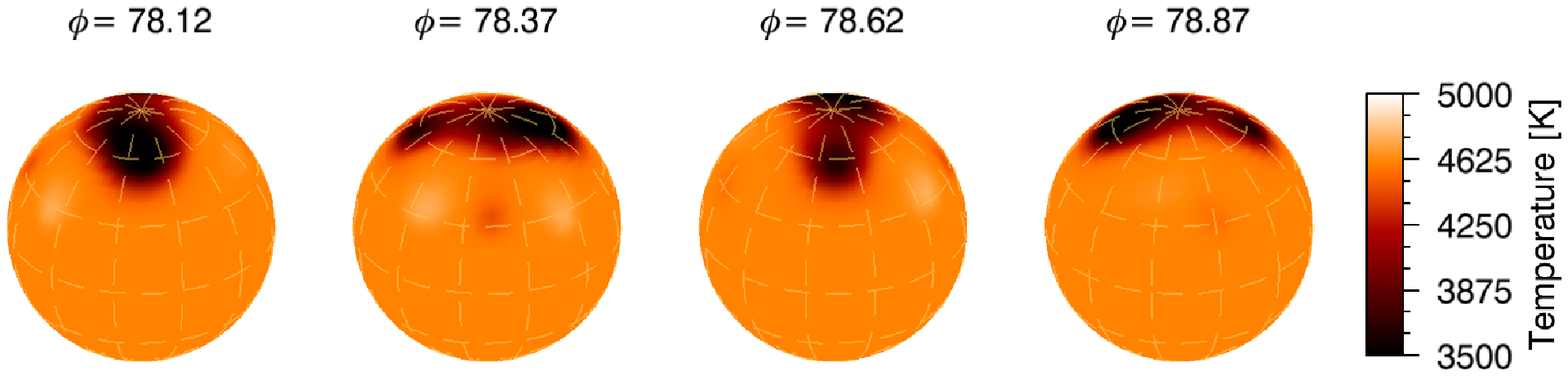}}
\end{minipage}\hspace{1ex}
\begin{minipage}{1.0\textwidth}
\captionsetup[subfigure]{labelfont=bf,textfont=bf,singlelinecheck=off,justification=raggedright,
position=top}
\subfloat[]{\includegraphics[width=250pt]{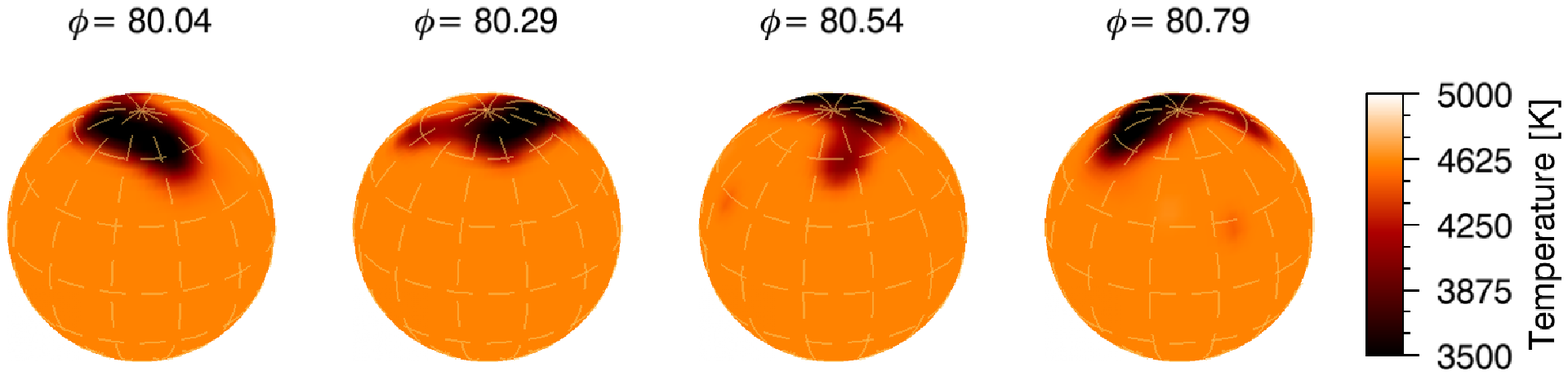}}
\end{minipage}\hspace{1ex}
\begin{minipage}{1.0\textwidth}
\captionsetup[subfigure]{labelfont=bf,textfont=bf,singlelinecheck=off,justification=raggedright,
position=top}
\subfloat[]{\includegraphics[width=250pt]{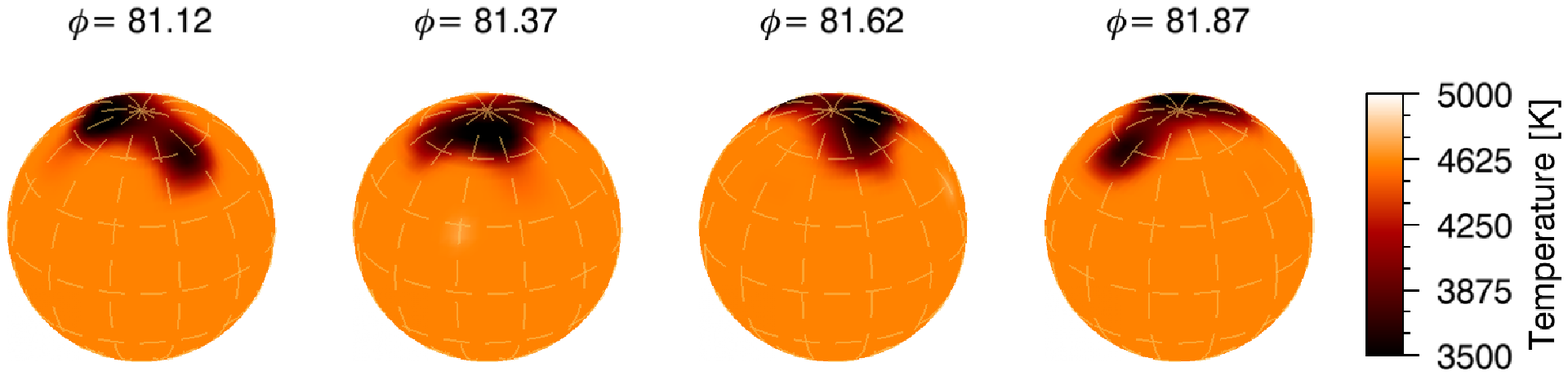}}
\end{minipage}\hspace{1ex}
\begin{minipage}{1.0\textwidth}
\captionsetup[subfigure]{labelfont=bf,textfont=bf,singlelinecheck=off,justification=raggedright,
position=top}
\subfloat[]{\includegraphics[width=250pt]{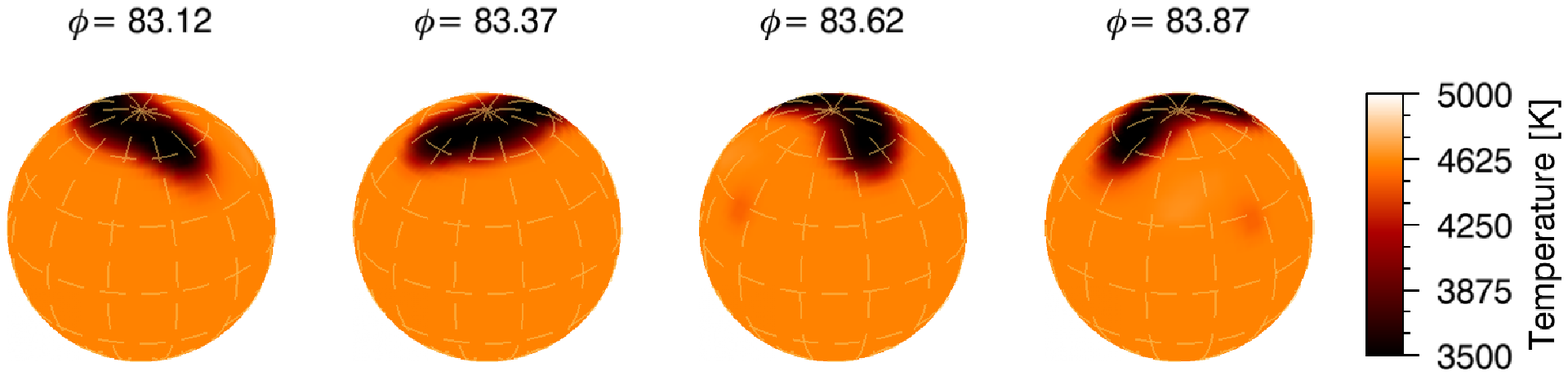}}
\end{minipage}\hspace{1ex}
\begin{minipage}{1.0\textwidth}
\captionsetup[subfigure]{labelfont=bf,textfont=bf,singlelinecheck=off,justification=raggedright,
position=top}
\subfloat[]{\includegraphics[width=250pt]{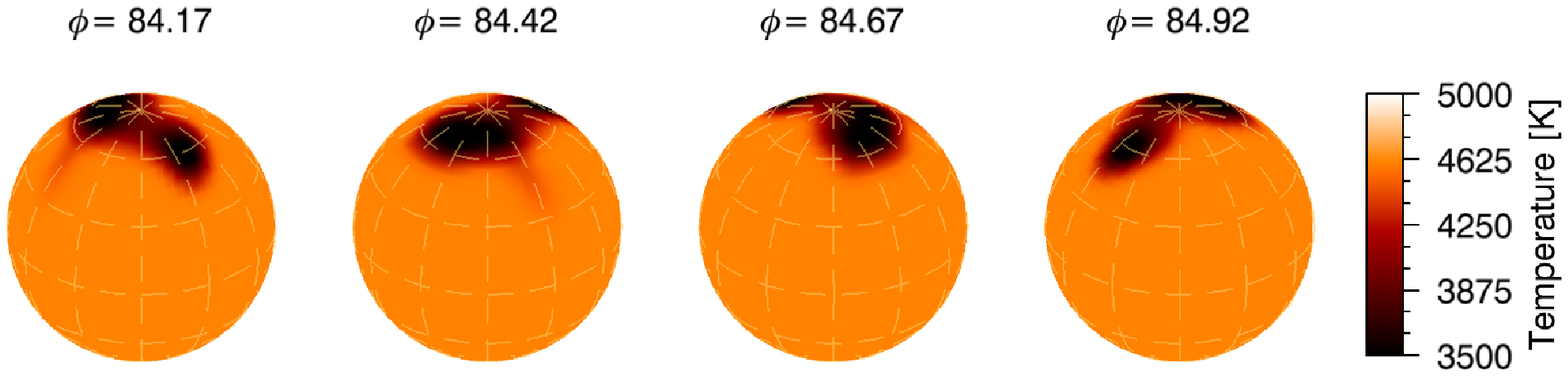}}
\end{minipage}\hspace{1ex}
\begin{minipage}{1.0\textwidth}
\captionsetup[subfigure]{labelfont=bf,textfont=bf,singlelinecheck=off,justification=raggedright,
position=top}
\subfloat[]{\includegraphics[width=250pt]{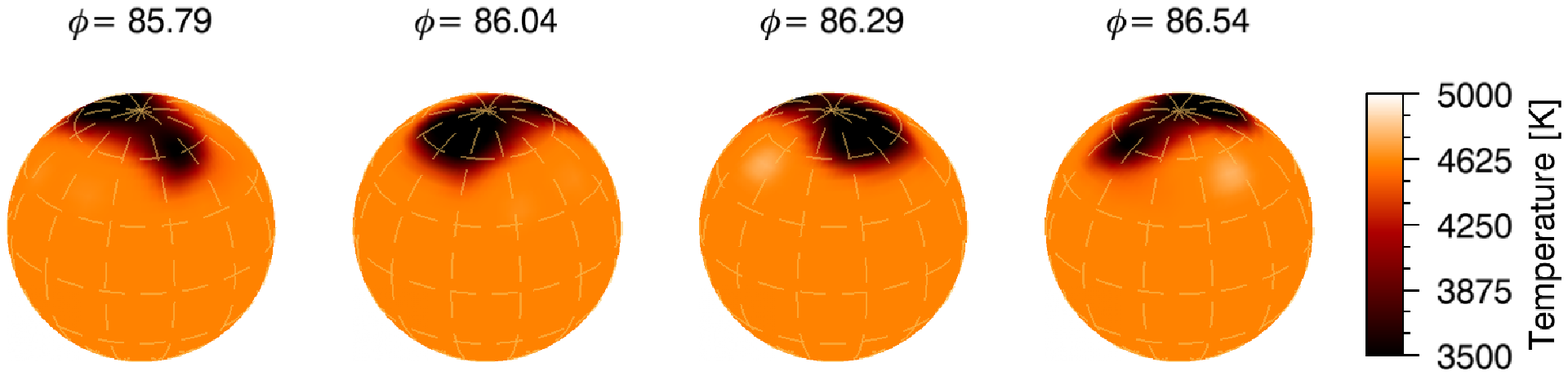}}
\end{minipage}\hspace{1ex}
\caption{Doppler images of XX~Tri for the observing season 2011/12. Otherwise as in
Fig.~\ref{fig:di_season_2}.}
\label{fig:di_season_6}
\end{figure}

\begin{figure}[!t]
\begin{minipage}{1.0\textwidth}
\captionsetup[subfigure]{labelfont=bf,textfont=bf,singlelinecheck=off,justification=raggedright,
position=top}
\subfloat[]{\includegraphics[width=250pt]{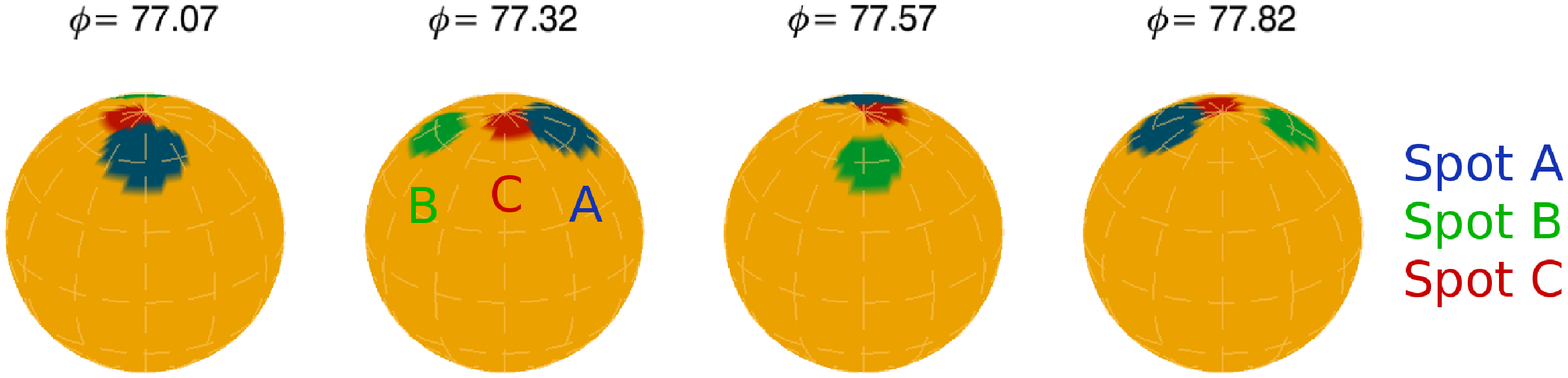}}
\end{minipage}\hspace{1ex}
\begin{minipage}{1.0\textwidth}
\captionsetup[subfigure]{labelfont=bf,textfont=bf,singlelinecheck=off,justification=raggedright,
position=top}
\subfloat[]{\includegraphics[width=250pt]{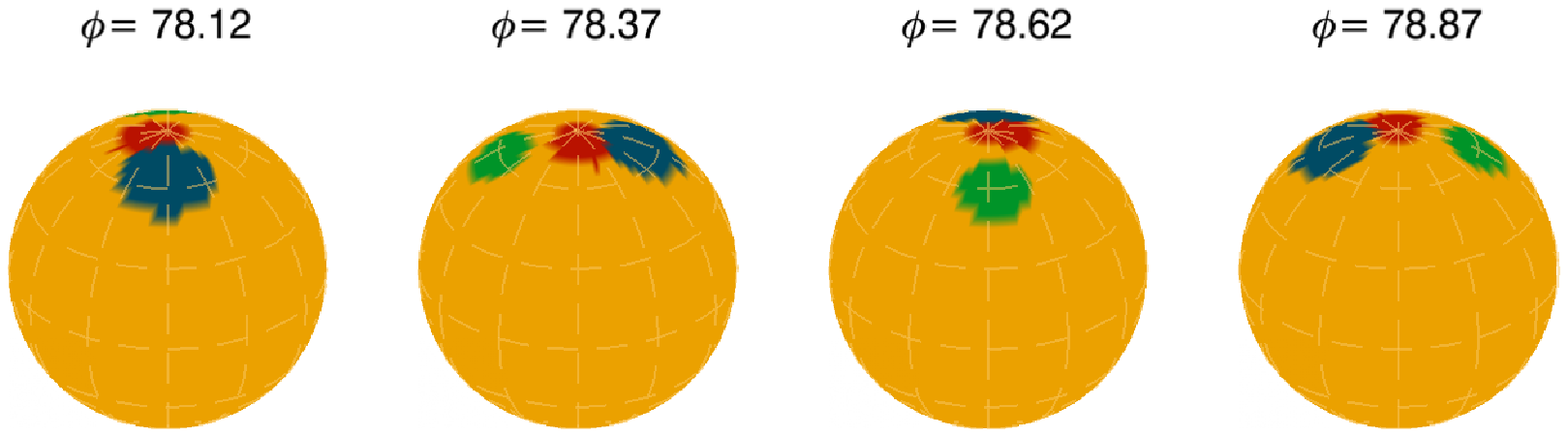}}
\end{minipage}\hspace{1ex}
\begin{minipage}{1.0\textwidth}
\captionsetup[subfigure]{labelfont=bf,textfont=bf,singlelinecheck=off,justification=raggedright,
position=top}
\subfloat[]{\includegraphics[width=250pt]{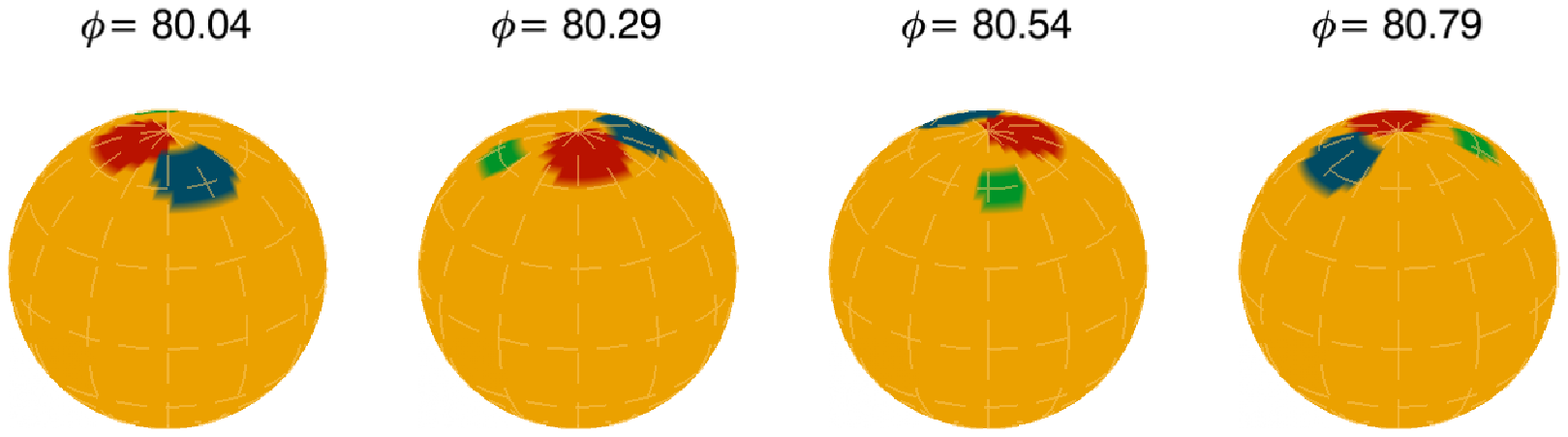}}
\end{minipage}\hspace{1ex}
\begin{minipage}{1.0\textwidth}
\captionsetup[subfigure]{labelfont=bf,textfont=bf,singlelinecheck=off,justification=raggedright,
position=top}
\subfloat[]{\includegraphics[width=250pt]{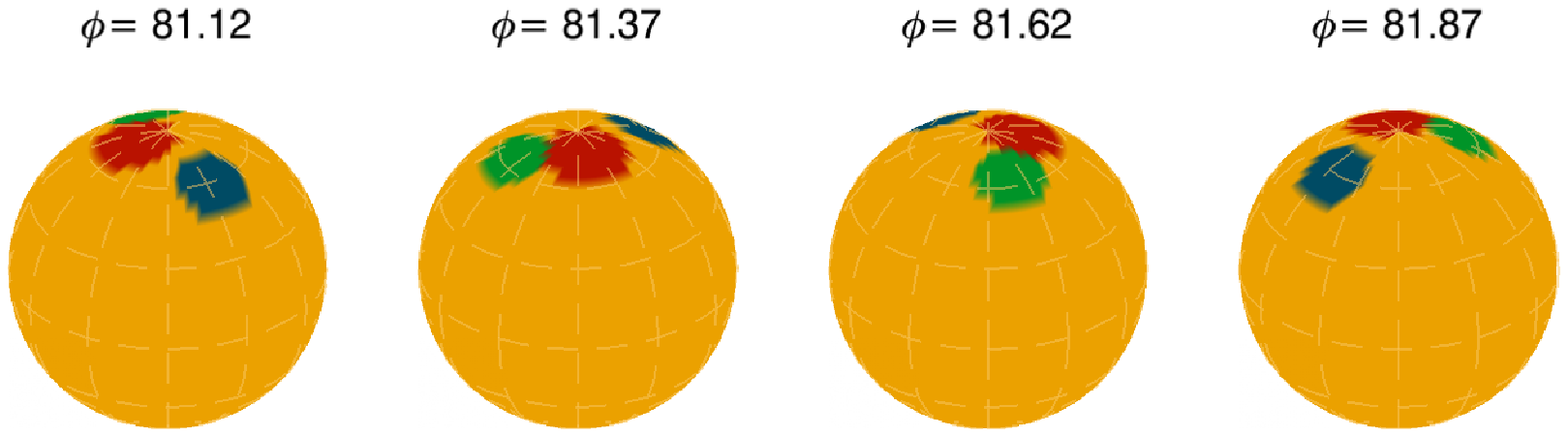}}
\end{minipage}\hspace{1ex}
\begin{minipage}{1.0\textwidth}
\captionsetup[subfigure]{labelfont=bf,textfont=bf,singlelinecheck=off,justification=raggedright,
position=top}
\subfloat[]{\includegraphics[width=250pt]{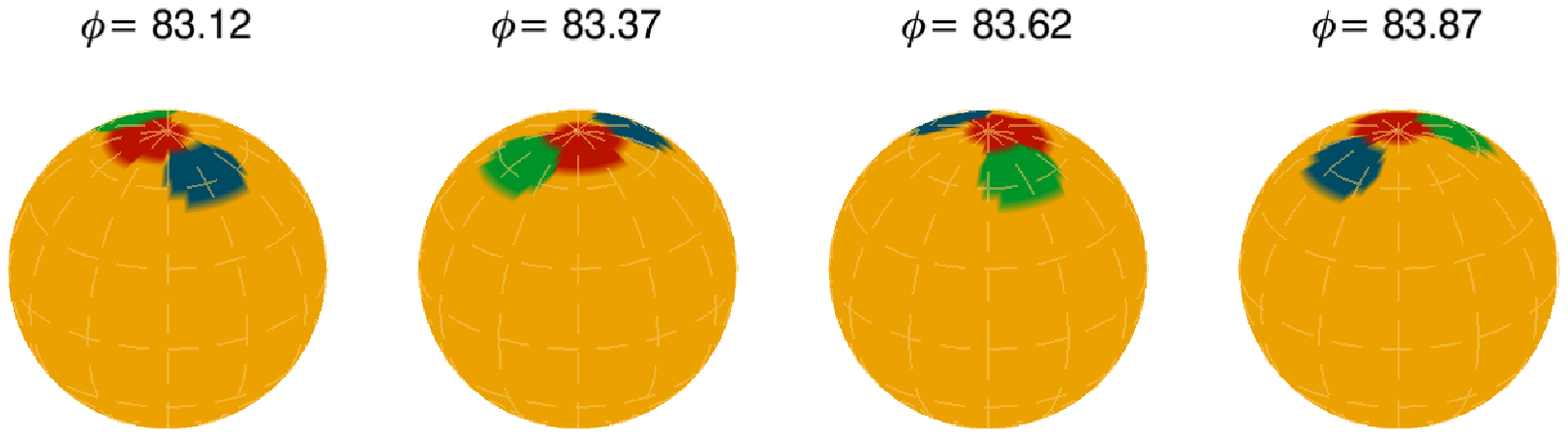}}
\end{minipage}\hspace{1ex}
\begin{minipage}{1.0\textwidth}
\captionsetup[subfigure]{labelfont=bf,textfont=bf,singlelinecheck=off,justification=raggedright,
position=top}
\subfloat[]{\includegraphics[width=250pt]{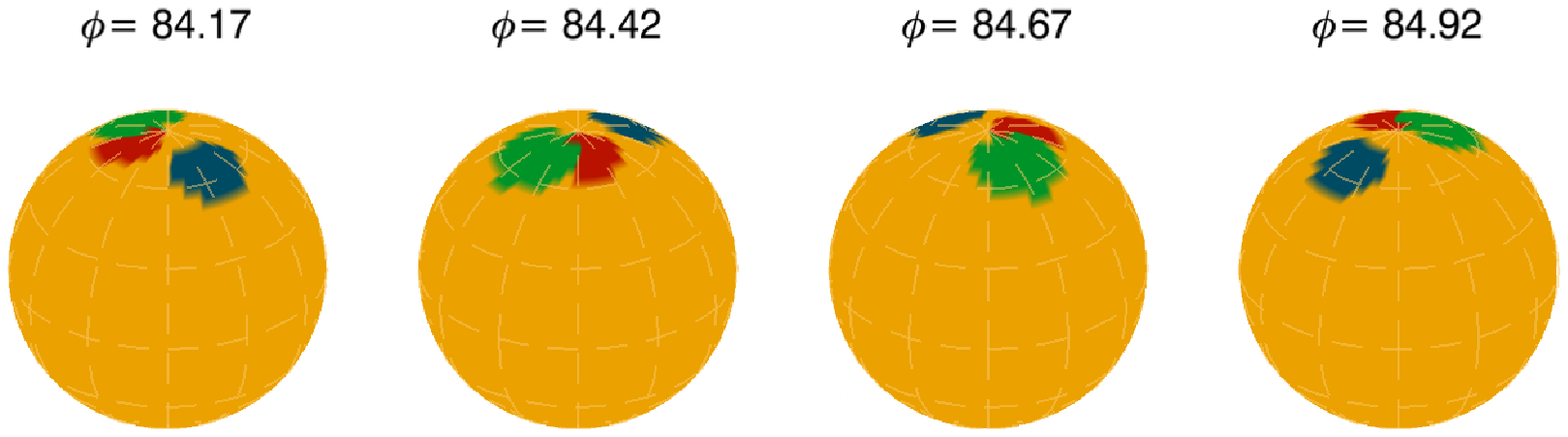}}
\end{minipage}\hspace{1ex}
\begin{minipage}{1.0\textwidth}
\captionsetup[subfigure]{labelfont=bf,textfont=bf,singlelinecheck=off,justification=raggedright,
position=top}
\subfloat[]{\includegraphics[width=250pt]{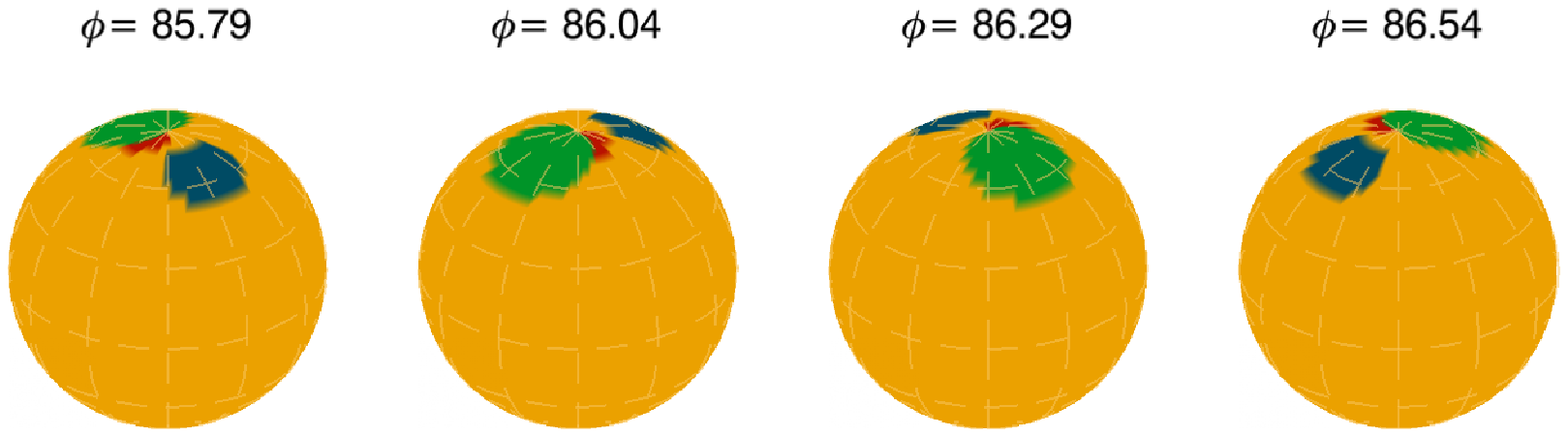}}
\end{minipage}\hspace{1ex}
\caption{Spot-model fits of the Doppler images in Fig.~\ref{fig:di_season_6}. Otherwise as in
Fig.~\ref{fig:di_season_2_fit}.}
\label{fig:di_season_6_fit}
\end{figure}

In Fig.~\ref{fig:di_season_6} seven almost consecutive Doppler images are shown, which cover around
ten rotations. In Fig.~\ref{fig:di_season_6_fit} the spot-model fits of the Doppler images are
shown.

\section{Line profiles of Doppler images for the seasons 2006/07 to 2011/12}

\begin{figure*}[!t]
\begin{minipage}{1.0\textwidth}
\captionsetup[subfigure]{labelfont=bf,textfont=bf,singlelinecheck=off,justification=raggedright,
position=top}
\subfloat[2006.58]{\includegraphics[width=90pt]{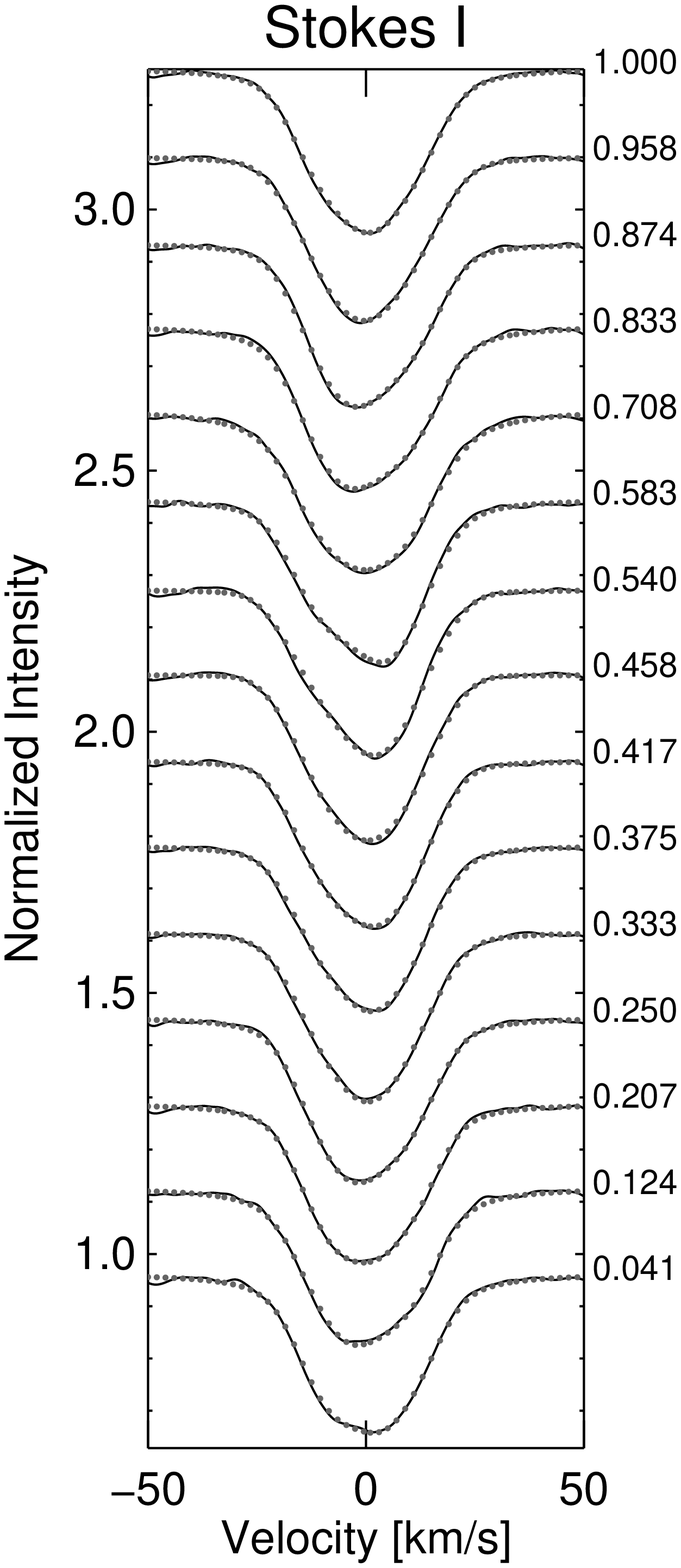}}
\subfloat[2006.64]{\includegraphics[width=90pt]{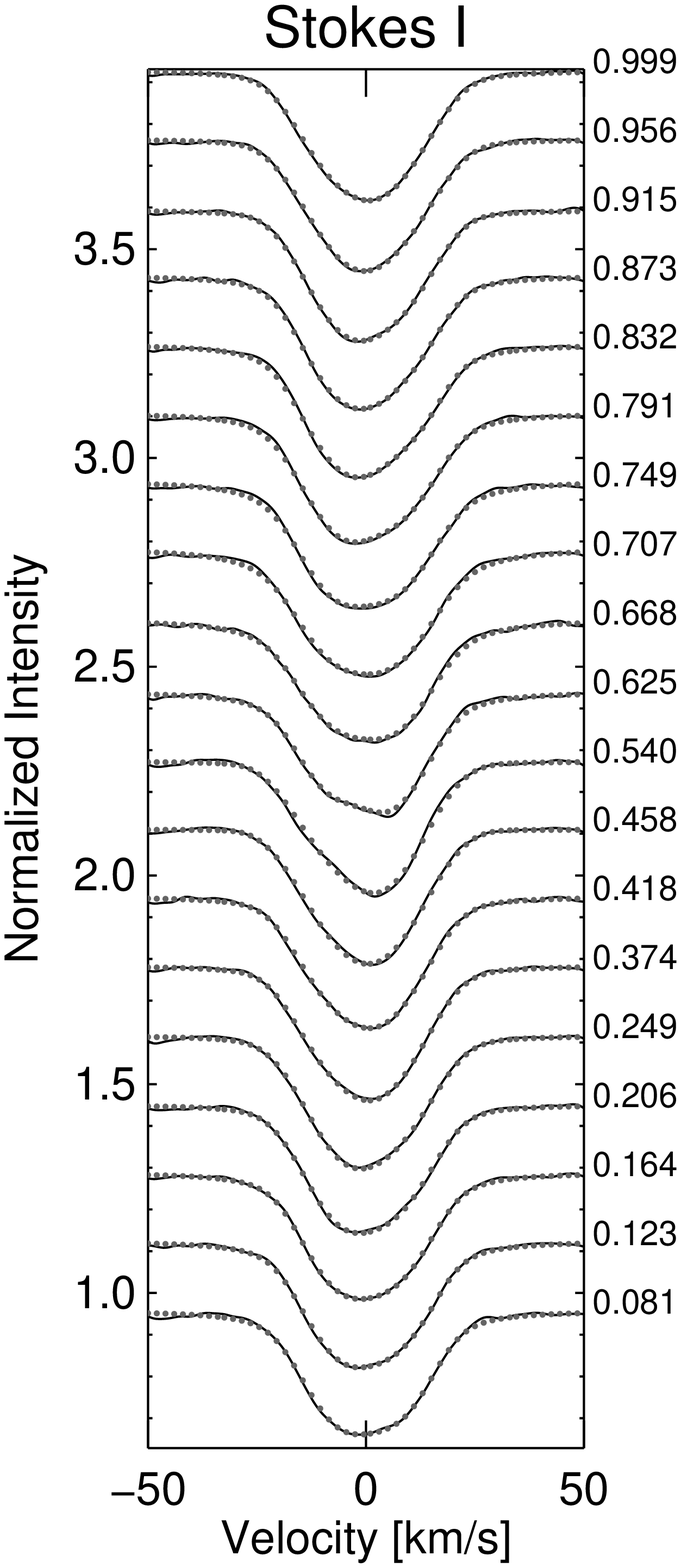}}
\subfloat[2006.71]{\includegraphics[width=90pt]{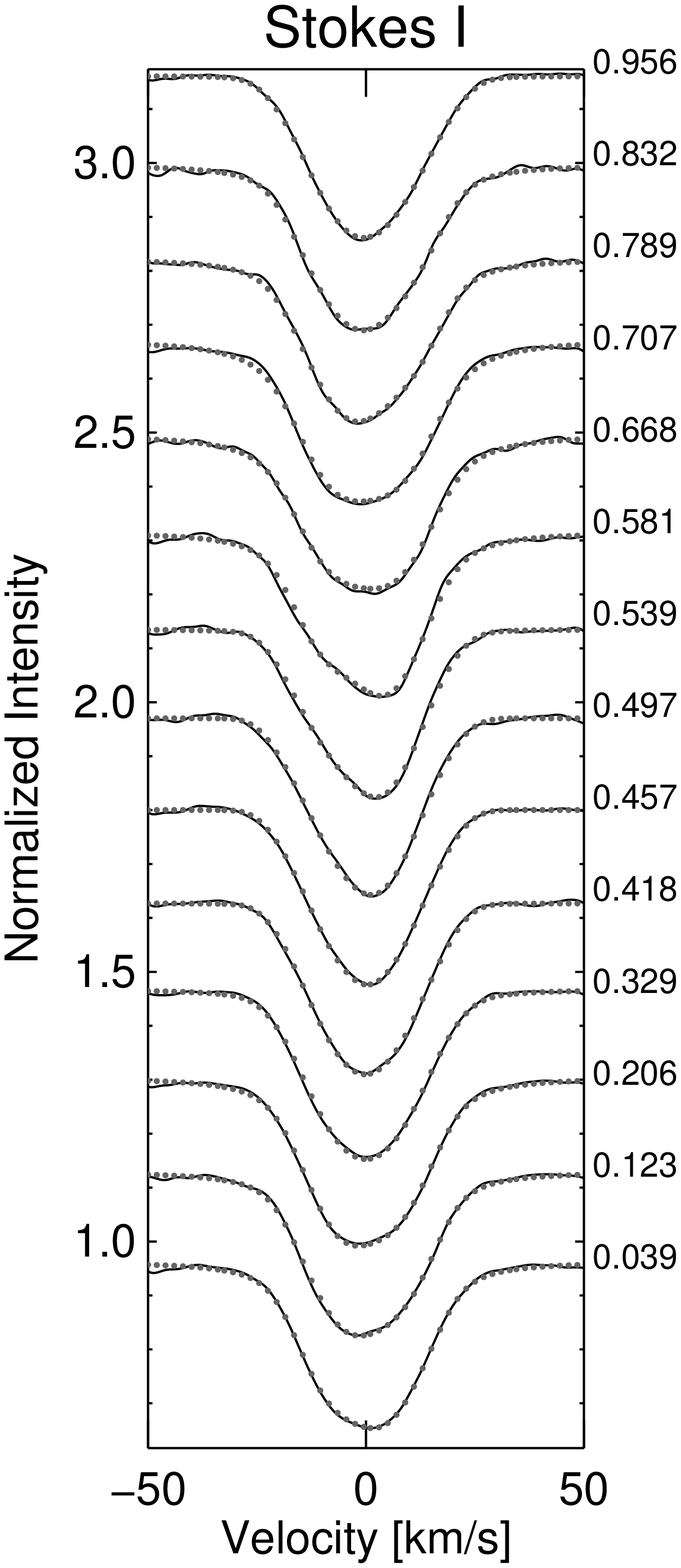}}
\subfloat[2006.78]{\includegraphics[width=90pt]{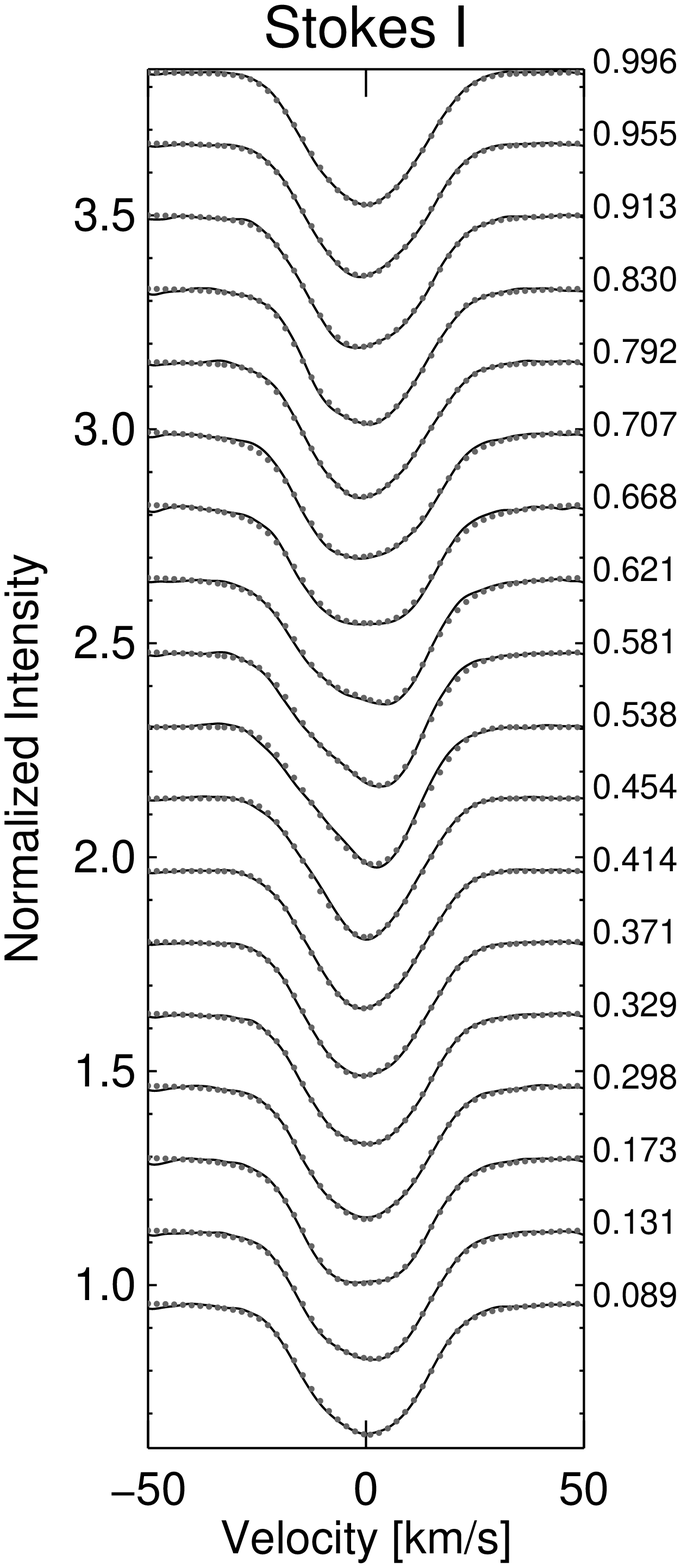}}
\subfloat[2006.91]{\includegraphics[width=90pt]{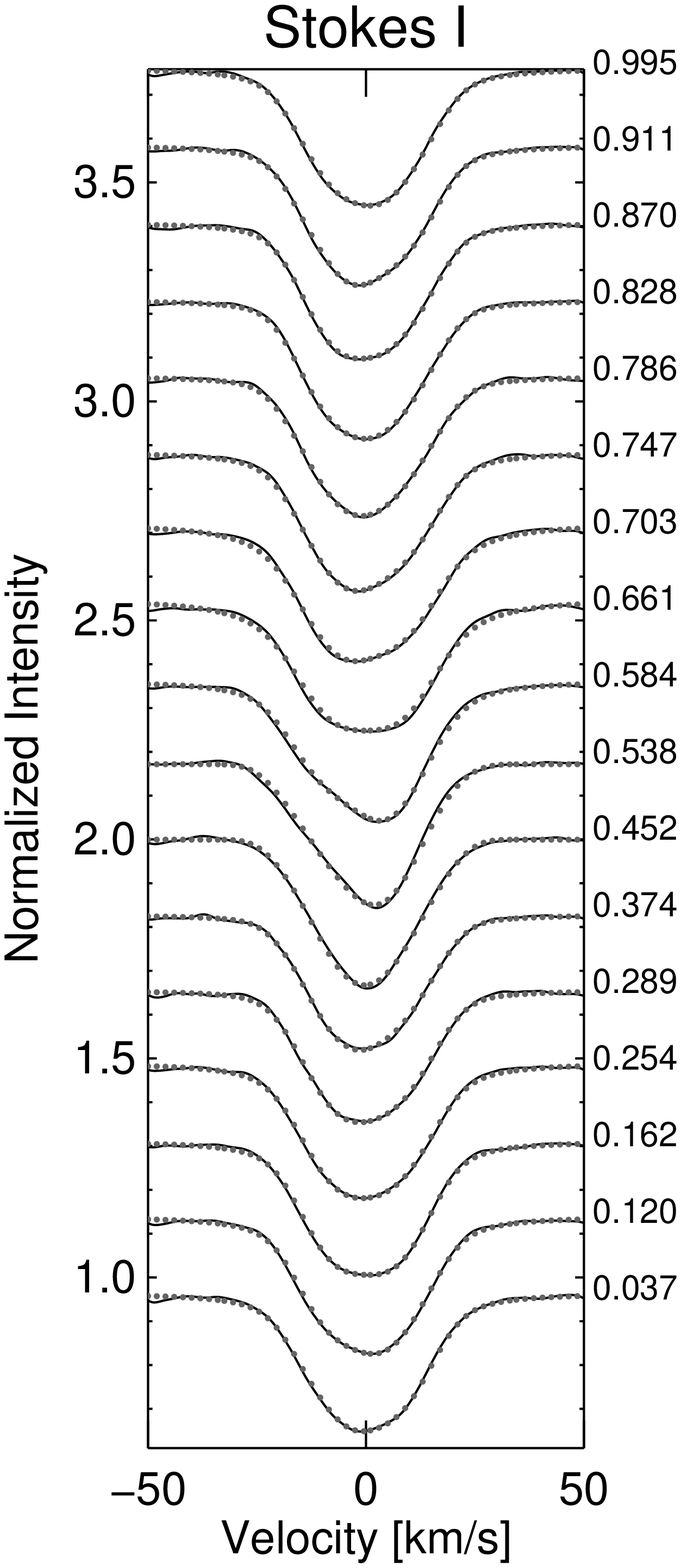}}
\subfloat[2007.01]{\includegraphics[width=90pt]{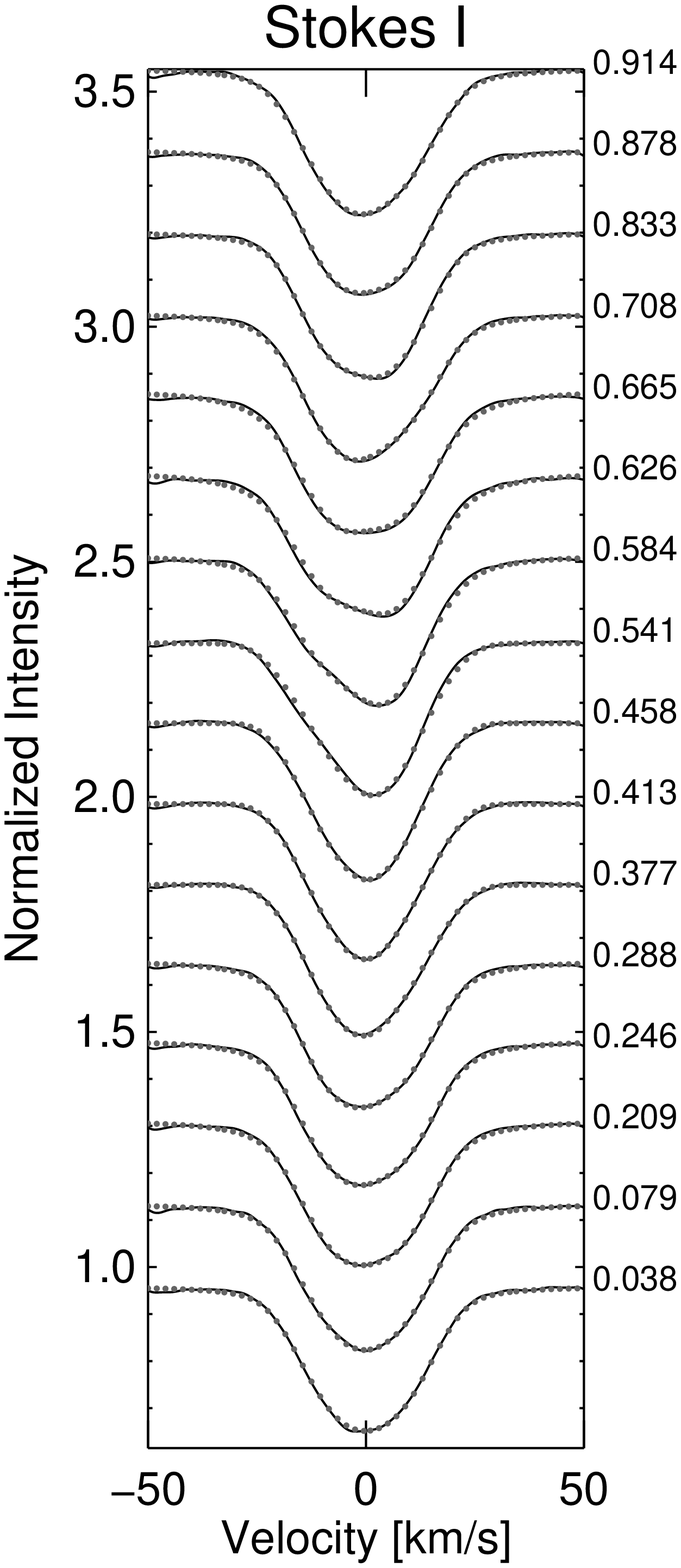}}
\end{minipage}\hspace{1ex}
\begin{minipage}{1.0\textwidth}
\captionsetup[subfigure]{labelfont=bf,textfont=bf,singlelinecheck=off,justification=raggedright,
position=top}
\subfloat[2007.14]{\includegraphics[width=90pt]{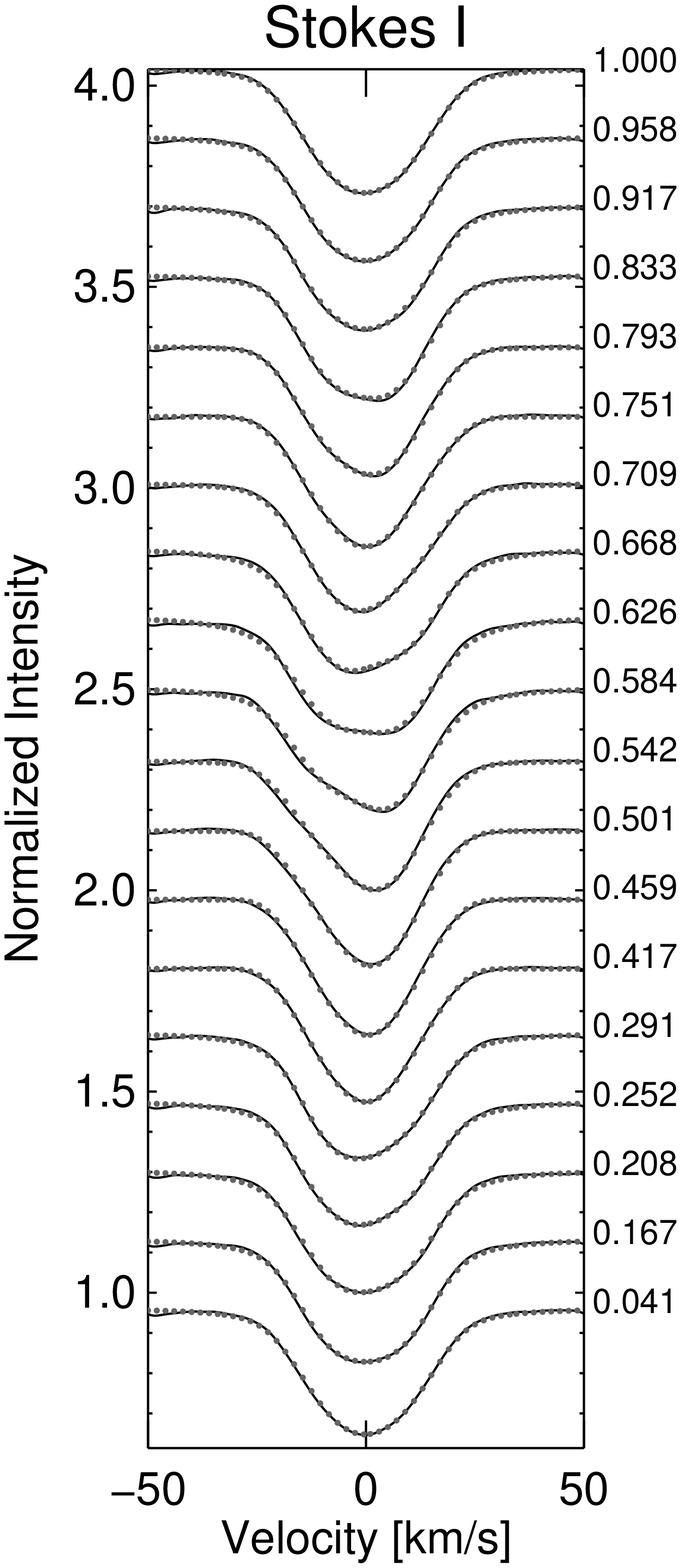}}
\subfloat[2007.67]{\includegraphics[width=90pt]{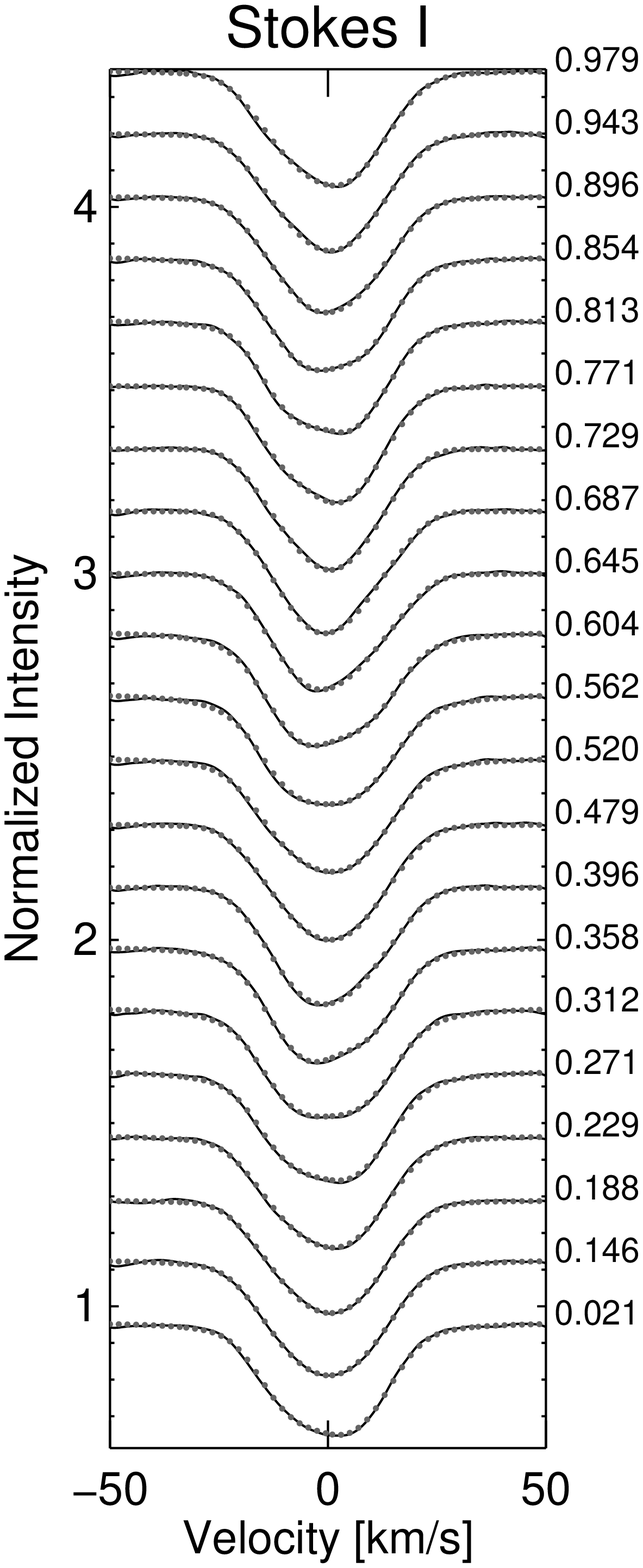}}
\subfloat[2007.73]{\includegraphics[width=90pt]{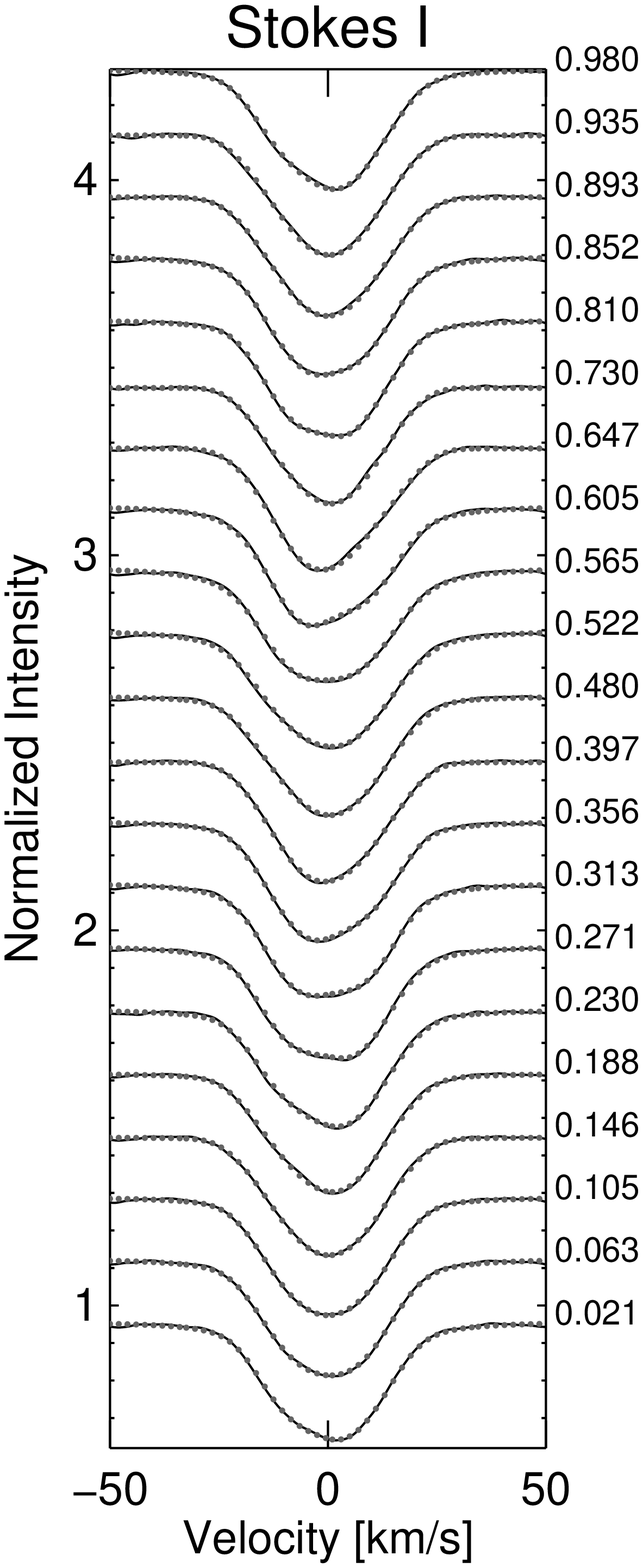}}
\subfloat[2007.87]{\includegraphics[width=90pt]{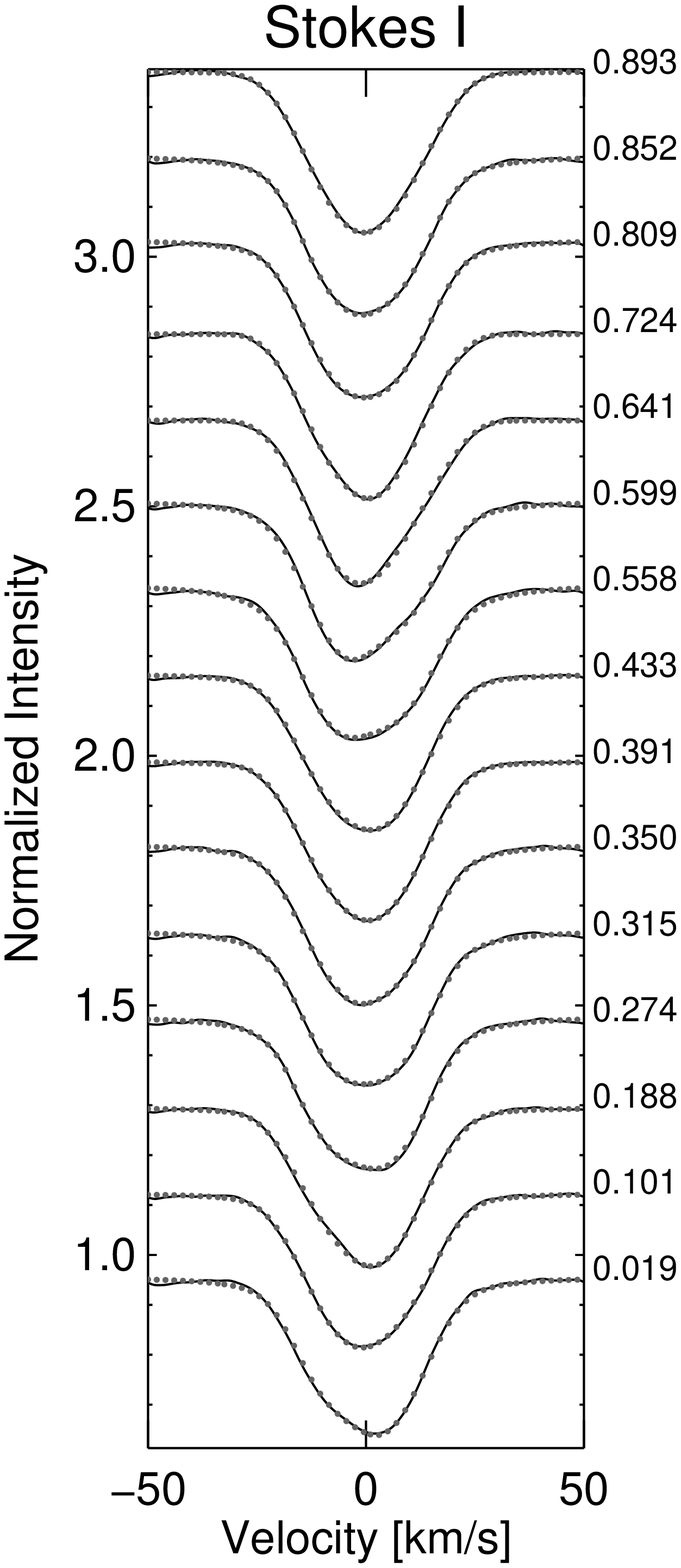}}
\subfloat[2008.05]{\includegraphics[width=90pt]{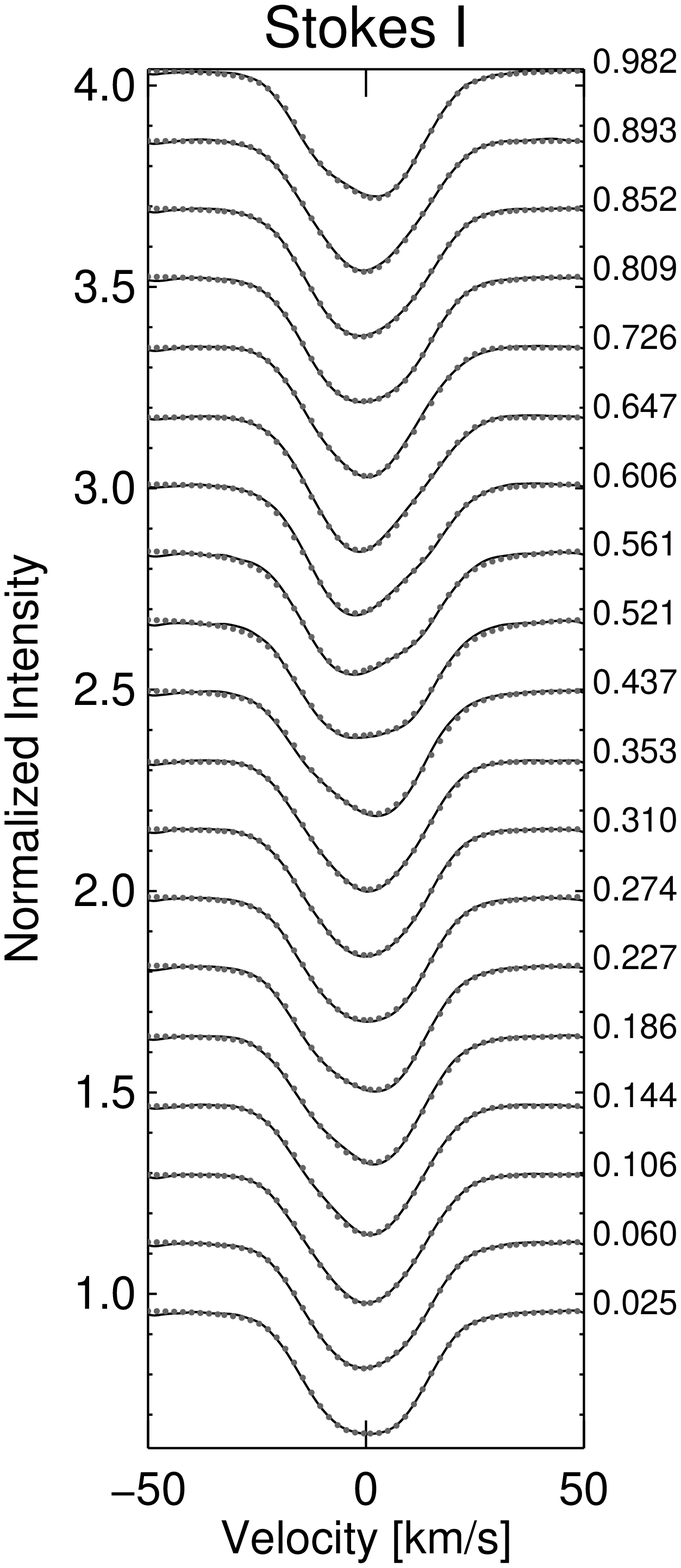}}
\subfloat[2008.12]{\includegraphics[width=90pt]{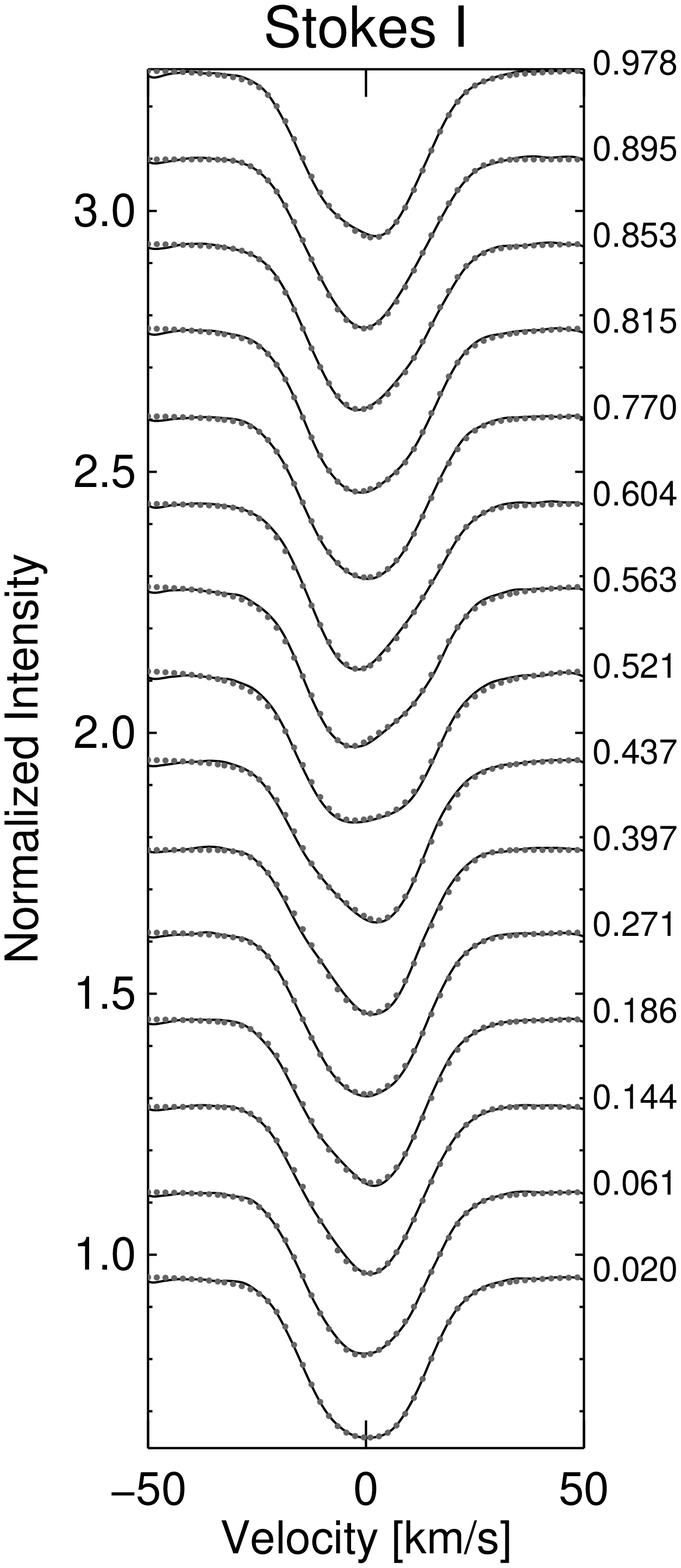}}
\end{minipage}\hspace{1ex}
\begin{minipage}{1.0\textwidth}
\captionsetup[subfigure]{labelfont=bf,textfont=bf,singlelinecheck=off,justification=raggedright,
position=top}
\subfloat[2008.53]{\includegraphics[width=90pt]{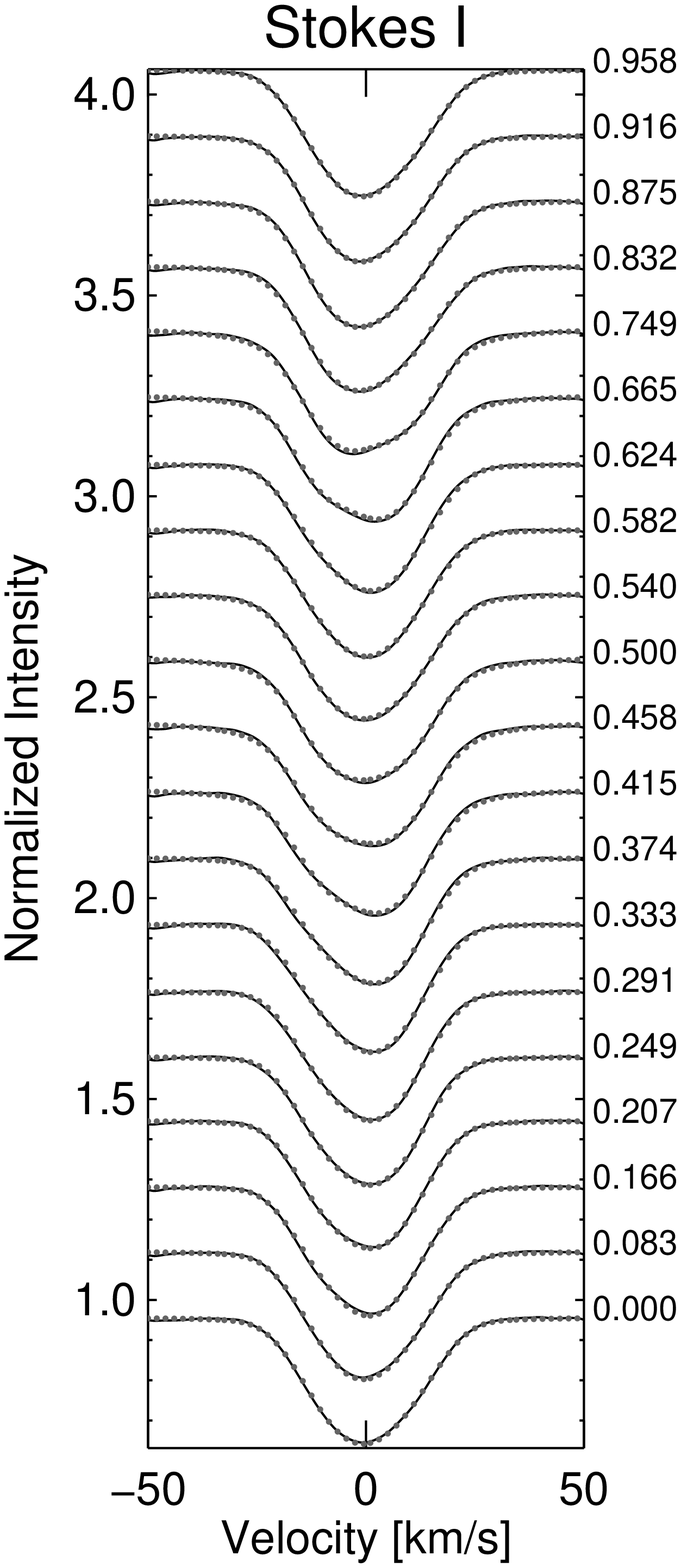}}
\subfloat[2008.60]{\includegraphics[width=90pt]{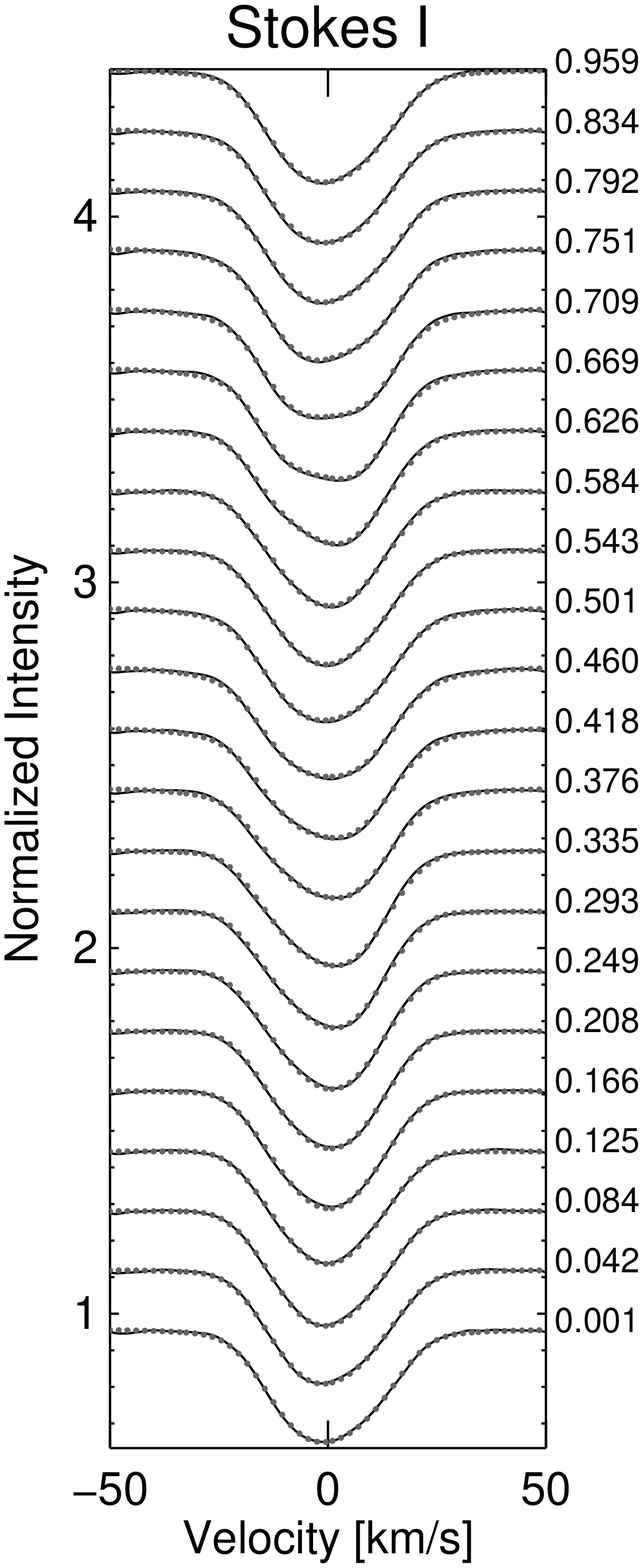}}
\subfloat[2008.73]{\includegraphics[width=90pt]{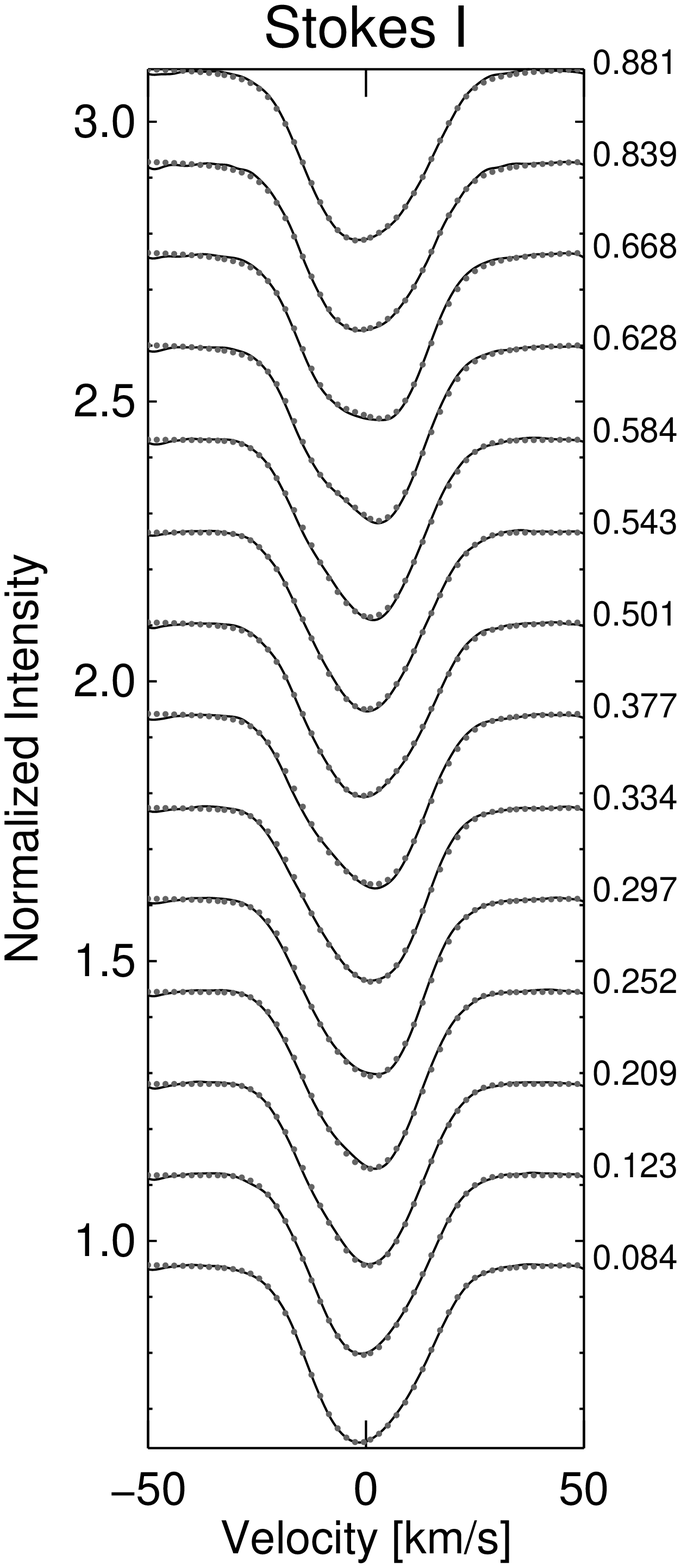}}
\subfloat[2008.80]{\includegraphics[width=90pt]{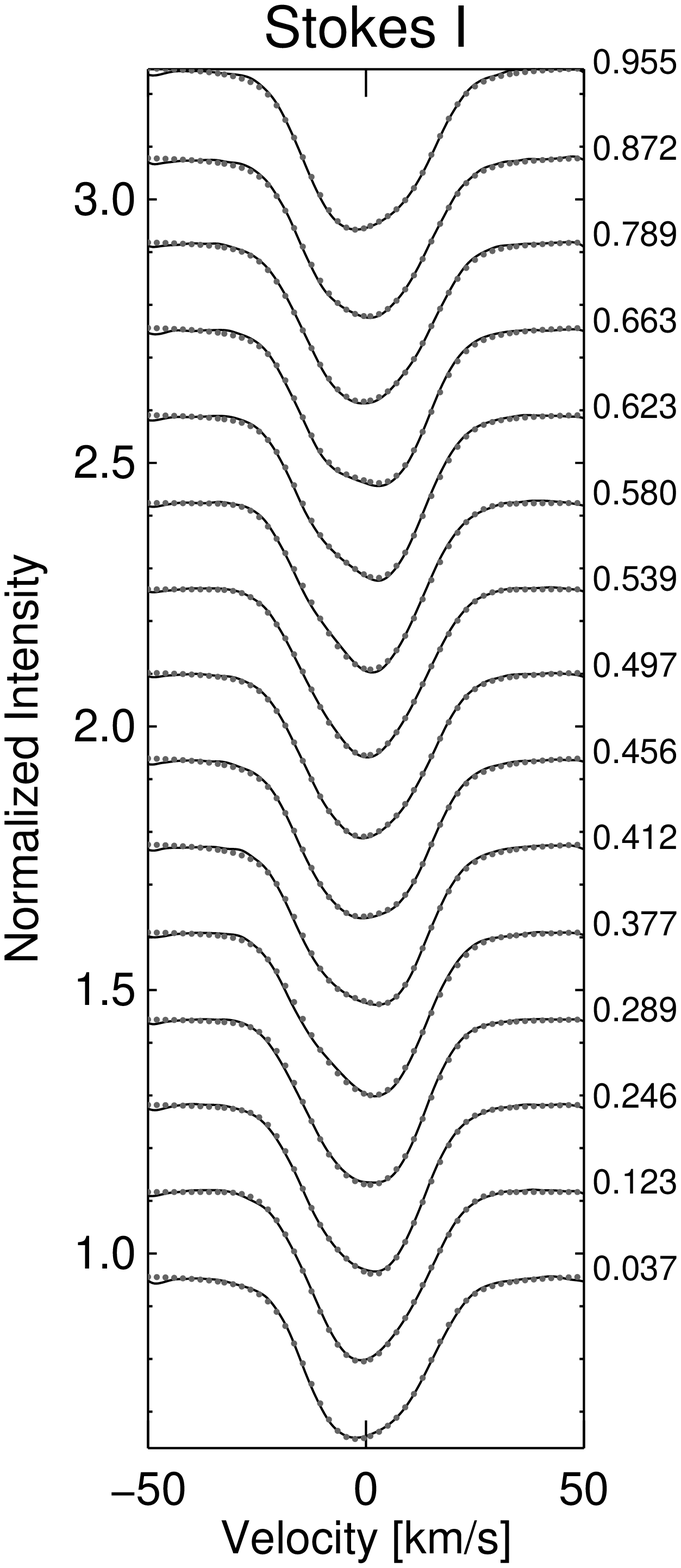}}
\subfloat[2008.88]{\includegraphics[width=90pt]{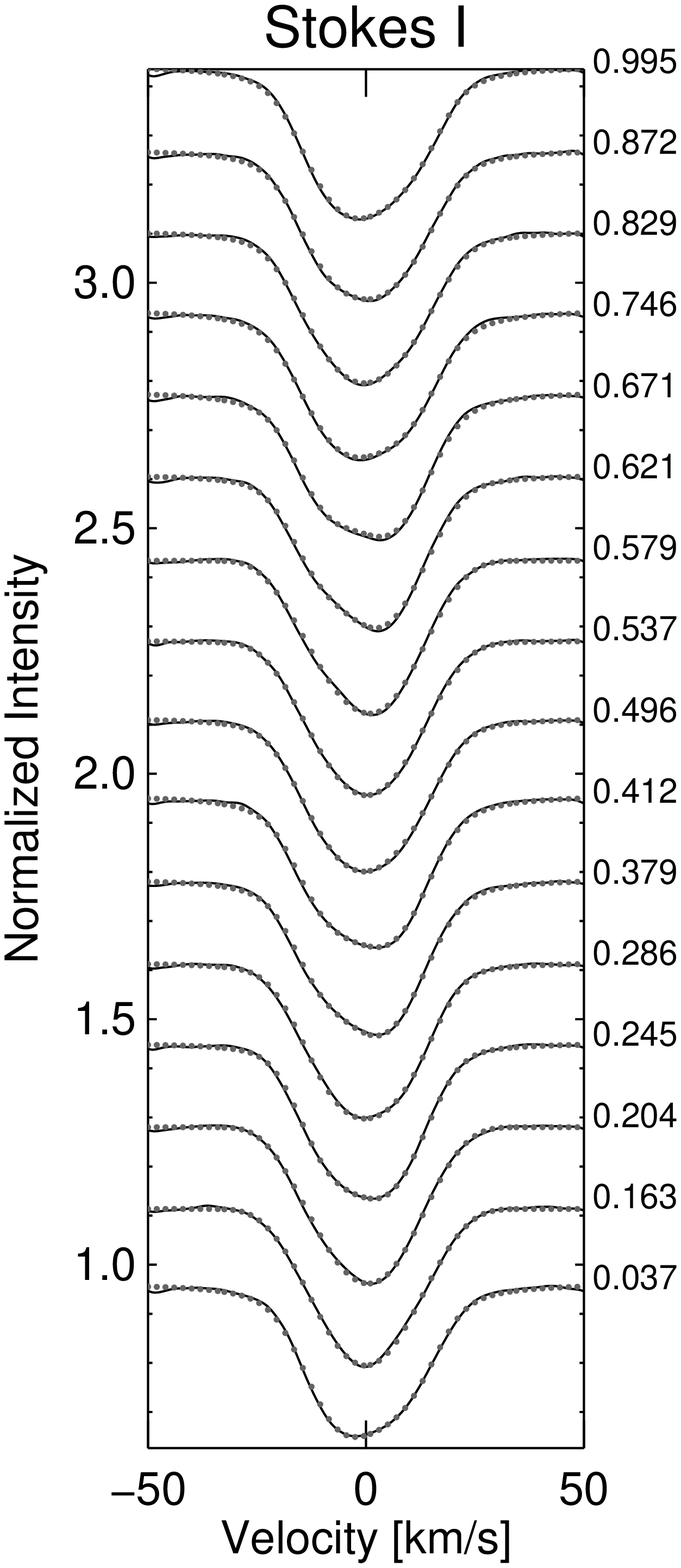}}
\subfloat[2008.94]{\includegraphics[width=90pt]{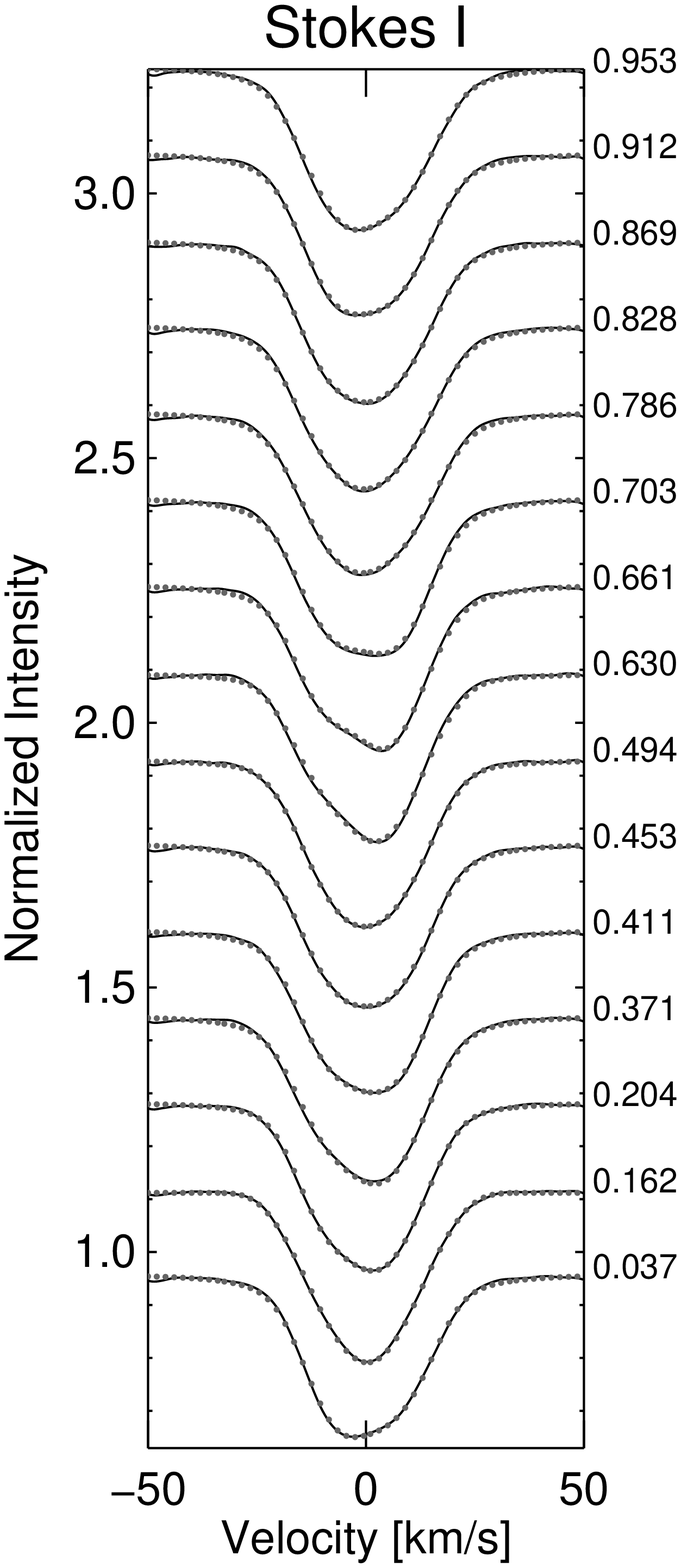}}
\end{minipage}\hspace{1ex}
\caption{Line profiles of Doppler images \#1-18. Each figure shows the observed (solid lines) and
inverted (dotted lines) line profiles for one Doppler image stating their mid times and the
respective phases. Rotation advances from bottom to the top. Their corresponding RMS-errors are
given in Table~\ref{tab:maps_log}.}
\label{fig:stokes_profiles_1}
\end{figure*}

\begin{figure*}[!t]
\begin{minipage}{1.0\textwidth}
\captionsetup[subfigure]{labelfont=bf,textfont=bf,singlelinecheck=off,justification=raggedright,
position=top}
\subfloat[2009.07]{\includegraphics[width=90pt]{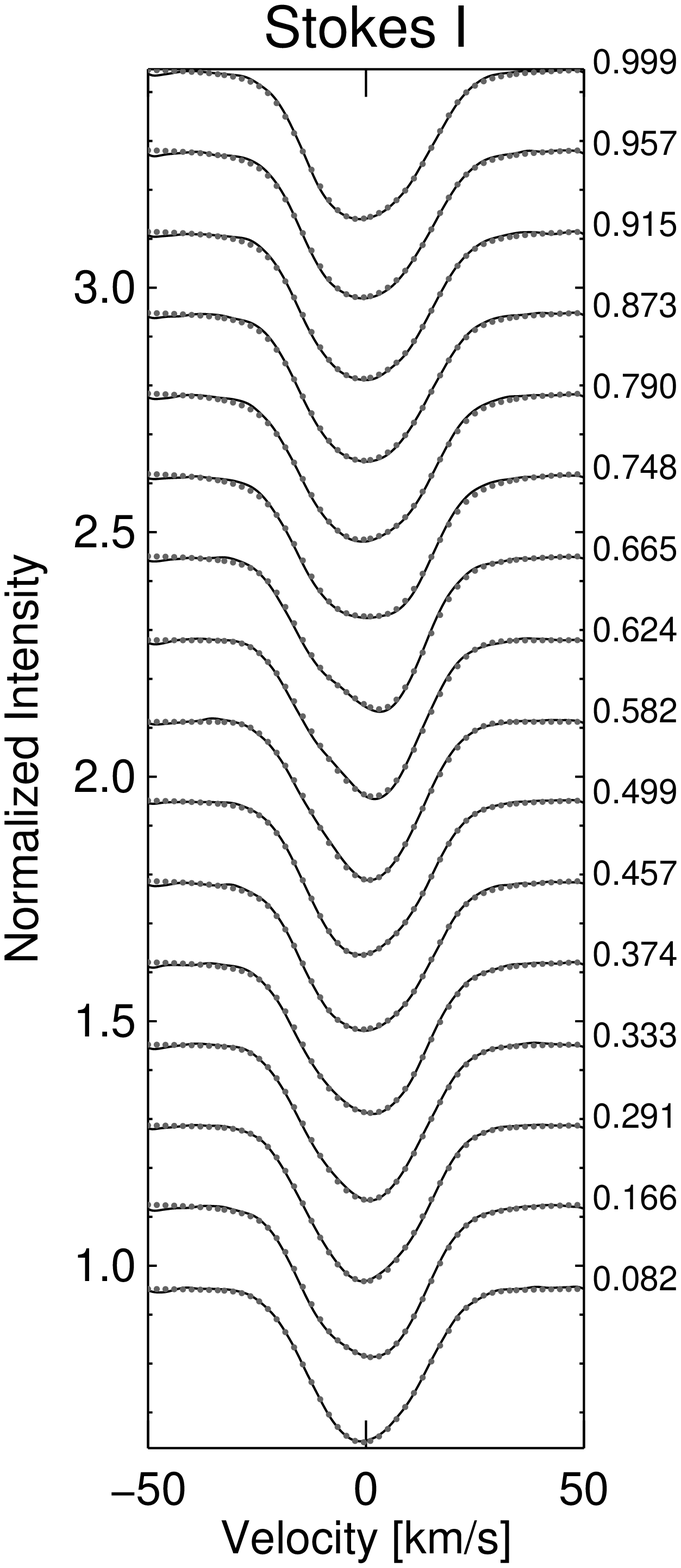}}
\subfloat[2009.60]{\includegraphics[width=90pt]{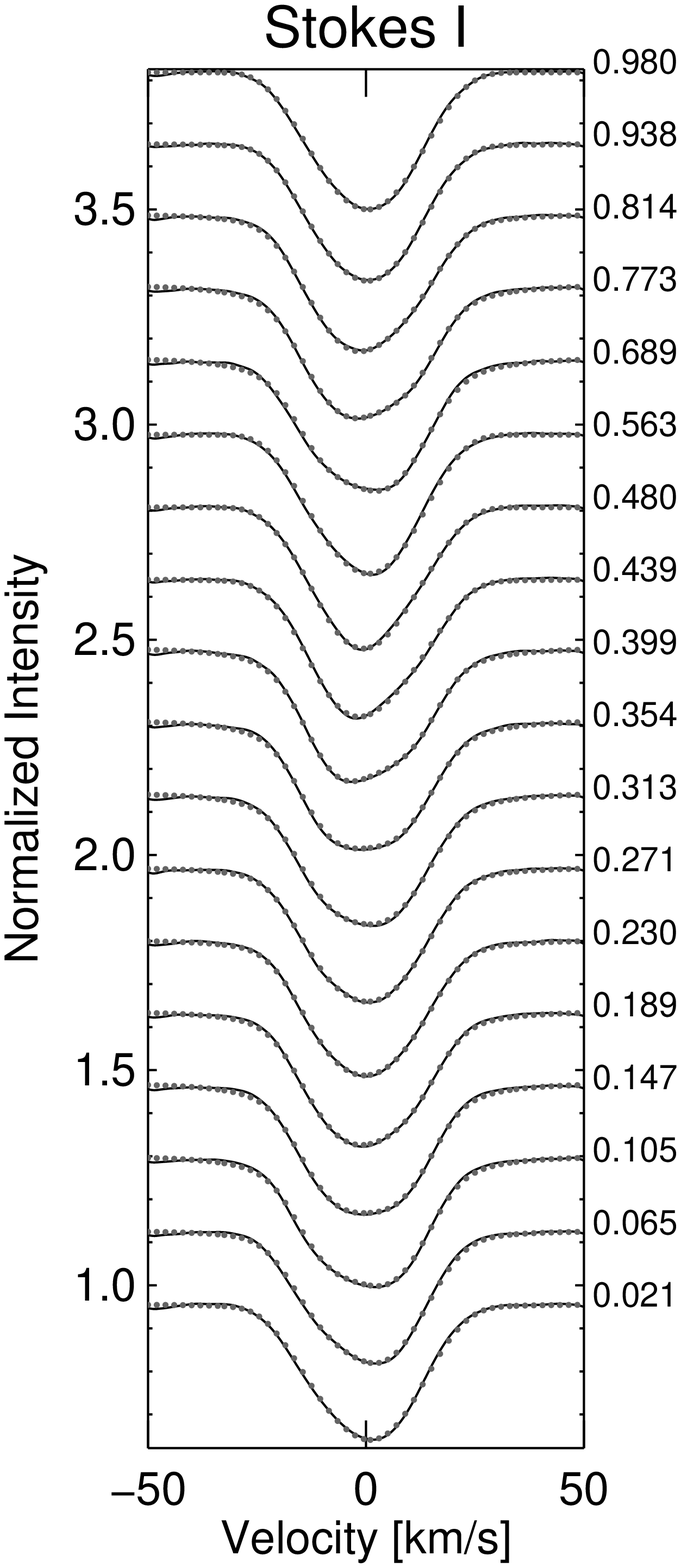}}
\subfloat[2009.67]{\includegraphics[width=90pt]{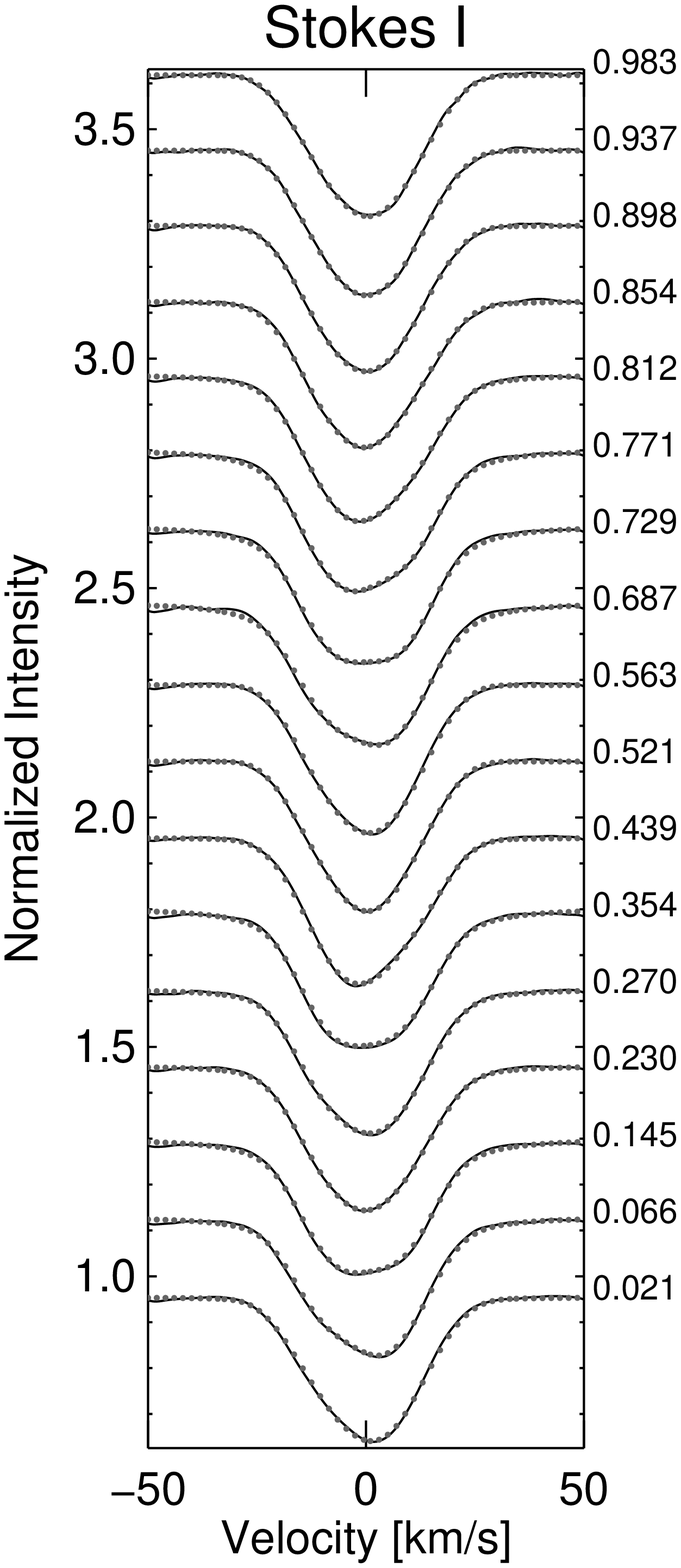}}
\subfloat[2009.86]{\includegraphics[width=90pt]{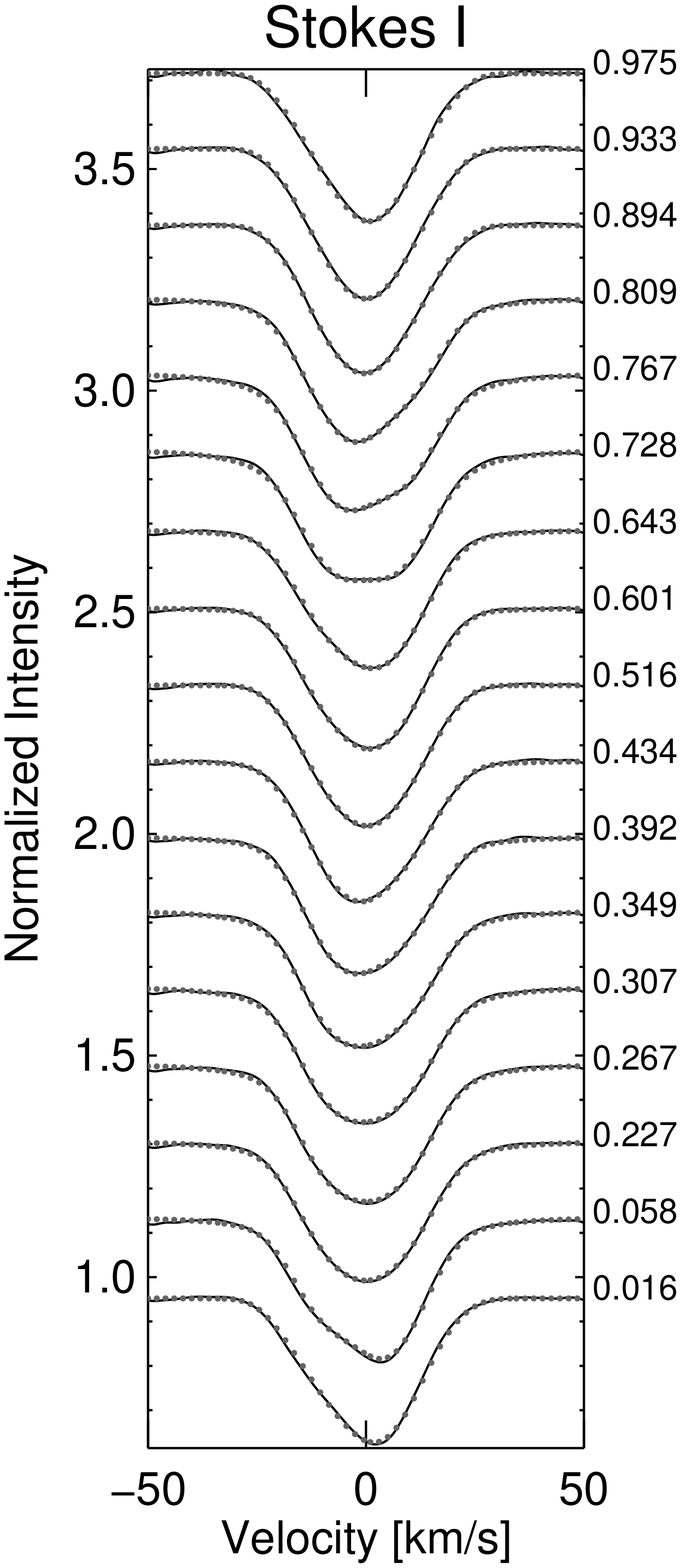}}
\subfloat[2009.93]{\includegraphics[width=90pt]{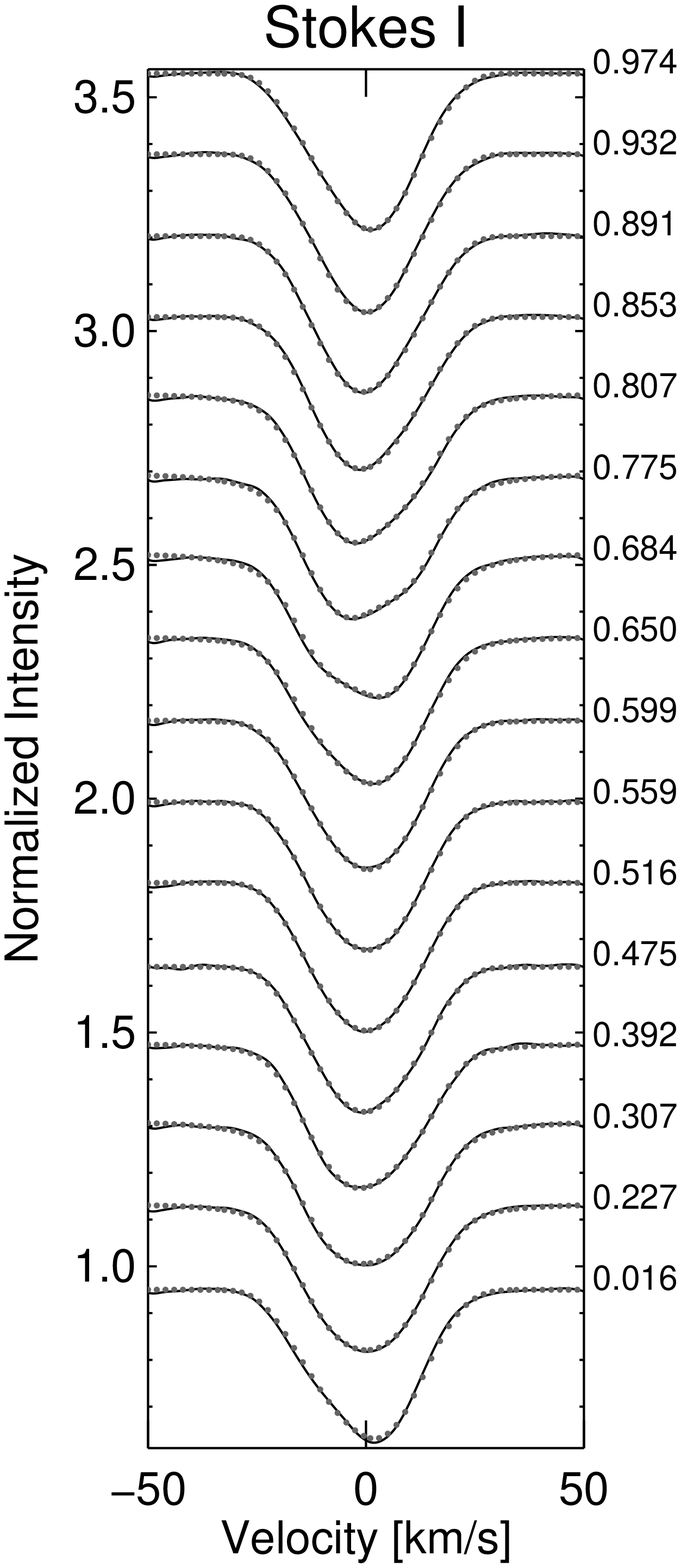}}
\subfloat[2010.06]{\includegraphics[width=90pt]{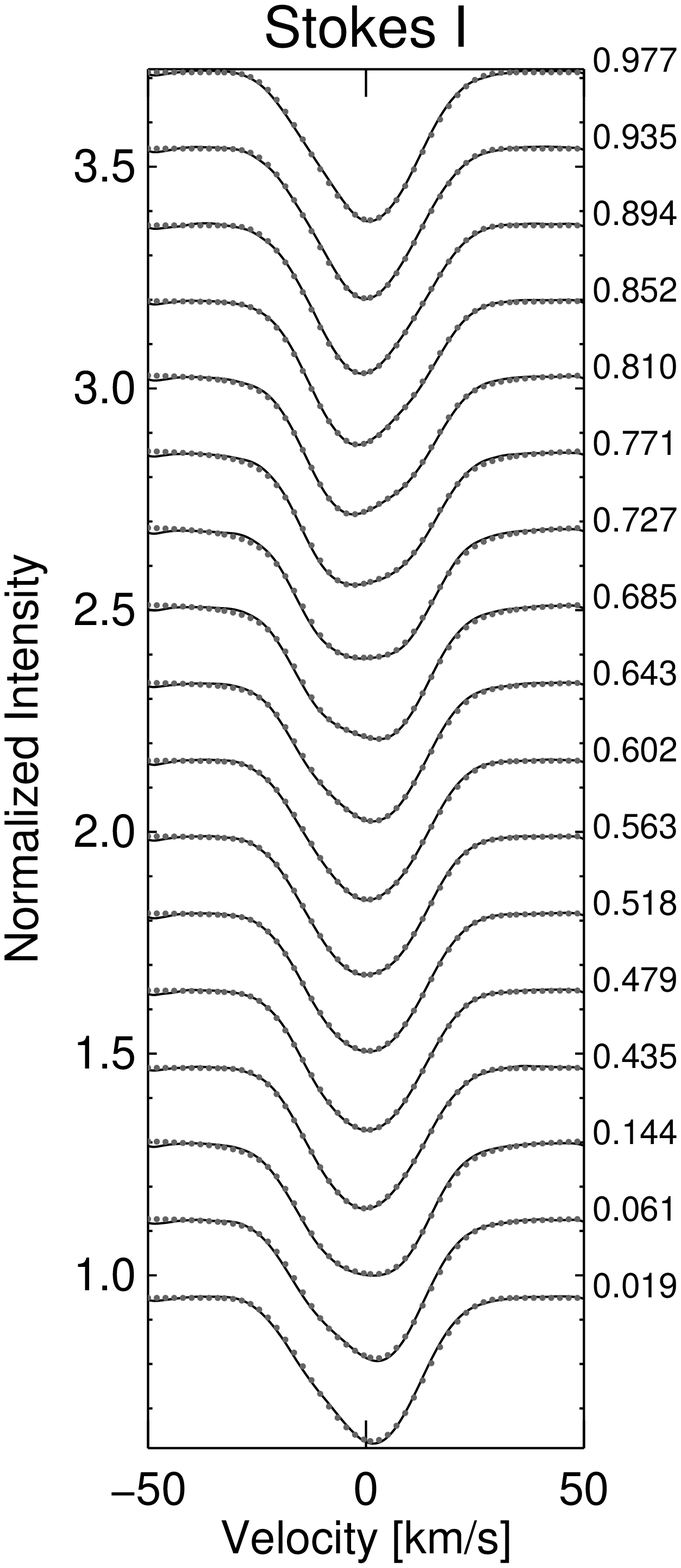}}
\end{minipage}\hspace{1ex}
\begin{minipage}{1.0\textwidth}
\captionsetup[subfigure]{labelfont=bf,textfont=bf,singlelinecheck=off,justification=raggedright,
position=top}
\subfloat[2010.59]{\includegraphics[width=90pt]{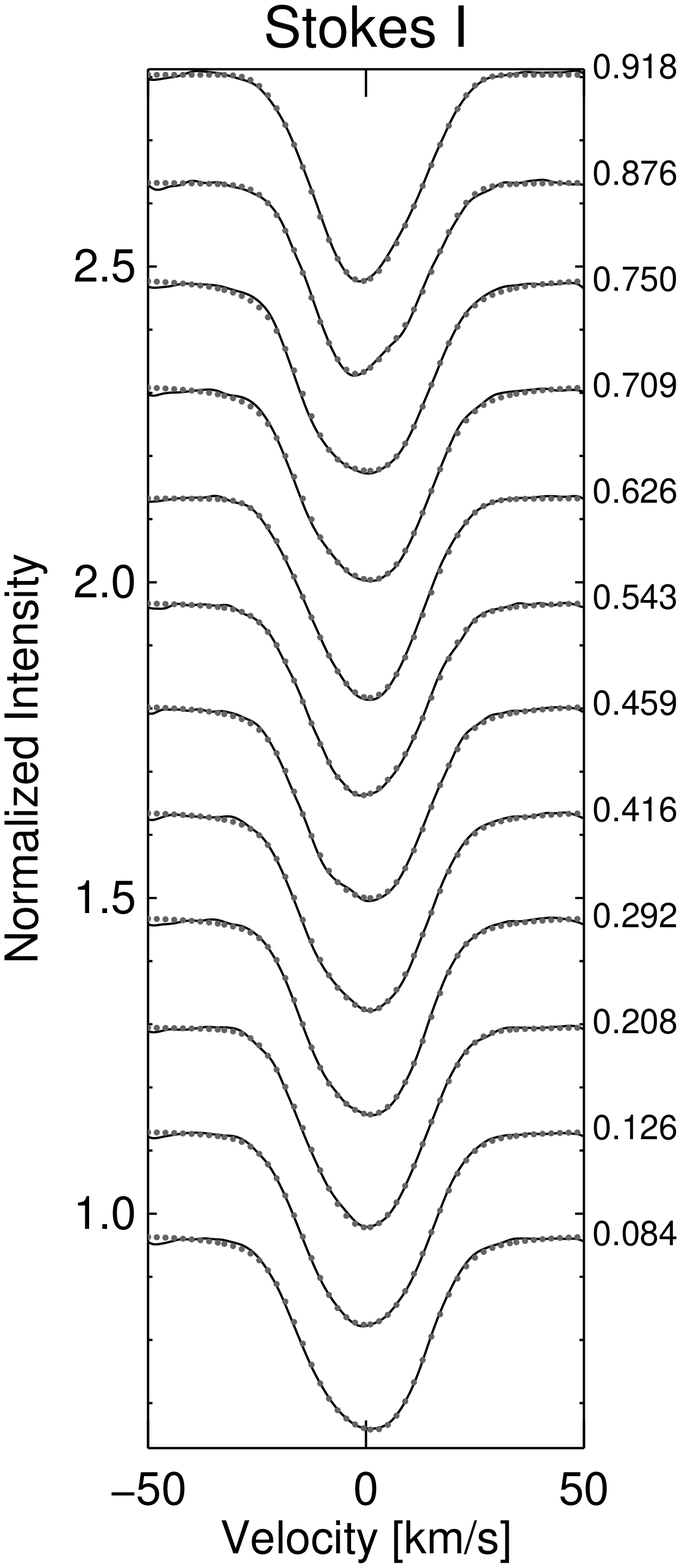}}
\subfloat[2010.67]{\includegraphics[width=90pt]{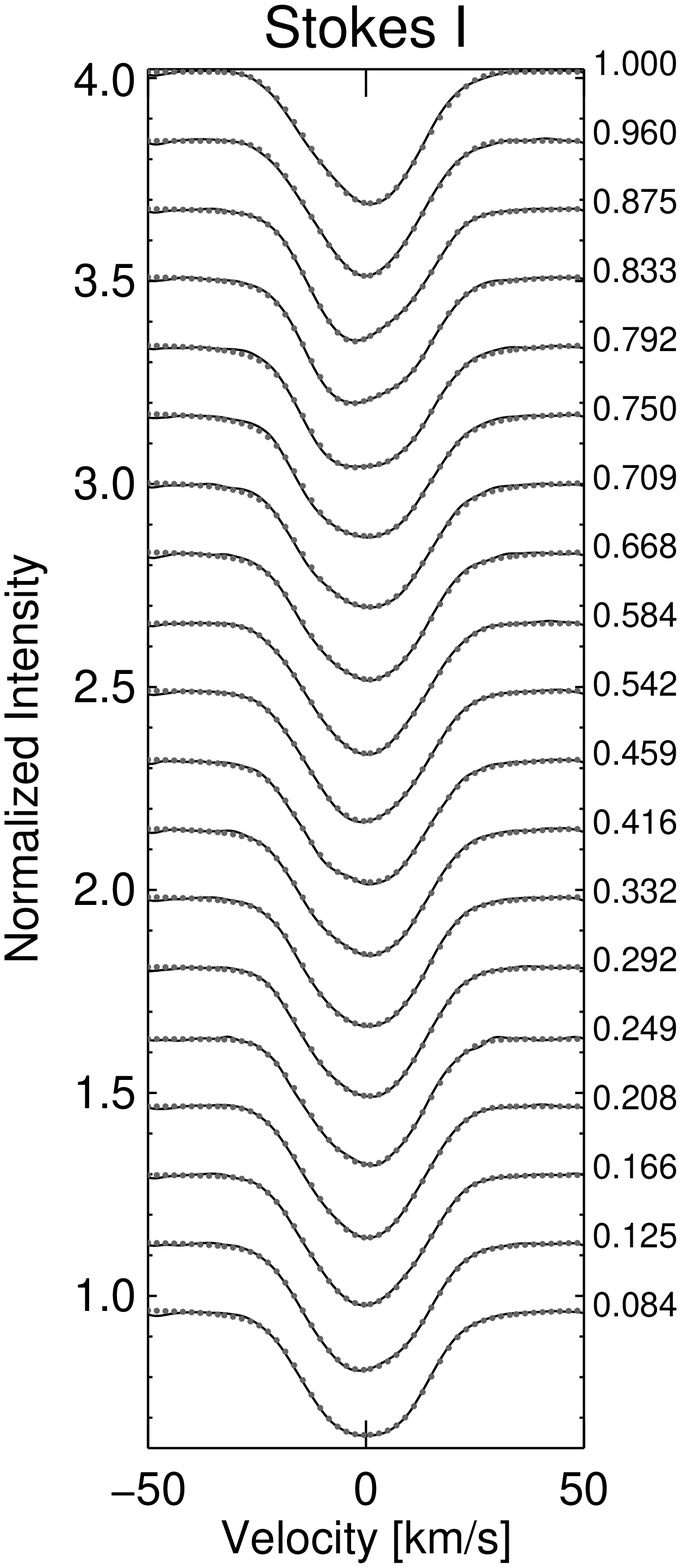}}
\subfloat[2010.74]{\includegraphics[width=90pt]{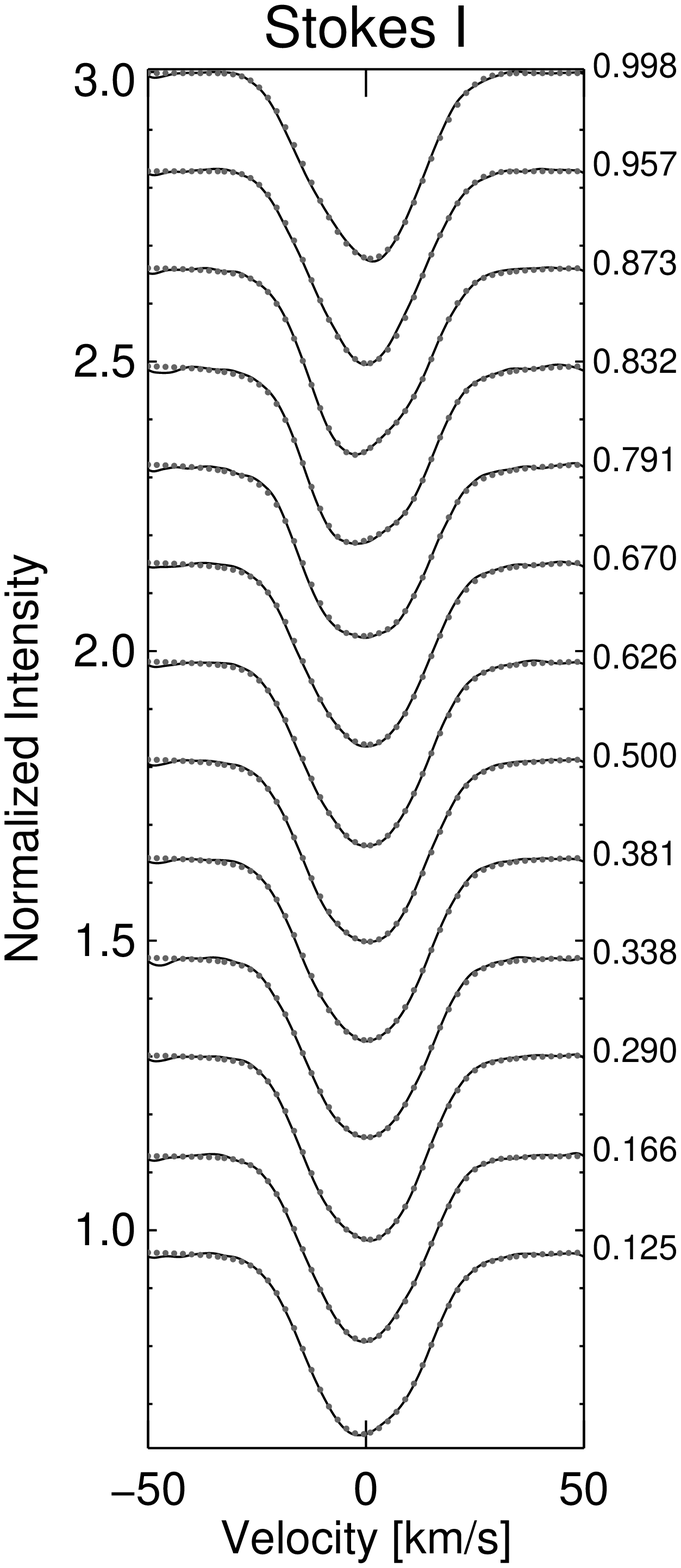}}
\subfloat[2010.84]{\includegraphics[width=90pt]{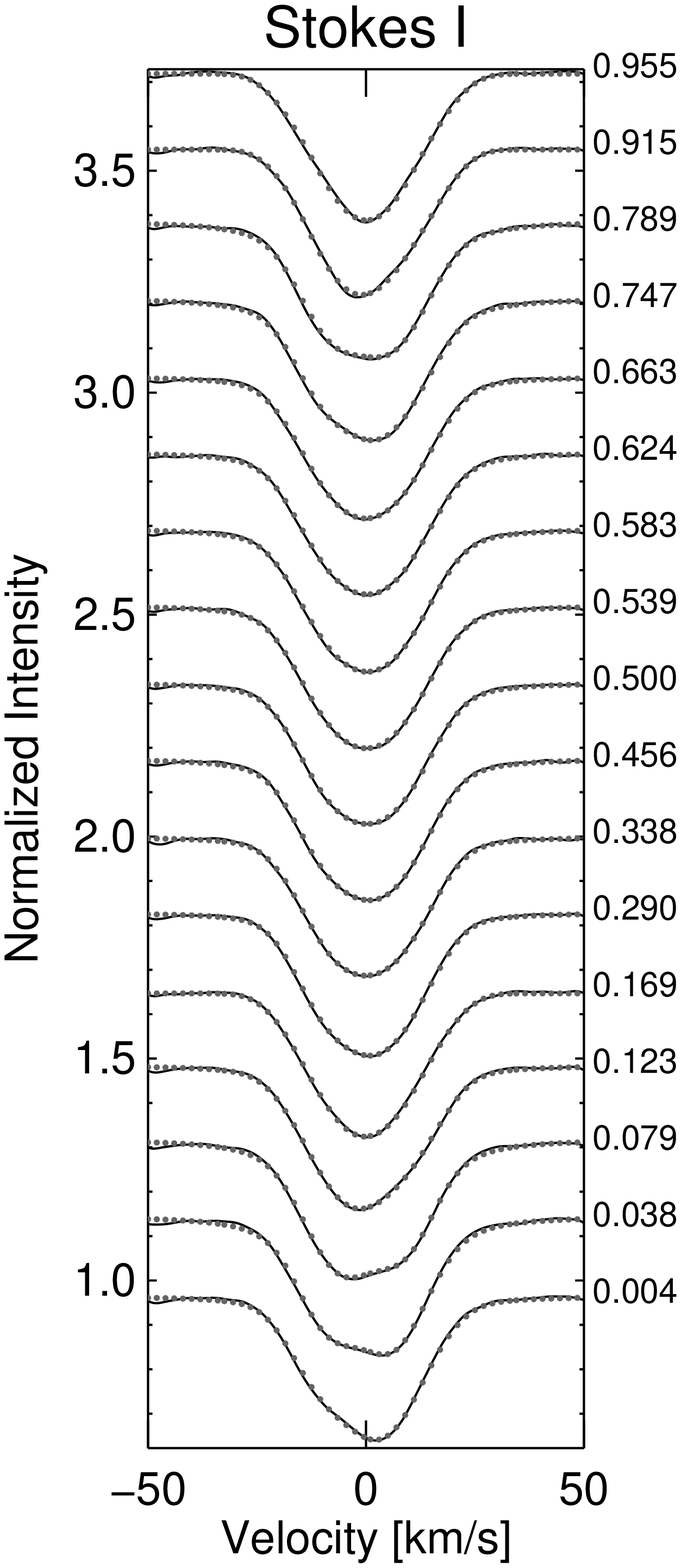}}
\subfloat[2011.12]{\includegraphics[width=90pt]{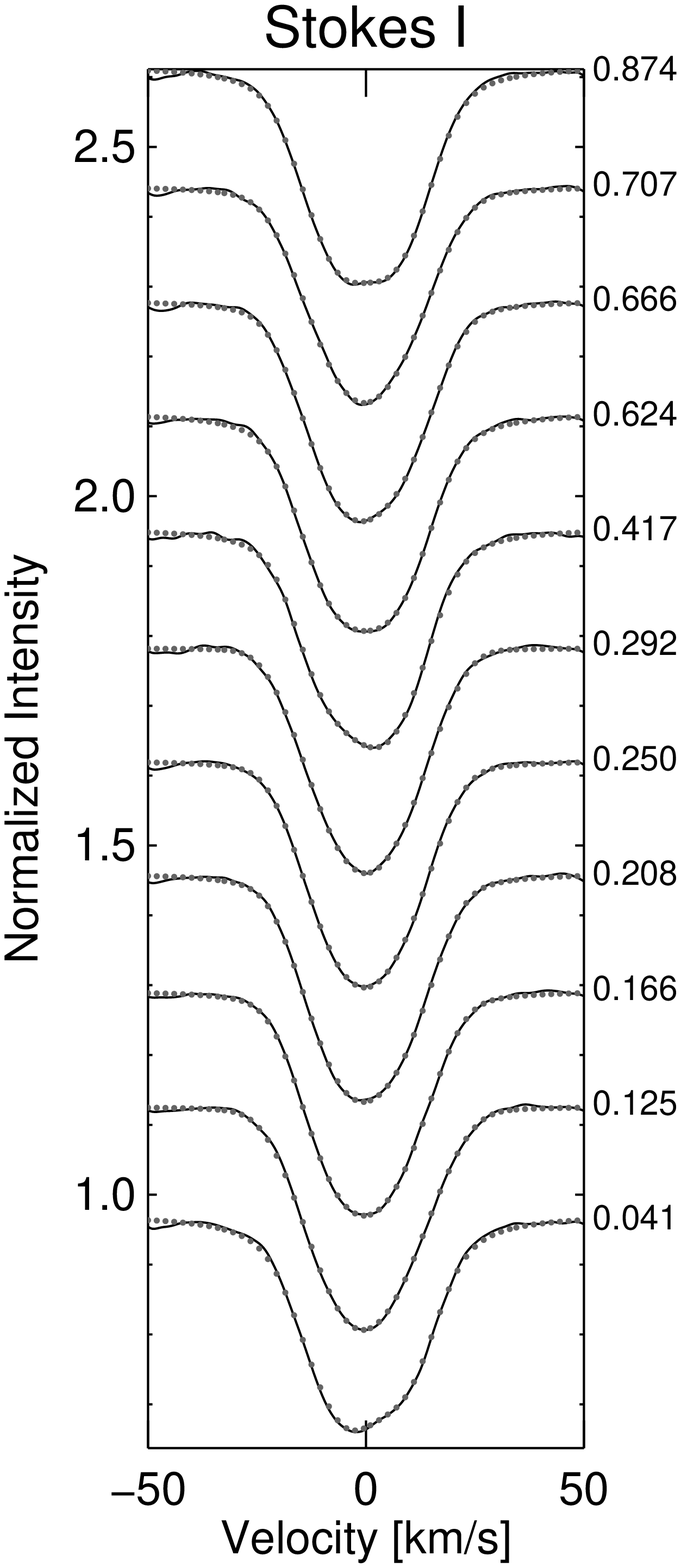}}
\subfloat[2011.61]{\includegraphics[width=90pt]{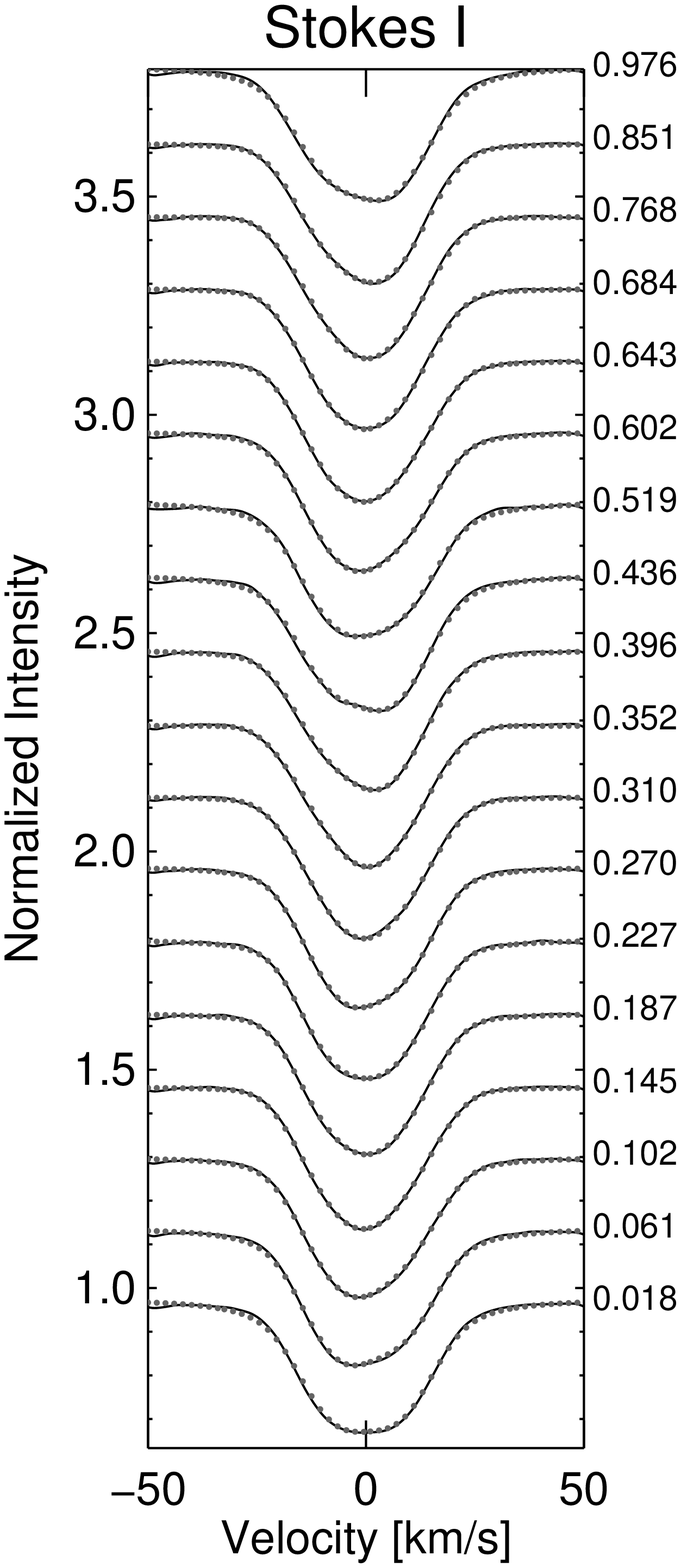}}
\end{minipage}\hspace{1ex}
\begin{minipage}{1.0\textwidth}
\captionsetup[subfigure]{labelfont=bf,textfont=bf,singlelinecheck=off,justification=raggedright,
position=top}
\subfloat[2011.68]{\includegraphics[width=90pt]{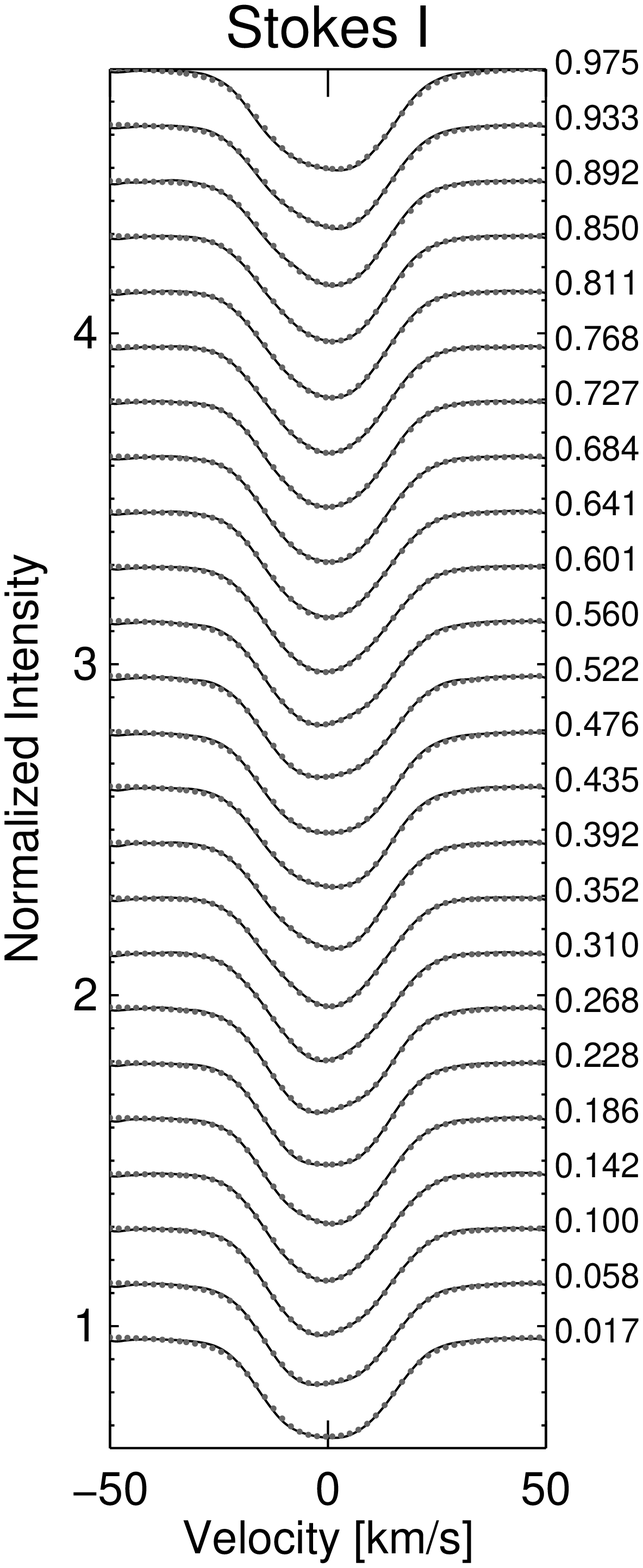}}
\subfloat[2011.81]{\includegraphics[width=90pt]{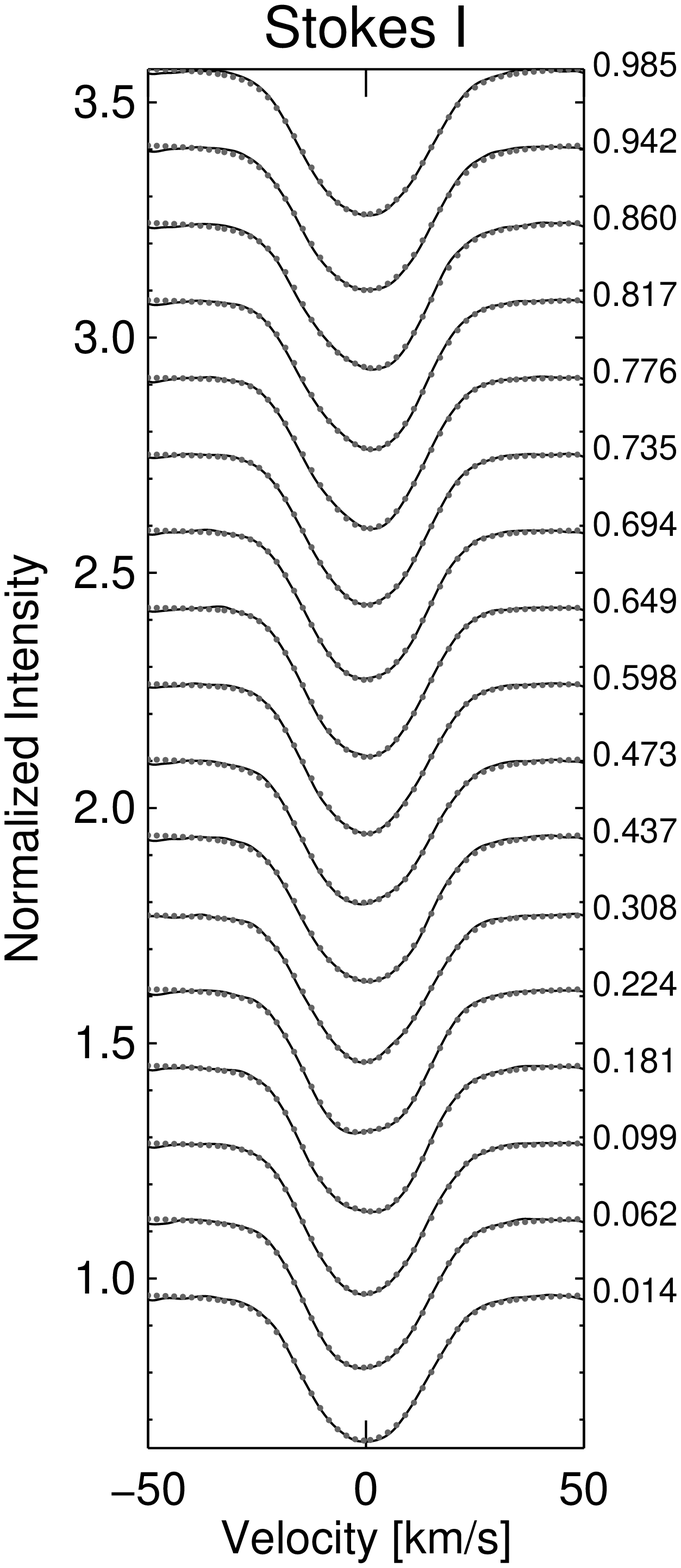}}
\subfloat[2011.88]{\includegraphics[width=90pt]{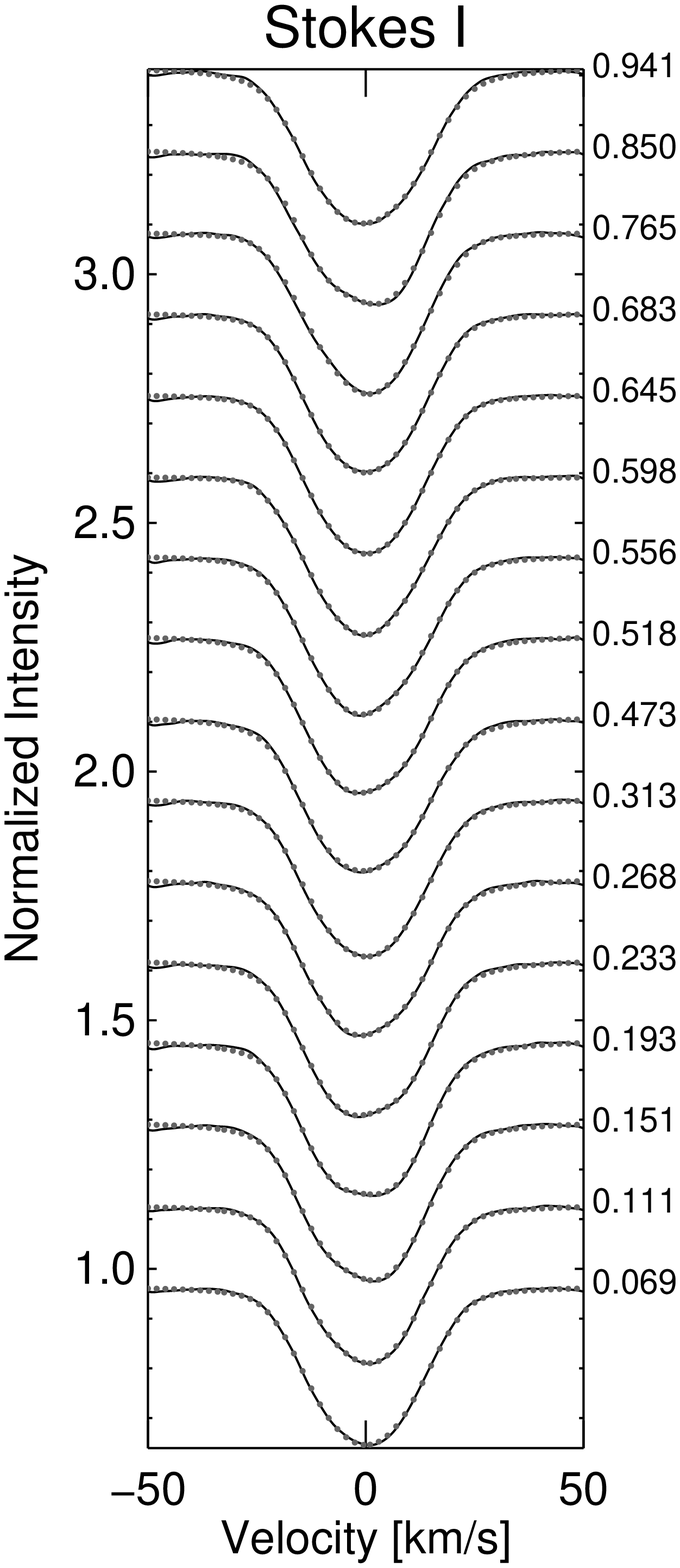}}
\subfloat[2012.01]{\includegraphics[width=90pt]{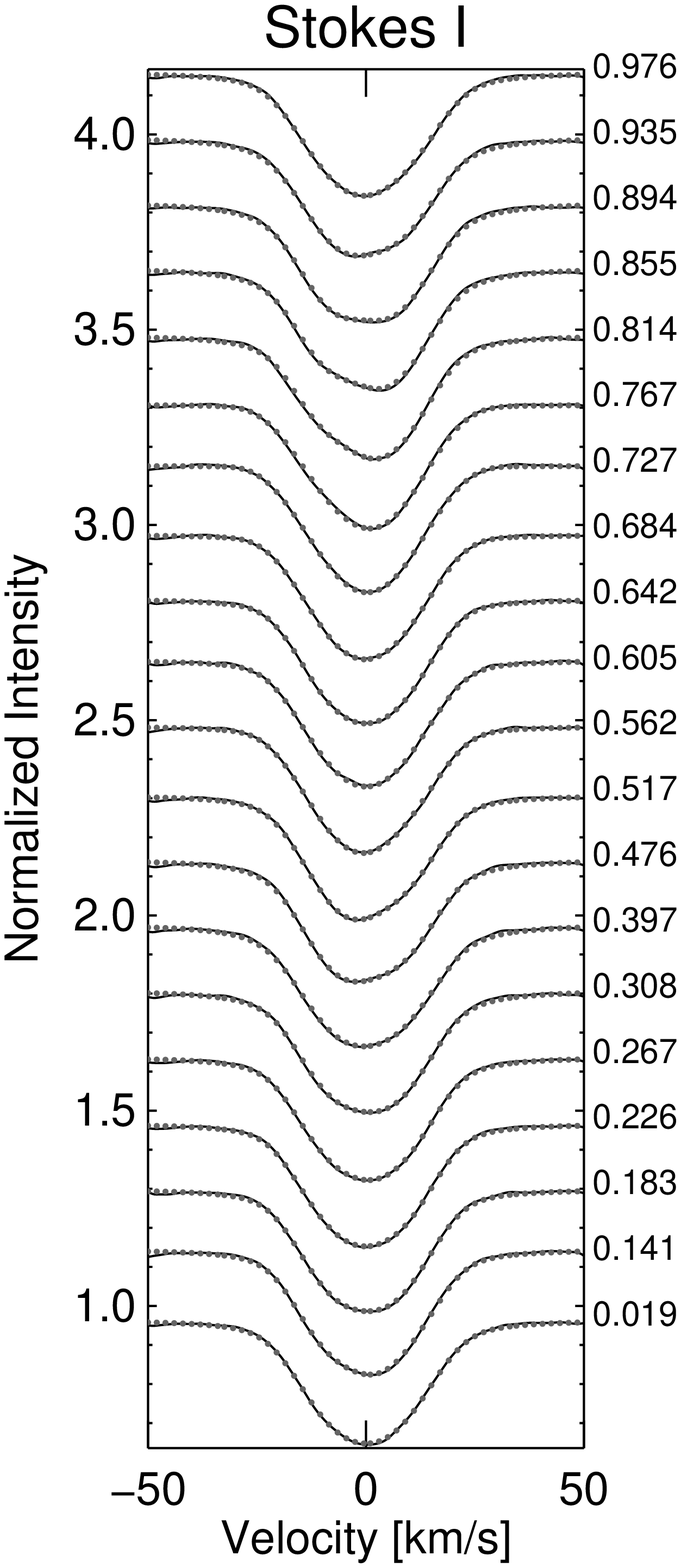}}
\subfloat[2012.08]{\includegraphics[width=90pt]{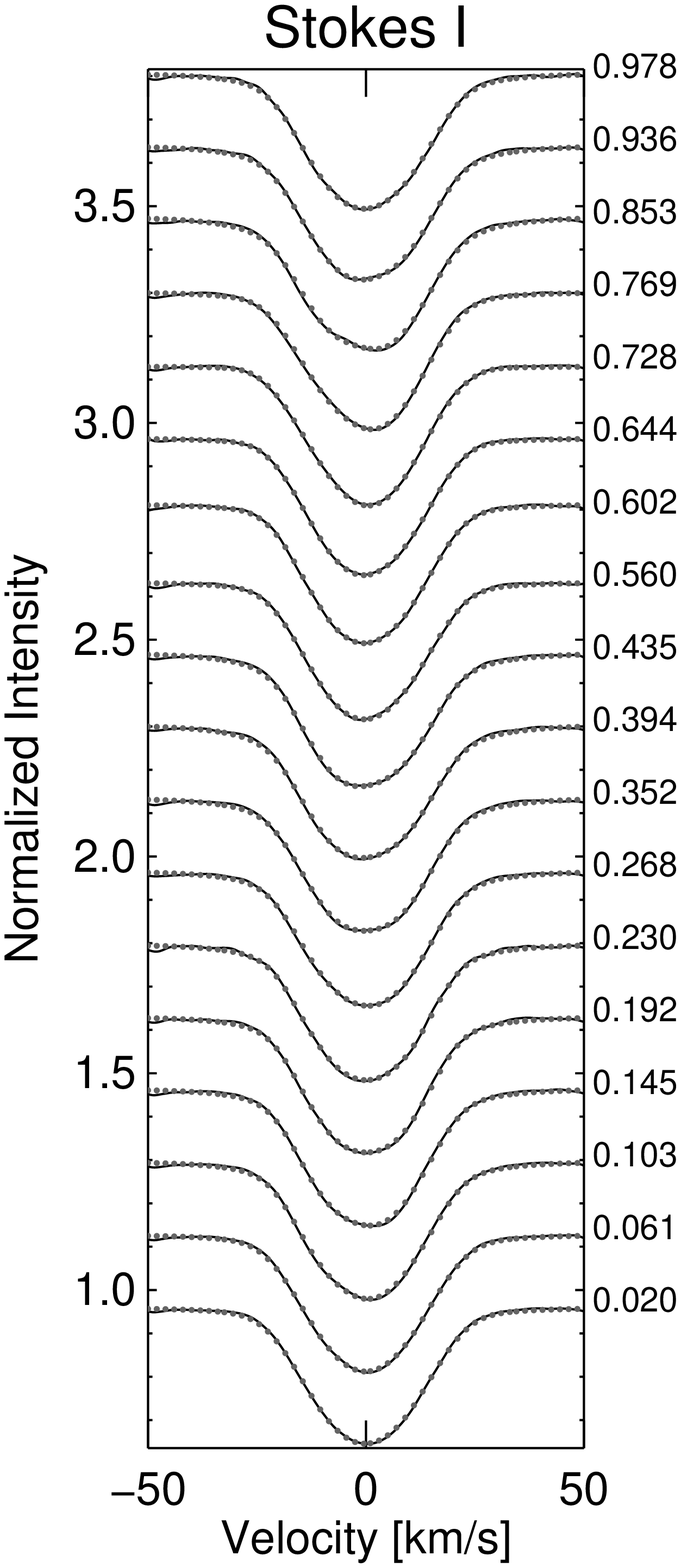}}
\subfloat[2012.19]{\includegraphics[width=90pt]{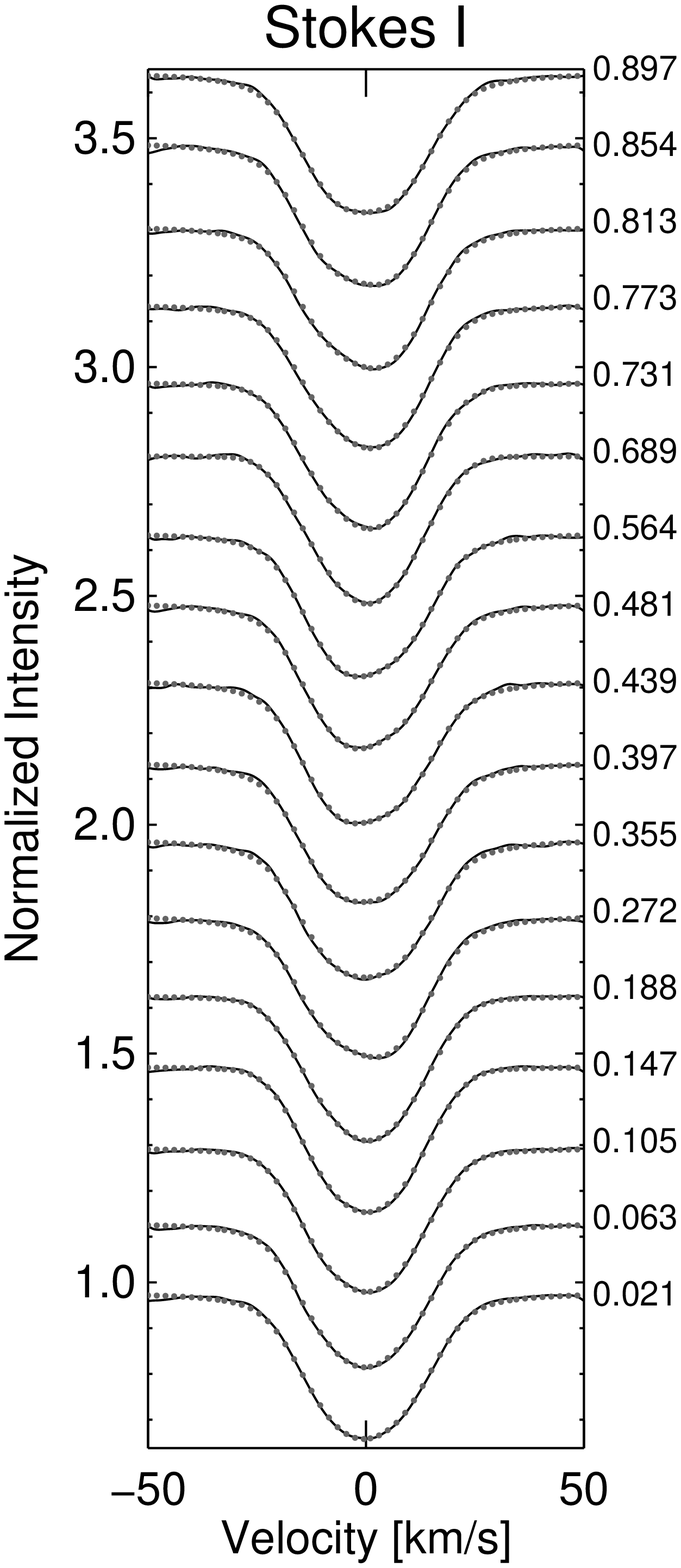}}
\end{minipage}\hspace{1ex}
\caption{Line profiles of Doppler images \#19-36. Otherwise as in Fig.~\ref{fig:stokes_profiles_1}.}
\label{fig:stokes_profiles_2}
\end{figure*}

Fig.~\ref{fig:stokes_profiles_1} and Fig.~\ref{fig:stokes_profiles_2} show the observed and inverted
line profiles of each Doppler image for all observational seasons.

\section{Phase coverage of Doppler images for the seasons 2006/07 to 2011/12}

\begin{figure*}[!t]
\begin{minipage}{1.0\textwidth}
\captionsetup[subfigure]{labelfont=bf,textfont=bf,singlelinecheck=off,justification=raggedright,
position=top}
\subfloat[]{\includegraphics[width=250pt]{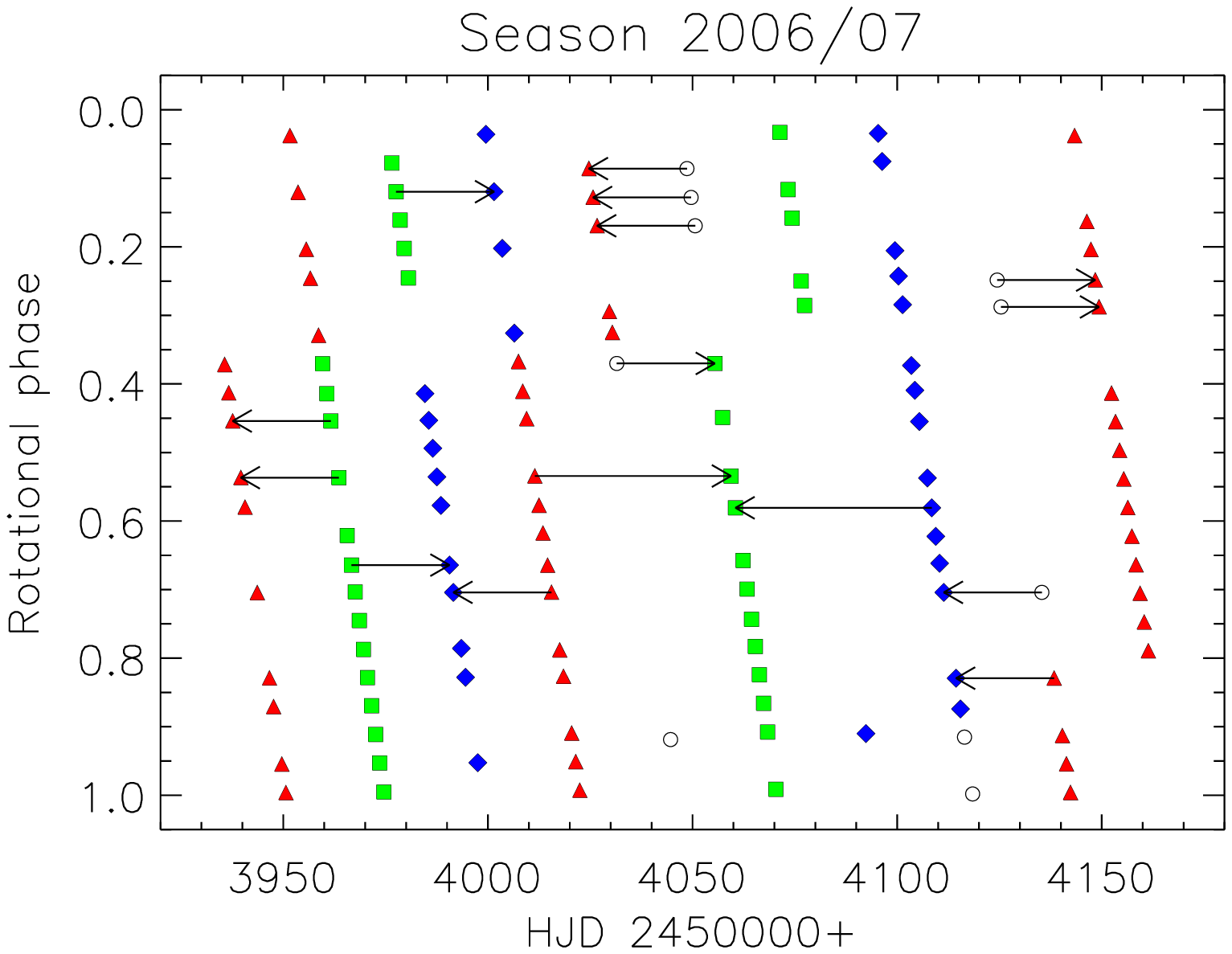}}
\subfloat[]{\includegraphics[width=250pt]{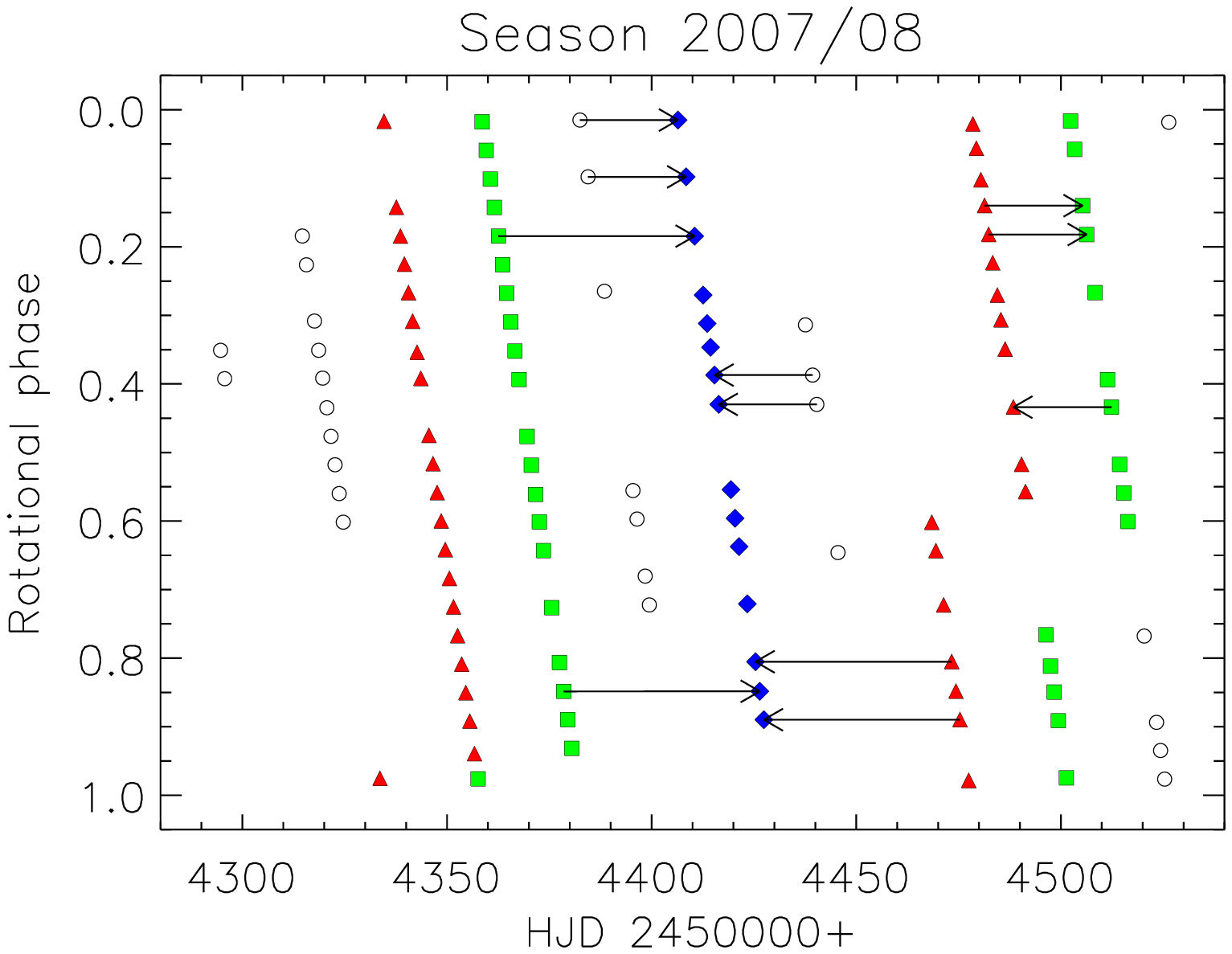}}
\end{minipage}\hspace{1ex}
\begin{minipage}{1.0\textwidth}
\captionsetup[subfigure]{labelfont=bf,textfont=bf,singlelinecheck=off,justification=raggedright,
position=top}
\subfloat[]{\includegraphics[width=250pt]{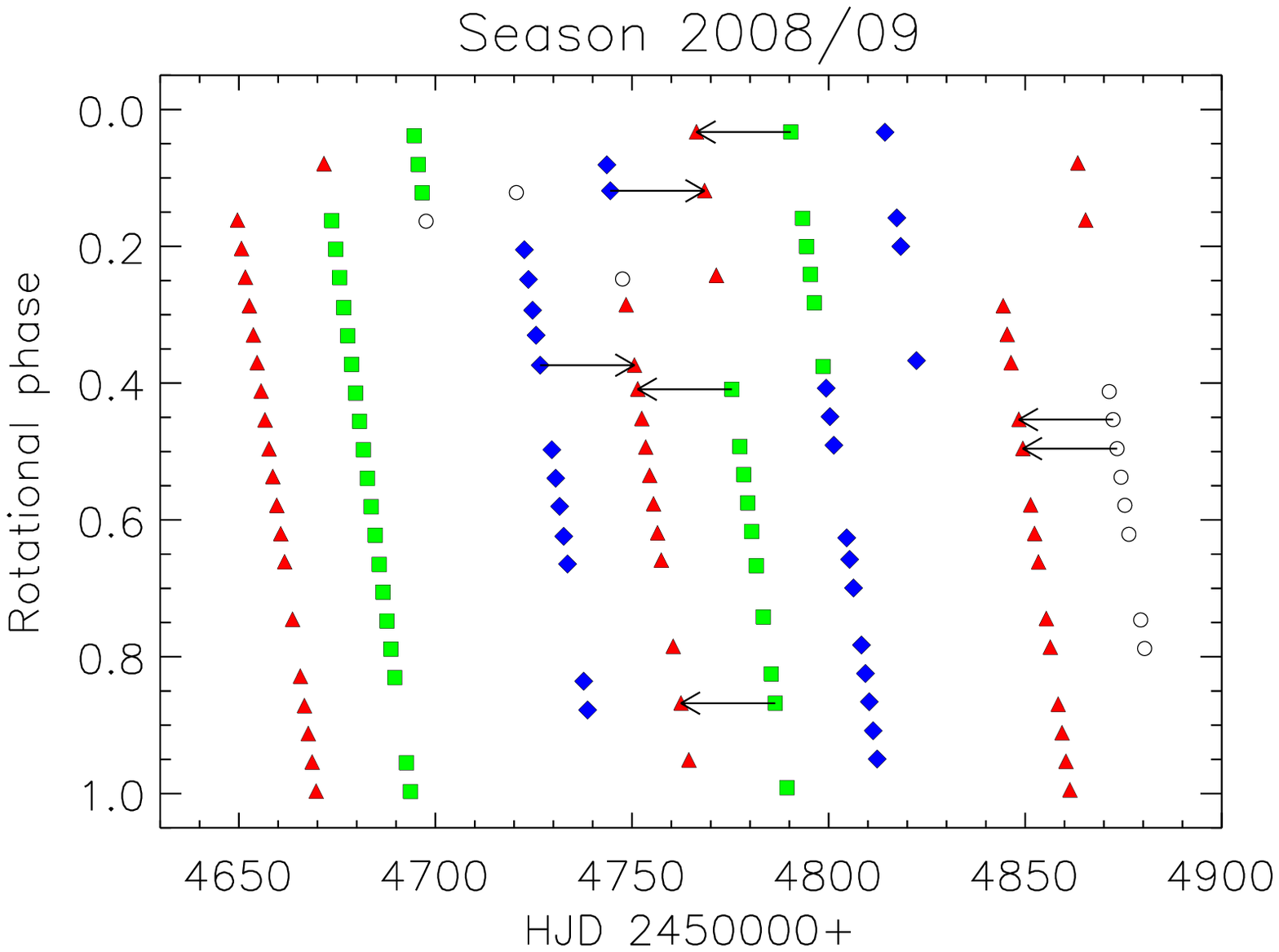}}
\subfloat[]{\includegraphics[width=250pt]{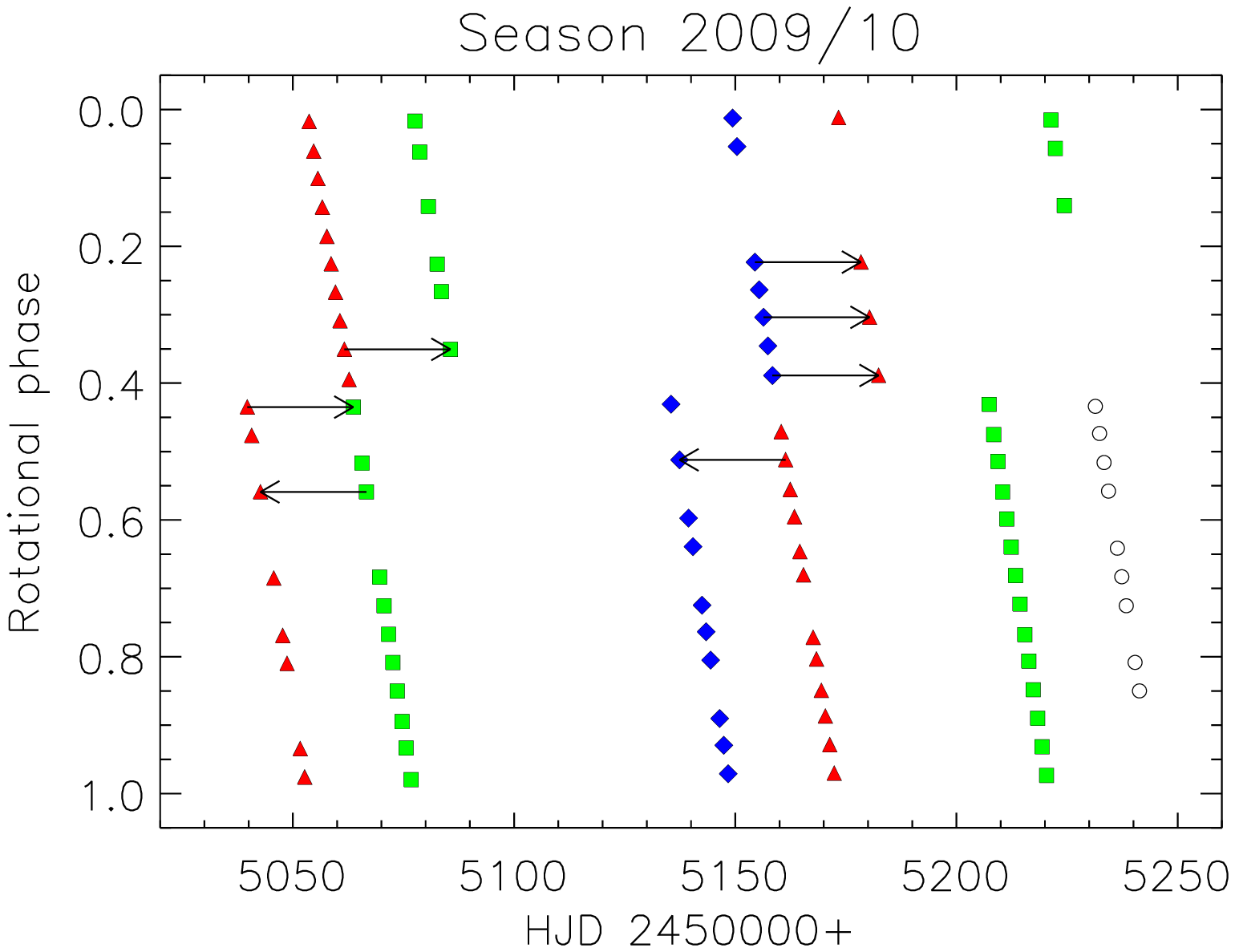}}
\end{minipage}\hspace{1ex}
\begin{minipage}{1.0\textwidth}
\captionsetup[subfigure]{labelfont=bf,textfont=bf,singlelinecheck=off,justification=raggedright,
position=top}
\subfloat[]{\includegraphics[width=250pt]{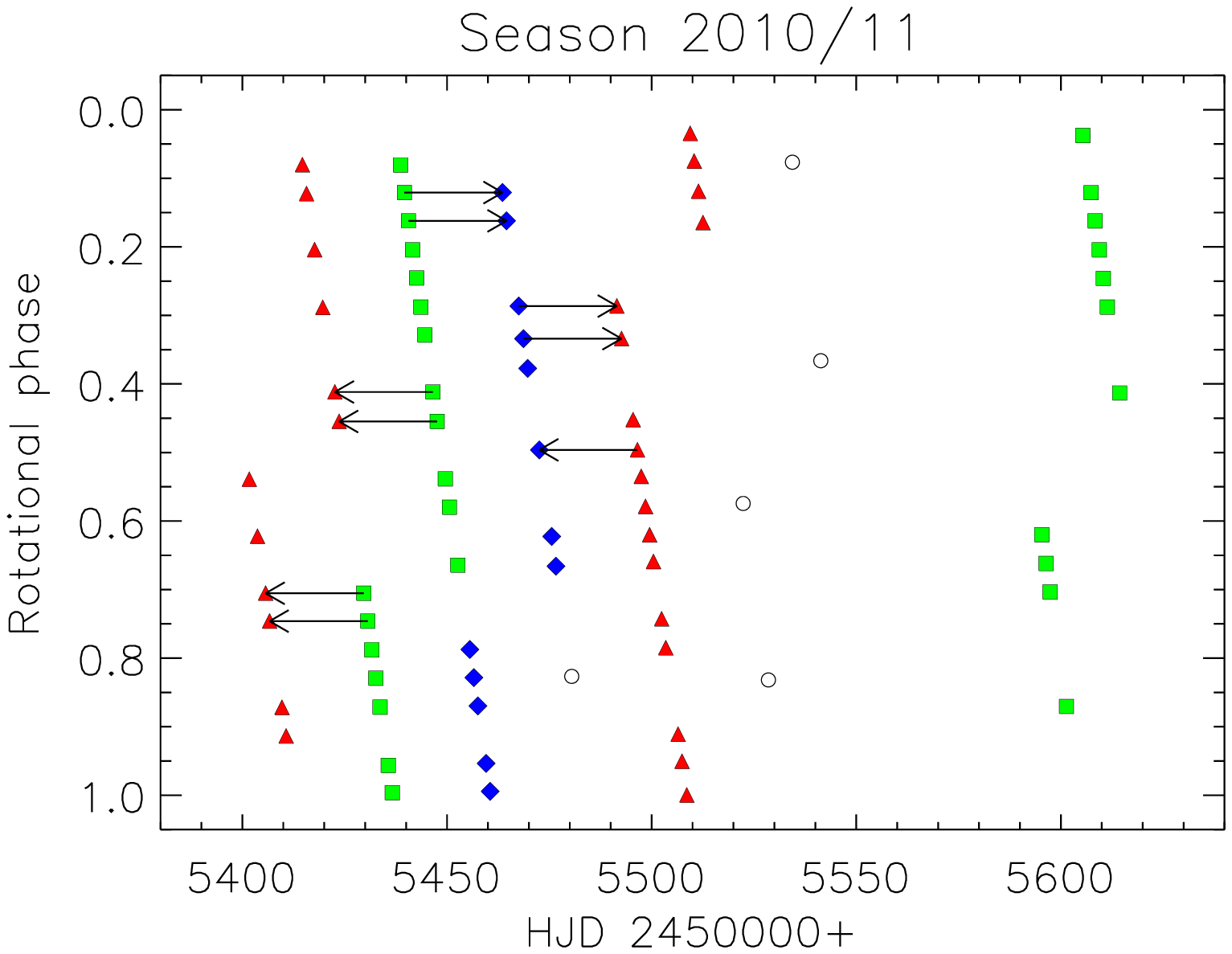}}
\subfloat[]{\includegraphics[width=250pt]{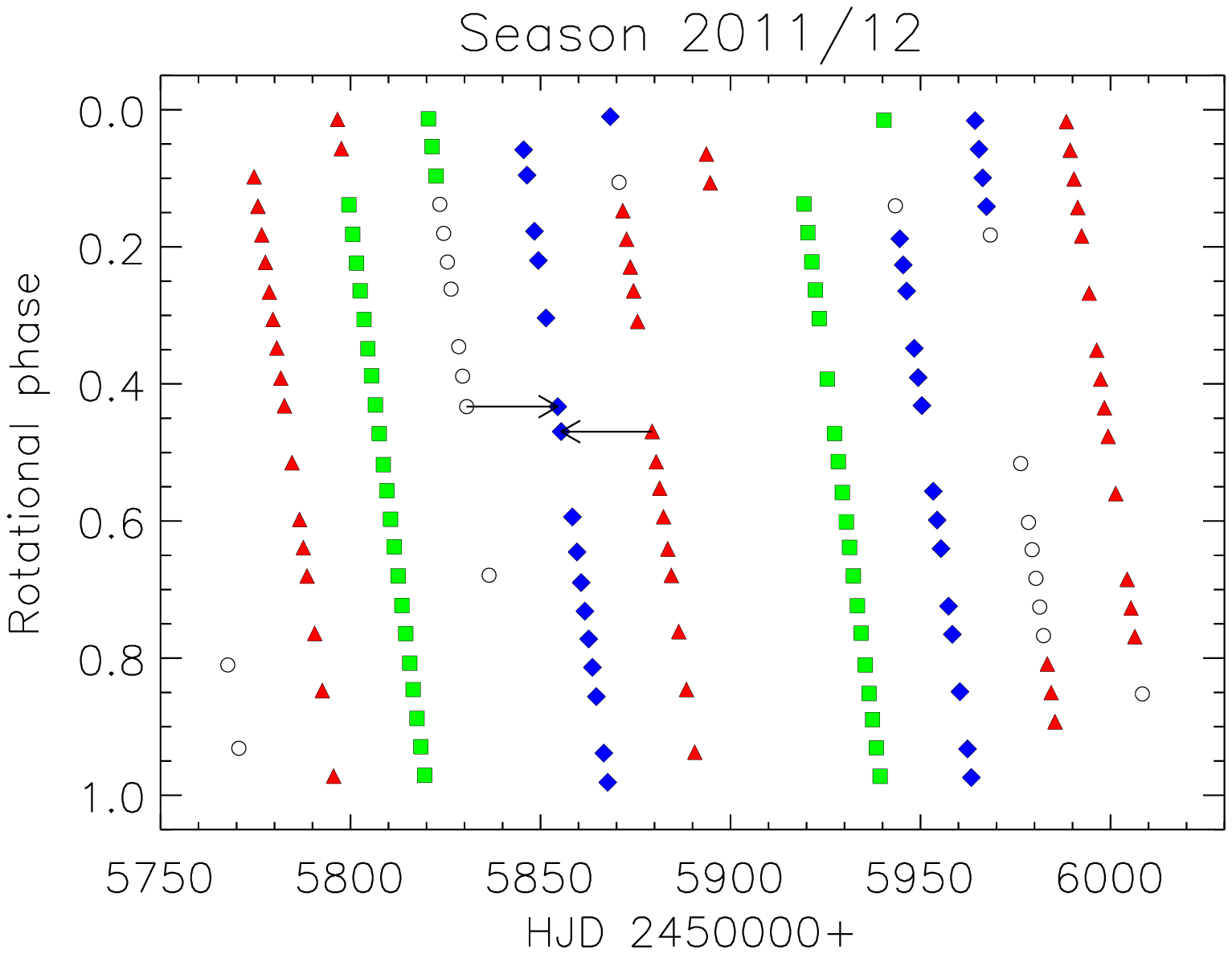}}
\end{minipage}\hspace{1ex}
\caption{Phase coverage of Doppler images from 2006 to 2012. Different filled (colored) symbols
represents the phases of each individual Doppler image, whereas not-filled circles represents
non-used spectra (except for gap filling). The arrows indicate the spectra, which were used to fill
up large observational gaps. Detailed information is given in Table~\ref{tab:maps_log}.}
\label{fig:phase_coverage}
\end{figure*}

Fig.~\ref{fig:phase_coverage} shows the phase coverage of each Doppler image for all observational
seasons.

\section{CCF-maps and DR-fits for the seasons 2006/07 to 2011/12}

\begin{figure*}[!t]
\begin{minipage}{1.0\textwidth}
\captionsetup[subfigure]{labelfont=bf,textfont=bf,singlelinecheck=off,justification=raggedright,
position=top}
\subfloat[]{\includegraphics[width=180pt]{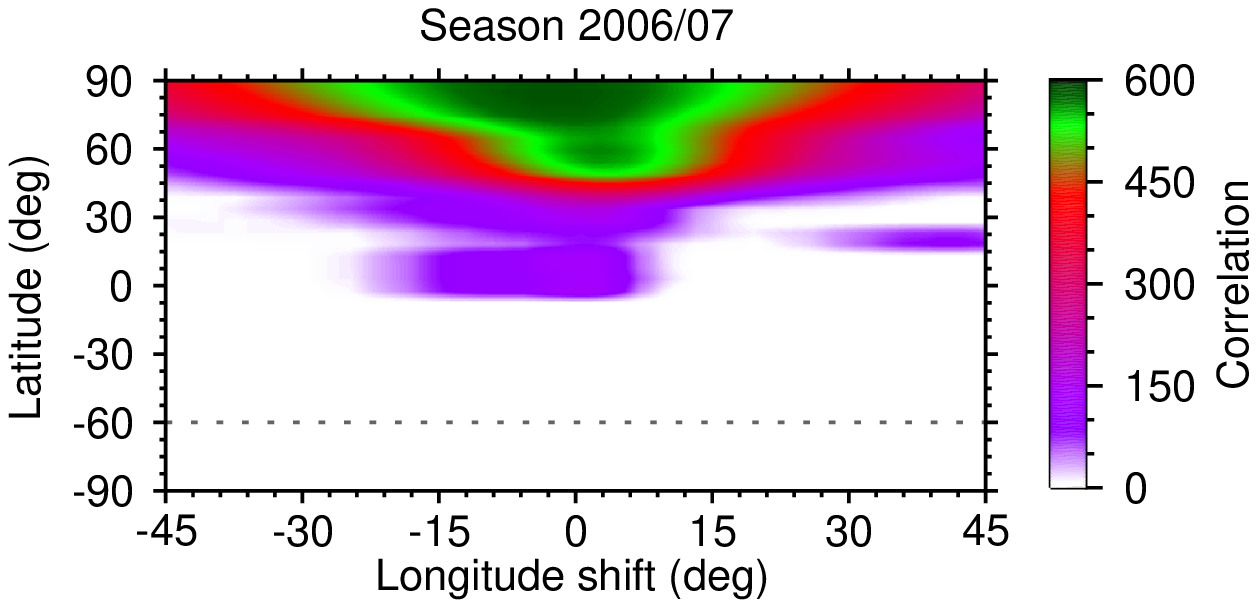}}
\subfloat[]{\includegraphics[width=180pt]{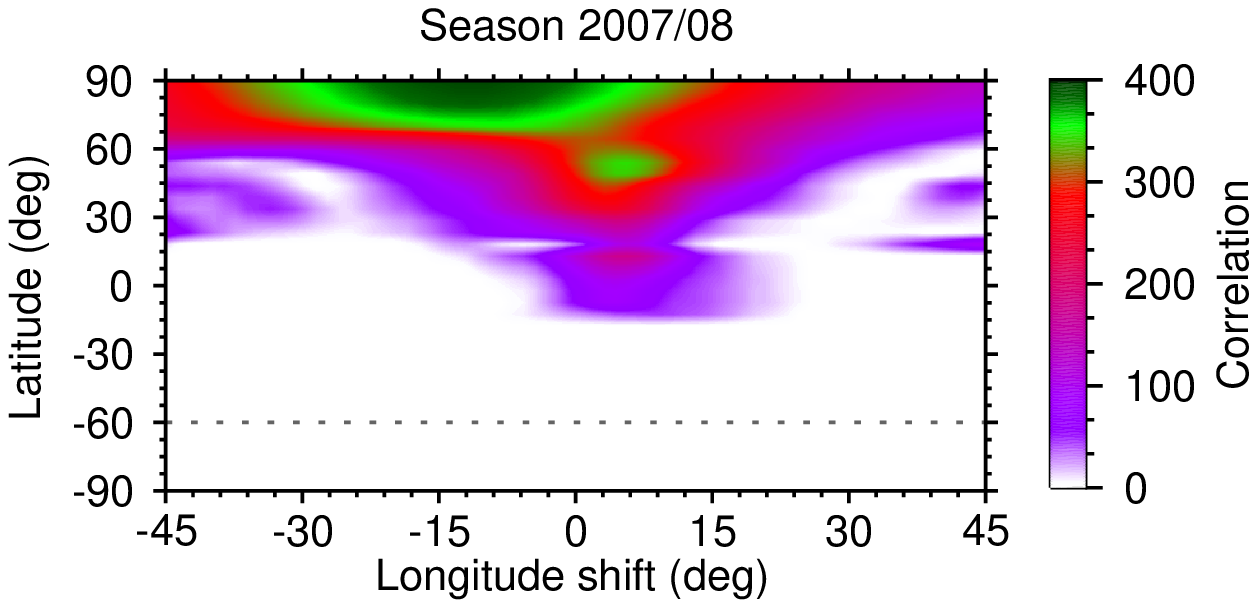}}
\subfloat[]{\includegraphics[width=180pt]{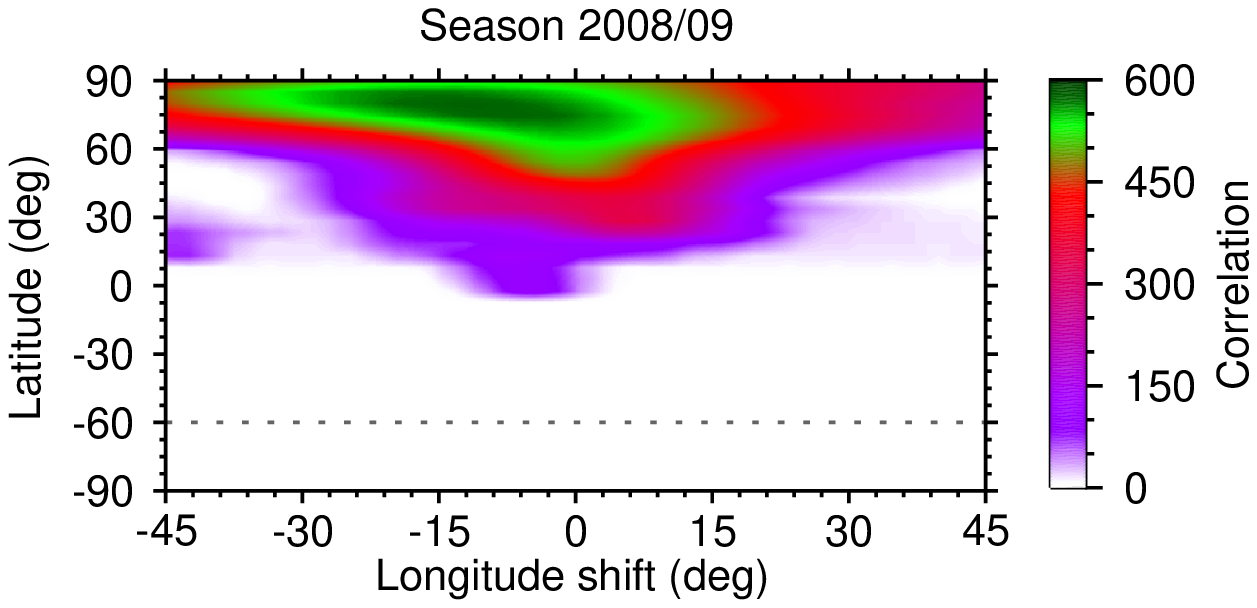}}
\end{minipage}\hspace{1ex}
\begin{minipage}{1.0\textwidth}
\captionsetup[subfigure]{labelfont=bf,textfont=bf,singlelinecheck=off,justification=raggedright,
position=top}
\subfloat[]{\includegraphics[width=180pt]{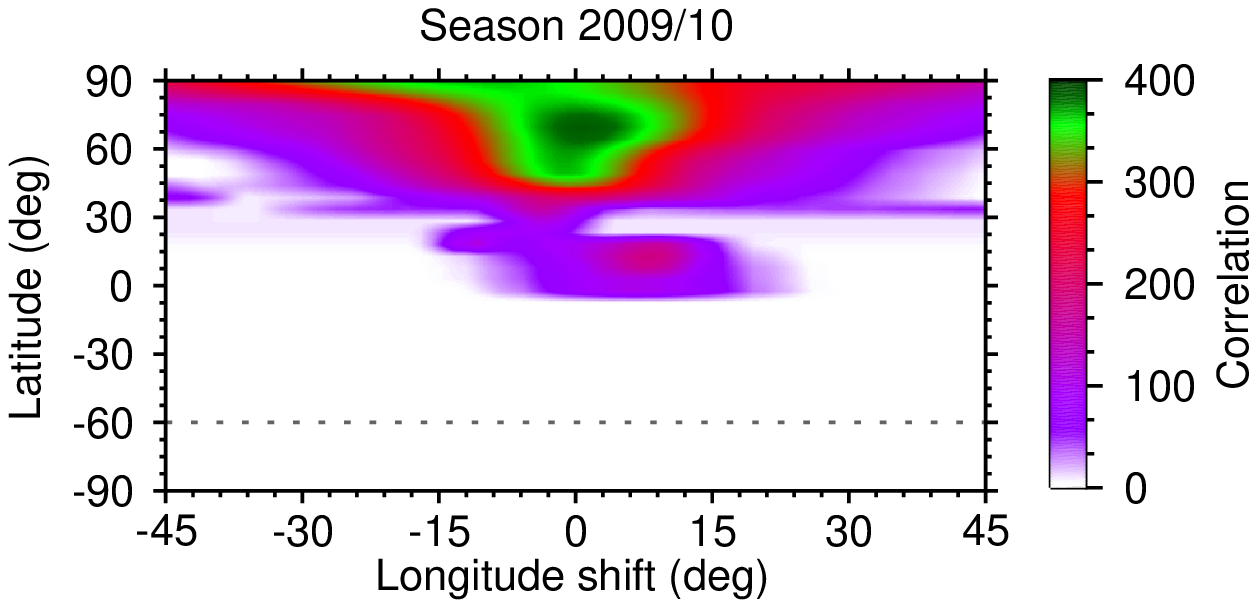}}
\subfloat[]{\includegraphics[width=180pt]{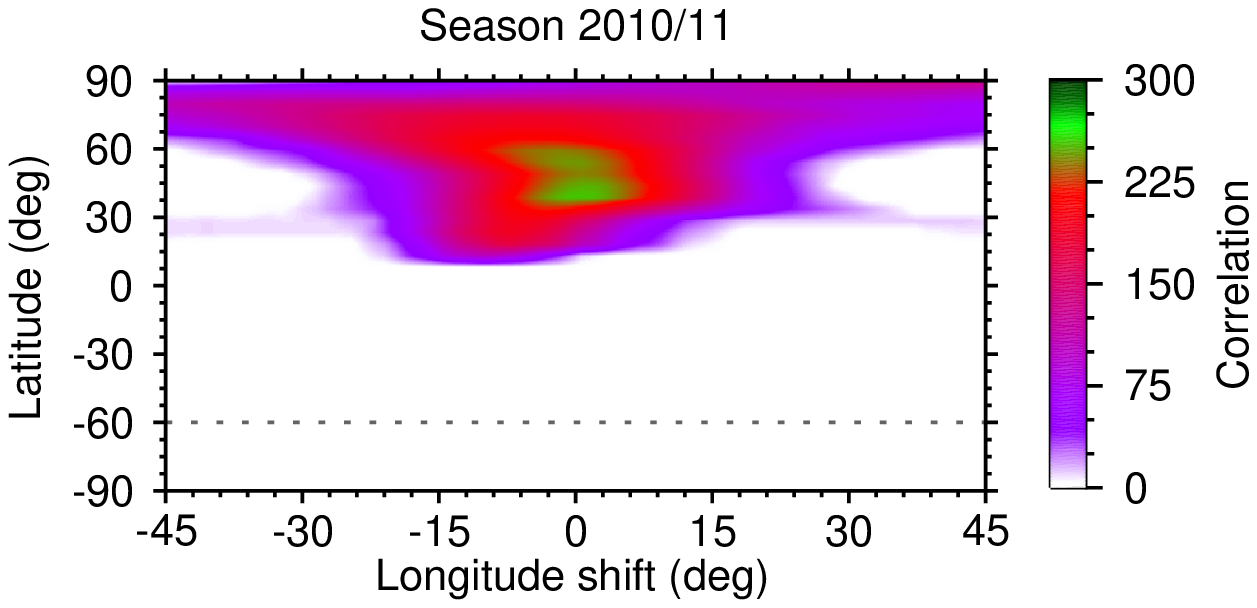}}
\subfloat[]{\includegraphics[width=180pt]{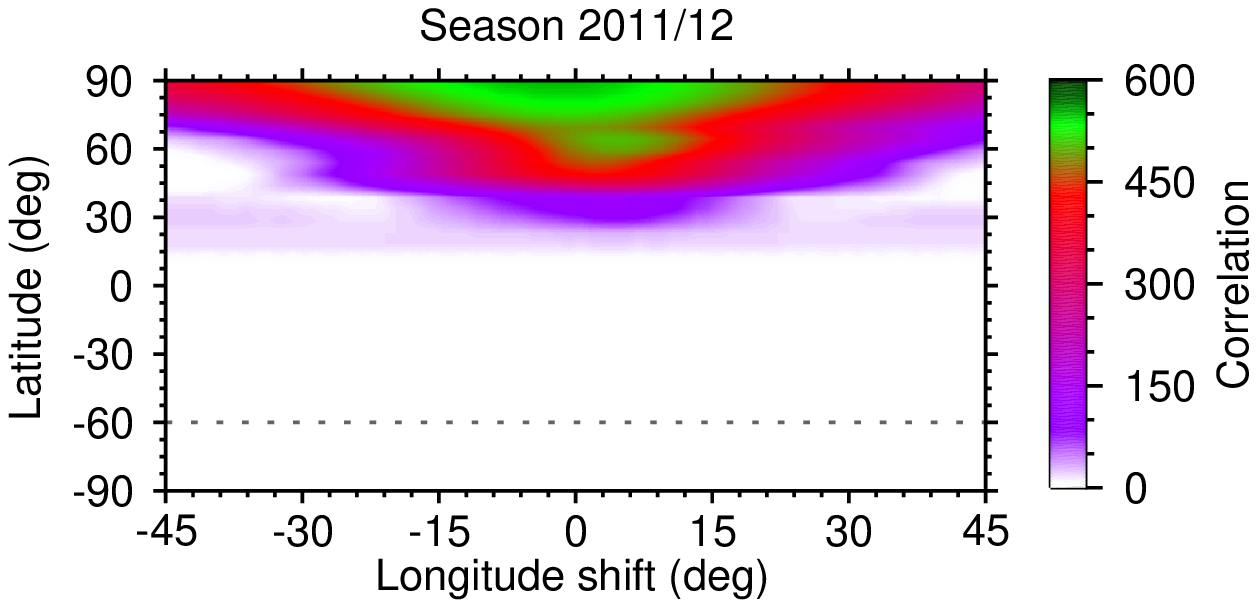}}
\end{minipage}\hspace{1ex}
\caption{Cross-correlation function maps from 2006 to 2012. Each map represents the average ccf map
for one observing season.}
\label{fig:ccf_maps_all}
\end{figure*}

\begin{figure*}[!b]
\begin{minipage}{1.0\textwidth}
\captionsetup[subfigure]{labelfont=bf,textfont=bf,singlelinecheck=off,justification=raggedright,
position=top}
\subfloat[]{\includegraphics[width=175pt]{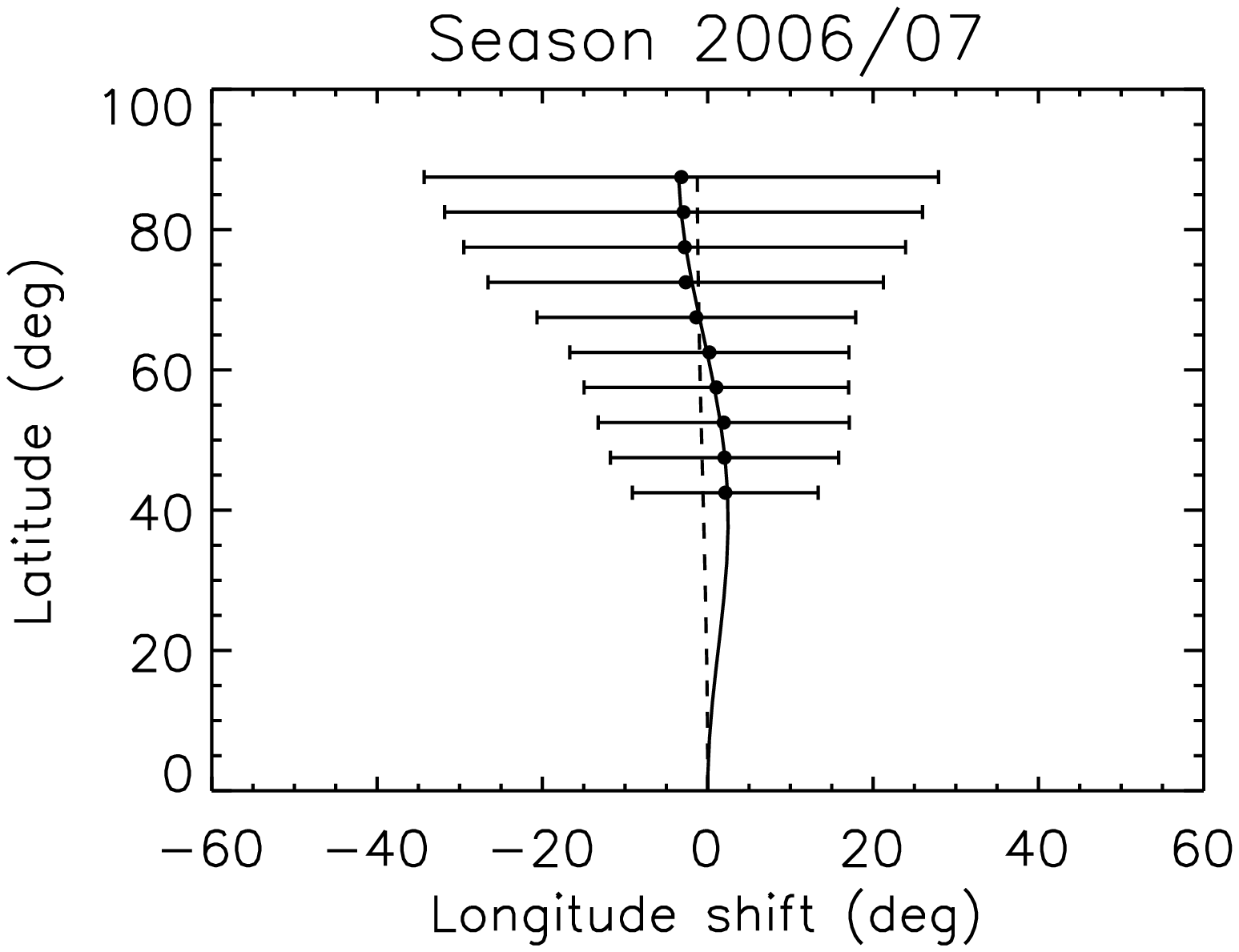}}
\subfloat[]{\includegraphics[width=175pt]{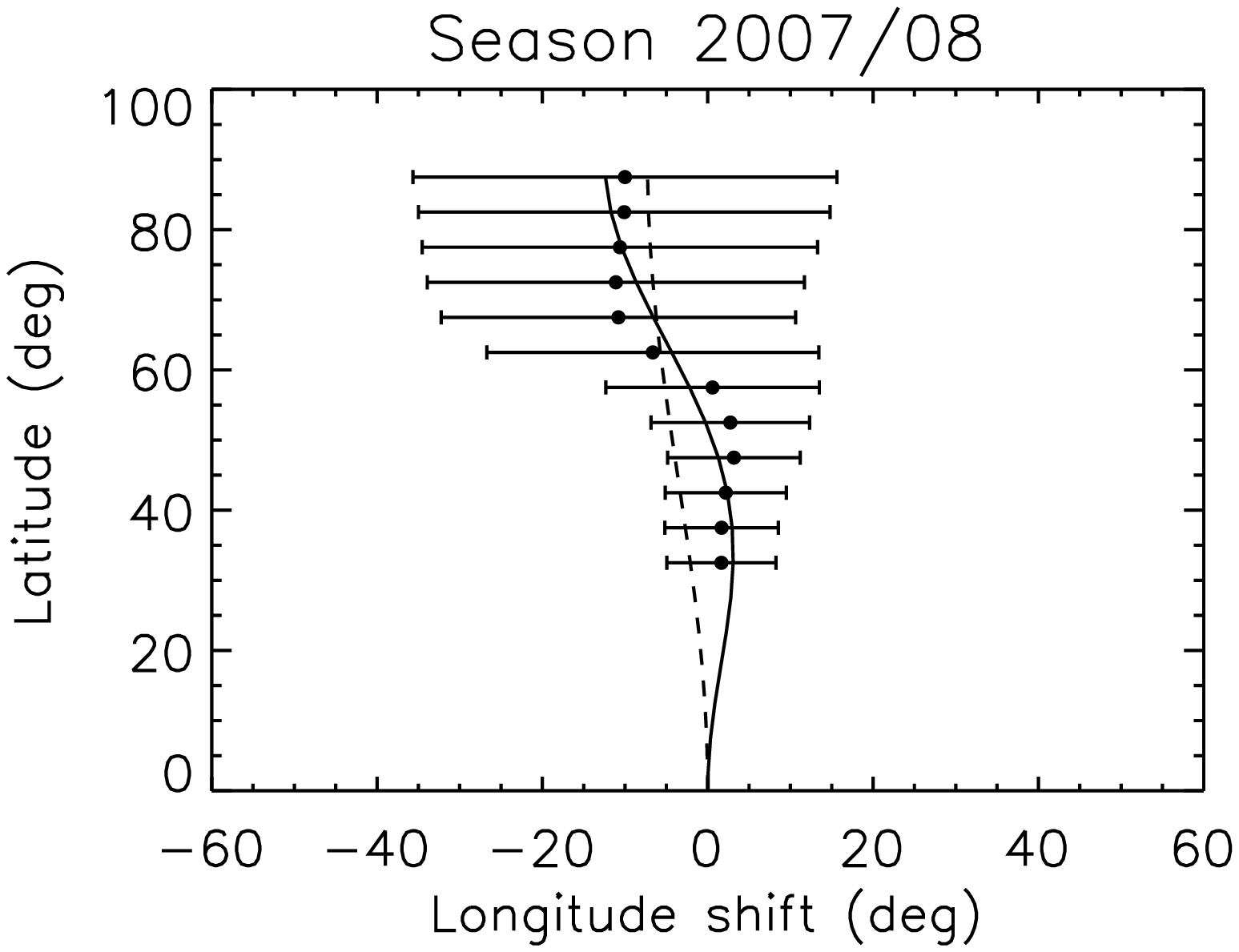}}
\subfloat[]{\includegraphics[width=175pt]{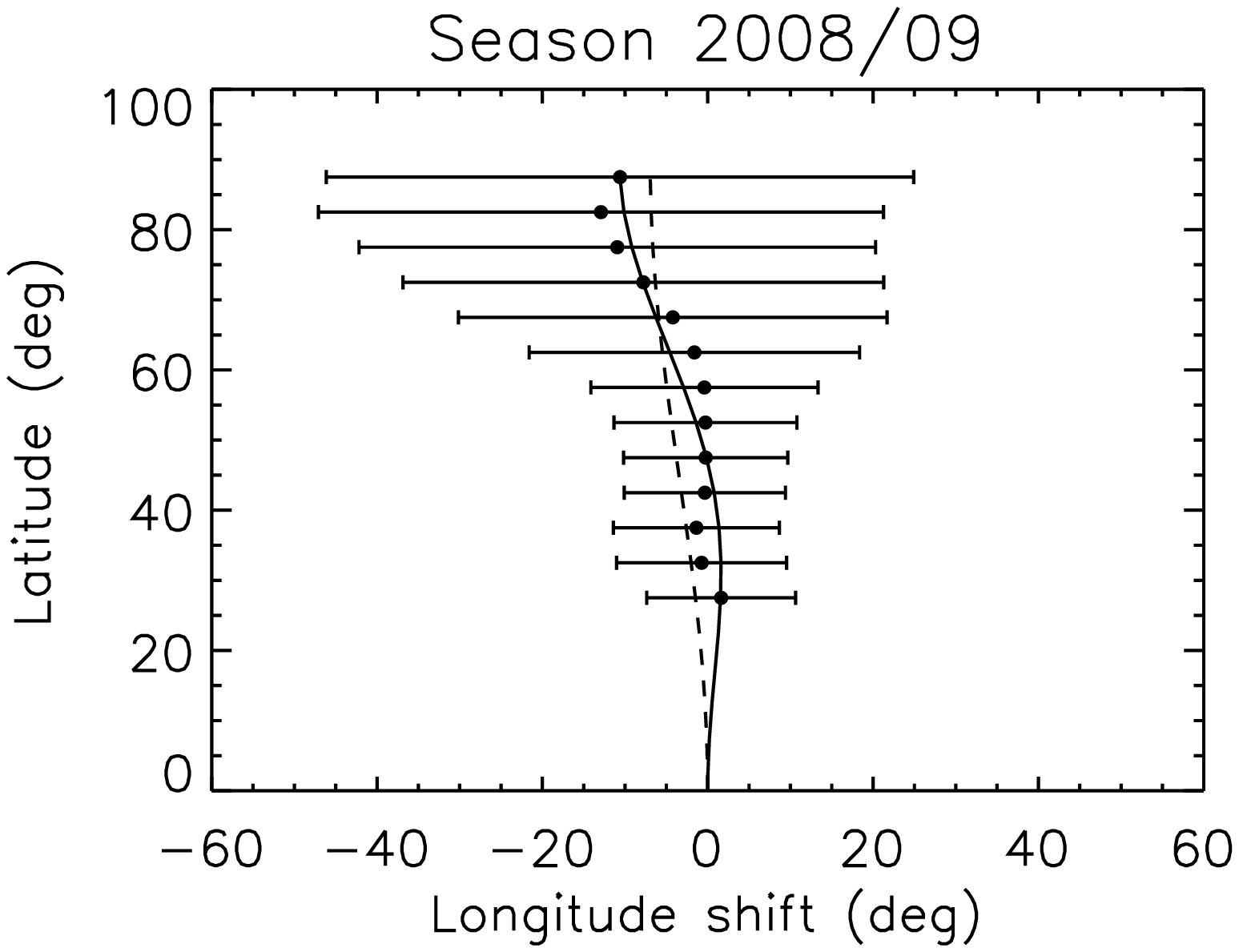}}
\end{minipage}\hspace{1ex}
\begin{minipage}{1.0\textwidth}
\captionsetup[subfigure]{labelfont=bf,textfont=bf,singlelinecheck=off,justification=raggedright,
position=top}
\subfloat[]{\includegraphics[width=175pt]{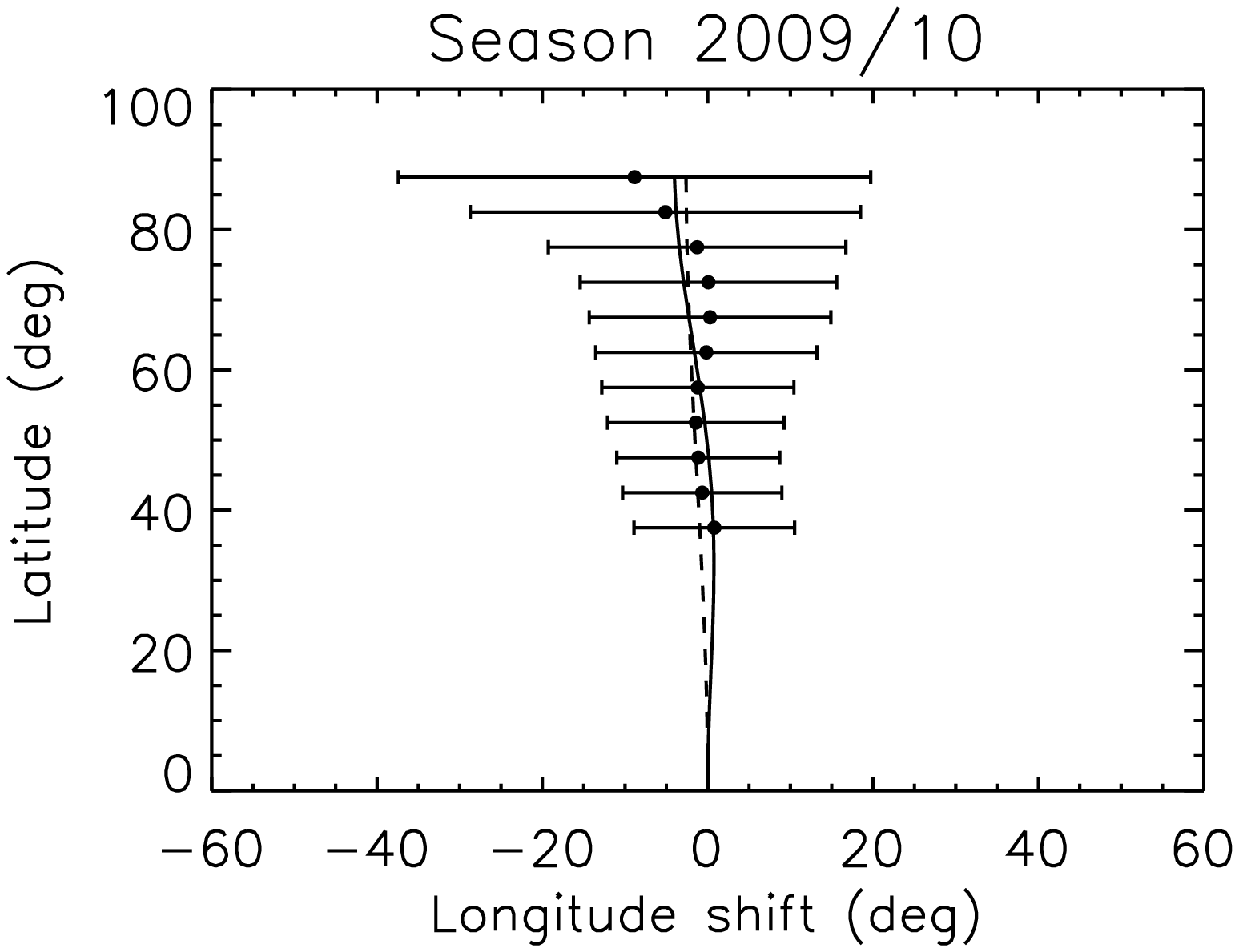}}
\subfloat[]{\includegraphics[width=175pt]{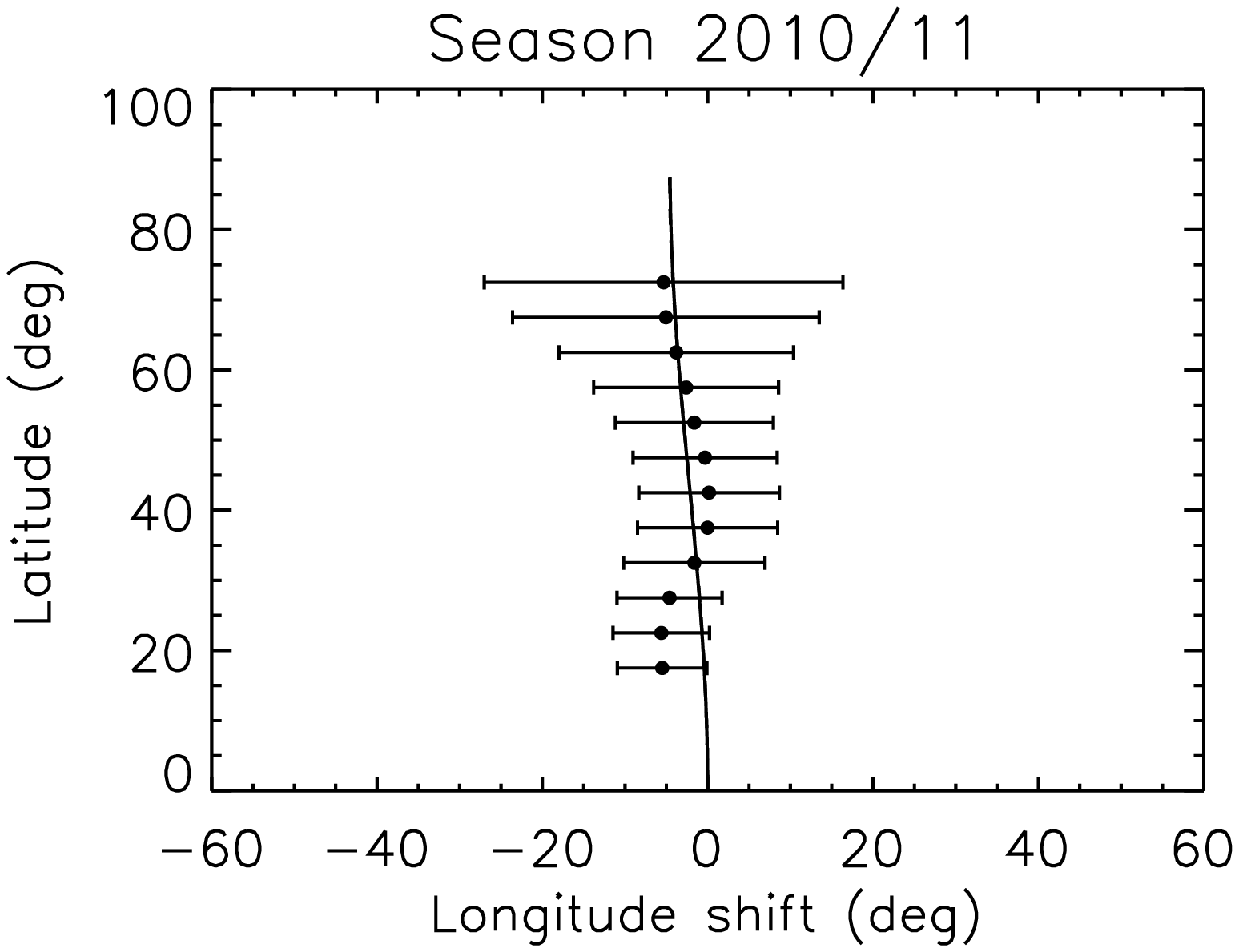}}
\subfloat[]{\includegraphics[width=175pt]{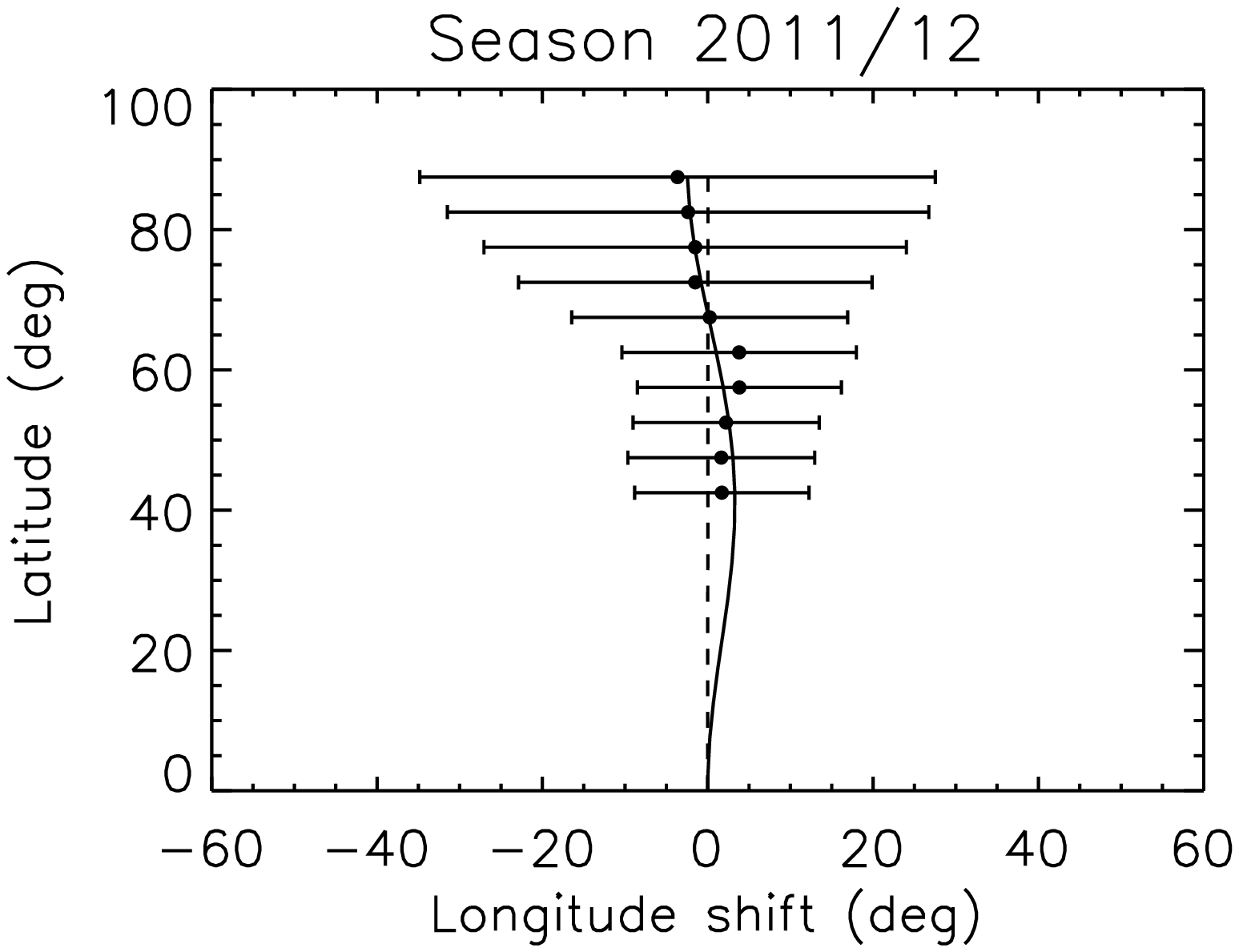}}
\end{minipage}\hspace{1ex}
\caption{Differential rotation signatures from 2006 to 2012. Analyzing the ccf maps in
Fig.~\ref{fig:ccf_maps_all} reveals a weak solar-like differential rotation. The dots are the
correlation peaks per $5^{\circ}$-latitude bin and their error bars are defined as the FWHMs of the
corresponding Gaussians. The dashed line represents a fit using Eq.~\ref{eqn:diff_rot_eqn_1},
whereas the solid line represents a fit using Eq.~\ref{eqn:diff_rot_eqn_2}. The parameters for each
fit are summarized in Table~\ref{tab:diff_rot_par}.}
\label{fig:diff_rot_fit_all}
\end{figure*}

Fig.~\ref{fig:ccf_maps_all} shows the ccf maps from each observing season. In
Fig.~\ref{fig:diff_rot_fit_all} the observed differential rotation pattern determined from the ccf
maps, together with the best fit of the differential rotation following Eq.~\ref{eqn:diff_rot_eqn_1}
and Eq.~\ref{eqn:diff_rot_eqn_2} are shown.

\section{Longitudinal spot distribution for the seasons 2006/07 to 2011/12}

\begin{figure*}[!t]
\begin{minipage}{1.0\textwidth}
\captionsetup[subfigure]{labelfont=bf,textfont=bf,singlelinecheck=off,justification=raggedright,
position=top}
\subfloat[]{\includegraphics[width=175pt]{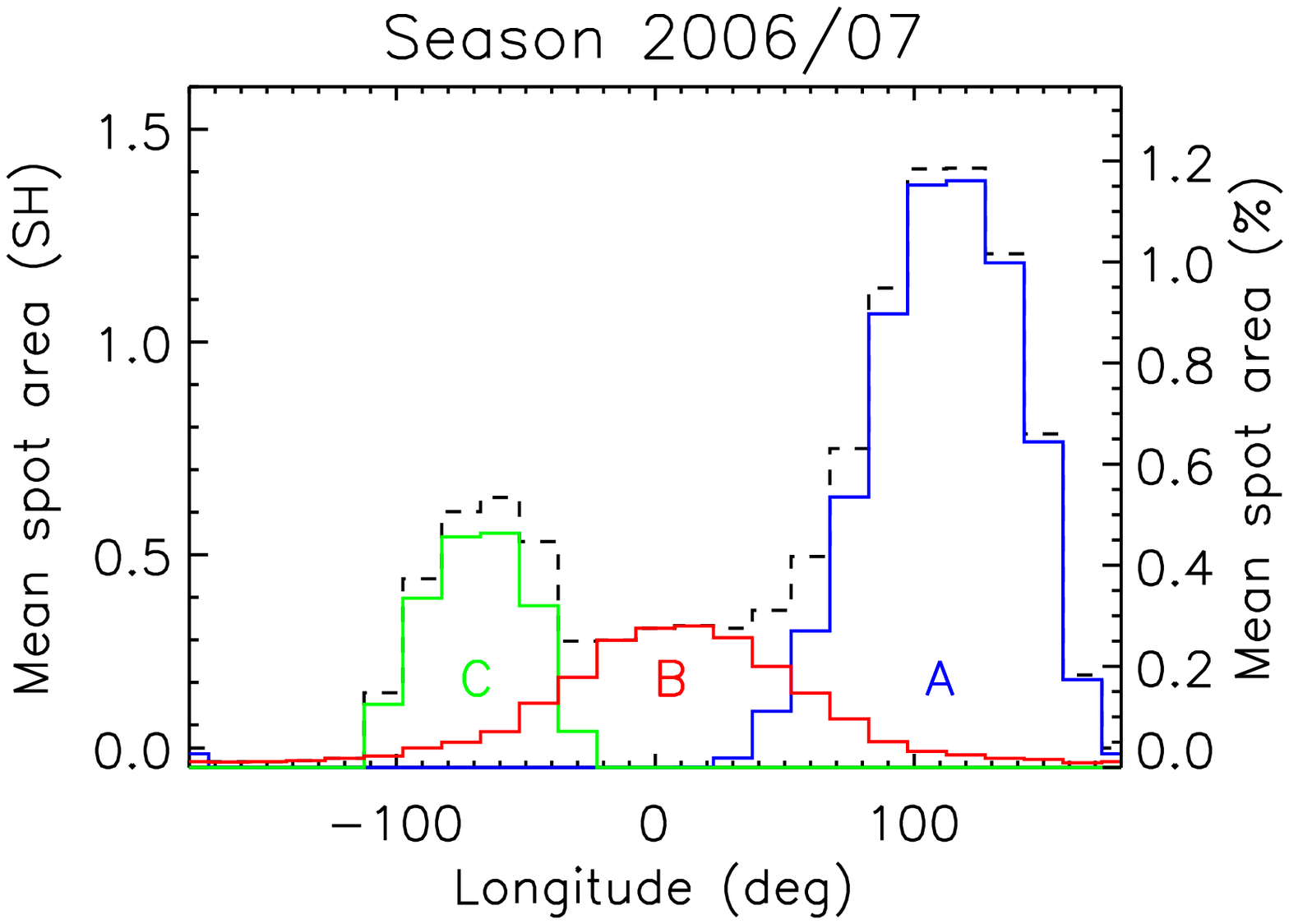}}
\subfloat[]{\includegraphics[width=175pt]{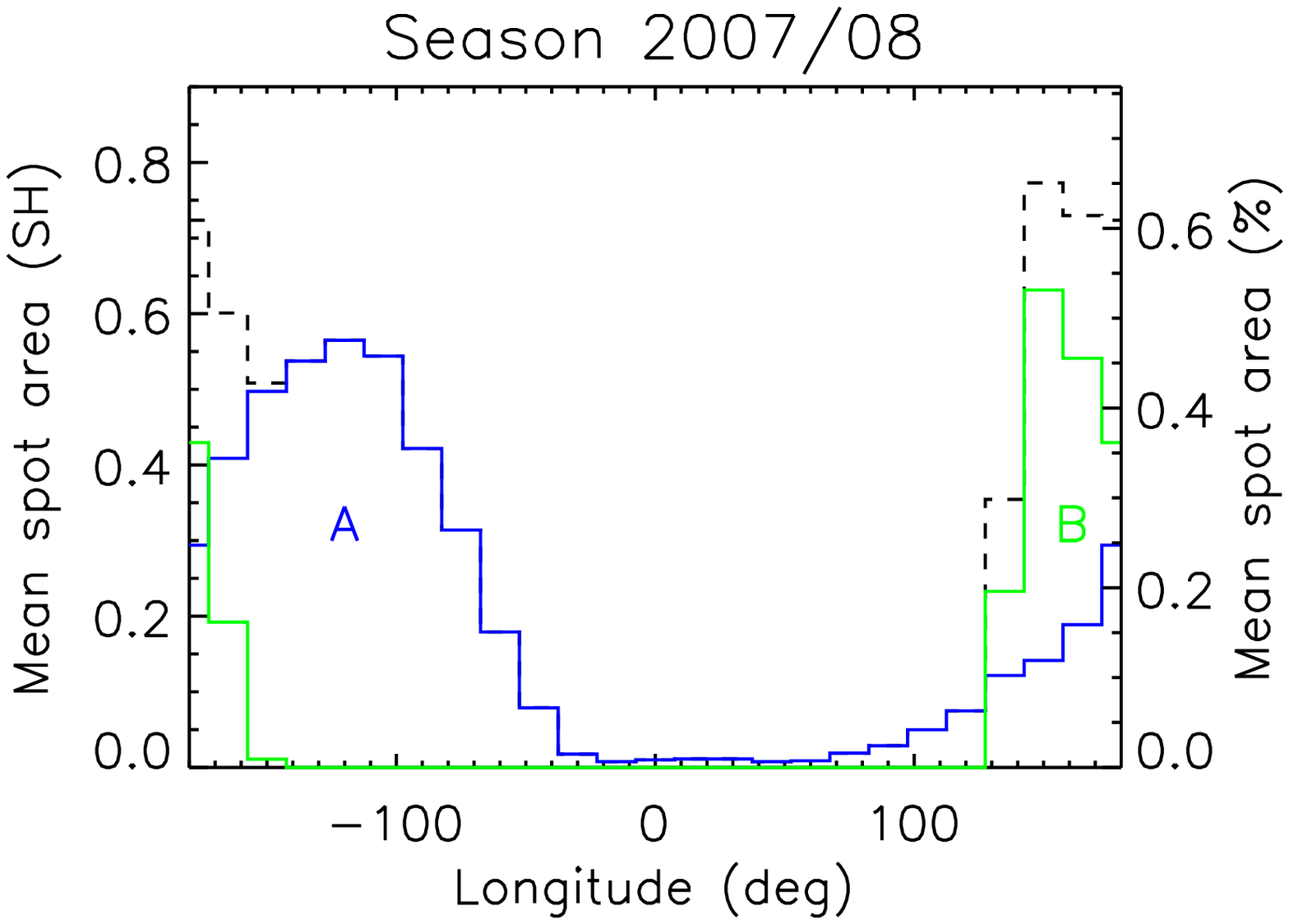}}
\subfloat[]{\includegraphics[width=175pt]{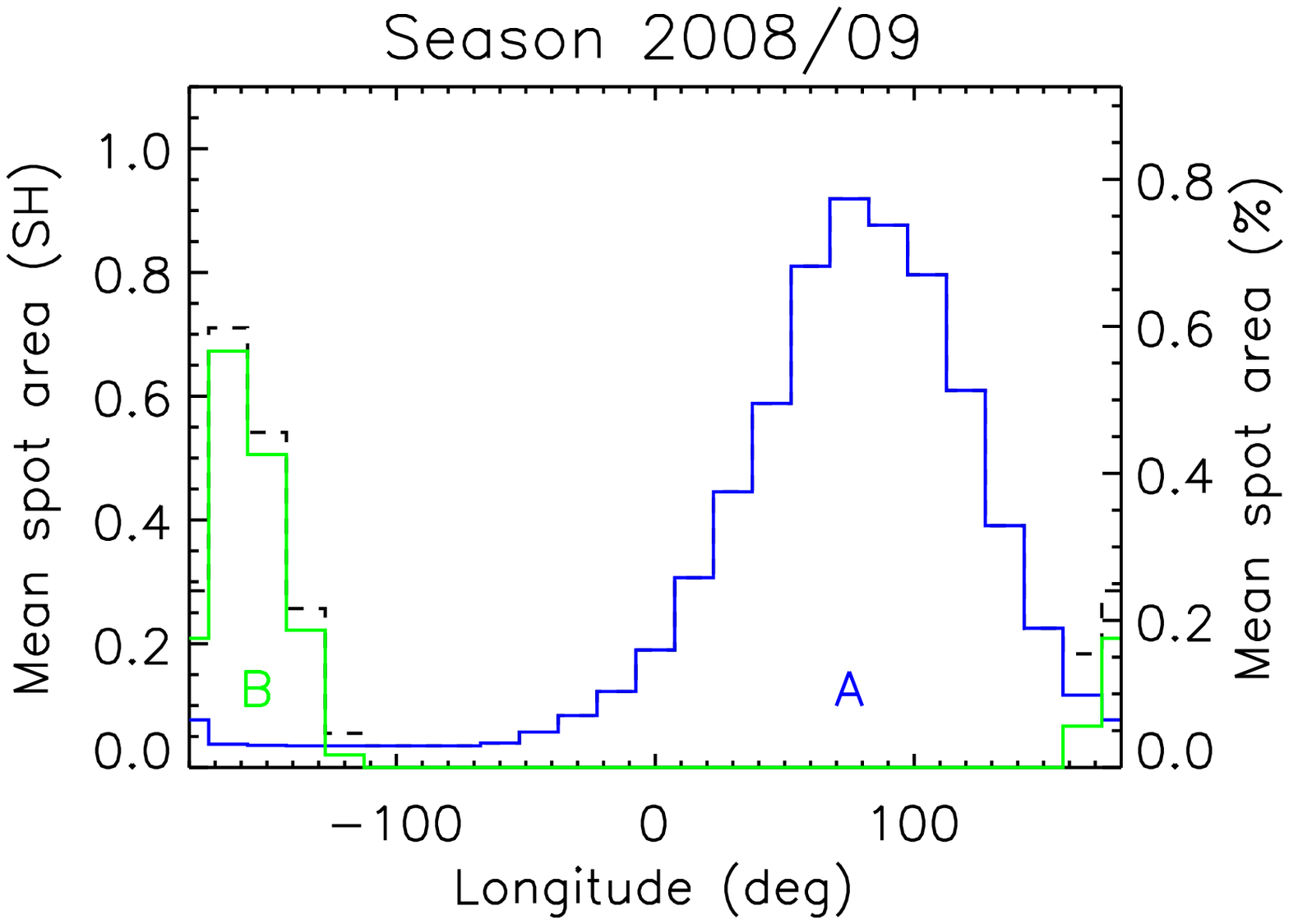}}
\end{minipage}\hspace{1ex}
\begin{minipage}{1.0\textwidth}
\captionsetup[subfigure]{labelfont=bf,textfont=bf,singlelinecheck=off,justification=raggedright,
position=top}
\subfloat[]{\includegraphics[width=175pt]{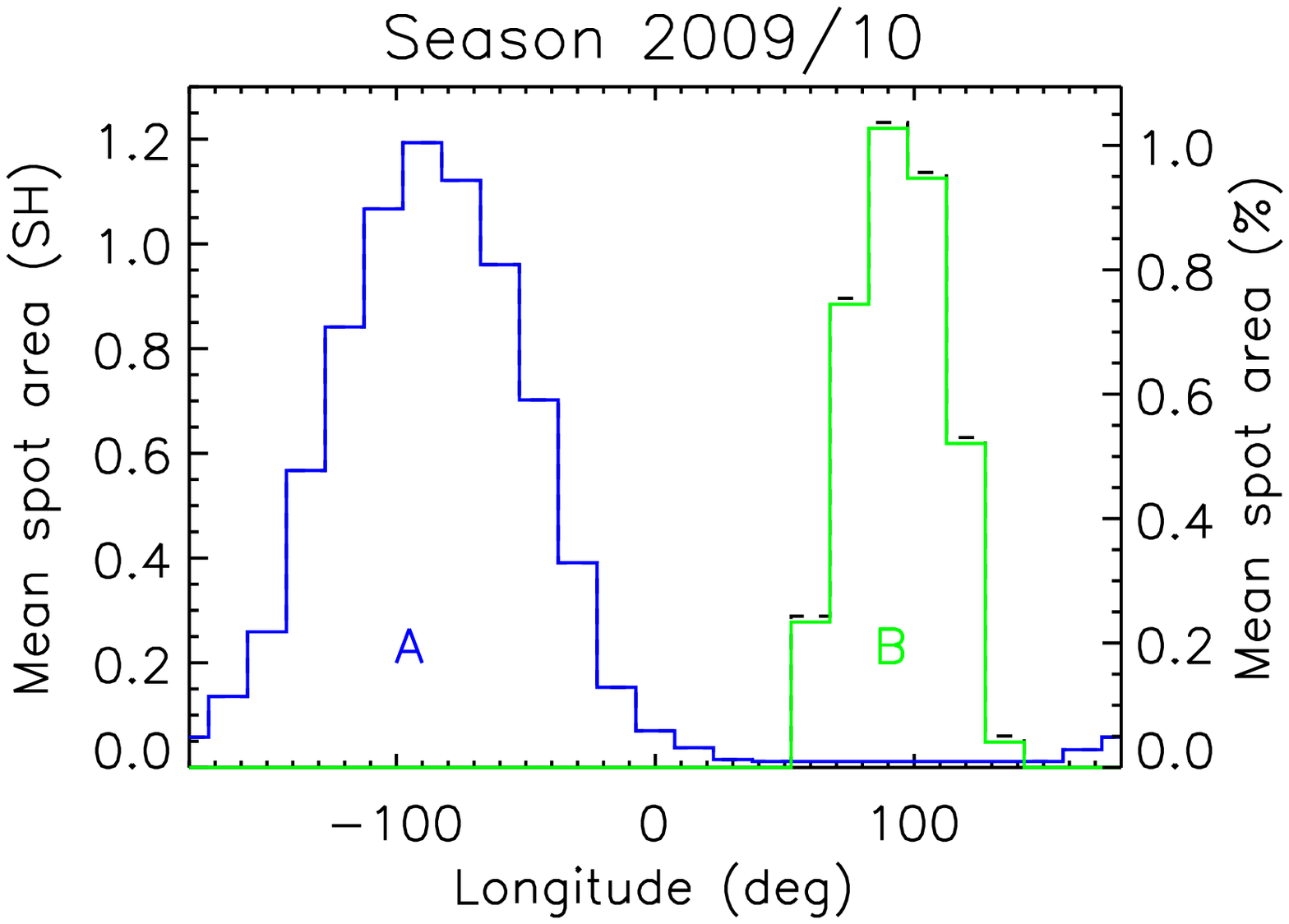}}
\subfloat[]{\includegraphics[width=175pt]{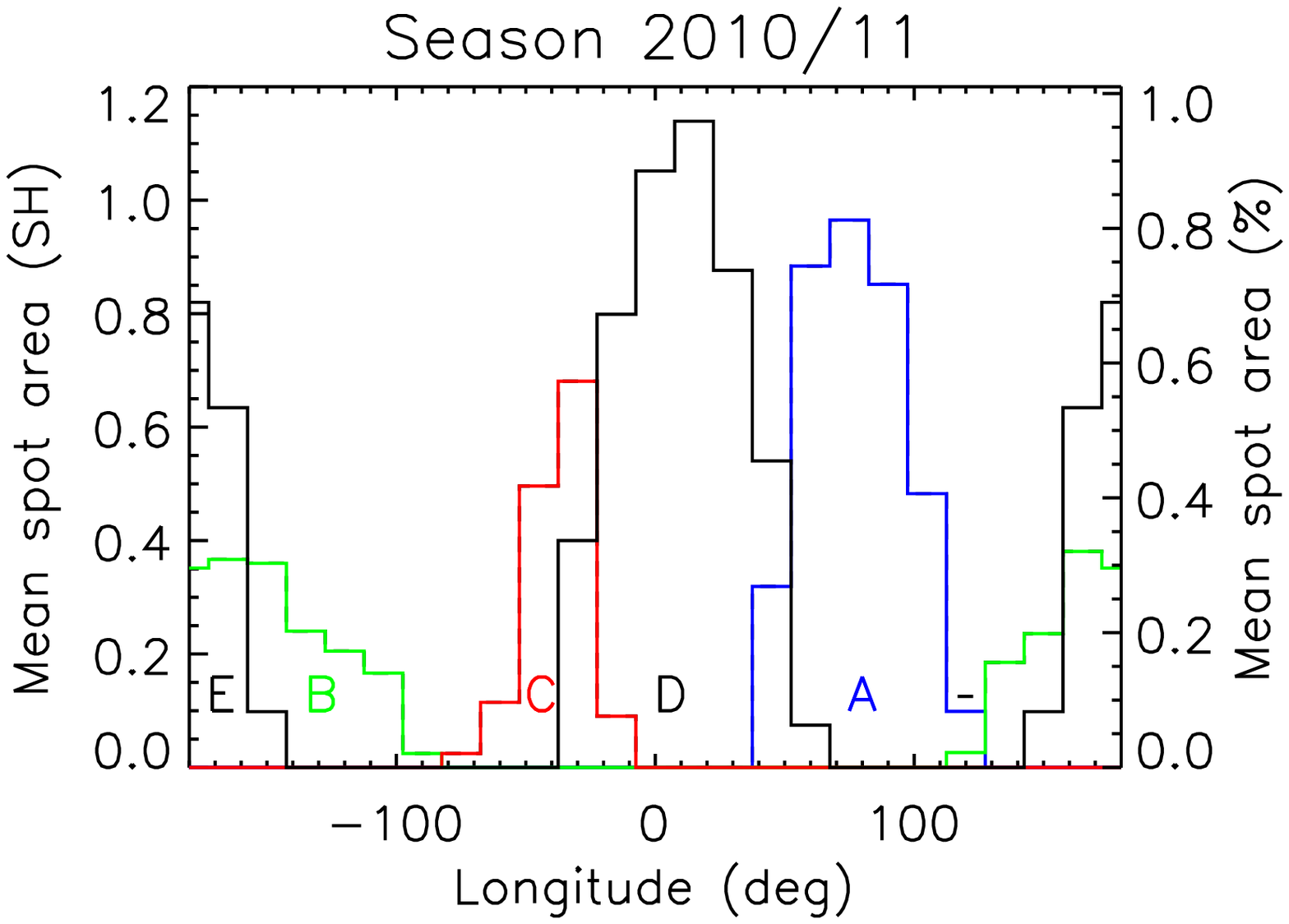}}
\subfloat[]{\includegraphics[width=175pt]{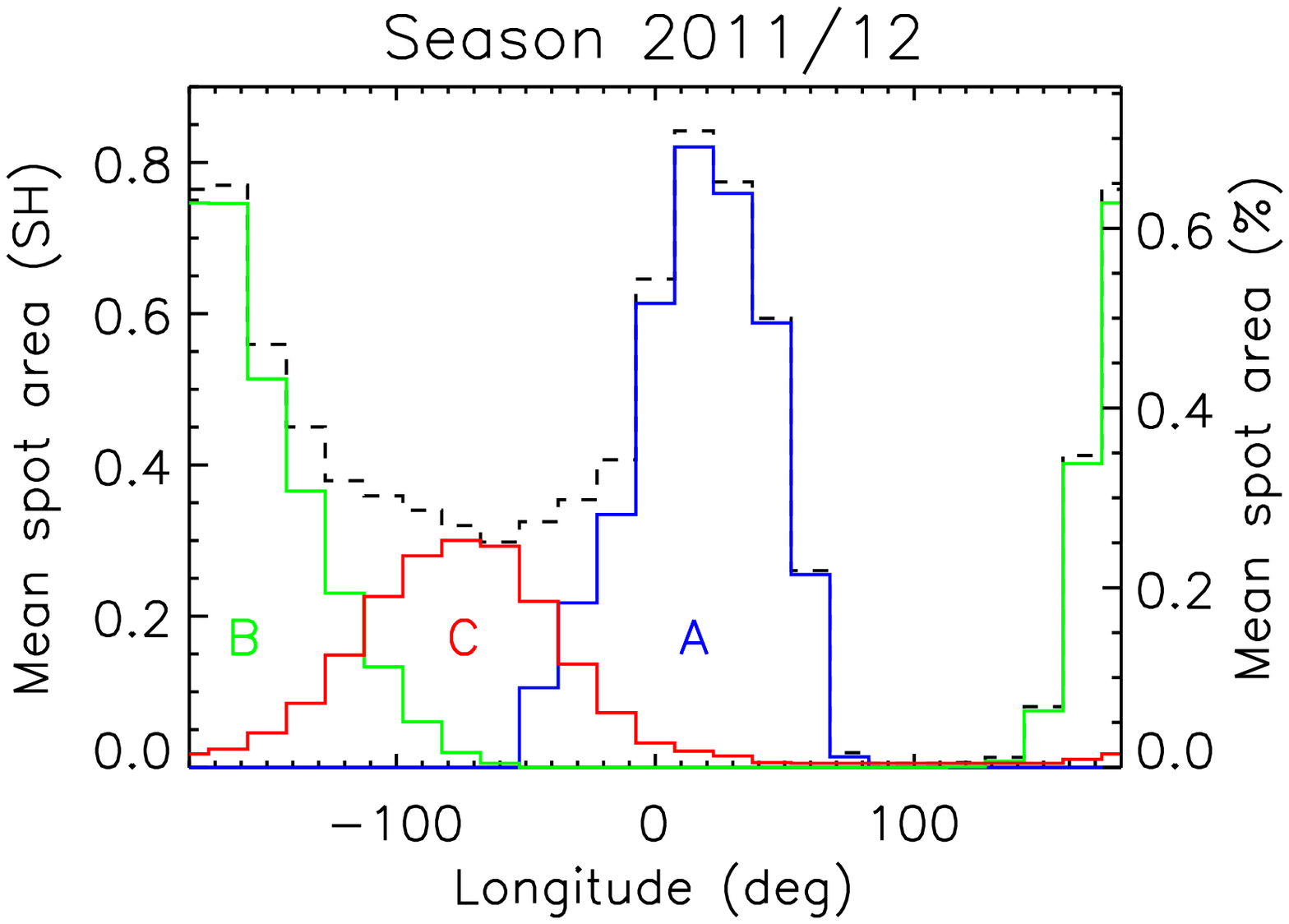}}
\end{minipage}\hspace{1ex}
\caption{Longitudinal spot area distribution on XX~Tri from 2006 to 2012. Shown are the seasonal
mean distributions of the individual spots (solid colored lines) from our spot-model fits for each
observing season. The black dashed line represents the total spotted area. The spot area is given in
solar hemispheres on the left axis (1~SH\;=\;3.05~$\mathrm{Gm^2}$) and relative to the total area
of a stellar hemisphere of XX~Tri on the right axis.}
\label{fig:spot_distribution_season_all}
\end{figure*}

Fig.~\ref{fig:spot_distribution_season_all} shows the mean distribution of the spot area from our
spot-model fits for each observing season.

\end{appendix}

\end{document}